\pdfoutput=1

\documentclass[12pt,a4paper]{article}

\usepackage{ifthen} 
\newboolean{pdflatex}
\setboolean{pdflatex}{true} 

\newboolean{articletitles}
\setboolean{articletitles}{true} 

\newboolean{uprightparticles}
\setboolean{uprightparticles}{false} 

\newboolean{inbibliography}
\setboolean{inbibliography}{false} 

\def\authorfont{\footnotesize}
\def\keywords#1{\par
	\vspace*{8pt}
	{\authorfont{\leftskip18pt\rightskip\leftskip
	\noindent{\it Keywords}\/:\ #1\par}}\par}

\usepackage{longtable} 
\usepackage{subfigure,rotating,multirow}
\usepackage[UKenglish]{isodate}
\newcommand{\tbl}[1]{\caption{#1}}

\usepackage{microtype}
\usepackage{lineno}  
\usepackage{xspace} 
\usepackage{caption} 

\usepackage{graphicx}  
\usepackage{color}
\usepackage{colortbl}
\graphicspath{{./figs/}} 

\usepackage{amsmath} 
\usepackage{amssymb}
\usepackage{amsfonts}
\usepackage{upgreek} 

\newcommand*\patchAmsMathEnvironmentForLineno[1]{%
\expandafter\let\csname old#1\expandafter\endcsname\csname #1\endcsname
\expandafter\let\csname oldend#1\expandafter\endcsname\csname
end#1\endcsname
 \renewenvironment{#1}%
   {\linenomath\csname old#1\endcsname}%
   {\csname oldend#1\endcsname\endlinenomath}%
}
\newcommand*\patchBothAmsMathEnvironmentsForLineno[1]{%
  \patchAmsMathEnvironmentForLineno{#1}%
  \patchAmsMathEnvironmentForLineno{#1*}%
}
\AtBeginDocument{%
\patchBothAmsMathEnvironmentsForLineno{equation}%
\patchBothAmsMathEnvironmentsForLineno{align}%
\patchBothAmsMathEnvironmentsForLineno{flalign}%
\patchBothAmsMathEnvironmentsForLineno{alignat}%
\patchBothAmsMathEnvironmentsForLineno{gather}%
\patchBothAmsMathEnvironmentsForLineno{multline}%
}

\usepackage{hyperref}    
\usepackage[all]{hypcap} 

\def\deltaLLXpi      { \Delta {\rm log} {\mathcal L} (X-\pi) \xspace}

\def\deltaLLCombXpi  { \Delta {\rm log} {\mathcal L}_{comb} (X-\pi) \xspace}
\def\deltaLLCombepi  { \Delta {\rm log} {\mathcal L}_{comb} (e-\pi) \xspace}
\def\deltaLLCombKpi  { \Delta {\rm log} {\mathcal L}_{comb} (K-\pi) \xspace}
\def\deltaLLCombmupi { \Delta {\rm log} {\mathcal L}_{comb} (\mu-\pi) \xspace}

\def\deltaLLcalo { \Delta {\rm log} {\mathcal L}^\text{CALO} (e-h) \xspace}
\def\deltaLLecal { \Delta {\rm log} {\mathcal L}^\text{ECAL} (e-h) \xspace}
\def\deltaLLhcal { \Delta {\rm log} {\mathcal L}^\text{HCAL} (e-h) \xspace}
\def\deltaLLps   { \Delta {\rm log} {\mathcal L}^\text{PS} (e-h) \xspace}

\def\eOverP {\mbox{\em E/pc}\xspace}

\def\sPlot {\mbox{\em sPlot}\xspace}

\def\RSens     {\ensuremath{R}\xspace}
\def\PhiSens   {$\varPhi$\xspace}

\def\neutroneq {\ensuremath{\rm \,n_{eq}}\xspace}

\def\degreesC  {\ensuremath{^{\circ}}\rm C\xspace}

\newcommand{\BuJpsiK}{\BuToJPsiK}
\newcommand{\Jpsi}{\jpsi}
\newcommand{\BdKpi}{\BdToKpi}
\newcommand{\BdDpi}{\BdToDpi}
\newcommand{\DKpi}{\ensuremath{\D^0\to K^-\pi^+}\xspace}
\newcommand{\DpKpipi}{\ensuremath{\Dp\to K^-\pi^+\pi^+}\xspace}
\newcommand{\Dst}{\ensuremath{D^{*+}}\xspace}
\newcommand{\DstDpi}{\ensuremath{D^{*+}\to D^0\pi^+}\xspace}

\newcommand{\etos}{\ensuremath{\epsilon^{\rm TOS}}\xspace}
\newcommand{\etis}{\ensuremath{\epsilon^{\rm TIS}}\xspace}

\def\lhcb {LHCb\xspace}
\def\ux85 {UX85\xspace}

\def\lhc {LHC\xspace}

\def\velo   {VELO\xspace}
\def\rich   {RICH\xspace}
\def\richone {RICH1\xspace}
\def\richtwo {RICH2\xspace}
\def\ttracker {TT\xspace}
\def\intr     {IT\xspace}
\def\introne  {IT1\xspace}
\def\st     {ST\xspace}
\def\ot     {OT\xspace}

\ifthenelse{\boolean{uprightparticles}}%
{

 \def\Pmu         {\ensuremath{\upmu}\xspace}

 \def\Ppi         {\ensuremath{\uppi}\xspace}

 \def\Ppsi        {\ensuremath{\uppsi}\xspace}

 \def\PDelta      {\ensuremath{\Delta}\xspace}
 \def\PXi      {\ensuremath{\Xi}\xspace}
 \def\PLambda      {\ensuremath{\Lambda}\xspace}
 \def\PSigma      {\ensuremath{\Sigma}\xspace}
 \def\POmega      {\ensuremath{\Omega}\xspace}
 \def\PUpsilon      {\ensuremath{\Upsilon}\xspace}

 \def\PB      {\ensuremath{\mathrm{B}}\xspace}
 
 \def\PD      {\ensuremath{\mathrm{D}}\xspace}

 \def\PJ      {\ensuremath{\mathrm{J}}\xspace}
 \def\PK      {\ensuremath{\mathrm{K}}\xspace}

 \def\PZ      {\ensuremath{\mathrm{Z}}\xspace}
 
 \def\Pb      {\ensuremath{\mathrm{b}}\xspace}
 \def\Pc      {\ensuremath{\mathrm{c}}\xspace}

 \def\Pi      {\ensuremath{\mathrm{i}}\xspace}

 \def\Pp      {\ensuremath{\mathrm{p}}\xspace}

 \def\Ps      {\ensuremath{\mathrm{s}}\xspace}

}
{

 \def\Pmu         {\ensuremath{\mu}\xspace}

 \def\Ppi         {\ensuremath{\pi}\xspace}

 \def\Ppsi        {\ensuremath{\psi}\xspace}
 
 \mathchardef\PDelta="7101
 \mathchardef\PXi="7104
 \mathchardef\PLambda="7103
 \mathchardef\PSigma="7106
 \mathchardef\POmega="710A
 \mathchardef\PUpsilon="7107
 
 \def\PB      {\ensuremath{B}\xspace}
 
 \def\PD      {\ensuremath{D}\xspace}

 \def\PJ      {\ensuremath{J}\xspace}
 \def\PK      {\ensuremath{K}\xspace}

 \def\PZ      {\ensuremath{Z}\xspace}
 
 \def\Pb      {\ensuremath{b}\xspace}
 \def\Pc      {\ensuremath{c}\xspace}

 \def\Pi      {\ensuremath{i}\xspace}

 \def\Pp      {\ensuremath{p}\xspace}

 \def\Ps      {\ensuremath{s}\xspace}

}

\def\mup        {\ensuremath{\Pmu^+}\xspace}
\def\mun        {\ensuremath{\Pmu^-}\xspace} 
\def\mumu       {\ensuremath{\Pmu^+\Pmu^-}\xspace}

\def\Z      {\ensuremath{\PZ^0}\xspace}

\def\squark    {\ensuremath{\Ps}\xspace}

\def\cquark    {\ensuremath{\Pc}\xspace}

\def\bquark    {\ensuremath{\Pb}\xspace}

\def\pion  {\ensuremath{\Ppi}\xspace}
\def\piz   {\ensuremath{\pion^0}\xspace}

\def\pip   {\ensuremath{\pion^+}\xspace}
\def\pim   {\ensuremath{\pion^-}\xspace}
\def\pipi  {\ensuremath{\pion^+\pion^-}\xspace}

\def\kaon  {\ensuremath{\PK}\xspace}
  \def\Kbar  {\kern 0.2em\overline{\kern -0.2em \PK}{}\xspace}

\def\Kz    {\ensuremath{\kaon^0}\xspace}
\def\Kzb   {\ensuremath{\Kbar^0}\xspace}
\def\KzKzb {\ensuremath{\Kz \kern -0.16em \Kzb}\xspace}
\def\Kp    {\ensuremath{\kaon^+}\xspace}
\def\Km    {\ensuremath{\kaon^-}\xspace}

\def\KpKm  {\ensuremath{\Kp \kern -0.16em \Km}\xspace}
\def\KS    {\ensuremath{\kaon^0_{\rm\scriptscriptstyle S}}\xspace}

\def\Kstarz  {\ensuremath{\kaon^{*0}}\xspace}

\def\Dbar    {\kern 0.2em\overline{\kern -0.2em \PD}{}\xspace}
\def\D       {\ensuremath{\PD}\xspace}

\def\Dz      {\ensuremath{\D^0}\xspace}
\def\Dzb     {\ensuremath{\Dbar^0}\xspace}
\def\DzDzb   {\ensuremath{\Dz {\kern -0.16em \Dzb}}\xspace}
\def\Dp      {\ensuremath{\D^+}\xspace}
\def\Dm      {\ensuremath{\D^-}\xspace}

\def\DpDm    {\ensuremath{\Dp {\kern -0.16em \Dm}}\xspace}

\def\Dstarp  {\ensuremath{\D^{*+}}\xspace}

\def\B       {\ensuremath{\PB}\xspace}
  \def\Bbar    {\kern 0.18em\overline{\kern -0.18em \PB}{}\xspace}

\def\Bz      {\ensuremath{\B^0}\xspace}

\def\Bu      {\ensuremath{\B^+}\xspace}

\def\Bp      {\ensuremath{\Bu}\xspace}

\def\Bd      {\ensuremath{\B^0}\xspace}
\def\Bs      {\ensuremath{\B^0_\squark}\xspace}
\def\BzBs    {\ensuremath{\B^0_{(\squark)}}\xspace}
\def\Bsb     {\ensuremath{\Bbar^0_\squark}\xspace}

\def\jpsi     {\ensuremath{{\PJ\mskip -3mu/\mskip -2mu\Ppsi\mskip 2mu}}\xspace}
\def\psitwos  {\ensuremath{\Ppsi{(2S)}}\xspace}

  \def\Y#1S{\ensuremath{\PUpsilon{(#1S)}}\xspace}
\def\OneS  {\Y1S}
\def\TwoS  {\Y2S}
\def\ThreeS{\Y3S}

\def\proton      {\ensuremath{\Pp}\xspace}

\def\L {\ensuremath{\PLambda}\xspace}

\def\Lc      {\ensuremath{\L_\cquark^+}\xspace}

\newcommand{\decay}[2]{\ensuremath{#1\!\to #2}\xspace}         
\def\ra                 {\ensuremath{\rightarrow}\xspace}
\def\to                 {\ensuremath{\rightarrow}\xspace}

\def\CP                {\ensuremath{C\!P}\xspace}

\def\BdToKstmm    {\decay{\Bd}{\Kstarz\mup\mun}\xspace}

\def\BsToJPsiPhi  {\decay{\Bs}{\jpsi\phi}\xspace}

\def\BuToJPsiK  {\decay{\Bu}{\jpsi\Kp}\xspace}
\def\BdToDpi  {\decay{\Bd}{\Dp\pim}\xspace}

\def\BdToKpi      {\decay{\Bd}{\Kp\pim}\xspace}

\def\AT#1     {\ensuremath{A_T^{#1}}\xspace}           

\def\Bsmm     {\decay{\Bs}{\mup\mun}\xspace}

\def\C#1      {\ensuremath{\mathcal{C}_{#1}}\xspace}                       
\def\Cp#1     {\ensuremath{\mathcal{C}_{#1}^{'}}\xspace}                    
\def\Ceff#1   {\ensuremath{\mathcal{C}_{#1}^{\mathrm{(eff)}}}\xspace}        
\def\Cpeff#1  {\ensuremath{\mathcal{C}_{#1}^{'\mathrm{(eff)}}}\xspace}       
\def\Ope#1    {\ensuremath{\mathcal{O}_{#1}}\xspace}                       
\def\Opep#1   {\ensuremath{\mathcal{O}_{#1}^{'}}\xspace}                    

\def\agamma     {\ensuremath{A_{\Gamma}}\xspace}

\newcommand{\unit}[1]{\ensuremath{\rm\,#1}\xspace}          

\newcommand{\tev}{\ensuremath{\mathrm{\,Te\kern -0.1em V}}\xspace}
\newcommand{\gev}{\ensuremath{\mathrm{\,Ge\kern -0.1em V}}\xspace}
\newcommand{\GeV}{\ensuremath{\mathrm{\,Ge\kern -0.1em V}}\xspace}
\newcommand{\mev}{\ensuremath{\mathrm{\,Me\kern -0.1em V}}\xspace}
\newcommand{\kev}{\ensuremath{\mathrm{\,ke\kern -0.1em V}}\xspace}
\newcommand{\ev}{\ensuremath{\mathrm{\,e\kern -0.1em V}}\xspace}
\newcommand{\gevc}{\ensuremath{{\mathrm{\,Ge\kern -0.1em V\!/}c}}\xspace}
\newcommand{\mevc}{\ensuremath{{\mathrm{\,Me\kern -0.1em V\!/}c}}\xspace}
\newcommand{\gevcc}{\ensuremath{{\mathrm{\,Ge\kern -0.1em V\!/}c^2}}\xspace}
\newcommand{\gevgevcccc}{\ensuremath{{\mathrm{\,Ge\kern -0.1em V^2\!/}c^4}}\xspace}
\newcommand{\mevcc}{\ensuremath{{\mathrm{\,Me\kern -0.1em V\!/}c^2}}\xspace}

\def\ma  {\ensuremath{{\rm \,m}^2}\xspace}
\def\cm   {\ensuremath{\rm \,cm}\xspace}

\def\mm   {\ensuremath{\rm \,mm}\xspace}

\def\mum  {\ensuremath{\,\upmu\rm m}\xspace}

\def\invnb {\ensuremath{\mbox{\,nb}^{-1}}\xspace}

\def\invpb {\ensuremath{\mbox{\,pb}^{-1}}\xspace}

\def\invfb   {\ensuremath{\mbox{\,fb}^{-1}}\xspace}

\def\ns   {\ensuremath{{\rm \,ns}}\xspace}

\def\fs   {\ensuremath{\rm \,fs}\xspace}

\def\mhz  {\ensuremath{{\rm \,MHz}}\xspace}
\def\khz  {\ensuremath{{\rm \,kHz}}\xspace}
\def\hz   {\ensuremath{{\rm \,Hz}}\xspace}

\def\invps{\ensuremath{{\rm \,ps^{-1}}}\xspace}

\def\neutroneq {\ensuremath{\rm \,n_{eq}}\xspace}

\newcommand{\chisq}{\ensuremath{\chi^2}\xspace}

\def\gsim{{~\raise.15em\hbox{$>$}\kern-.85em
          \lower.35em\hbox{$\sim$}~}\xspace}
\def\lsim{{~\raise.15em\hbox{$<$}\kern-.85em
          \lower.35em\hbox{$\sim$}~}\xspace}

\def\sqs   {\ensuremath{\protect\sqrt{s}}\xspace}

\def\pt         {\mbox{$p_{\rm T}$}\xspace}
\def\invpt      {\mbox{$1/p_{\rm T}$}\xspace}
\def\et         {\mbox{$E_{\rm T}$}\xspace}

\def\degrees{\ensuremath{^{\circ}}\xspace}

\def\mrad{\ensuremath{\rm \,mrad}\xspace}

\def\murad  {\ensuremath{\,\upmu\rm rad}\xspace}

\newcommand{\muvis}{\ensuremath{\mu_\text{vis}}}

\def\nonn {\ensuremath{\rm {\it{n^+}}\mbox{-}on\mbox{-}{\it{n}}}\xspace}
\def\ponn {\ensuremath{\rm {\it{p^+}}\mbox{-}on\mbox{-}{\it{n}}}\xspace}

\def\tell1  {TELL1\xspace}
\def\ukl1   {UKL1\xspace}

\def\cfourften     {\ensuremath{\rm C_4 F_{10}}\xspace}
\def\cffour        {\ensuremath{\rm CF_4}\xspace}

\newcommand{\eg}{\mbox{\itshape e.g.}\xspace}
\newcommand{\ie}{\mbox{\itshape i.e.}}

\def\DeltaThetaC  {\ensuremath{\Delta\theta_C}\xspace}
\def\SigmaThetaC  {\ensuremath{\sigma(\theta_C)}\xspace}
\def\RichDzeroKPi {\ensuremath{\D^0 \rightarrow K^- \pi^+}\xspace}
\def\RichDstarDPi {\ensuremath{\D^{*+}\rightarrow \D^0 \pi^+}\xspace}
\def\RichPPPPmumu {\ensuremath{p \thinspace p \rightarrow p \thinspace p \thinspace \mu^+ \mu^-}\xspace}
\def\Npe          {\ensuremath{N_{\rm pe}}\xspace}
\def\DeltaLLKPi   {\ensuremath{\rm \Delta log \mathcal{L}(K-\pi)}\xspace}

\usepackage{cite} 
\usepackage{mciteplus}

\usepackage{units}

\begin{document}

\renewcommand{\thefootnote}{\fnsymbol{footnote}}
\setcounter{footnote}{1}


\begin{titlepage}
\pagenumbering{roman}

\vspace*{-1.5cm}
\centerline{\large EUROPEAN ORGANIZATION FOR NUCLEAR RESEARCH (CERN)}
\vspace*{1.5cm}
\hspace*{-0.5cm}
\begin{tabular*}{\linewidth}{lc@{\extracolsep{\fill}}r}
\ifthenelse{\boolean{pdflatex}}
{\vspace*{-2.7cm}\mbox{\!\!\!\includegraphics[width=.14\textwidth]{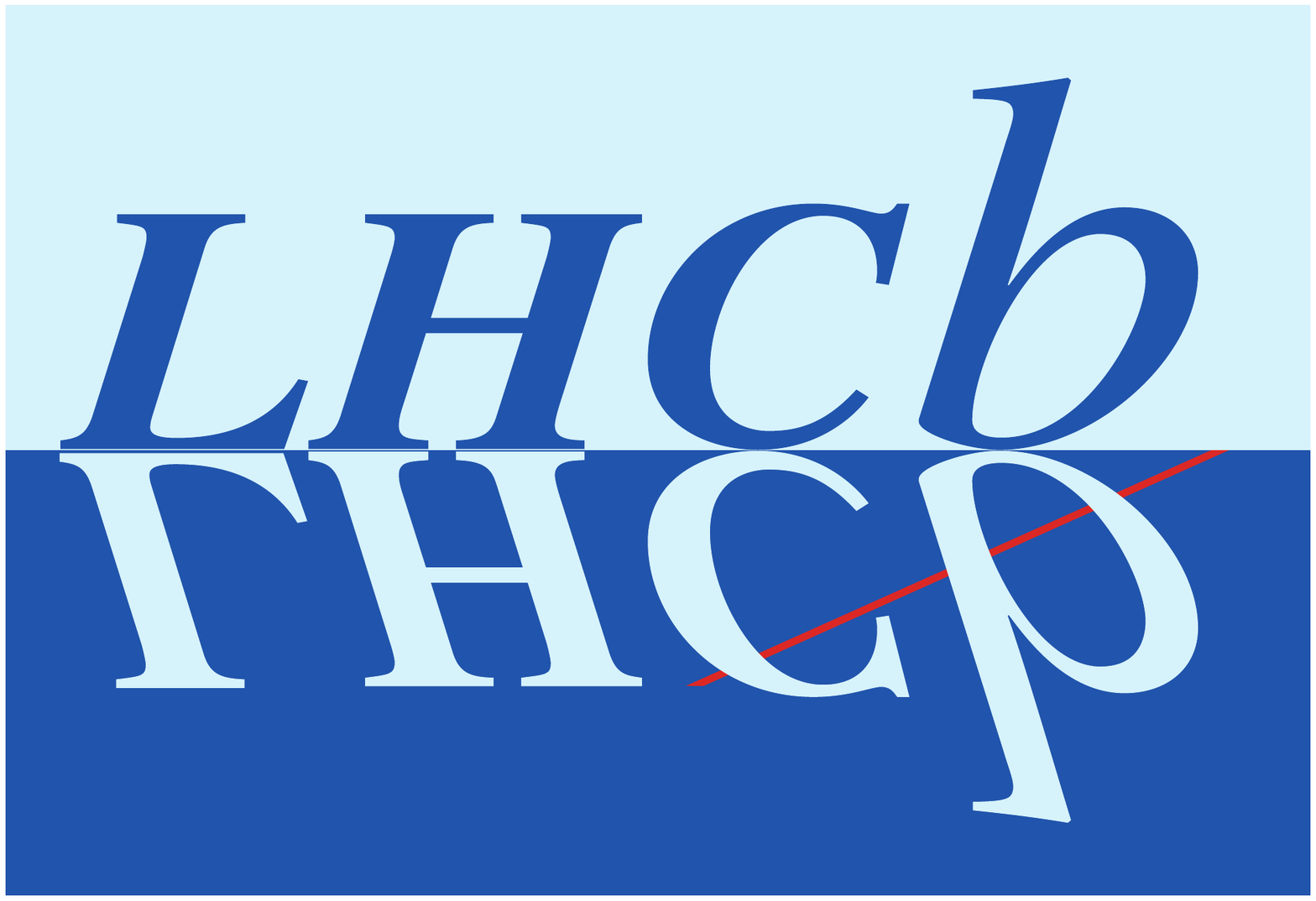}} & &}%
{\vspace*{-1.2cm}\mbox{\!\!\!\includegraphics[width=.12\textwidth]{lhcb-logo.eps}} & &}%
\\
 & & CERN-PH-EP-2014-290 \\  
 & & LHCb-DP-2014-002 \\  
 & & $19^{\mathrm{th}}$ December 2014 \\ 
 & & \\
\end{tabular*}

\vspace*{3.0cm}

{\bf\boldmath\huge
\begin{center}
LHCb Detector Performance
\end{center}
}

\vspace*{2.0cm}

\begin{center}
The LHCb collaboration\footnote{Authors are listed at the end of this paper.}
\end{center}

\vspace{\fill}

\begin{abstract}

\noindent
The LHCb detector is a forward spectrometer at the Large Hadron Collider (LHC)
at CERN. The experiment is designed for precision measurements of \CP{}
violation and rare decays  of beauty and charm hadrons. In this paper the
performance of the various LHCb sub-detectors  and the trigger system are
described, using data taken from 2010 to 2012.  It is shown that the design
criteria of the experiment have been met. The excellent performance of the
detector has allowed the LHCb collaboration to publish a wide range of physics
results, demonstrating LHCb's unique role, both as a  heavy flavour experiment
and as a general purpose detector in the forward region. 

\keywords{Large detector systems for particle and astroparticle physics; Particle tracking
detectors; Gaseous detectors; Calorimeters; Cherenkov detectors; Particle
identification methods; Detector alignment and calibration methods; Trigger; LHC }

\end{abstract}



\vspace*{1.0cm}

{\footnotesize 
\centerline{\it Published in Int. J. Mod. Phys. A 30, 1530022 (2015)}
\vspace*{1.0cm}
\centerline{\copyright~CERN on behalf of the \lhcb collaboration, license \href{http://creativecommons.org/licenses/by/4.0/}{CC-BY-4.0}.}}

\vspace*{2mm}

\end{titlepage}


\newpage
\setcounter{page}{2}
\mbox{~}
\newpage

\renewcommand{\thefootnote}{\arabic{footnote}}
\setcounter{footnote}{0}


\pagestyle{plain} 
\setcounter{page}{1}
\pagenumbering{arabic}


\cleardoublepage
\tableofcontents
\cleardoublepage

\section{Introduction}
\label{sec:introduction}

\subsection{Physics goals of the LHCb experiment}

LHCb is a dedicated heavy flavour physics experiment at the LHC. Its main goal is
to search for indirect evidence of new physics in \CP violation and rare decays of
beauty and charm hadrons,
by looking for the effects of new particles in processes that are precisely predicted in the 
Standard Model (SM) and by utilising the distinctive flavour structure of the SM
with no tree-level flavour-changing neutral currents.  
Quark mixing in the SM is described by the Cabibbo-Kobayashi-Maskawa (CKM)
matrix~\cite{Cabibbo:1963yz,*Kobayashi:1973fv}, 
which has a single source of \CP violation.
Since the level of \CP violation in weak interactions cannot explain the
matter-antimatter asymmetry in the universe~\cite{Sakharov:1967dj}, 
new sources of \CP violation beyond the SM are needed. 
The effect of such new sources might be seen in heavy flavour physics, where
many models of new physics produce contributions that change the expectation
values of the \CP violating phases or the branching fractions of rare decays. Some
models even predict decay modes that are forbidden in the SM. 
To examine such possibilities, \CP violation and rare decays of hadrons
containing $b$ and $c$ quarks must be studied with large data samples, using
many different decay modes. 

Thanks to the large beauty and charm production cross-section at the
\lhc~\cite{LHCb-PAPER-2010-002,LHCb-PAPER-2012-041}, 
the LHCb experiment collected $\sim 10^{12}$ heavy flavour decays during 2011 and 2012.
Despite these large yields, at the LHC centre-of-mass energies of $\sqrt{s}$ = 7--8\tev 
the charm and beauty cross-sections are approximately a factor 10 and 200
smaller than the total cross-section, respectively. To separate the decays of interest from
the background, both displaced vertex and high transverse momentum signatures are exploited.
Excellent vertex resolution is required to measure impact parameters and to
achieve a good decay time resolution, which is essential to resolve \Bs flavour
oscillations and to reject various sources of background. Good momentum and
invariant mass resolution are important to minimise combinatorial background and
resolve heavy-flavour decays with kinematically similar topologies. Charged
particle identification is essential in any flavour physics programme, for
instance to isolate suppressed decays and for $b$-quark flavour
tagging. Detection of photons, in addition to charged particles, allows the
reconstruction of rare radiative decays and more common decays with a \piz or an
$\eta$ meson in the final state. Finally, to benefit from the high event rate at the \lhc, a
high-bandwidth data acquisition system and a robust and selective trigger system are required.

LHCb has various advantages over the $e^+e^-$ $B$ factories, including a higher
cross-section, a larger boost and the fact that all species of $b$ hadrons are
produced. Less attractive characteristics of the LHC environment are the
generally increased background 
levels encountered, inherent to hadronic collisions, which result in a
number of experimental compromises, such as reduced $b$ flavour
tagging efficiency and the difficulty in reconstructing final states with
missing or neutral particles. 
Despite these challenges, the results~\cite{LHCb-PAPER-2012-031} obtained
from data taken between 2010 and 2013 (LHC Run I) have clearly established LHCb 
as the next generation flavour physics experiment.
Thanks to efficient charged particle tracking and dedicated triggers for lepton,
hadron and photon signatures, LHCb has the world's largest sample of
exclusively reconstructed charm and beauty decays.
With these samples, LHCb has already made many key results, such as
the first evidence for the rare decay 
$\Bs \to \mumu$~\cite{LHCb-PAPER-2012-043,LHCb-PAPER-2013-046} 
and measurements of angular distributions in the
$\Bz \to \Kstarz \mumu$ decay~\cite{LHCb-PAPER-2013-019,LHCb-Paper-2013-037},
which are particularly sensitive to deviations from the SM. 
Another example is the measurement of the \CP violating phase ($\phi_s$) in the
interference between mixing and decay of $\Bs$ mesons, where the value predicted
within the SM is small, but much larger values are possible in new physics
models. LHCb has measured this phase with results 
that are at present consistent with the SM within the
uncertainties~\cite{LHCb-PAPER-2014-019,LHCb-PAPER-2014-059}.  
The measurement of the angle $\gamma$ of the Unitarity Triangle from $\B \to \D
K$ decays is
a crucial component in the determination of the parameters of the CKM quark mixing matrix.
The $\gamma$ results from LHCb~\cite{LHCb-PAPER-2013-020,LHCB-CONF-2014-004}
already dominate the global averages. 
In the charm sector, one of the most interesting observables (\agamma)
is the difference in the inverse effective
lifetimes between \Dz and \Dzb decays.
The most precise measurement of \agamma to date has been presented by
LHCb~\cite{LHCb-PAPER-2013-054}. 
These are just some of the results from LHCb that have made a significant impact
on the flavour physics landscape. 

The physics output of LHCb extends well beyond this core programme. Examples of
other topics include: measurements of the
production of electroweak gauge bosons in the forward kinematic region,
uniquely covered by the LHCb acceptance~\cite{LHCb-PAPER-2012-008,LHCb-PAPER-2014-033};
measurements of the properties of newly discovered exotic
hadrons~\cite{LHCb-PAPER-2013-001,LHCb-PAPER-2014-014}; 
searches for lepton number and lepton flavour
violation~\cite{LHCb-PAPER-2011-038,LHCb-PAPER-2013-014}, 
and measurements of heavy quarkonia in 
proton-lead collisions~\cite{LHCb-PAPER-2013-052,LHCb-PAPER-2014-015}.
These illustrate the wide variety in electroweak and QCD topics covered
by the LHCb experiment 
and establish LHCb as a general purpose detector in the forward region at a
hadron collider. 

In the remainder of this introduction an overview of the LHCb detector is given,
together with a summary of the data-taking periods and the operating conditions.
Thereafter, the paper discusses charged particle reconstruction, vertexing and
decay-time resolution in 
Section~\ref{sec:charged-particle-reco}, neutral particle reconstruction in
Section~\ref{sec:neutral-particle-id} and 
particle identification in Section~\ref{sec:charged-particle-id}. The performance
results shown are indicative, and depend on the specific requirements set by a
physics analysis, for example to achieve high efficiency or high purity.
Section~\ref{sec:trigger} discusses the trigger and the paper concludes with a
short summary in Section~\ref{sec:Conclusion-Outlook}.

\subsection{Overview of the experimental setup}

\begin{figure}[ht]
\begin{center}
 \includegraphics[width=130mm]{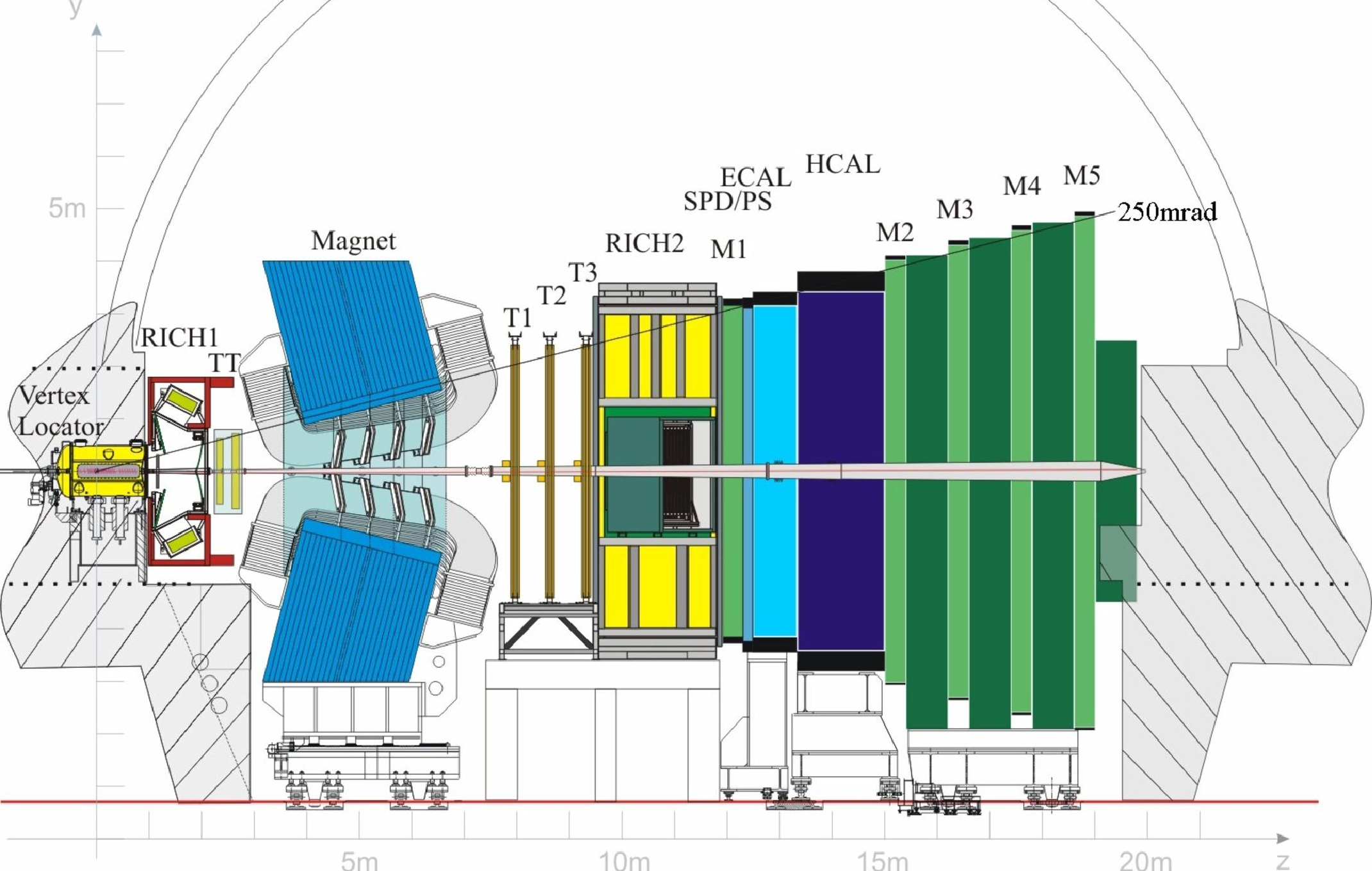}
 \caption{View of the LHCb detector~\protect\cite{LHCb-TDR-009}.}
 \label{fig:lhcb_layout}
\end{center}
\end{figure}

LHCb is a single-arm spectrometer with a forward angular coverage from approximately 15\mrad
to 300 (250)\mrad in the bending (non-bending) plane \cite{Alves:2008zz}.
The choice of the detector geometry is
driven by the fact that at high energies production of the $b$- and
$\overline{b}$-hadrons is highly correlated, such that they are
predominantly produced in the same forward or backward cone.
The layout of the LHCb spectrometer is shown in Figure~\ref{fig:lhcb_layout}.
Most detector subsystems are assembled in two halves, which can be moved out 
horizontally for assembly and maintenance purposes, as well as to provide access
to the beam-pipe. 
They are referred to as the detector A- and C-sides.
A right-handed coordinate system is defined with $z$ along the beam axis into
the detector, $y$ vertical and $x$ horizontal.
Cylindrical polar coordinates ($r$,$\phi$,$z$) are also used, as appropriate.

The spectrometer magnet, required for the momentum measurement of charged particles,
is a warm dipole magnet providing an integrated field of about 4\,Tm, 
which deflects charged particles in the horizontal plane.
The field of the spectrometer magnet also has an impact on the trajectory of the
LHC beams.
Three dipole magnets are used to compensate for this effect and to ensure a
closed orbit for the beams~\cite{LHCEvans:2008}. 

The tracking system consists of the VErtex LOcator (\velo), situated around the
interaction region inside a vacuum tank, and four planar tracking stations:
the Tracker Turicensis (TT) 
upstream of the dipole magnet, and tracking stations T1--T3 downstream of the magnet.
Silicon microstrips are used in TT and the region close to the beam-pipe (Inner Tracker, IT)
of stations T1--T3, whereas straw tubes are employed for the outer parts (Outer Tracker, OT).
Charged particles require a minimum momentum of $1.5\,\gevc$ to reach the tracking stations, T1--T3.

The \velo contains 42 silicon modules arranged along the beam,
each providing a measurement of the $r$ (\RSens sensors) and $\phi$ (\PhiSens sensors) coordinates.
The pitch within a module varies from
38\mum at the inner radius of 8.2\mm, increasing linearly to 102\mum at the
outer radius of 42\mm.
For detector safety, the \velo modules are retracted by 29\mm in the horizontal
direction during injection of the \lhc 
beams and are subsequently moved back, using a fully
automated procedure once stable conditions have been declared.
From the declaration of stable beams the VELO takes,
on average, 210 seconds to close. 
During LHC Run I approximately 750 closing procedures were performed.

The TT and IT detectors use silicon microstrip sensors with a strip pitch of
183\mum and 198\mum, respectively. 
The TT is about 150\cm wide and 130\cm high, with a total active area of around 8\ma. 
The IT covers a 120\cm wide and 40\cm high cross-shaped region in the centre of
the three tracking stations T1--T3. 
The total active area of the IT is approximately 4\ma. 
Each of the tracking stations has four detection layers in an \mbox{$x$ - $u$ -
  $v$ - $x$} arrangement with vertical strips in each of the two $x$ layers, and strips rotated
by a stereo angle of $-5\degrees$ and $+5\degrees$ in the $u$ and $v$ layers, respectively.

The Outer Tracker is a drift-tube gas detector consisting of approximately 200
gas-tight straw-tube modules 
with drift-time read-out. Each module contains two staggered layers of
drift-tubes with an inner diameter of $4.9\mm$. 
As a counting gas, a mixture of Argon (70\%), CO$_2$ (28.5\%) and O$_2$ (1.5\%)
is chosen to guarantee a drift time below $50\ns$ and a spatial resolution of $200\mum$.
As for the IT part of T1--T3, the OT has four layers arranged in an \mbox{$x$ -
  $u$ - $v$ - $x$} geometry. 
The total active area of a station is $597\cm \times 485\cm$.

Charged hadron identification in the momentum range from 2 to 100\,GeV/c
is achieved by two Ring Imaging Cherenkov detectors (RICH1 and RICH2) read out
by Hybrid Photon Detectors (HPDs). The upstream detector, RICH1, covers the low
momentum charged particle range from about 2 to 60\,GeV/c and uses Aerogel and
C${_4}$F$_{10}$ as radiators, while the downstream detector, RICH2, covers the
high momentum range from about 15\,GeV/c to 100\,GeV/c, using a CF${_4}$
radiator. RICH1 has a wide acceptance, covering the LHCb acceptance from
$\pm$25\mrad to $\pm$300\mrad (horizontal) and $\pm$250\mrad (vertical), while
RICH2 has a limited angular acceptance of $\pm$15\mrad to $\pm$120\mrad 
(horizontal) and $\pm$100\mrad (vertical).

The calorimeter system is composed of a Scintillating Pad Detector (SPD), a Preshower (PS),
a shashlik type electromagnetic calorimeter (ECAL) and a hadronic calorimeter (HCAL). 
It provides the identification of electrons, photons and hadrons as well as the
measurement of their energies and positions, 
and selects candidates with high transverse energy for the first trigger level (L0). 
The SPD improves the separation of electrons and photons.
A 15\mm lead converter with a thickness of 2.5 radiation lengths (${X}_0$) is placed between
the planes of rectangular scintillating pads of the SPD and the PS.
The background from charged pions is reduced by a measurement of
the longitudinal partitioning of the electromagnetic shower in the PS detector 
and the main section of ECAL. 
The ECAL is made of a sampling scintillator/lead structure with a total thickness of
25 $X_0$. The calorimeter system has a variable lateral segmentation which takes
into account the variation in hit density of two orders of magnitude over the calorimeter surface.
A segmentation into three different sections has been chosen for the ECAL 
with a corresponding projective geometry for the SPD and PS detectors, meaning that all of 
their transverse dimensions scale with the distance from the interaction point.
The outer dimensions match projectively those of the tracking system,
while the square hole around the beam-pipe approximately limits the inner
acceptance to projective polar angles $\theta_{x,y} > 25$\mrad. 
The hadron calorimeter (HCAL) is a sampling device made from iron and scintillating tiles,
as absorber and active material, respectively. The special feature of this sampling structure
is the orientation of the scintillating tiles which run parallel to the beam-axis.
Given the dimensions of the hadronic showers, the HCAL is segmented into two
zones with different lateral dimensions.
The thickness of the HCAL is limited to 5.6 nuclear interaction lengths
($\lambda_i$) due to space constraints.

The muon detection system provides muon identification and contributes 
to the L0 trigger of the experiment. It is composed of five stations (M1--M5) of rectangular
shape equipped predominantly with Multi Wire Proportional Chambers (MWPC),
except in the highest rate region of M1, 
where triple Gas Electron Multiplier (GEM) detectors are used.
The full system comprises 1380 chambers and covers a total area of 435\ma.
Stations M2 to M5 are placed downstream of the calorimeters and are interleaved with 80\cm thick iron absorbers
to select penetrating muons. The minimum momentum that a muon must have to traverse the five stations
is approximately $6\,\gevc$. The total absorber thickness, including the
calorimeters, is approximately 20 $\lambda_i$. 
Station M1 is placed in front of the calorimeters and is used to improve the
\pt measurement in the trigger. The geometry of the five stations is projective, with
each station divided into four regions, R1 to R4, with increasing distance from the beam axis.
The linear dimensions of the regions R1, R2, R3, R4, and their segmentation scale in the ratio 1:2:4:8.
With this geometry, the channel occupancies are comparable in each of 
the four regions of a given station.

The LHCb trigger system consists of two levels. The first level is implemented in hardware and
is designed to reduce the event rate from the nominal LHC bunch crossing rate of
40\mhz to a maximum of 1.1\mhz. The complete 
detector is then read out and the data is sent to the High Level Trigger (HLT)
implemented on the Event Filter Farm (EFF),  
which had about 30000 processing cores in 2012. The HLT is a software trigger,
running a simplified version of the offline event reconstruction to accommodate
the more stringent CPU time requirements. 

\subsection{Data taking periods and operating conditions}

\begin{figure}
\begin{center}
\includegraphics[width=140mm]{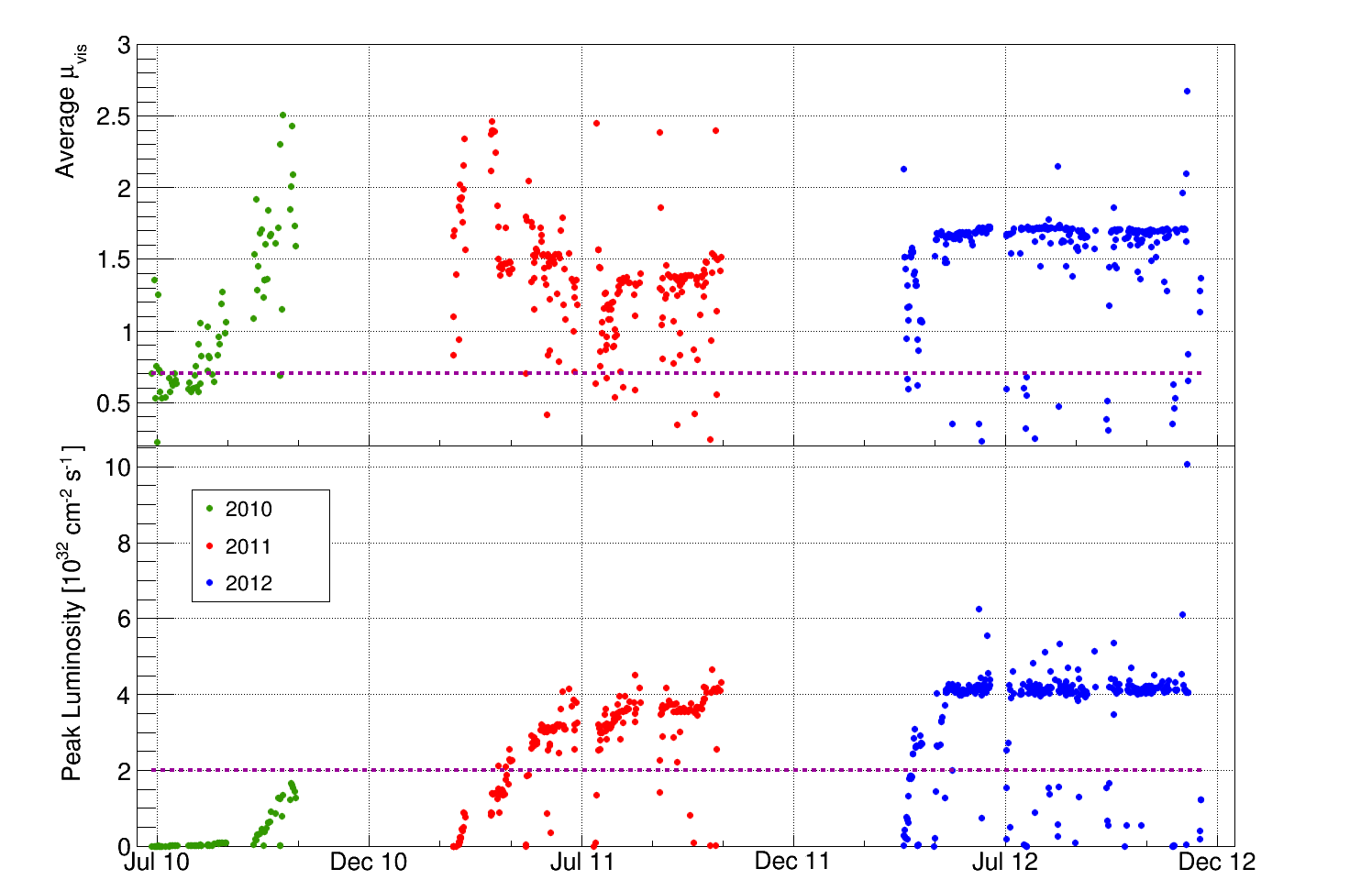}
\caption{Average number of visible interactions per bunch crossing ('pile-up', top) and
  instantaneous luminosity (bottom) at the LHCb interaction point in the period
  2010-2012. The dotted lines show the design values.} 
\label{fig:oper_conditions}
\end{center}
\end{figure}

At the end of 2009, LHCb recorded its first $pp$ collisions at the injection
energy of the LHC, $\sqs=0.9\tev$.
These data have been used to finalise the commissioning of the sub-detector
systems and the reconstruction software,  
and to perform a first alignment and calibration of the tracking, calorimeter and
particle identification (PID) systems. In this period, the \velo was left in the open
position, due to the larger aperture required at lower beam energies.

During 2010 the operating conditions changed rapidly due to the
ramp-up of the LHC luminosity. A critical parameter for LHCb
performance is the pile-up \muvis{}, defined as the average
number of visible interactions per beam-beam
crossing~\cite{LHCb-PAPER-2011-015}. The evolution of the LHCb
operating conditions during LHC Run I is shown in
Figure~\ref{fig:oper_conditions}.  Starting with luminosities
$\sim 10^{28}\mathrm{\,cm^{-2}s^{-1}}$ and almost no pile-up, the
luminosity reached $10^{32}\mathrm{\,cm^{-2}s^{-1}}$ with
$\muvis \approx 2.5$.

While the highest luminosity in 2010 was already 75\% of the LHCb design luminosity,
the pile-up was much larger than the design value due to the low number of bunches in the machine. 
It was demonstrated that the trigger and reconstruction work efficiently under such harsh conditions 
with increased detector occupancy due to pile-up, and that the physics output was not compromised.

\begin{figure}
\begin{center}
\includegraphics[width=135mm]{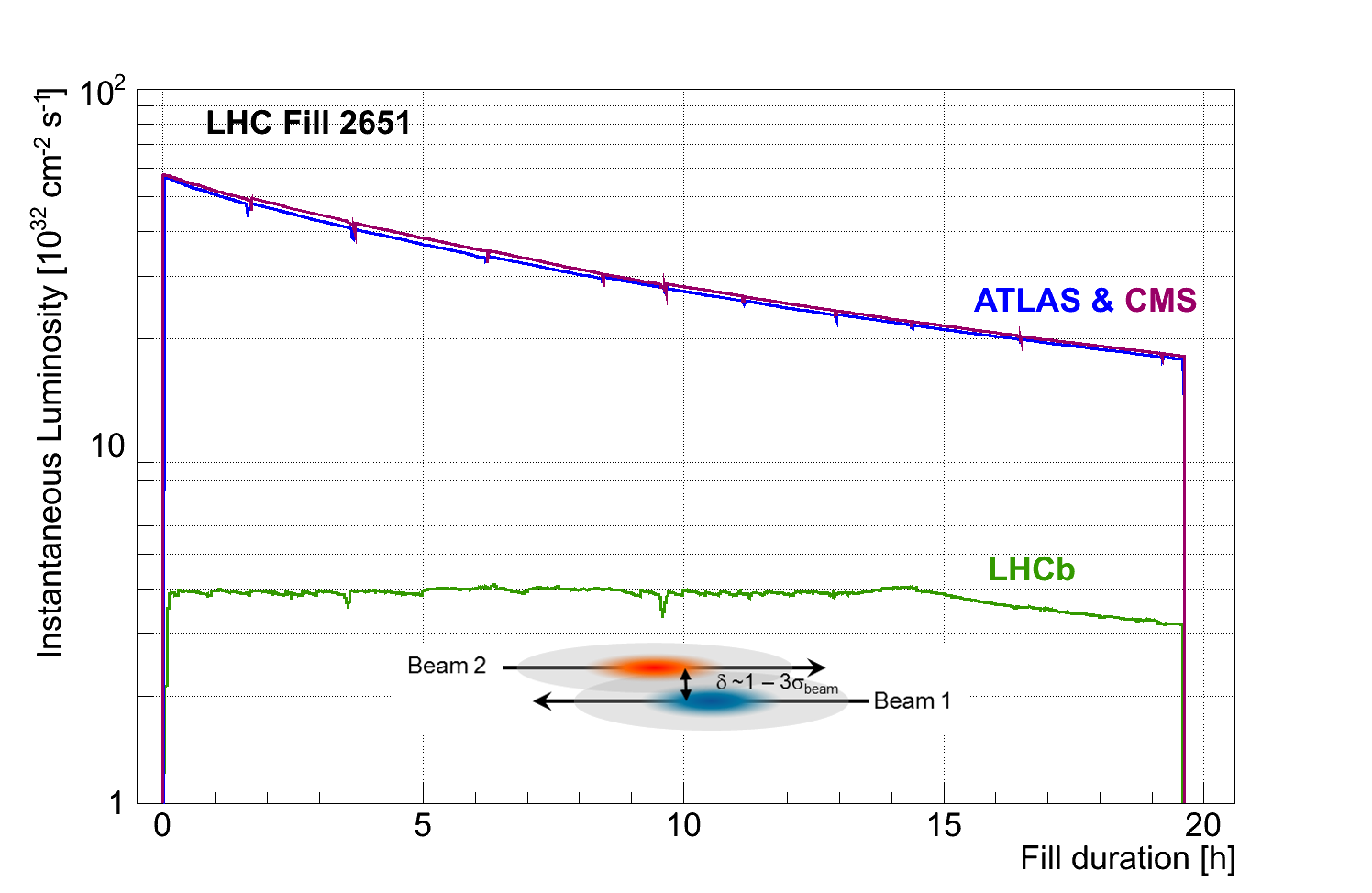}
\caption{Development of the instantaneous luminosity for
  ATLAS, CMS and LHCb during LHC fill 2651. After ramping to the desired value
  of $4 \times 10^{32}\mathrm{cm^{-2}s^{-1}}$ for LHCb, the luminosity is kept stable
  in a range of 5\% for about 15 hours by adjusting the transversal beam overlap.
  The difference in luminosity towards the end of the fill between ATLAS, CMS and LHCb
  is due to the difference in the final focusing at the collision points, commonly referred to
  as the beta function, $\beta^*$.}
\label{fig:lumi-levelling}
\end{center}
\end{figure}

The LHC beam energy was 3.5\tev during 2010 and 2011.
In the first part of the 2011 data taking the number of bunches in the machine
increased in several steps 
to about 1300, the maximum possible with $50\,$ns bunch spacing.
Due to the larger number of bunches the pile-up over the year could be reduced,
while LHCb took the majority 
of the data at a luminosity of $3.5\times 10^{32}\mathrm{\,cm^{-2}s^{-1}}$. This
was 1.75 times more 
than the design luminosity of $2\times 10^{32}\mathrm{\,cm^{-2}s^{-1}}$, as
shown in Figure~\ref{fig:oper_conditions}. 
In 2011 a luminosity levelling procedure was introduced at the LHCb
interaction point. By  adjusting the transverse overlap of the beams
at LHCb, the instantaneous luminosity could be kept
stable to within about 5\% during a fill, as illustrated in
Figure~\ref{fig:lumi-levelling}. 
For this particularly long fill, a maximum overlap with head-on beams was
reached only after 15 hours. The luminosity levelling procedure
minimises the effects of luminosity decay, allowing to maintain the same trigger
configuration during a fill and to reduce systematic uncertainties due to
changes in the detector occupancy. 

In 2012 the LHC beam energy was increased to 4\tev. 
LHCb took data at a luminosity of $4\times 10^{32}\mathrm{\,cm^{-2}s^{-1}}$,
twice the LHCb design luminosity. 
The LHC delivered stable beams for about 30\% of the operational year. An effort
was made in 2012 to use more 
efficiently the processing power available in the Event-Filter-Farm (EFF), which
otherwise would have been idle during 70\% of the time.
The mechanism put in operation defers 
a fraction of the HLT processing to the inter-fill time, typically several
hours, between the LHC collision periods. 
In this approach about 20\% of the L0 accepted events during data-taking are
temporarily saved on the local disks 
of the EFF nodes and are processed only after the end of stable beams.
This deferred triggering method allowed LHCb to increase the data sample
available for physics analysis. 

\begin{figure}
\begin{center}
\includegraphics[width=135mm]{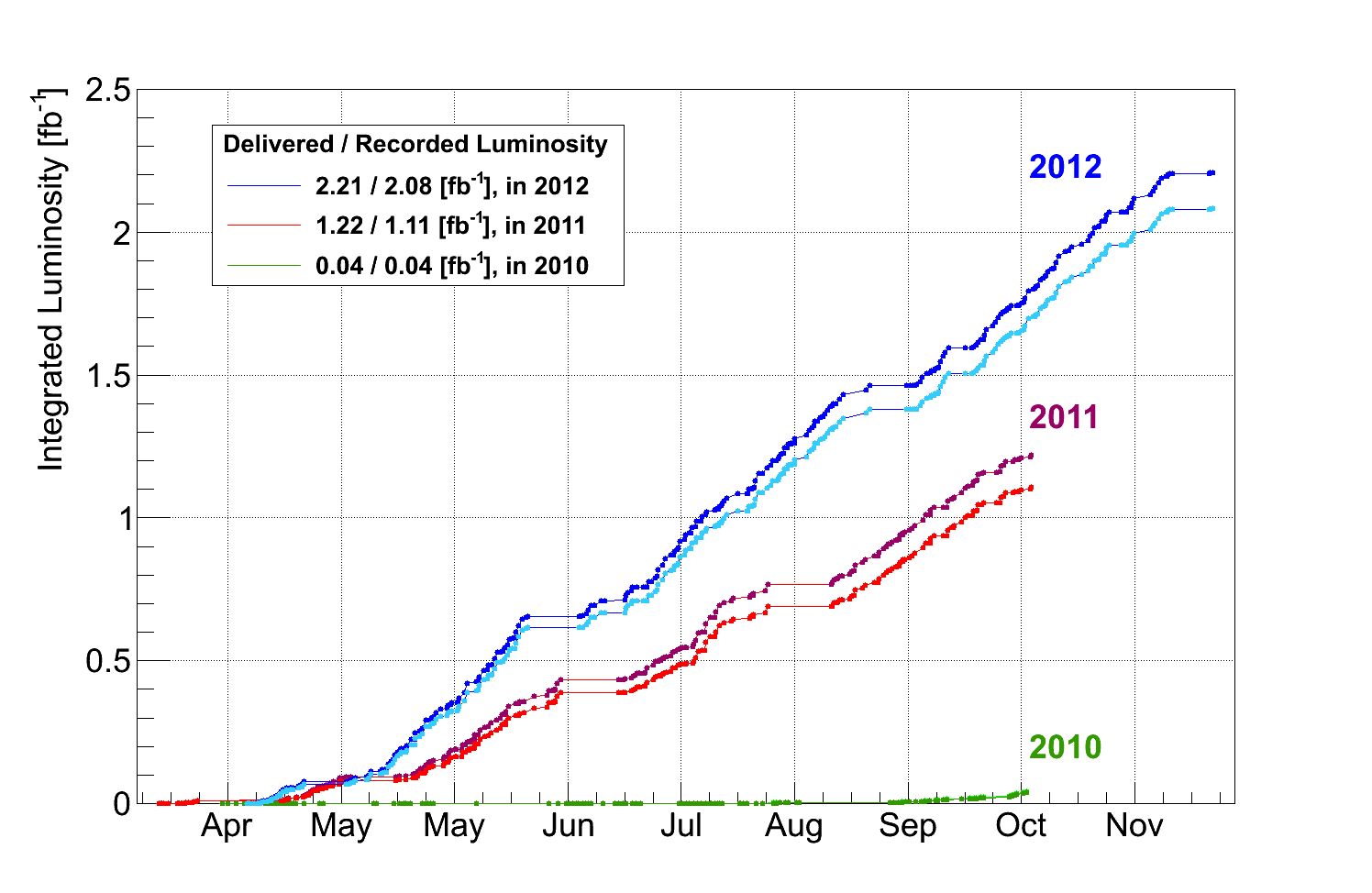}
\caption{Integrated luminosity in LHCb during the three years of LHC Run I.
  The figure shows the curves for the delivered (dark coloured lines)
  and recorded (light coloured lines) integrated luminosities.}
\label{fig:int_luminosity}
\end{center}
\end{figure}

The integrated luminosity recorded by LHCb was 38\invpb in 2010, 1.11\invfb in
2011 and 2.08\invfb in 2012. The evolution of the integrated luminosity for the
years 2010 to 2012 is shown in Figure~\ref{fig:int_luminosity}. 

Luminosity calibrations were carried out with the LHCb detector for the various
centre-of-mass energy $\sqrt{s}$  
at which data has been taken. Both the "van der Meer scan" and "beam-gas
imaging" luminosity calibration methods  
were employed~\cite{LHCb-PAPER-2014-047}. 
For proton-proton interactions at $\sqrt{s}$ = 8\tev a relative precision of the
luminosity calibration of  
1.47\% was obtained using van der Meer scans and 1.43\% using beam-gas imaging,
resulting in a combined  
precision of 1.12\%. Applying the calibration to the full data set determines
the luminosity with a precision  
of 1.16\%. This represents the most precise luminosity measurement achieved so
far at a bunched-beam hadron collider. 

The average operational efficiency, defined as the ratio of recorded over
delivered luminosity, 
was 93\% during LHC Run~I, reaching 95\% on average in 2012.
The inefficiency contains two irreducible sources. The first one is the
detector-safety procedure for the VELO closing, 
amounting to 0.9\%, which is in line with expectations. The second originates
from non-conformities in the implementation 
of the read-out protocol of some sub-detector front-end systems and introduces
2.4\% of dead-time at 1\mhz read-out frequency. 
The remaining 3.6\% is related to short technical problems with the sub-detector
electronics or the central read-out system. 
About 99\% of the recorded data is used for physics analyses.

After a short pilot run in 2012, the LHC delivered for the first time
proton-lead collisions in January and February 2013. 
The beam energy of the proton beam was 4\tev, while the corresponding nucleon
energy of the lead beam was 1.58\tev, 
corresponding to a centre-of-mass energy of 5\tev. The LHC delivered collisions
with both protons and lead nuclei as the clockwise, and anti-clockwise beams,
which made it possible for LHCb to collect data in the forward and backward
direction of proton-lead collisions. 
The integrated recorded luminosity during the proton-lead run was 1.6\invnb.

Since the LHCb magnet deflects positive and negative particles in opposite
directions in the $x-z$ plane, 
a difference in performance of the left and right sides of the detector leads to
charge detection asymmetries. 
To reach its design sensitivity in \CP{} violation measurements, LHCb aims to
control such detection asymmetries 
to a precision of $10^{-3}$ or better. This is achieved by changing the
direction of the magnetic field regularly 
and then combining data sets with different polarity to cancel left-right
asymmetries. In \mbox{Run I} the polarity 
of the magnet was inverted about two times per month, such that smoothly varying
changes in data-taking conditions or detector performance would not jeopardise
the cancellation.

The LHCb operation with both field polarities leads to different
effective crossing angles between the two beams, in particular when the beam
crossing is performed in the horizontal plane, 
as it was the case in 2010 and 2011. The effective total crossing angles varied
between about 40\murad and 1040\murad 
for the two spectrometer polarities.
During 2012 the beam crossing was performed in the vertical plane. Together with
the deflection caused by the LHCb 
spectrometer magnet this led to more similar total effective crossing angles of
about $\pm$ 470\murad in the horizontal plane 
for the two spectrometer polarities, respectively, and of $\pm$ 200\murad in the
vertical plane. 
However, the physics performance of the experiment has not been affected by the
various beam crossing scenarios mentioned here.

\section{Charged particle reconstruction}
\label{sec:charged-particle-reco}

The trajectories of charged particles inside the LHCb detector are reconstructed
using dedicated tracking detectors; The \velo detector encompassing the interaction region,
the TT stations before the spectrometer magnet and the T1-T3 stations further
downstream. By determining the deflection of the charged particles
after traversing the magnetic field, their momentum can be determined. The high
spatial resolution of the \velo enables a precise determination of the particle's
flight direction close to the primary
interaction point, resulting in a good vertex resolution.

\subsection{Hit efficiencies and hit resolutions of the tracking detectors}

The hit efficiencies and hit resolutions of the different tracking detectors are
discussed in the following sections. Hit efficiencies in general exceeding 99\%
were achieved, more than sufficient for an efficient track reconstruction. The hit
resolutions of all tracking detectors are as expected from test-beam
measurements. Hit occupancies for the 2011 data taking conditions, although
running at much higher luminosity and pile-up than originally planned, 
are well within acceptable levels, only mildly affecting the track finding
efficiency and rate of wrongly reconstructed trajectories. 

\subsubsection{Vertex Locator}
\label{sec:velohitresolution}

The overall performance of the \velo is described in detail 
in Reference~\cite{LHCb-DP-2014-001}.
A summary of the hit efficiency, hit resolution,
occupancy and radiation damage given below.

The \velo hit efficiency is evaluated by two methods.
The cluster finding efficiency~\cite{Parkes:1074928} is determined by removing
each sensor in turn from the track reconstruction, extrapolating the tracks to
this sensor and searching for a hit around the intercept
point. Alternatively, the
channel occupancy spectra is analysed to identify strips with
a substantially lower or higher number of hits than the average.
The two methods are in agreement. At the end of \lhc Run~I, the
occupancy method identified $0.6\%$ inefficient strips and $0.02\%$
noisy strips in the detector, these numbers are effectively identical
to those at the start of operations in 2010.

\begin{figure}[!tb]
  \vspace{1mm}
  \begin{center}
    \resizebox{\textwidth}{!}{
      \includegraphics[width=0.42\textwidth]{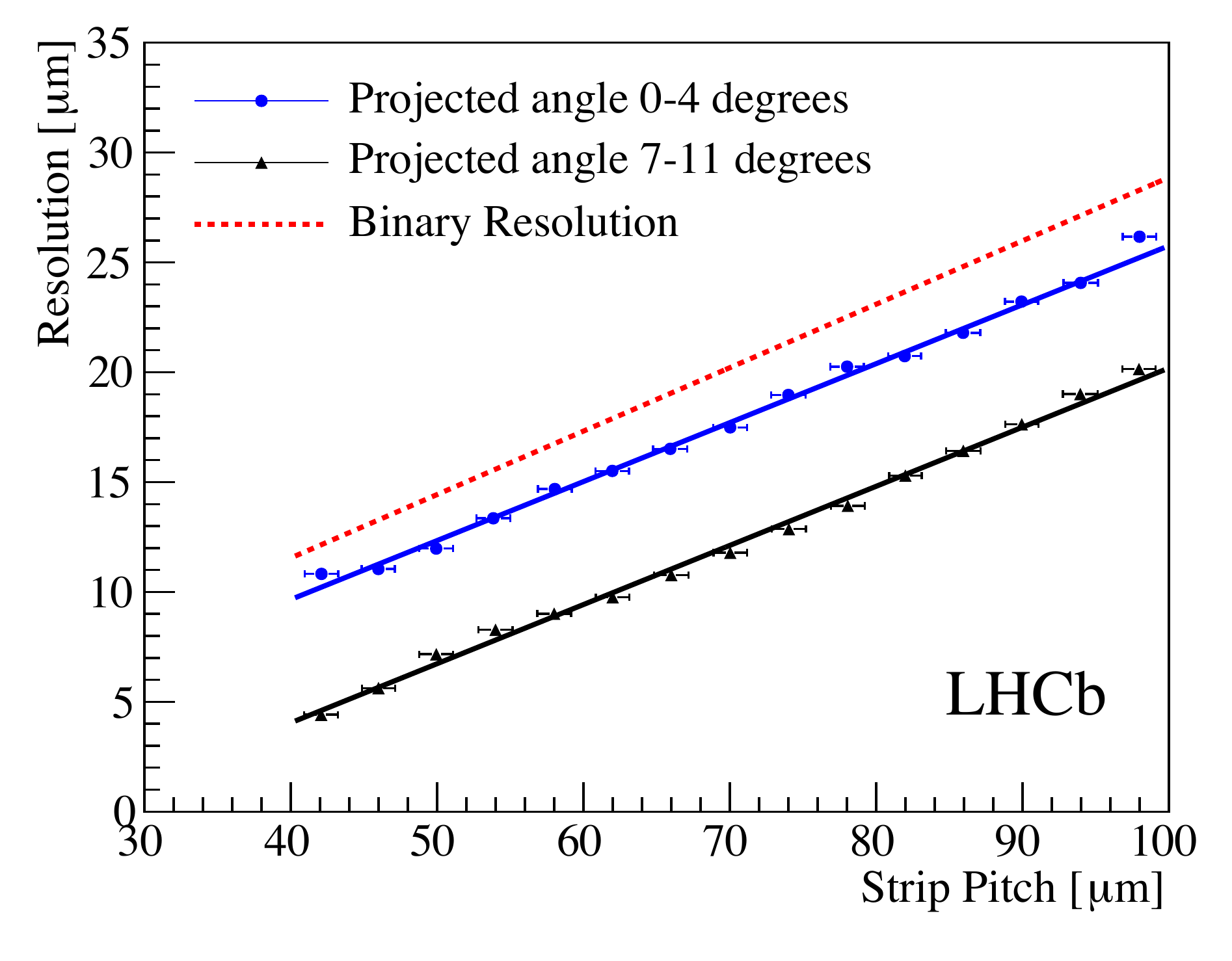}
      \includegraphics[width=0.4\textwidth]{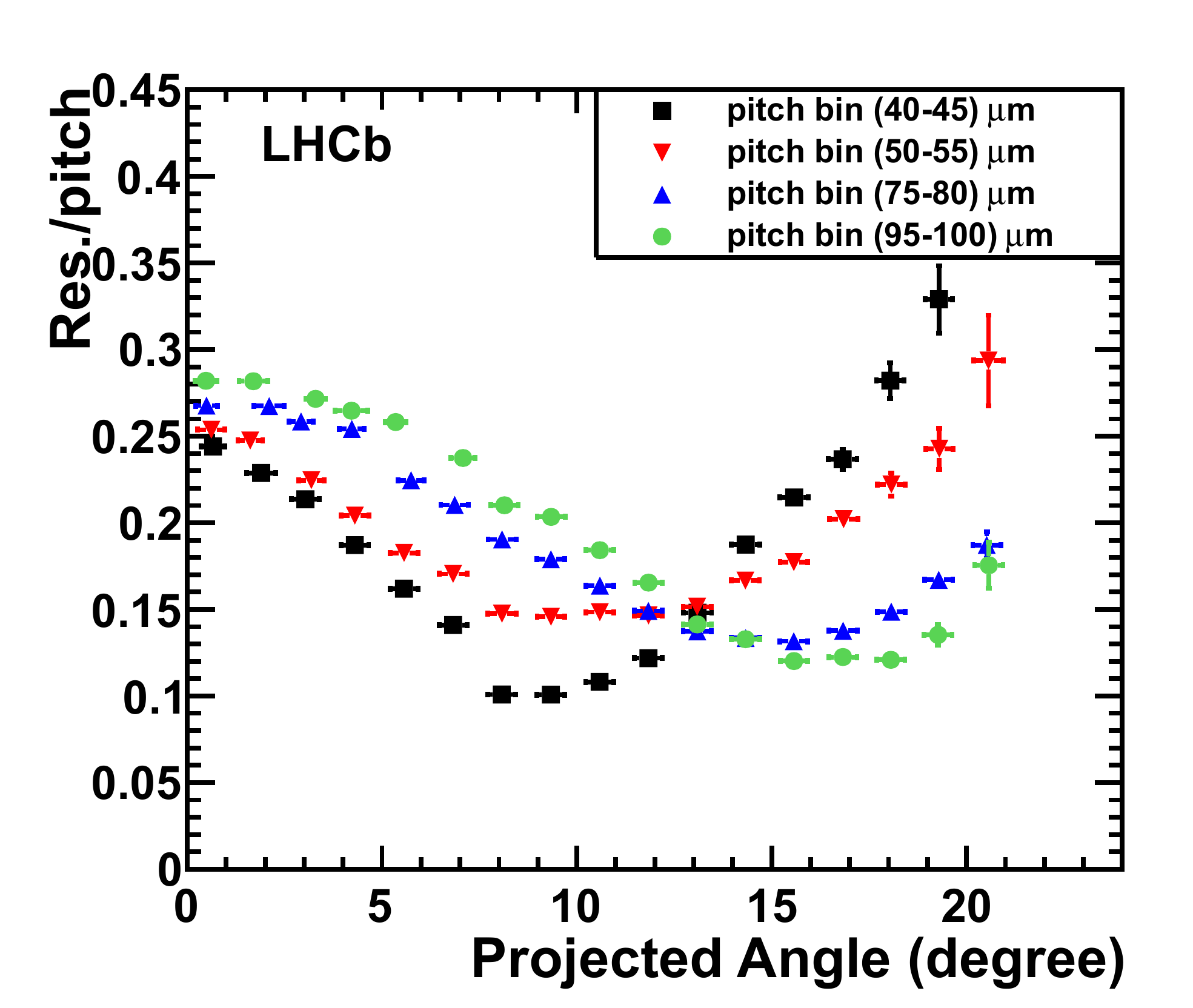}
    }
    \caption{The \velo hit resolution as a function of the
     inter-strip pitch (left) evaluated with 2010 data  for the $R$
     sensors. Results are shown for two projected angle ranges and the
     expected resolution of a single-hit binary system is indicated
     for comparison. Resolution divided by pitch as a function
   of the track projected angle for four different strip pitches (right).}
    \label{figVELORes}
  \end{center}
  \vspace{1mm}
\end{figure}

The hit resolution in silicon devices depends on the inter-strip
read-out pitch and the charge sharing between strips. The charge
sharing varies with the operational bias voltage and the projected angle of the
track. The bias voltage was 150\unit{V} throughout the physics data taking in
2010 to 2013. The projected angle \cite{Parkes:1074928} provides information on
the number of strips that the particle crosses while it traverses the thickness
of the silicon sensor. Initially the resolution
improves with increasing angle due to the charge sharing between strips,
allowing more accurate interpolation of the hit position. The optimal resolution
is obtained when the tracks cross the width of one strip
when traversing the 300\mum thickness of the sensor. For the \velo
the optimal projected angle varies between about 7\degrees at the lowest
inter-strip pitch of 40\mum to about 18\degrees for the 
largest 100\mum pitch strips, as shown in Figure~\ref{figVELORes}~(right). 
Above the optimal angle the resolution begins to
deteriorate due to the fluctuations in the charge on the strips and because 
the signal to noise ratio on individual strips may drop below the clustering threshold.

The \velo reads out analogue pulse-height information from the strips,
and this information is used offline to calculate the cluster position
\cite{Parkes:1074928} 
using the weighted average of the strip ADC values. The resolution of the sensors
is determined from the residual between the extrapolated position of
the fitted track and the measured cluster position.
The use of the evaluated cluster position in the track fit gives rise to a bias in the
residual, for which a correction is applied.

The resolution is determined as a function of the strip pitch
and of the projected angle.  For each bin, the resolution is determined
from the width of the fit of a Gaussian function to the distribution of the
corrected residuals. The resolution is evaluated using tracks that have hits in
the tracking stations behind the magnet and hence for which the momentum
measurement is available. The tracks are required to have a momentum greater
than 10\gevc to reduce the dependence of the estimation on the multiple
scattering effect, and a number of other track quality criteria are applied to
reject fake tracks. The results are presented here for the \RSens sensor. The
\PhiSens sensor results are compatible with those of the \RSens sensor
but the almost radial geometry of the \PhiSens sensor
strips means that tracks primarily have small projected angles.

The measured hit resolution has a linear dependence on the strip pitch in
projected angle bins, as shown in Figure~\ref{figVELORes}~(left).
The best hit precision measured is around 4\mum for an optimal projected angle of
8\degrees and the minimum pitch of approximately 40\mum.

The detector occupancy is a key parameter in the performance of the
pattern recognition and tracking algorithms. The cluster
occupancy was measured to be around $0.5\%$ in randomly triggered events
and $1\%$ in events passing the high level trigger, for data with a
\muvis{} = 1.7. The pitch of the strips on the sensors increases with radius, 
keeping the local occupancy values to within $25\%$ of these typical values.
The occupancy from
noise is negligible compared with that from particles; in the absence
of circulating beams the occupancy is below $0.01\%$.

The proximity of the \velo sensors to the LHC $pp$ collisions
results in the sensors receiving a significant radiation dose. A
study of the observed effects is available in Reference~\cite{LHCb-DP-2012-005}. 
During \lhc Run~I the sensors have been exposed to a range of fluences up
to  a maximum value of $1.8 \times 10^{14}\,1 \mev$ neutron
equivalents $/ \cm^2$ (\neutroneq) at the radius of the inner strip of
8.2\mm.

The current drawn from a silicon sensor increases linearly with fluence.
The sensor current is composed of two dominant sources, referred to as
{\it bulk} and {\it surface} currents. Studying the current as a
function of the temperature allows the two sources to be separated,
and dedicated data is taken to allow this study to be performed.
At the operational sensor temperature of approximately
$-7\degreesC$, the average rate of sensor current increase is 18\,$\upmu$A per
\invfb, in agreement with predictions~\cite{LHCb-DP-2012-005}.  

\begin{figure}[!tb]
   \begin{center}
     \resizebox{0.8\textwidth}{!}{
       \includegraphics[width = 0.49\textwidth]{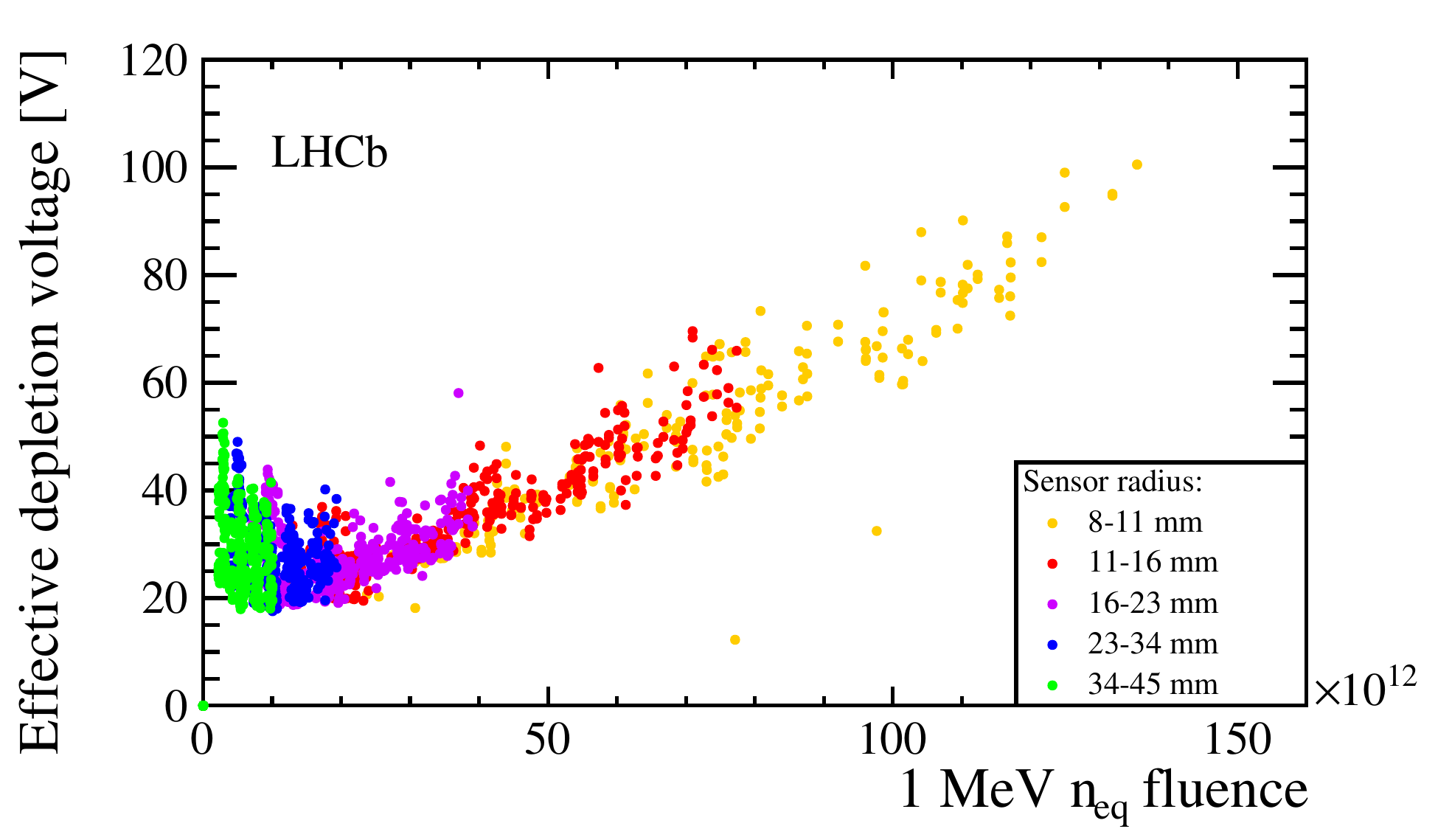}
     }
 \caption{ The effective depletion voltage versus fluence for all \velo
 sensors up to the end of LHC Run~1 at 3.4\invfb delivered integrated luminosity.}
 \label{figVELOEDV}
   \end{center}
\end{figure}

Dedicated data are taken around three times a year to study the charge
collection of the \velo as a function of the bias voltage of the
sensors. The bias voltage required to extract a fixed fraction of the
maximum charge can then be determined. From this measurement the
`effective depletion voltage' can be
determined~\cite{LHCb-DP-2012-005}, and this is shown as a function of fluence in
Figure~\ref{figVELOEDV}. Each sensor contributes multiple points to this
plot in each data sample as the sensors are divided in the analysis into radial
regions 
that have received similar fluences, as denoted by the different colours in the
figure. The ${\rm{\it n}}$-bulk sensors undergo
space-charge sign inversion under irradiation, and hence their
depletion voltage initially reduces with irradiation. This continues until type
inversion 
occurs, after which it increases with further irradiation. The first observation
of \nonn sensor 
space-charge-sign-inversion at the LHC was made during
2011~\cite{LHCb-DP-2012-005}, occurring 
at a fluence of a round $15 \times 10^{12}\,1\mev\neutroneq$.  The effective
depletion voltage at the maximal fluence at the end of \lhc Run~I was
approximately 100\,V, and followed the expectation. The current detector
is predicted to deliver an acceptable physics performance until the
end of \lhc Run~II with an operating voltage below 500\,V.

\subsubsection{Silicon Tracker}
\label{sec:tracking:st}

The Tracker Turicensis (TT) and the Inner Tracker (IT) are constructed
from \ponn silicon microstrip detectors. 
The TT sensors are 9.64\cm wide, 9.44\cm long and 500\mum thick.  
The TT modules have read-out sectors with one, two, three or four sensors bonded
together, and are arranged such that 
the single-sensor sectors are closest to the beam-pipe in the region with the
highest flux of particles.   
The sensors in the IT are 7.6\cm wide, 11\cm long and are either 320\mum or
410\mum thick.  
Two 410\mum thick sensors are bonded together for the IT modules on either side
of the beam-pipe while the modules above  
and below the beam-pipe use one 320\mum sensor.  
In total, there are 280 (336) read-out sectors with 512 (384) strips in the TT (IT).

The cluster finding efficiency of the detectors depends on the fraction of
working channels and the intrinsic hit efficiency of the silicon sensor. The
number of working channels is affected by problems with the read-out, and the
masking of dead or noisy strips found during the calibration of the detector.
The fraction of working channels varied during data taking.  The
luminosity-weighted average of the fraction of working channels during Run~I is
calculated to be $99.7\%$ and $98.6\%$ for the TT and IT, respectively. Repairs
can be made to the TT read-out during short technical stops whereas problems
with the IT read-out can only be fixed during the LHC shutdowns at the end of
each year.  Two read-out sectors were disabled in the IT as they could not be
properly configured. 

\begin{figure}[!tb]
  \centering\includegraphics[width=0.7\textwidth]{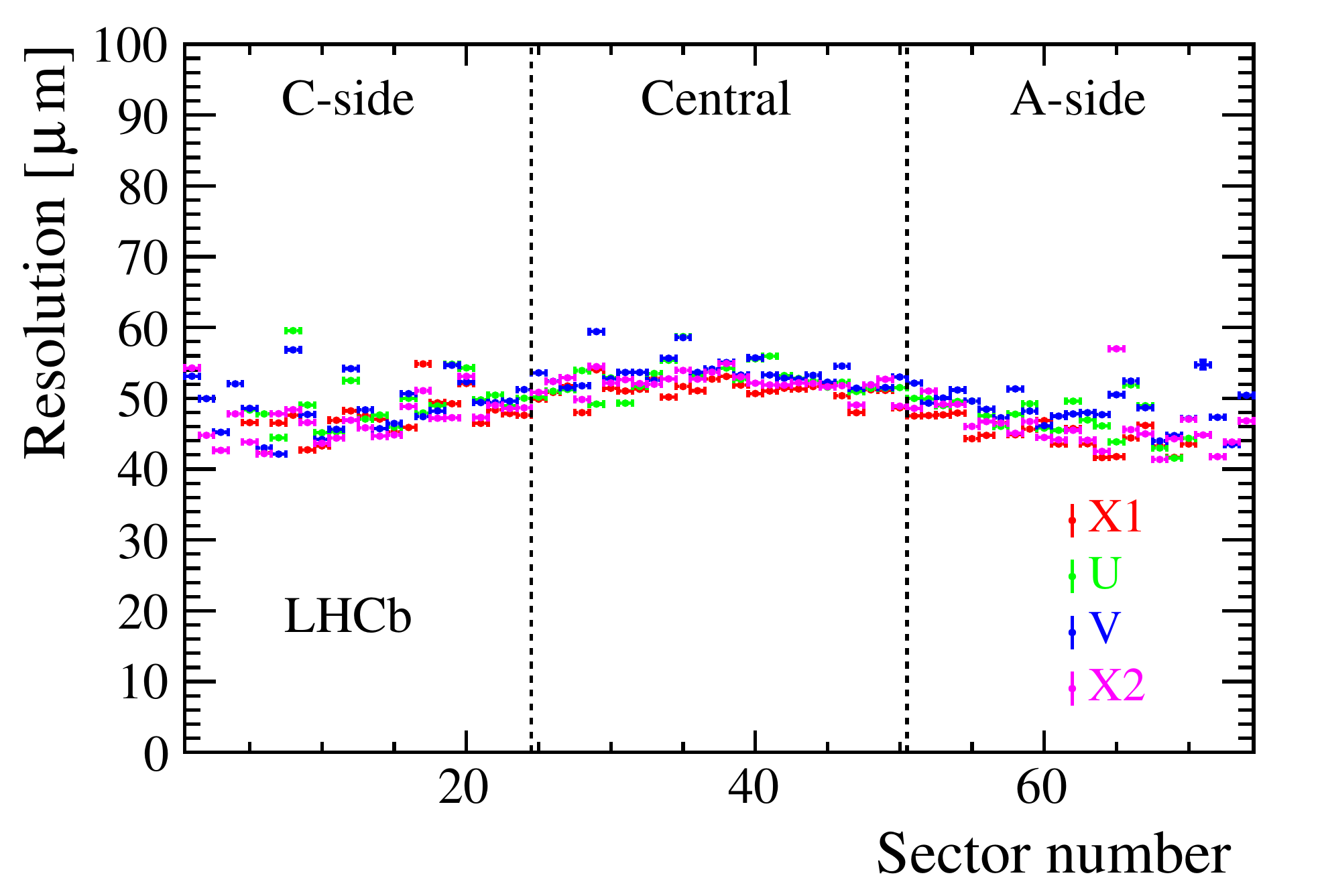}
  \caption{Hit resolution measured for all modules in the TT.  The sector number
  corresponds approximately to the $x$-direction. The resolution improves in the
  outer regions of the ``A-side'' and ``C-side'' regions where there is more
  charge sharing due to the larger track angle.  It is almost constant in the
  sectors in the ``Central'' region where the occupancy is highest.  The labels
  $X1$, $U$, $V$ and $X2$  correspond to the four detection layers arranged with
  an $(x-u-v-x)$ geometry in the TT box.}  
  \label{fig:ResolutionVsSectorTT}
\end{figure}

The intrinsic hit efficiency of the silicon sensors can be measured using
reconstructed tracks to probe whether or not the expected hits on a track are
found.  The efficiency is defined as the ratio between the number of hits found
and the number of hits expected for a given sector.  The measurement uses
daughter tracks from clean samples of {\decay{\jpsi}{\mumu}} decays.  The method
looks for hits in a window around the intersection point between a track and
each sensor on the track where a hit is expected.  The tracks are required to
have momentum greater than 10\gevc to reduce the effect of multiple scattering.
Additional cuts are placed on the track quality to minimise the effect of fake
tracks on the efficiency measurement.  The efficiency is calculated relative to
the number of working channels \ie~hits are not expected to be found when a
channel or group of channels is disabled.  The overall hit efficiency is
determined to be greater than 99.7\% and 99.8\% for TT and IT, respectively.

\begin{figure}[!tb] \centering\includegraphics[width=0.9\textwidth]{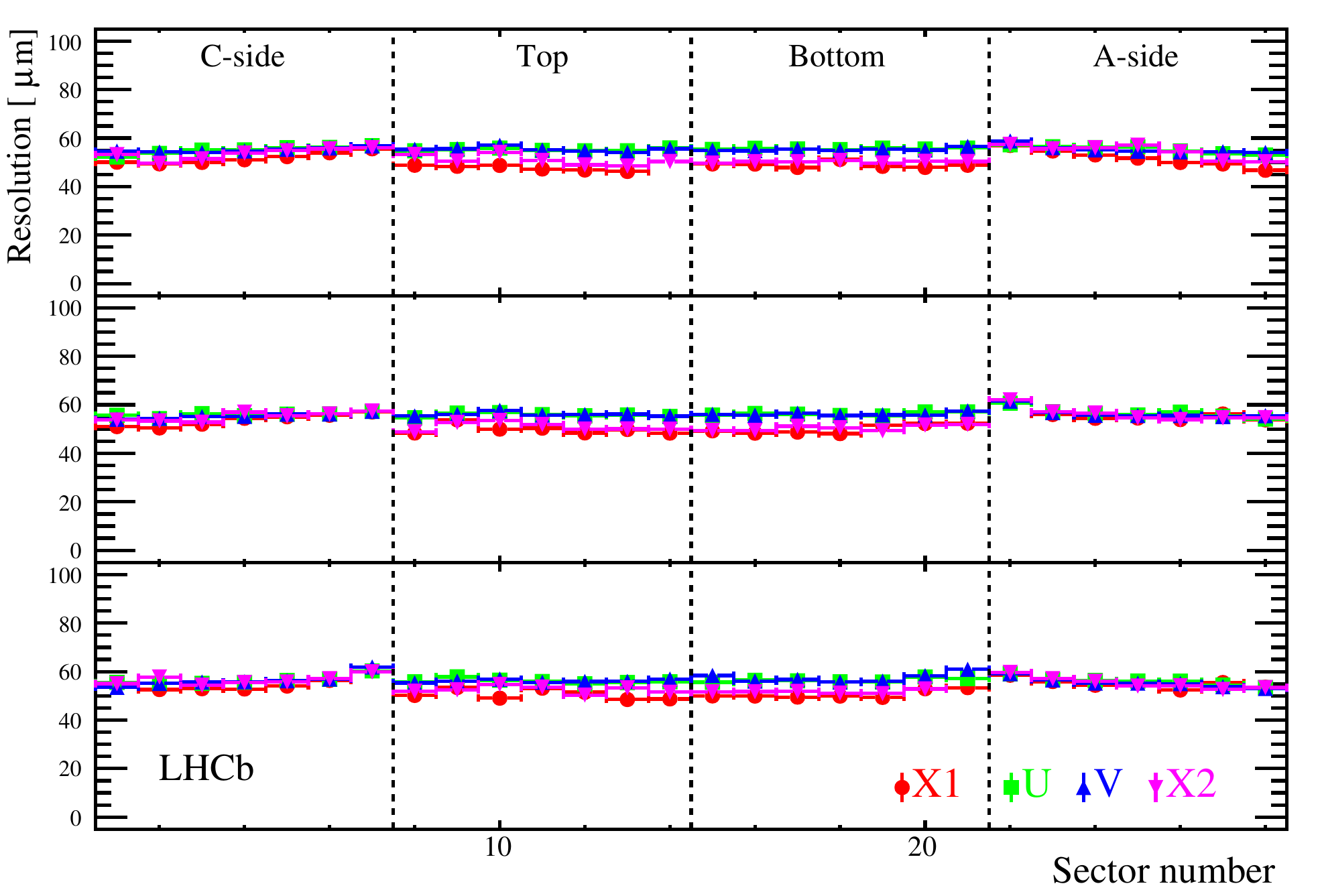}
  \caption{Hit resolution measured for modules in IT1 (bottom), IT2 (middle) and
    IT3 (top).  The sector number corresponds approximately to the
    $x$-direction.  The resolution in the 1-sensor sectors in the boxes above
    (Top) and below (Bottom) the beam-pipe are constant.  The resolution
    improves for the 2-sensor sectors in the A- and C-side boxes with increasing
    distance from beam-pipe where the track angle is typically larger. The
    labels $X1$, $U$, $V$ and $X2$ correspond to the four detection layers
    arranged with an $(x-u-v-x)$ geometry in each box.} 
  \label{fig:ResolutionVsSectorIT}
\end{figure}

The hit resolution is determined from the residuals between the measured hit
position and the extrapolated track position.    The unbiased residual is
calculated by removing the hit from the track fit and calculating the distance
between the hit and the extrapolated track position.   The resolution is given
by the spread of the unbiased residual distribution after correcting for the
uncertainty in the track parameters. The hit resolution measured using the 2011
data is 52.6\mum for the TT and 50.3\mum for the IT. The resolution measured
using the 2012 data is shown as a function of the sector number in
Figure~\ref{fig:ResolutionVsSectorTT} for the four TT layers and in
Figure~\ref{fig:ResolutionVsSectorIT} for the IT.  The resolution is worse in
the central regions closest to the beam-pipe where the track angles are smallest
and, consequently, where there is the least amount of charge sharing between
strips.

\begin{table}[!tb]
\centering
\tbl{Summary of the hit efficiency and resolution measurements made using 2011
and 2012 data.  Results are also shown for simulated events.} 
{\begin{tabular}{c l c c c c}
\hline
 Detector & Measurement & 2011 Data & 2012 Data & 2011 MC & 2012 MC \\
 \hline 
 \multirow{2}{*}{TT} & Hit efficiency & 99.7\% & 99.8\% & 99.9\% & 99.9\% \\
  & Hit resolution & 52.6\mum & 53.4\mum & 47.8\mum & 48.0\mum \\
  \hline
 \multirow{2}{*}{IT} & Hit efficiency & 99.8\% & 99.9\% & 99.9\% & 99.9\% \\
 & Hit resolution & 50.3\mum & 54.9\mum & 53.8\mum & 53.9\mum \\
 \hline
 \end{tabular}
\label{tab:ST_summary}}
\end{table}

The measurements of the hit efficiency and the resolution are summarised in
Table~\ref{tab:ST_summary} for the 2011 and 2012 data taking periods.  The
results are compared with the expectation from simulations for 2011 and 2012
data taking conditions, respectively.  The measured hit resolutions are in
agreement with those expected from simulation.  The small differences observed
can be partially explained by the remaining misalignment of the modules. The
measured hit efficiency is well above 99\% in all cases.

The particle density falls significantly as the distance from the beam-pipe is
increased.  The occupancy in each of the read-out sectors was estimated using a
data sample containing events randomly selected after the Level-0 trigger with
$\mu=1.7$.  The average occupancy in the TT varies between 1.9\%  for the
sectors closest to be beam-pipe compared to 0.2\% for the outermost modules.
Similarly, the average occupancy was found to vary between 1.9\% and 0.2\% for
the IT sectors.

\subsubsection{Outer Tracker}
\label{sec:tracking:ot}

\begin{figure}[!tb]
\vspace{-1mm}
  \centering
  \begin{picture}(350,150)(0,0)
     \put(  0,10){\includegraphics[scale=0.32]{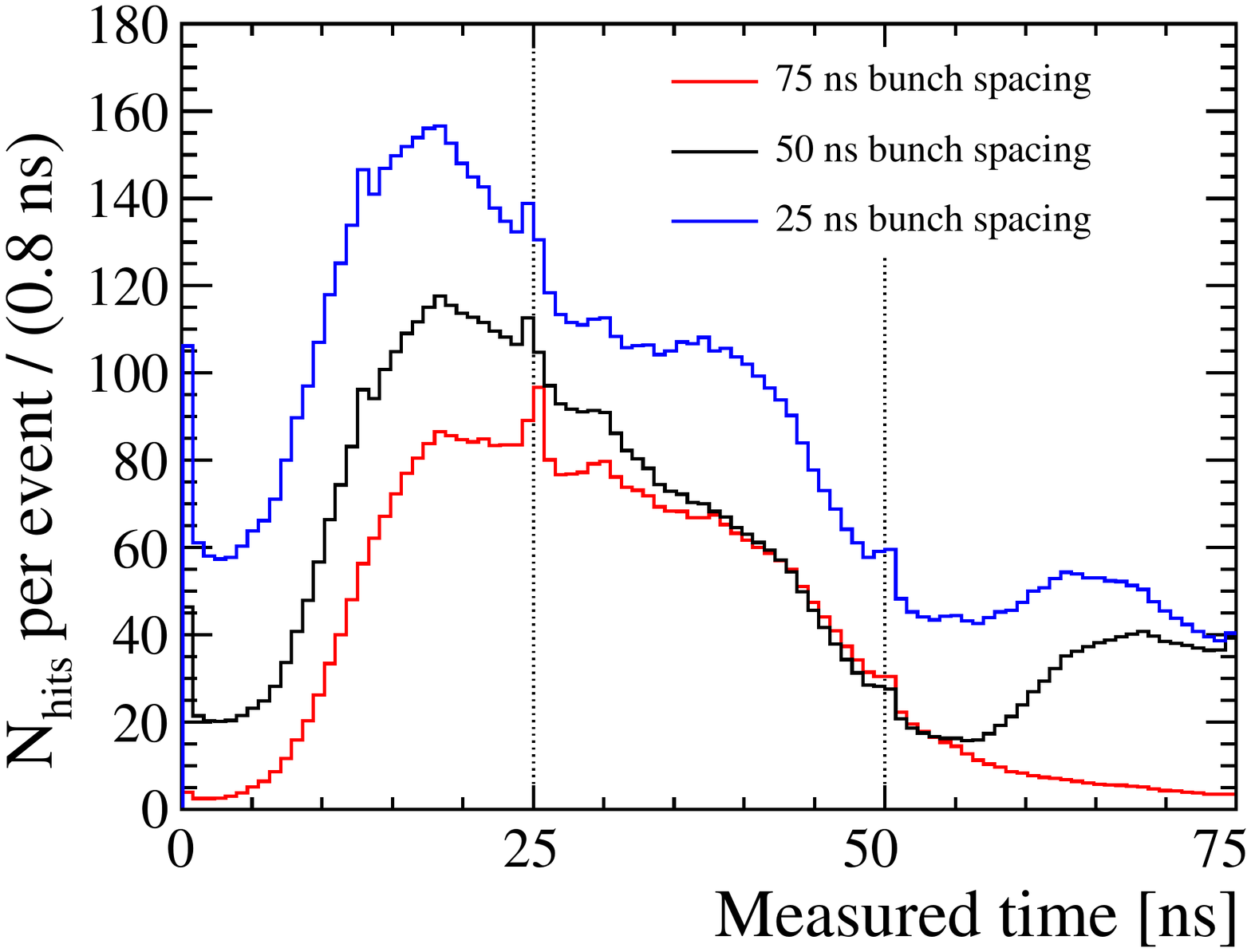}}
     \put(190,15){\includegraphics[scale=0.32]{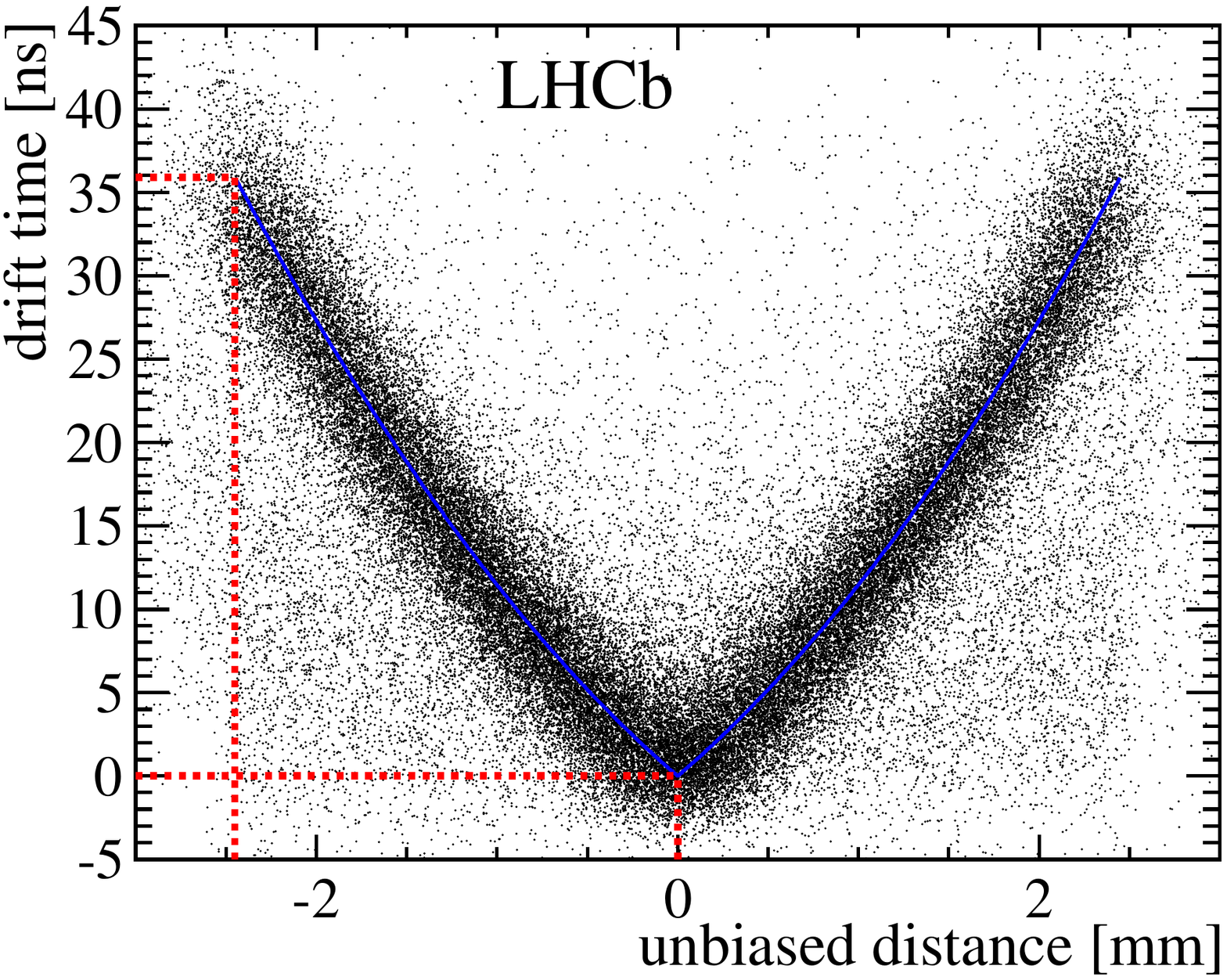}}
     \put(30,130){LHCb}
  \end{picture}      
\caption{Drift time distribution (left) for the modules located closest to the
  beam (``M8''). Drift time versus distance relation (right) where the
  red-dotted lines indicate the centre and the edge of the straw, corresponding
  to drift times of 0 and 36\,ns, respectively~\protect\cite{LHCb-DP-2013-003}.}
\label{fig:ot:drifttime}
\vspace{-1mm}
\end{figure}

The outer parts of the tracking stations T1--T3 are equipped with a straw-tube
detector (OT)~\cite{LHCb-DP-2013-003}. Charged particles traversing the straw-tubes
will ionise the gas along their trajectory. The drift-times of the ionisation
electrons to the wire located at the centre of the straw  are measured with
respect to the beam crossing signal. The distribution of the recorded
drift-time,  which is proportional to the distance of the particle trajectory to
the wire, is shown in Figure~\ref{fig:ot:drifttime}~(right). The calibration of
the drift-time to distance  relation~\cite{LHCb-DP-2013-003} has been done on data.  

The maximum drift time in the straw tubes is about 35\,ns, but to account for
variations in the time-of-flight of the particles, the signal propagation time
through the wire, and variations in time offset constants in the electronics,
three bunch crossings are read out upon a positive L0 trigger on the first bunch
crossing, corresponding to a time window of 75\,ns.   

\begin{figure}[!tb]
\vspace{-3mm}
  \centering
  \setlength{\unitlength}{0.1\textwidth}
  \begin{picture}(10,4)
    \put(0,0){\includegraphics[bb=150 180 740 430, width=\textwidth]{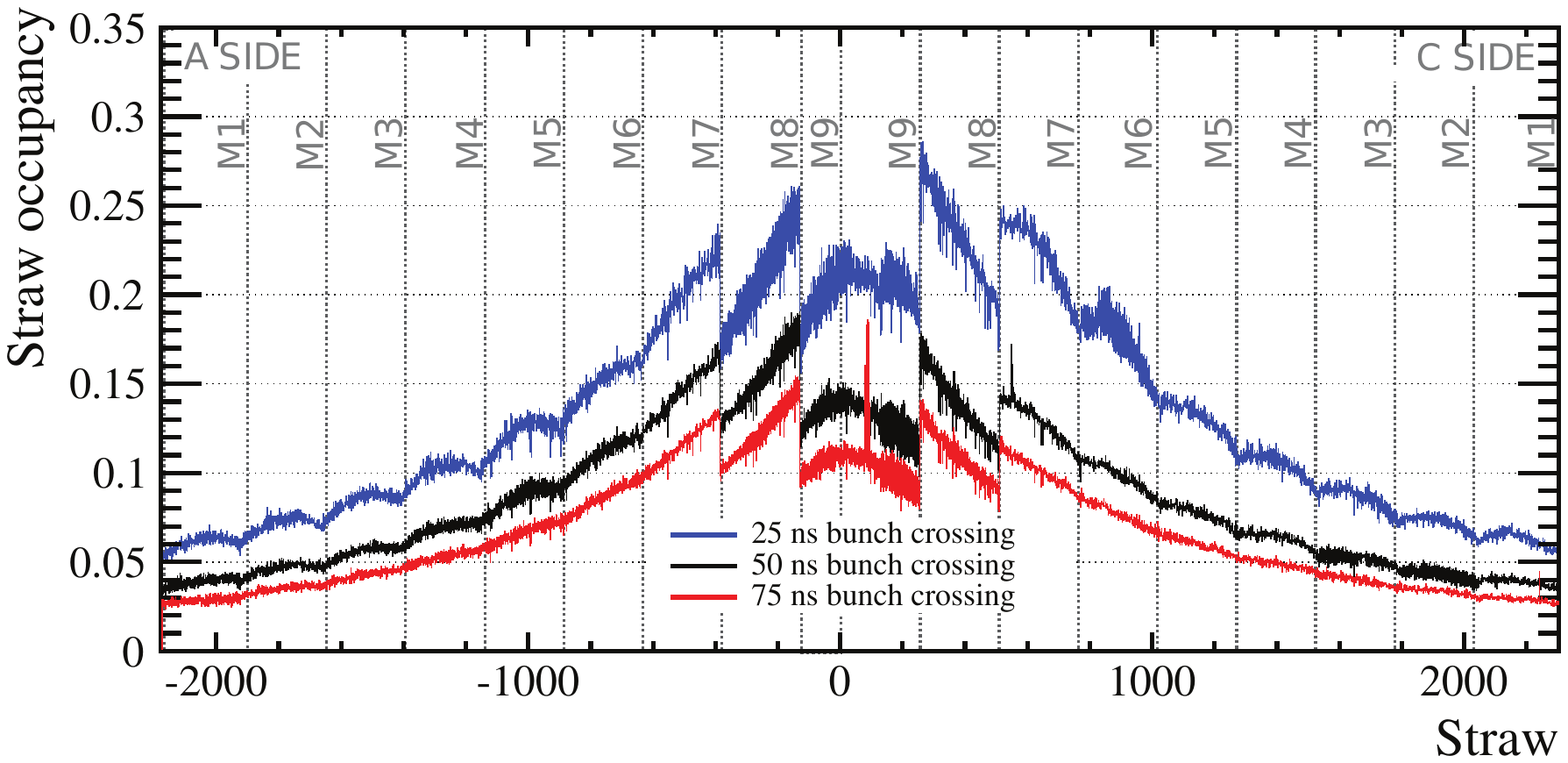}}
    \put(2.15,3.65){LHCb}
  \end{picture}
\caption{Straw occupancy for (red) $75\,\mathrm{ns}$, (black) $50\,\mathrm{ns}$
  and (blue) $25\,\mathrm{ns}$ bunch-crossing spacing, for comparable pile-up
  conditions~\protect\cite{LHCb-DP-2013-003}. The modules are indicated by 'M',
  and contain 256 straws each. The width of the module is 340\,mm.} 
\label{fig:ot:ocupmod}
\vspace{-3mm}
\end{figure}

During Run I, the LHC was operating predominantly in either 75\,ns or 50\,ns bunch
spacing schemes. A short running period in 2012 with 25\,ns bunch spacing was
also performed, allowing a study of the detector performance under these
conditions to be undertaken. The contribution from earlier and later bunch
crossings is visible in the drift time spectrum, see Figure~\ref{fig:ot:drifttime}.
These additional hits from different bunch crossings 
increase the occupancy as shown in Figure~\ref{fig:ot:ocupmod}.
The occupancy for the most central modules is reduced with respect to the
neighbouring modules, as these modules are located further away from the beam,
in the vertical direction. The straws with highest average occupancy for typical
running conditions in 2011 (\ie~50\,ns bunch spacing conditions and about 1.4
visible overlapping events) amounts to about 17\%. This increases to about 25\%
for 25\,ns bunch spacing conditions with on average 1.2 overlapping events.
The average pile-up conditions in 2012 were slightly different, corresponding to
about 1.8 visible overlapping events, resulting in a higher multiplicity
compared to 2011. 

\begin{figure}[!tb]
 \centering\begin{picture}(350,180)(0,0)
  \put(15,10){\includegraphics[scale=0.5]{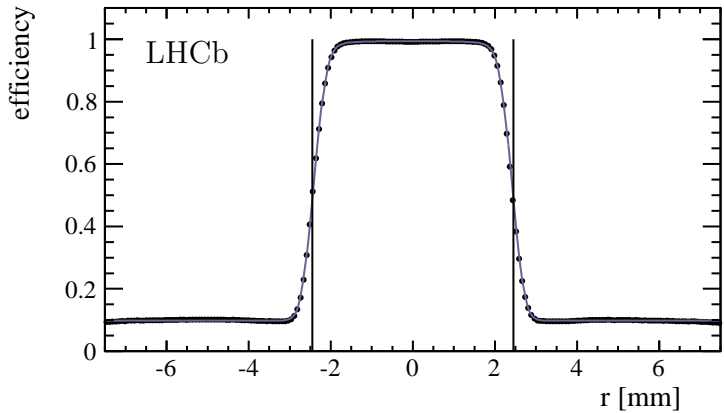}}
  \put(70,145){LHCb}
 \end{picture}	
\caption{Example of the OT efficiency profile as a function of the distance
  between the extrapolated track position and the centre of the straw for hits
  in the detector modules on either side of the beam-pipe (type
  M7)~\protect\cite{LHCb-DP-2013-003}. The vertical bars represent the edges of
  the straw cell.} 
\label{fig:ot:cellprofile}
\end{figure}

A scan of the hit efficiency as a function of the predicted distance between the
expected hit and the centre of the considered straw is performed on 2011
and 2012 data. An efficiency profile of the detector single cell is thus
obtained, an example of which is shown in Figure~\ref{fig:ot:cellprofile}.  The
efficiency drops close to the cell edges due to two effects. First, the
probability for ionisation to occur decreases for shorter path lengths inside
the straw.  Secondly, a fraction of the hits are positioned outside the straw
volume due to the uncertainty on the track extrapolation.
The average single cell efficiency for tracks in the central half of the 
straw, closer than 1.25\,mm to the wire, amounts to 99.2\%.  
Radiation damage could in principle lead to a decrease in signal amplitude.
This was monitored during the 2011 and 2012 running periods
and no degradation is observed~\cite{LHCb-DP-2012-001}.

The single hit resolution is determined by comparing the
predicted hit position from the track with the hit position
obtained from the drift-time. The hit under study is not used in the
reconstruction of the extrapolated 
track, in order not to bias the resolution determination.
The resulting single hit resolution is 205\mum, close to the design value of 200\mum.
Only tracks with a momentum larger than 10\gevc are used, and the 
residual is corrected for the uncertainty in the track parameters, caused by
effects such as multiple scattering.

\subsubsection{Muon system}
\label{sec:tracking:muon}

\begin{figure}[tb]
  \begin{center}
    \includegraphics[width = 0.95\textwidth]{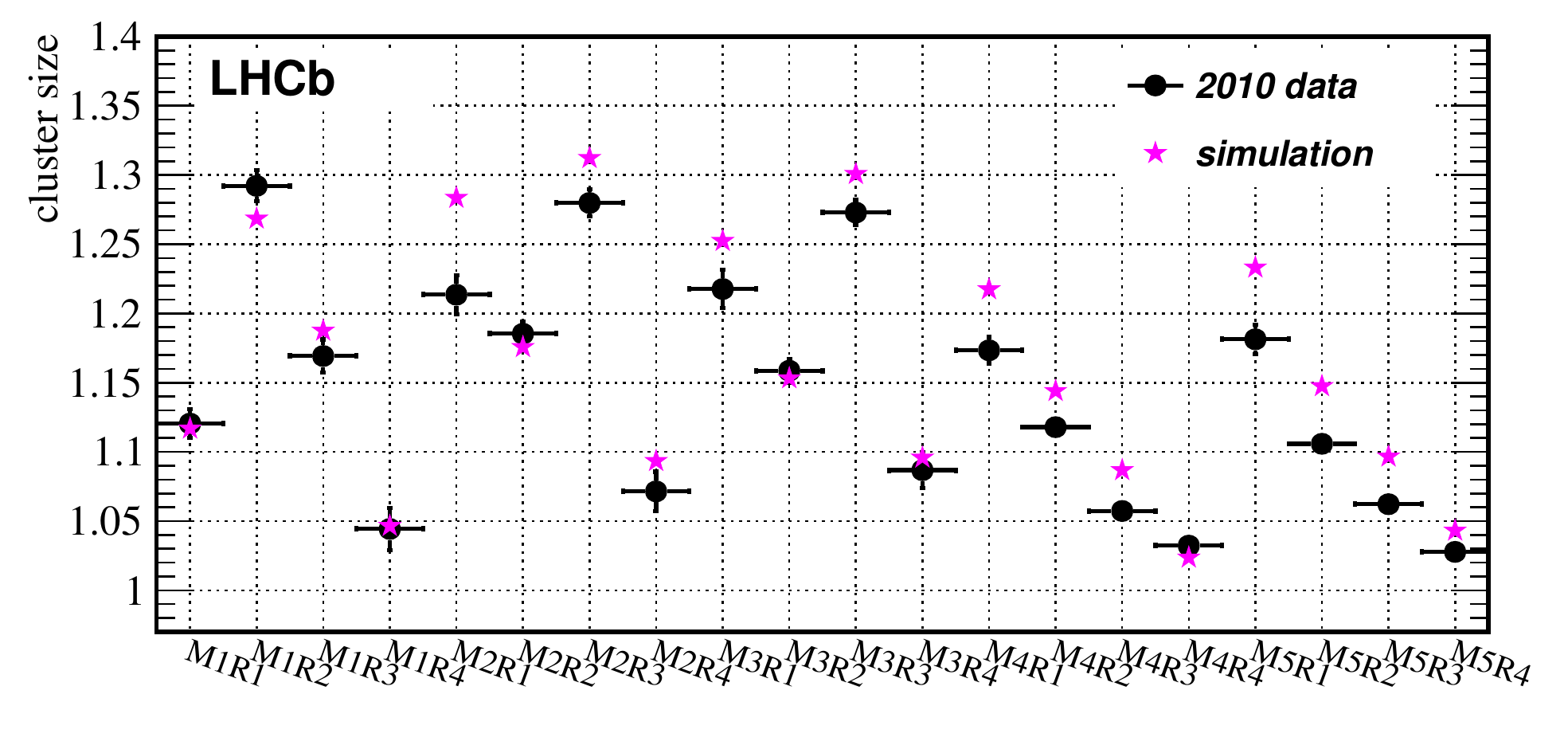}
    \caption{Average cluster size in each detector region for
      data and simulation~\protect\cite{LHCb-DP-2012-002}.  
      The labels refer to the stations (M1 to M5) and to the
      four regions with different granularity used in each station (from the
      innermost R1 to the outermost R4) . Only isolated muon tracks are used,
    and angular effects are corrected for.}
    \label{fig:muon_clsize}
  \end{center}
\end{figure}

To discriminate muons against the abundant hadronic background,
muon candidates are formed from aligned hits in each of the five
stations. Since \lhcb aims at a trigger efficiency for muons larger than 95\%,
the average efficiency of each muon station must  exceed 99\%.
To meet this stringent requirement, a redundant design was chosen for the muon
chambers, consisting of four active layers per chamber for the M2--M5 stations,
and two for the M1 station~\cite{Alves:2008zz}. Chambers are operated with a gas
gain providing a signal detection efficiency, for the logical OR of the different
layers, well above the required 99\%. A conflicting demand, also
dictated by  the L0 trigger algorithm, is to minimise
cross-talk between channels. The cluster size of muon
track hits, defined as the average number of adjacent pads fired by an
isolated muon track, is measured using 2010 data. The result depends on
the station (M1 to M5) and region (from the innermost R1 to the
outermost R4), since twenty chamber types of different size and
granularity are used. As shown in Figure~\ref{fig:muon_clsize}, the cluster size
values observed in the data are in reasonable agreement with the simulation and
are sufficient to meet the L0 trigger requirements~\cite{Aslanides:2002bpa}.

Due to the redundant design, all of the 1380 muon chambers were continuously
operating during the whole data taking.  
The few cases of a broken MWPC or GEM detector layer only caused a locally
limited reduction of efficiency.  
Dead detector channels were only due to faulty components in the
read-out chain, and never affected more than 0.2\% of the total detector surface.
Their effect on the muon trigger efficiency is estimated to be less than 1\%.
The other main sources of inefficiencies are incorrect time reconstruction
and dead-time of the read-out electronics.

Since the signals must be detected within the 25\,ns LHC time gate around a
bunch crossing, the detector time resolution is required to be smaller than
4.5\,ns. The 122,112 physical channels were aligned in time with an accuracy of
1\,ns using samples of cosmic rays~\cite{LHCb-DP-2011-001} 
and tracks from the first $pp$ collisions in the detector~\cite{LHCb-DP-2012-002}.
The timing performance is measured from special calibration runs
where events triggered by the calorimeters were acquired in a 125\,ns wide
gate around the triggered collision. A high-purity sample of muon
candidates is obtained by reconstructing track segments from aligned muon
detector hits in all of the five stations, and matching such segments
with high-quality tracks reconstructed by the tracking detectors. 
The time resolution of muon detector hits associated to
these tracks is measured to be between 2.5 and 4\,ns, depending on the
detector region~\cite{LHCb-DP-2012-002}. The inefficiency due to tails in the time
measurement is estimated by counting the fraction of muon tracks
having one or more hits outside the 25\,ns time gate.
In each of the time alignment runs acquired during the data taking, such
inefficiency is found to never exceed 1.2\%.

The total hit efficiency of the muon chambers is measured using muon
candidates in events triggered independently of the muon detector during the
normal data taking.
The efficiency for each station is estimated by searching hits around
the position predicted by the segment reconstructed using only the
other four stations, which must have a good matching with a high-quality track.
For station M1, which is located upstream of the
calorimeter system, candidate muons are also required to originate from a \jpsi
decay. Tracks close to the known dead channels
are removed from the sample. The contribution of background hits
accidentally matching the candidate track is subtracted using a
statistical model.

\begin{figure}[tb]
  \begin{center}
    \includegraphics[width = 0.92\textwidth]{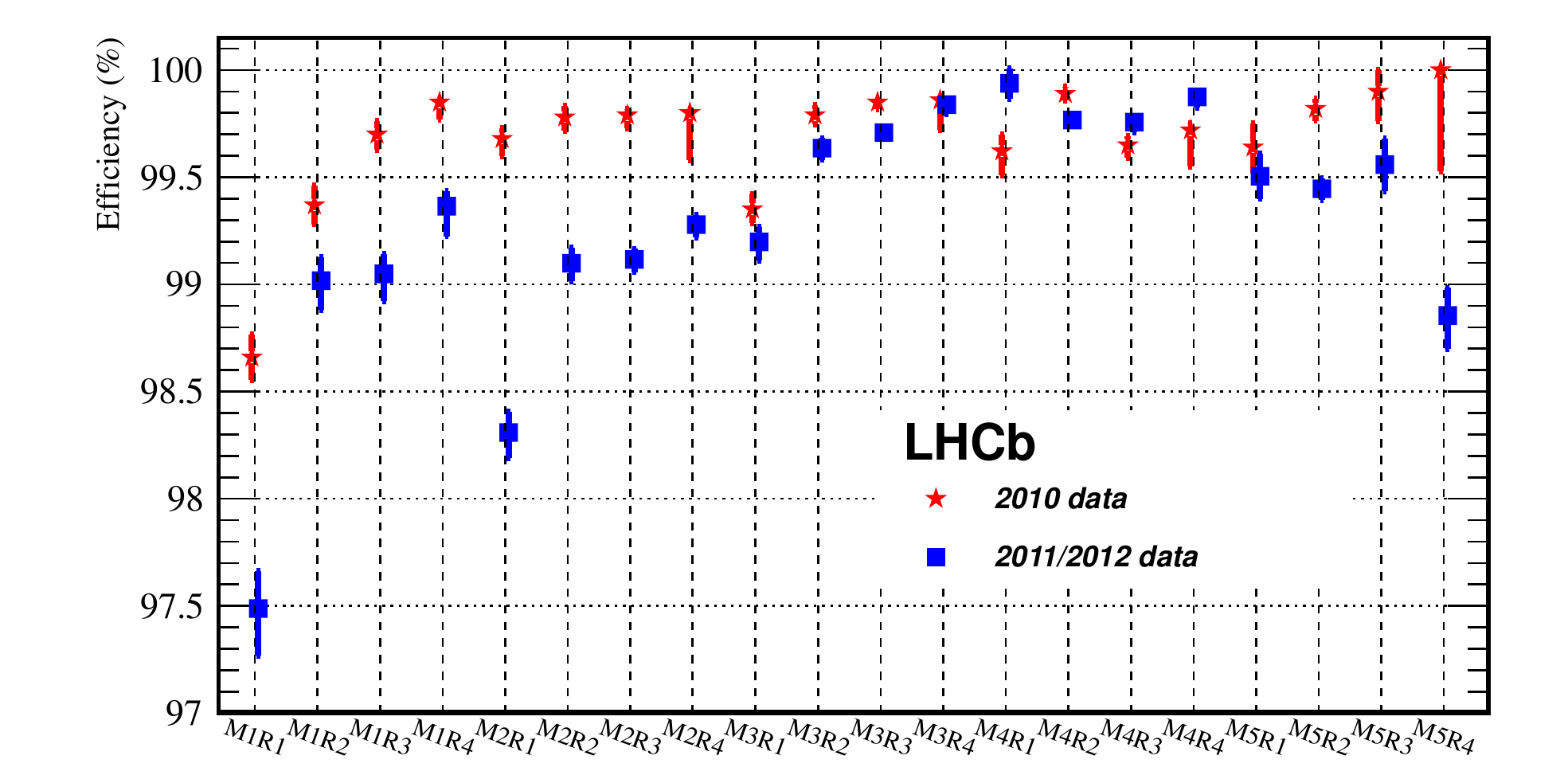}
    \caption{Average measured hit efficiency, in percent, for the different
    regions of the muon detector. Statistic and systematic uncertainties are
    added in quadrature. The effect of the few known dead channels is not
    included. Measurement in the 2010 and 2011/2012 data taking periods are
    shown separately due to different pile-up conditions.}
    \label{fig:muon_eff}
  \end{center}
\end{figure}

The resulting efficiencies, measured separately for the twenty chamber
types, and for the 2010 and 2011-2012 data taking conditions, are shown in
Figure~\ref{fig:muon_eff}. The 2010 values are compatible with the
inefficiencies due to incorrect time reconstruction~\cite{LHCb-DP-2012-002}.
The larger inefficiency observed in 2011 and 2012 is due to
the different beam conditions, with 50\,ns bunch spacing and higher
luminosity, causing a non-negligible  dead-time of the read-out chain.

The dead-time of the front-end read-out chips varies
from 50 to 100\,ns, depending on the region and on the signal
amplitude. This affects in particular the inner regions
having the highest channel occupancy, reaching average values
of 2.5\% in M1R1 and  0.6\% in M2R1 for the 2012 data taking.
A second source of dead time is the finite length of the digital output signals,
18 to 25\,ns, depending on the region and the data taking period.
In order to reduce the  number of off-detector read-out
channels, these signals are formed  from the logical OR of several
contiguous physical channels. The occupancy of these logical channels is thus
larger than the occupancy of physical channels, and can lead to measurable
dead-time effects, even in the outer detector regions. This happens in particular
for station M5, which is affected by spurious hits due to
back-scattering from the beam-line elements located behind the detector.
Since the detector was operated at twice the nominal luminosity of  $2\times
10^{32}\mathrm{\,cm^{-2}s^{-1}}$, the dead-time effect is larger than
originally expected. Nonetheless, most regions meet the 99\% efficiency
requirement.
Taking into account the combined response of the five stations, the detector
is found to provide muon identification for trigger and offline
reconstruction with an efficiency larger than 95\%.

\subsection{Track reconstruction}
\label{sec:tracking}

The trajectories of the charged particles traversing the tracking system are
reconstructed from hits in the \velo, TT, IT and OT detectors.
Depending on their paths through the spectrometer, the following track types
are defined, as illustrated in Figure~\ref{fig:tracktypes}:

\begin{figure}[!tb]
\begin{center}
\includegraphics[width=0.7\textwidth]{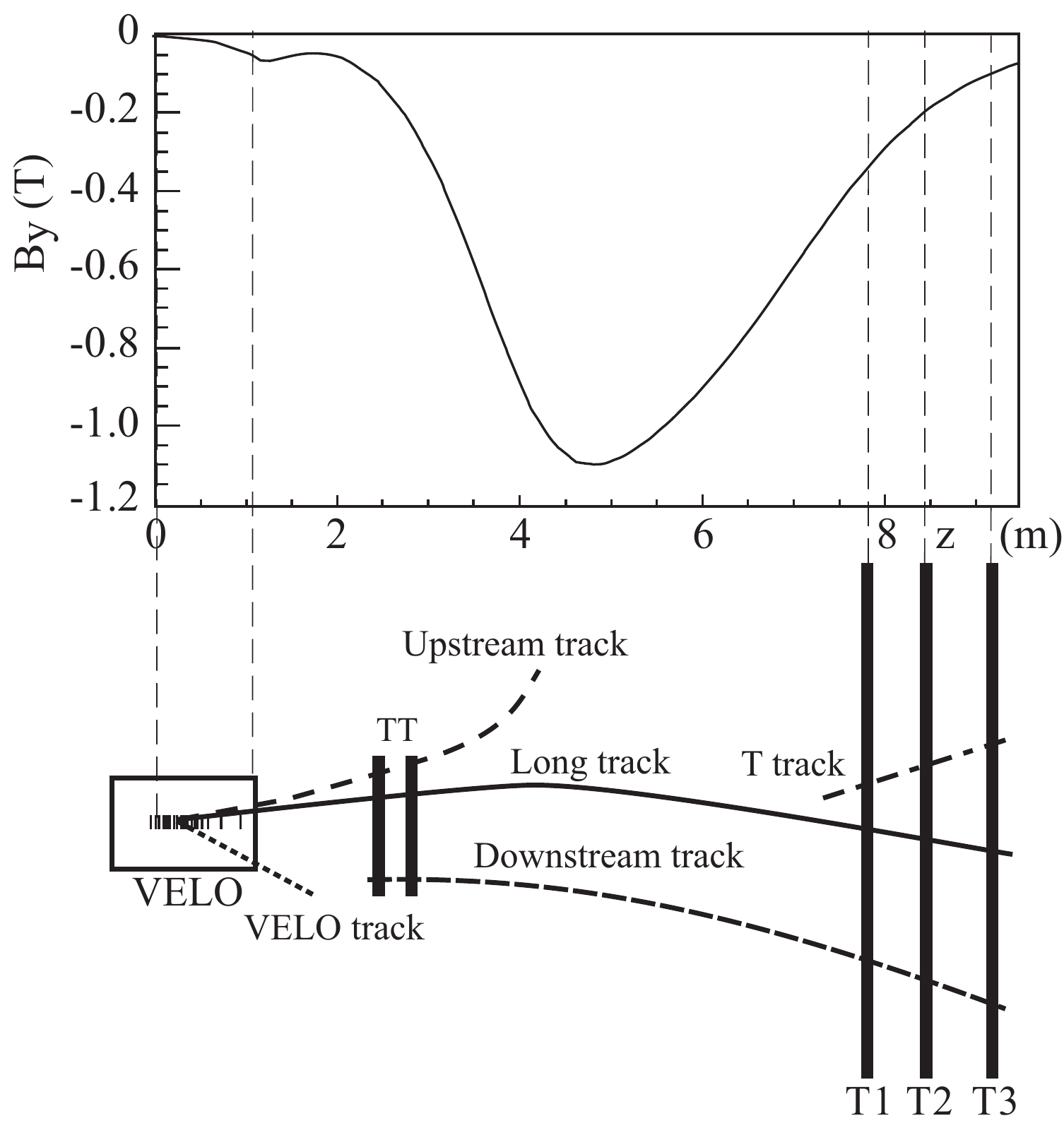}
\caption{A schematic illustration of the various track types
  \protect\cite{Alves:2008zz}:  long, upstream, downstream, \velo and T
  tracks. For reference the main $B$-field component ($B_y$) is plotted above as
  a function of the $z$ coordinate.}
\label{fig:tracktypes}
\end{center}
\end{figure}

\begin{itemize}
  
\item {\bf Long tracks} traverse the full tracking system. They have hits in
both the \velo and the T stations, and optionally in TT. As they traverse
the full magnetic field they have the most precise momentum estimate and
therefore are the most important set of tracks for physics analyses.

\item {\bf Upstream tracks} pass only through the \velo and TT stations. In
general their momentum is too low to traverse the magnet and reach the T
stations. However, they pass through the \richone detector and may generate
Cherenkov photons if they have $p>1\gevc$. They are therefore also used to understand
backgrounds in the particle identification algorithm of the \rich.

\item {\bf Downstream tracks} pass only through the TT and T stations. They
  are important for the reconstruction of long lived particles, such as \KS and
  \L, that decay outside the \velo acceptance.

\item {\bf \velo tracks} pass only through the \velo and are typically
large-angle or backward tracks, which are useful for the primary vertex
reconstruction.

\item {\bf T tracks} pass only through the T stations. They are
typically produced in secondary interactions, but are still useful during the
treatment of \richtwo data for particle identification.

\end{itemize}

The long track reconstruction starts with a search in the \velo for straight
line trajectories\cite{LHCb-PUB-2011-001,LHCb-2007-013}. To be reconstructed as
\velo tracks, traversing particles must provide at least three hits in the
\RSens sensors  and three hits in the \PhiSens sensors. Then, there are two
complementary algorithms to add information from the downstream tracking
stations to these \velo tracks. In the first algorithm, the forward
tracking\cite{LHCb-2007-015}, the \velo tracks are combined with information
from the T stations. The momentum of a particle and its trajectory through the
detector are fully determined from the information provided by the \velo and a
single T station hit. Further hits in the T stations are then searched along
this trajectory to find the best possible combination of hits defining the long
track. In the second algorithm, called track
matching\cite{LHCb-2007-020,LHCb-2007-129}, the VELO tracks are combined with
track segments found after the magnet in the T stations, using a standalone
track finding algorithm\cite{LHCb-2008-042}. In order to form such a track
segment, particles traversing the T stations need to provide at least one hit in
the $x$ layers and one in the stereo layers in each of the three stations. The
candidate tracks found by each algorithm are then combined, removing duplicates,
to form the final set of long tracks used for analysis. Finally, hits in the TT
consistent with the extrapolated trajectories of each track are added to improve
their momentum determination. 

Downstream tracks are found starting with T tracks, extrapolating them through
the magnetic field and searching for corresponding hits in the
TT\cite{LHCb-2007-026,sascha}. 
Upstream tracks are found by extrapolating \velo
tracks to the TT where matching hits are then added in a procedure similar
to that used by the downstream tracking. At least three TT hits are required to
be present by these algorithms\cite{LHCb-2007-010}.

\begin{figure}[!tb]
\begin{center}
\includegraphics[width=0.68\textwidth]{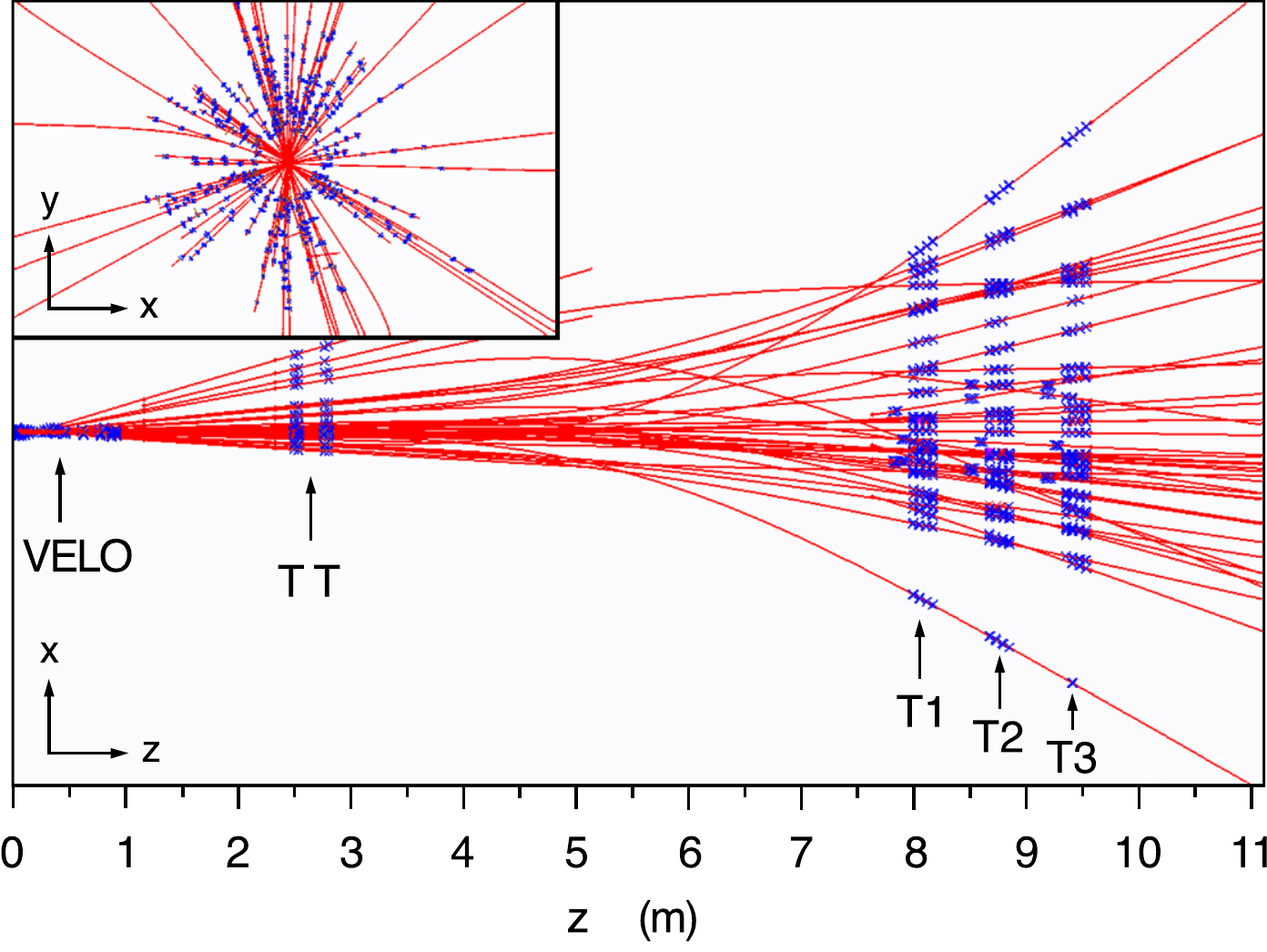}
\caption{Display of the reconstructed tracks and assigned hits in an event in
  the $x$-$z$ plane \protect\cite{Alves:2008zz}.
  The insert shows a zoom into the \velo region in the $x$-$y$ plane.}
\label{fig:trackevent}
\end{center}
\end{figure}

In a final step, the tracks are fitted using a Kalman
filter~\cite{fruhwirth,vanTilburg:2005ut}. The fit takes into account multiple
scattering and corrects for energy loss due to ionisation. The $\chi^2$ per
degree of freedom of the fit is used to determined the quality of the
reconstructed track. After the fit, the reconstructed track is represented by
state vectors ($x$, $y$, $dx/dz$, $dy/dz$, $q/p$) which are specified at given
$z$-positions in the experiment.  If two or more tracks have many hits in
common, only the one with most hits is kept. Figure~\ref{fig:trackevent} shows
the tracks reconstructed in a typical event.

Mis-reconstructed (fake) tracks are those that do not correspond to the
trajectory of a real charged particle. Due to the large extrapolation distance
in traversing the magnet, most of these fake tracks originate from wrong
associations between \velo tracks and tracks in the T stations. The fraction of
fake tracks in minimum bias events is typically around $6.5\%$, increasing to
about $20\%$ for large multiplicity events\cite{LHCb-PAPER-2013-070}.
This fake rate is significantly reduced, at the cost of a small drop in
efficiency, with a neural network classifier which uses as input the result of
the track fit, the track kinematics and the number of measured hits in the
tracking stations versus the number of expected hits.

\subsubsection{Track finding efficiency}

The tracking efficiency is defined here as the probability that
the trajectory of a charged particle that has passed through the full tracking
system is reconstructed. In particular it does not account for interactions with
the material, decays in flight and particles that fly outside of the detector
acceptance.

The efficiency is measured using a tag-and-probe technique with
$\jpsi\to\mup\mun$ decays.  In this method one of the daughter particles, the
``tag'' leg, is fully reconstructed, while the other particle, the ``probe''
leg, is only partially reconstructed. The probe leg should carry enough momentum
information such that the \jpsi invariant mass can be reconstructed with a
sufficiently high resolution. The tracking efficiency is then obtained by
matching the partially reconstructed probe leg to a fully reconstructed long
track. If a match is found the probe leg is defined as efficient. In the trigger
and offline selection of the $\jpsi$ candidates, no requirements are set on the
particle used for the probe leg to avoid biases on the measured efficiency.

\begin{figure}[!tb]
  \begin{center}
    \includegraphics[width=0.49\textwidth]{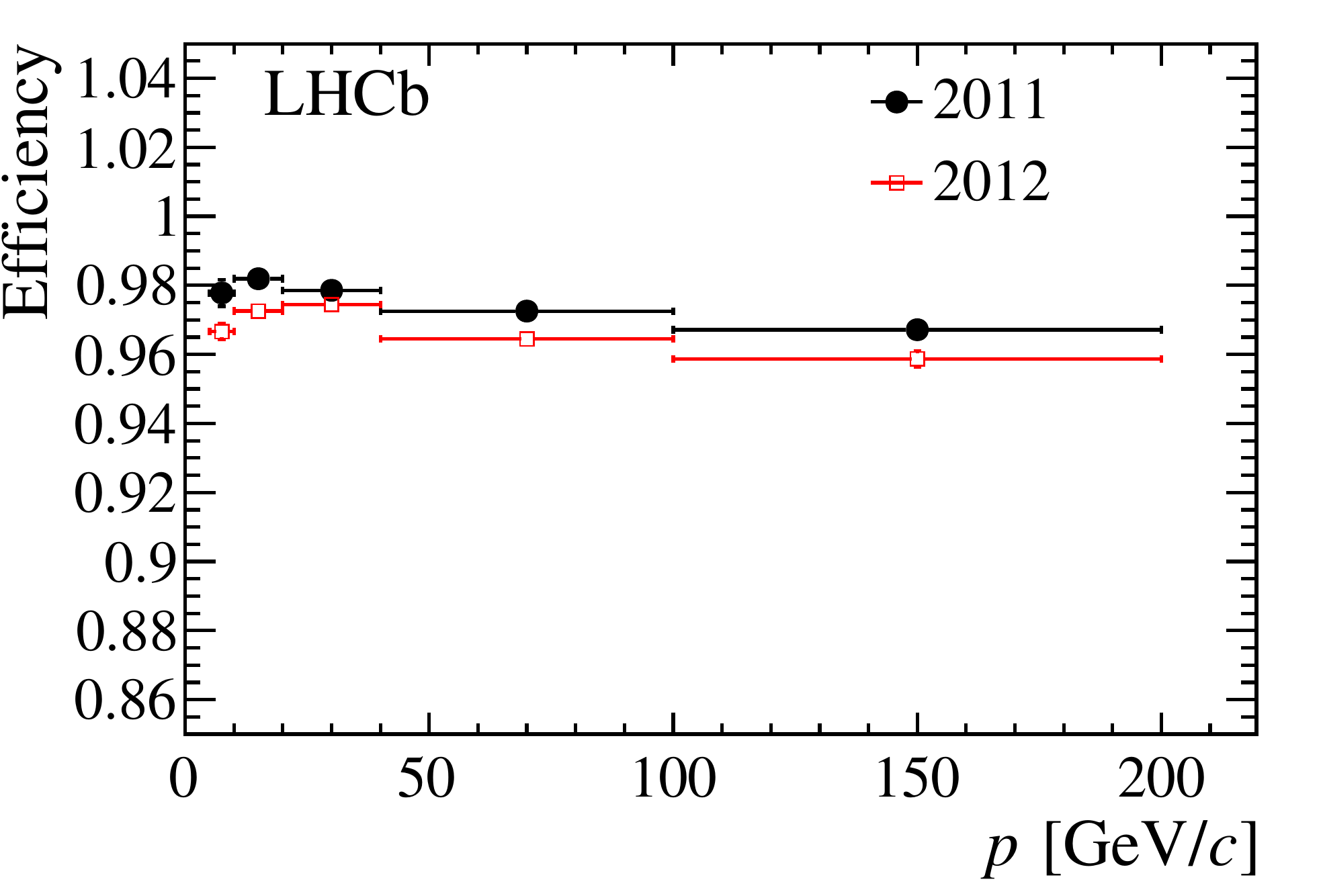}
    \includegraphics[width=0.49\textwidth]{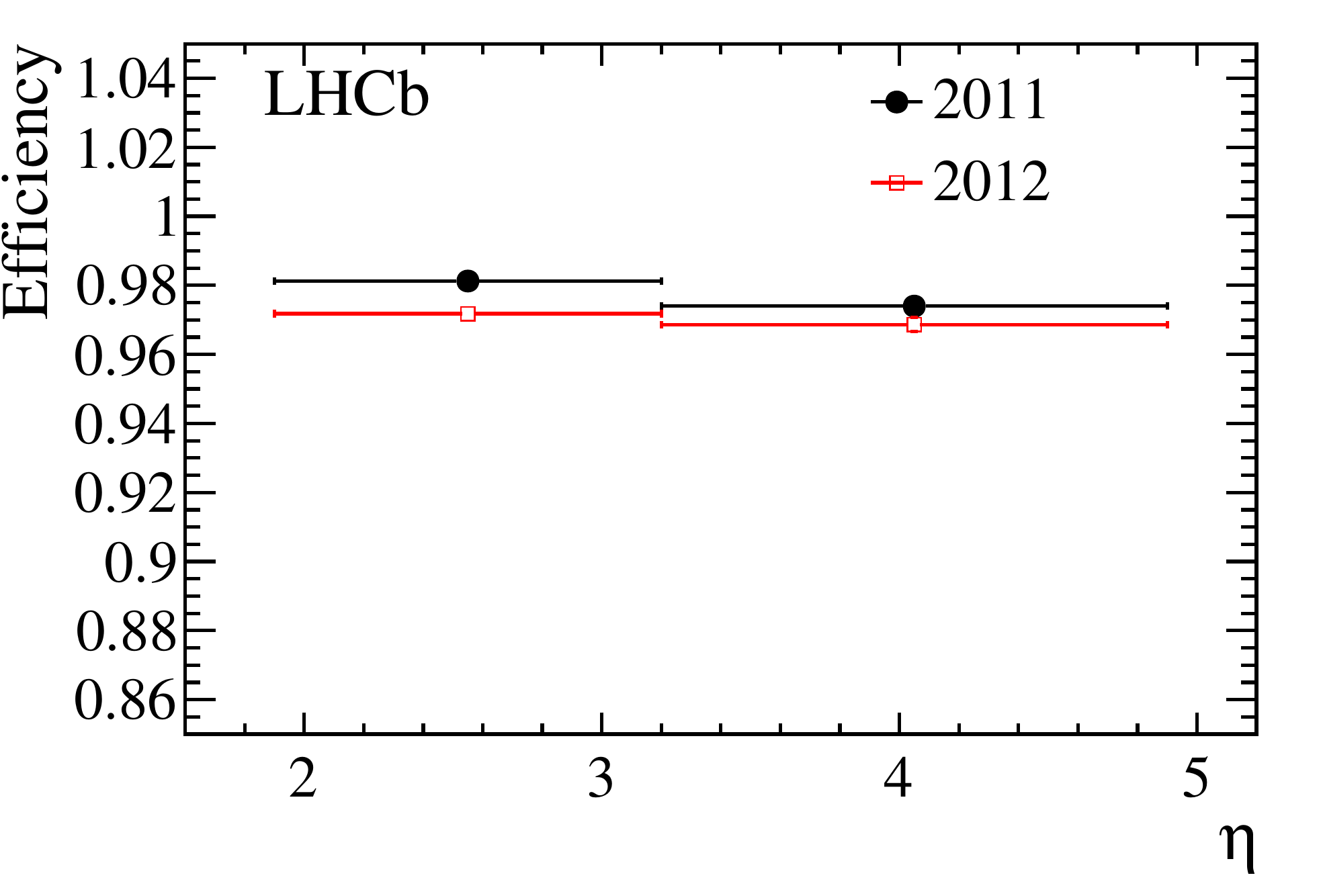}
    \includegraphics[width=0.49\textwidth]{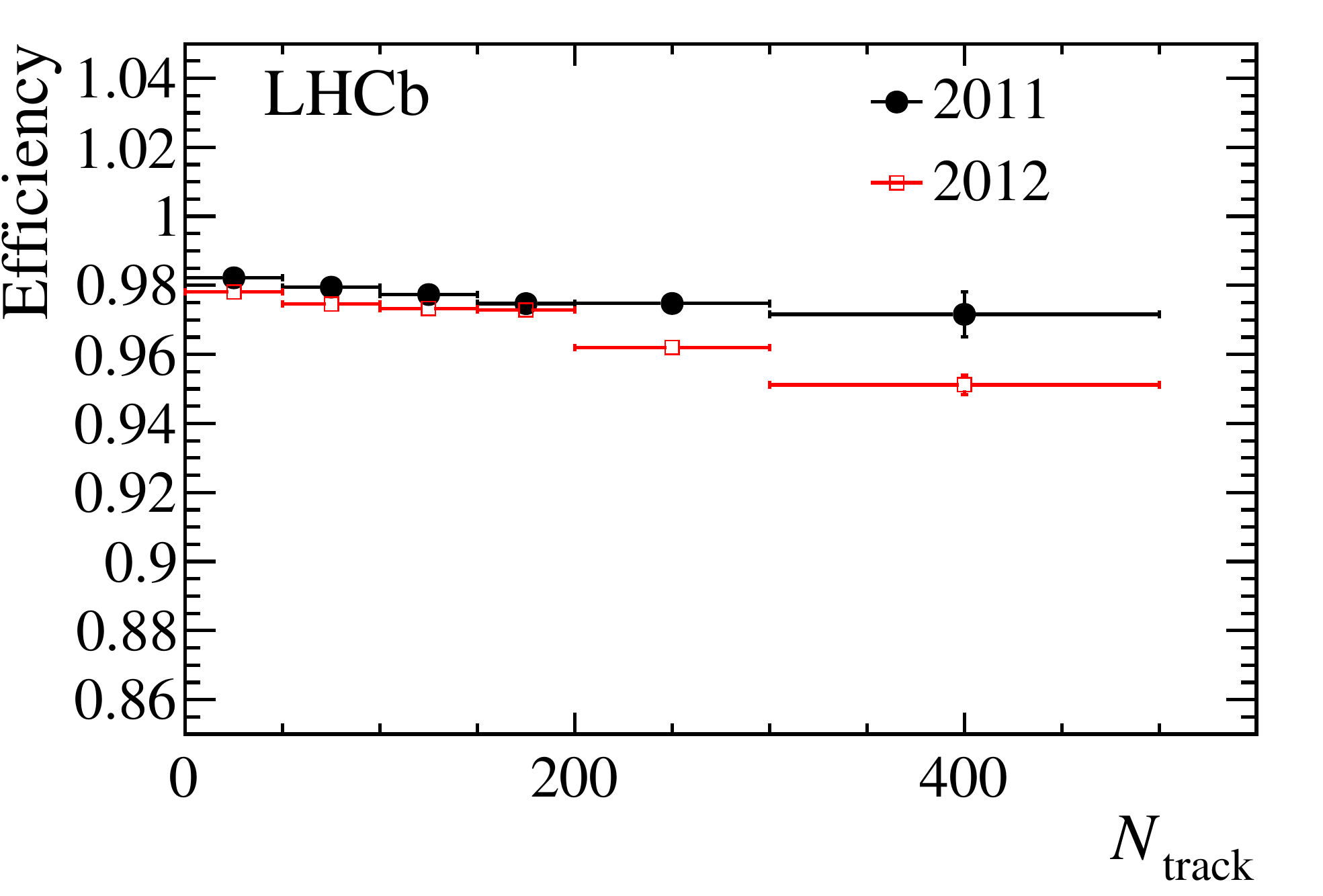}
    \includegraphics[width=0.49\textwidth]{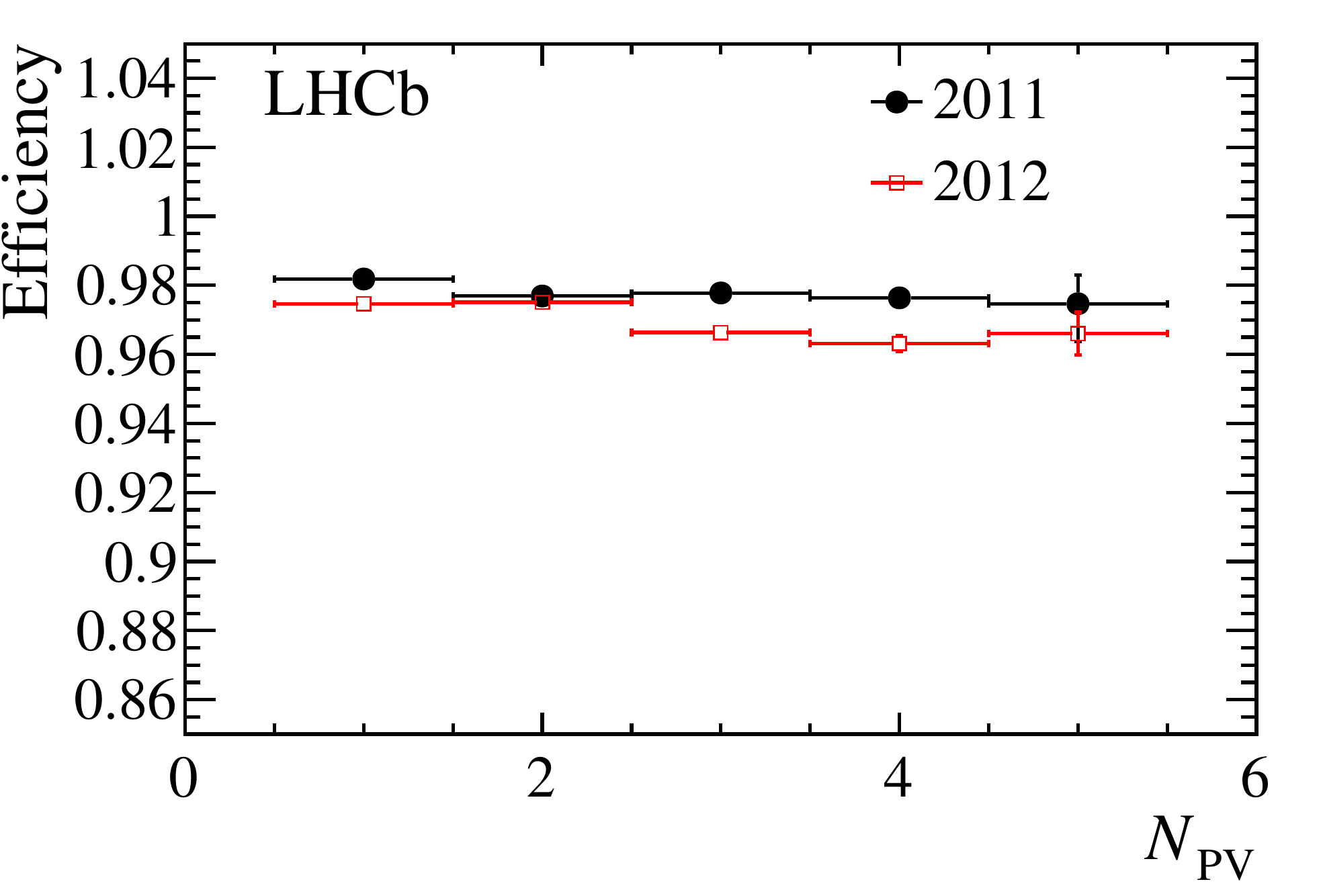}
  \end{center}
  \caption{\small Tracking efficiency as function of the
    momentum, $p$, the pseudorapidity, $\eta$, the total number of tracks in the
    event, $N_{\rm track}$, and the number of reconstructed primary vertices,
    $N_{\rm PV}$ \protect\cite{LHCb-DP-2013-002}.
    The error bars indicate the statistical uncertainty.}
  \label{fig:effLong2011}
\end{figure}

Two different tag-and-probe methods\cite{LHCb-PUB-2011-025,LHCb-DP-2013-002} are
used to measure the efficiency for long tracks.  The overall efficiency depends
on the momentum spectrum of the tracks and the track multiplicity of the event.
The tracking efficiency is shown in Figure~\ref{fig:effLong2011} as a function of
the absolute momentum, $p$, of the pseudorapidity, $\eta$, of the total number
of tracks in the event, $N_{\rm track}$, and of the number of reconstructed
primary vertices, $N_{\rm PV}$. The performance in the 2012 data is slightly
worse, which is partially due to the higher hit multiplicity at the higher
centre-of-mass energy. As can be seen, the average efficiency is above $96\%$ in the
momentum range $5\gevc < p < 200\gevc$ and in the pseudorapidity range $2 < \eta
< 5$, which covers the phase space of LHCb. Only in high multiplicity events 
($N_{\rm track}>200$) it is slightly less than $96\%$. 
The track reconstruction efficiency has been shown to be well reproduced in
simulated events \cite{LHCb-DP-2013-002}.

\subsubsection{Mass and momentum resolution}
\label{sec:momentumresolution}

\begin{figure}[!tb]
  \begin{center}
    \includegraphics[width=0.65\textwidth]{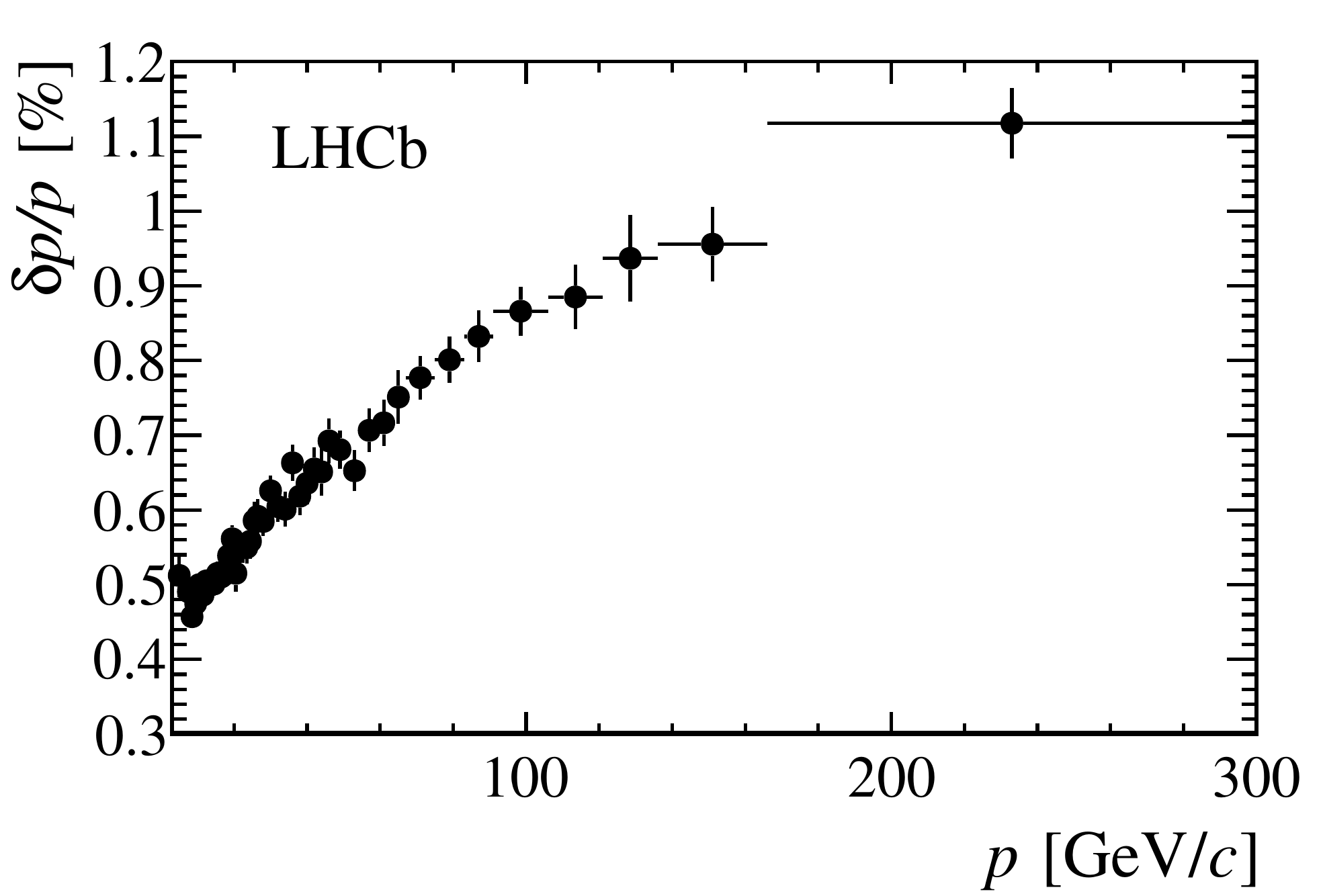}
  \end{center}
  \caption{\small Relative momentum resolution versus momentum for long tracks
    in data obtained using \jpsi decays.}
  \label{fig:dppVsp}
\end{figure}

The momentum resolution for long tracks in data is extracted using
$\jpsi\to\mup\mun$ decays. The mass resolution of the \jpsi is primarily defined
by the momentum resolution of the two muons. Neglecting the muon masses
and considering decays where the two muons have a similar momentum, the
momentum resolution, $\delta p$, can be approximated as:
\begin{linenomath}
\begin{equation}
  \left(\frac{\delta p}{p}\right)^2 = 2 \left(\frac{\sigma_m}{m}\right)^2
  - 2\left(\frac{p\,\sigma_{\theta}}{m\,c\,\theta}\right)^2 \ ,
\end{equation}
\end{linenomath}
where $m$ is the invariant mass of the \jpsi candidate and $\sigma_m$ is the
Gaussian width obtained from a fit to the mass distribution. The second term is
a correction for the opening angle, $\theta$, between the two muons, where
$\sigma_{\theta}$ is the per-event error on $\theta$ which is obtained from the
track fits of the two muons. Figure~\ref{fig:dppVsp} shows the relative momentum
resolution, $\delta p/p$, as a function of the momentum, $p$. The momentum
resolution is about 5 per mille for particles below $20\gevc$, rising to about 8
per mille for particles around $100\gevc$.

\begin{figure}[!tb]
  \begin{center}
    \includegraphics[width=0.49\textwidth]{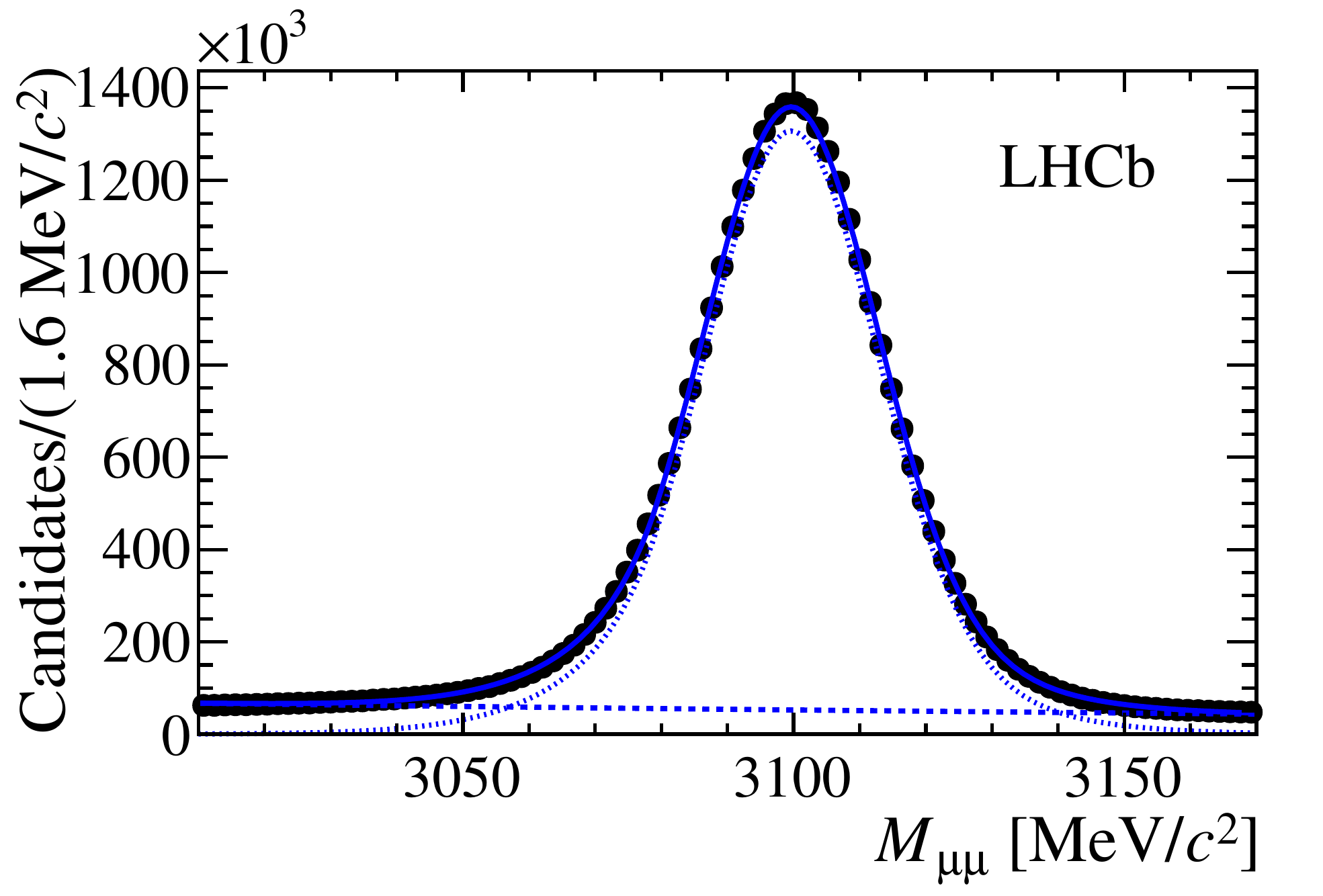}
    \includegraphics[width=0.49\textwidth]{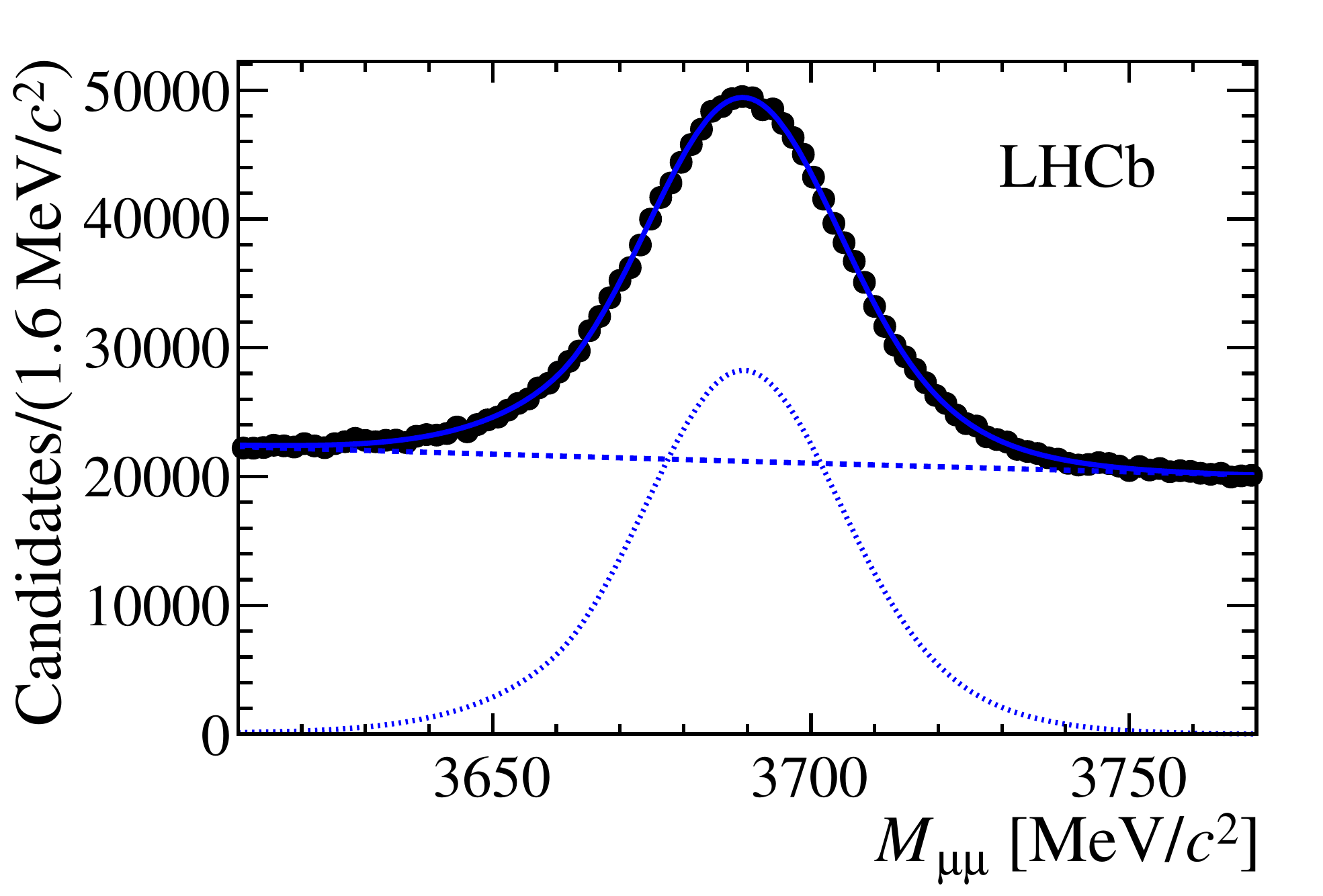}
    \includegraphics[width=0.49\textwidth]{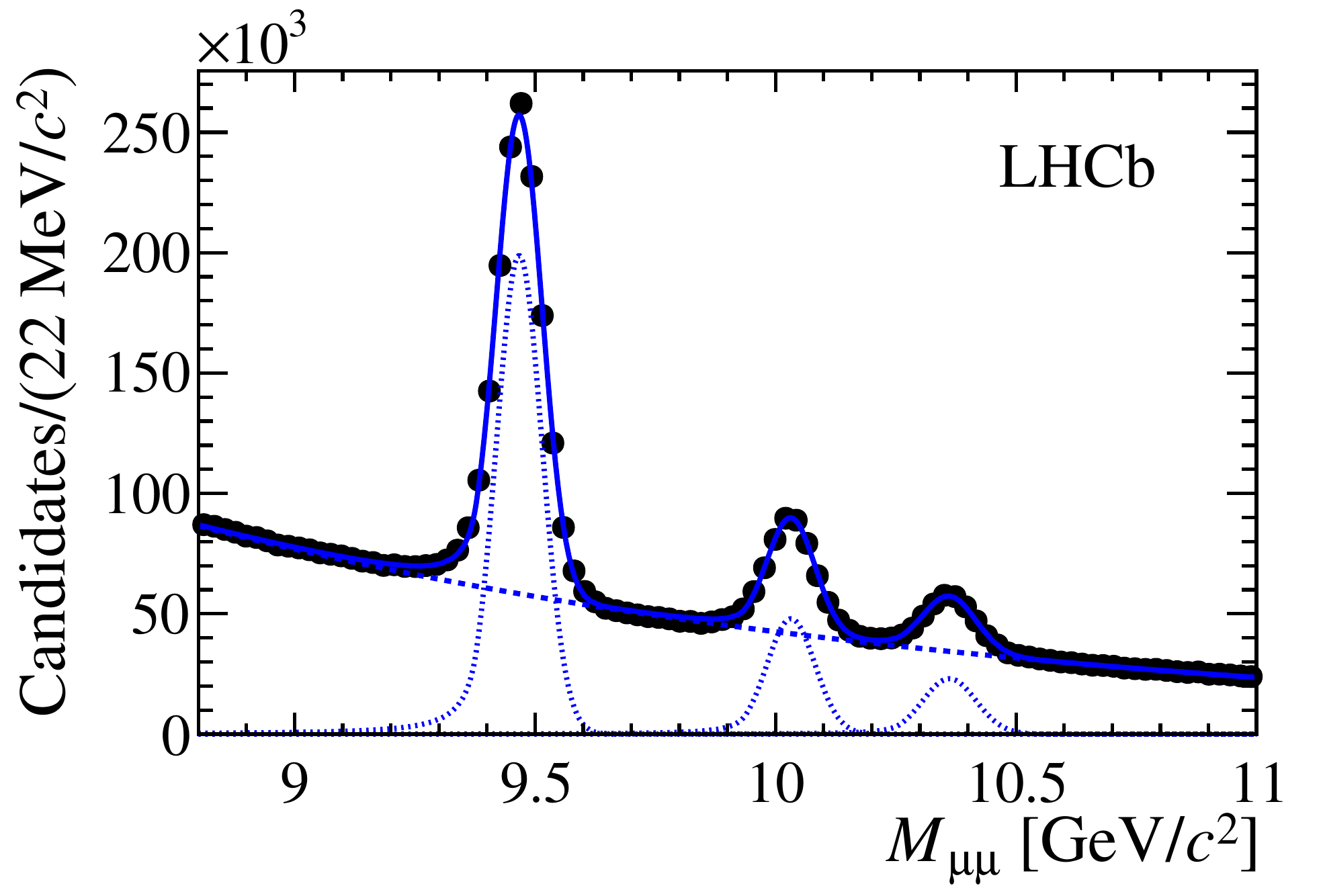}
    \includegraphics[width=0.49\textwidth]{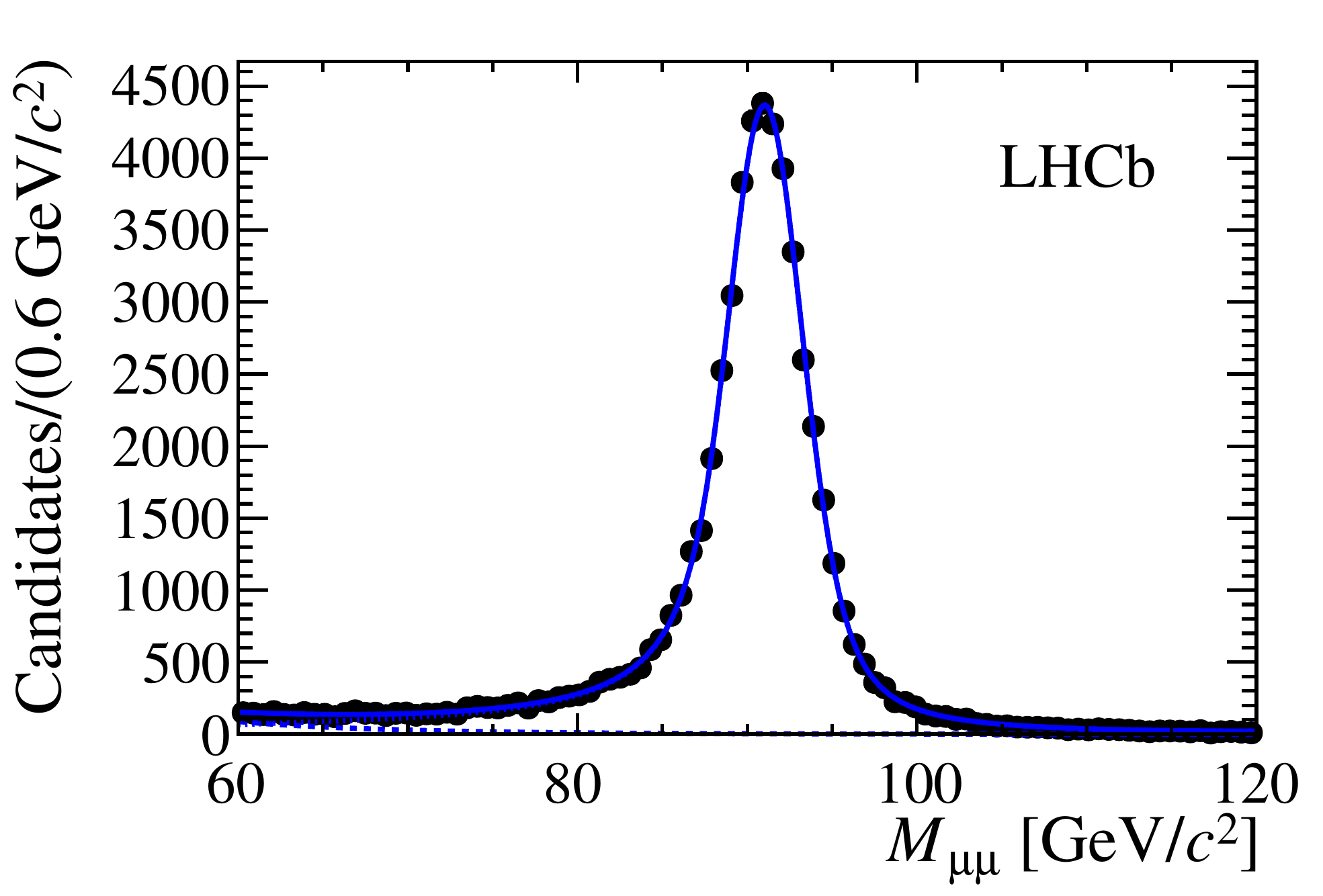}
  \end{center}
  \caption{\small Mass distributions for (top left) \jpsi, (top right) \psitwos,
    (bottom left) \OneS, \TwoS and \ThreeS, and (bottom right) \Z
    candidates. The shapes from the mass fits are superimposed, indicating the
    signal component (dotted line), the background component (dashed line) and
    the total yield (solid line).}
  \label{fig:massDistributions}
\end{figure}

The mass resolution is compared for six different dimuon resonances: the \jpsi,
\psitwos, \OneS, \TwoS and \ThreeS mesons, and the \Z boson. These resonances
are chosen as they share the same topology and exhibit a clean mass peak. A
loose selection is applied to obtain the invariant mass distributions, as shown
in Figure~\ref{fig:massDistributions}.

\begin{figure}[!tb]
  \begin{center}
    \includegraphics[width=0.49\textwidth]{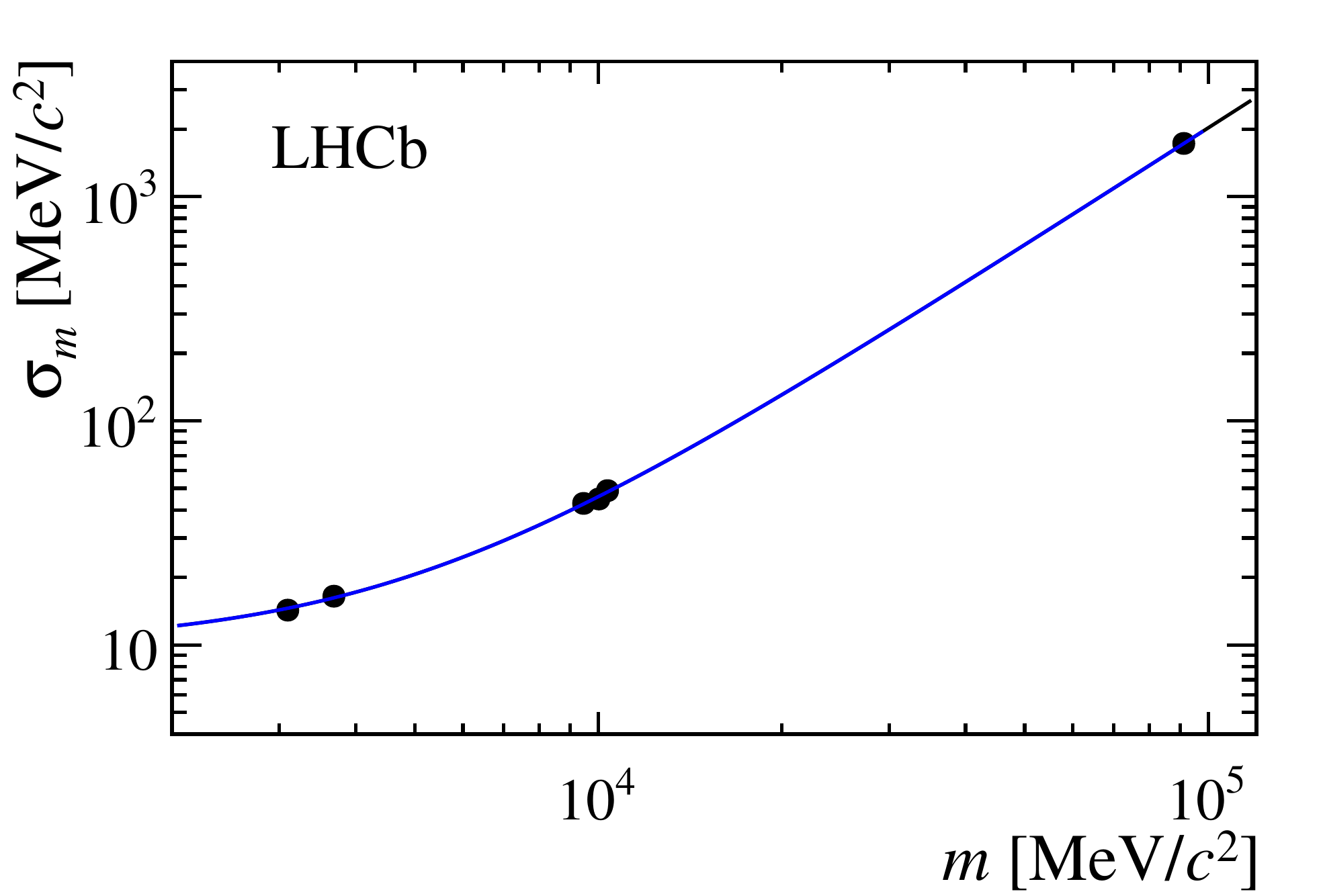}
    \includegraphics[width=0.49\textwidth]{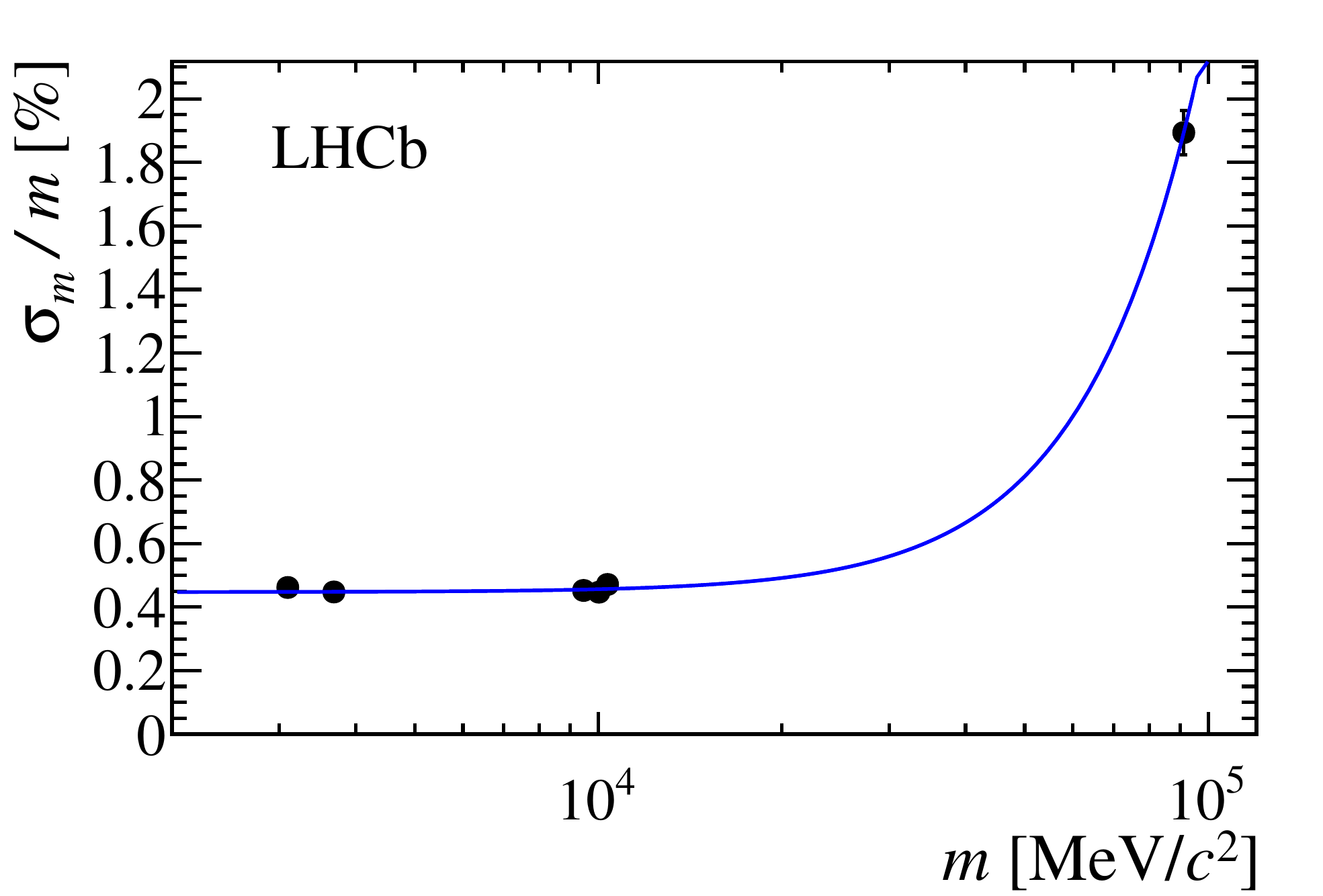}
  \end{center}
  \caption{\small Mass resolution ($\sigma_m$) (left) and relative mass
    resolution (right) as a function of the mass ($m$) of the dimuon resonance. The
    mass of the muons can be neglected in the invariant mass calculation of
    these resonances. The mass resolution is obtained from a fit to the mass
    distributions. The superimposed curve is obtained from an empirical
    power-law fit through the data points.}
  \label{fig:massDependence}
\end{figure}

The momentum scale is calibrated using large samples of $\jpsi\to\mup\mun$ and
$\Bp\to\jpsi\Kp$ decays, as is done for the precision measurements of
\bquark-hadron and \D meson masses~\cite{LHCb-PAPER-2011-035,
  LHCb-PAPER-2012-028, LHCb-PAPER-2012-048, LHCb-PAPER-2013-011}. By comparing
the measured masses of known resonances with the world average
values~\cite{PDG2014}, a systematic uncertainty of $0.03\%$ on the momentum
scale is obtained. As shown in Figure~\ref{fig:dppVsp} the momentum resolution
depends on the momentum of the final-state particles, and therefore the mass
resolution is not expected to behave as a pure single Gaussian. Nevertheless, a
double Gaussian function is sufficient to describe the observed mass
distributions. Final-state radiation creates a low-mass-tail to the left side of
the mass distribution, which is modelled by an additional power-law tail. To
describe the 
\Z mass distribution, a single Gaussian function with power-law tail is
convolved with a Breit-Wigner function, where the natural width is fixed to
$2495.2\mevcc$~\cite{PDG2014}. In all cases, an exponential shape models the
background. The results from the fits are overlaid in
Figure~\ref{fig:massDistributions}. The overall mass resolution is calculated as
the root mean square of the double Gaussian function. The mass resolution obtained
from the fits are shown in Table~\ref{tab:massRes}. The uncertainties are
statistical only. Figure~\ref{fig:massDependence} shows the mass resolution and
relative mass resolution versus the mass of the resonance. It can be seen that
the relative mass resolution, $\sigma_m/m$, is about 5 per mille up to the
$\PUpsilon$ masses.

\begin{table}[!tb]
\centering
\tbl{Mass resolution for the six different dimuon resonances.}
{\begin{tabular}{lcc}\hline
      Resonance & Mass resolution $(\mevcc)$ \\
      \hline
      \jpsi      & $14.3\pm0.1$  \\
      \psitwos   & $16.5\pm0.4$  \\
      \OneS      & $42.8\pm0.1$  \\
      \TwoS      & $44.8\pm0.1$    \\
      \ThreeS    & $48.8\pm0.2$    \\
      \Z         & $1727\pm64$     \\
      \hline
    \end{tabular}
    \label{tab:massRes}}
\end{table}

\subsection{Spatial alignment of the tracking detectors}

The alignment of the \lhcb{} tracking detector uses information from
optical and mechanical surveys and from reconstructed charged particle
trajectories. To ensure adequate tracking performance, the position and
orientation of detector elements in the global reference frame must be known
with an accuracy significantly better than the single hit resolution. Since
\lhcb{} is a forward spectrometer, the requirements in terms of absolute units
of distance are different for the different coordinate axes: tracks are less
sensitive to displacements of elements in the $z$ direction compared
to equally sized displacements in $x$ and $y$. Similarly, rotations
around the $z$ axis are more important than those around the $x$ and
$y$ axis.

Although the final alignment precision is obtained with reconstructed
tracks, a precise survey is indispensable both as a starting point for
the track-based alignment and to constrain degrees of freedom
to which fitted track trajectories are insensitive. For example, the
knowledge of the $z$ scale of the vertex detector originates solely
from the pre-installation survey. Ultimately this is what limits, for
example, certain measurements such as the $\Bs$ oscillation frequency.

Several methods have been deployed for track-based alignment in
\lhcb{}. One technique used for the \velo divides the
alignment in three stages, corresponding to different detector
granularity~\cite{Viret:2008jq,Gersabeck:2008jr}. The relative
alignment of each \PhiSens{} sensor with respect to the \RSens{}
sensor in the same module is performed by fitting an analytical form
to the residuals as a function of $\phi$. The relative alignment of
the modules within each \velo{} half are obtained with a $\chi^2$
minimisation based on an implementation of Millepede
method~\cite{Blobel:2002ax}. The relative alignment of one \velo half
with respect to the other half is also based on the Millepede method.
It is performed using a track sample crossing the overlap region
between the two halves and with a $\chi^2$ minimisation that exploits
the difference in the position between primary vertices reconstructed
in both halves.  Similar approaches based on Millepede have been
considered for the \ot~\cite{Deissenroth:thesis} and \intr.

The implementations of the Millepede algorithm in LHCb use a
simplified model of the track, ignoring the effects of the magnetic
field, multiple scattering and energy loss. These effects are
accounted for in the default LHCb track fit, which is based on a
Kalman filter. Therefore, another global $\chi^2$ minimisation that
uses the default track fit has been
implemented~\cite{Hulsbergen:2008yv,Amoraal:thesis}. The algorithm can
align all tracking detectors simultaneously. The correct treatment of
magnetic field and material effects facilitates the use of relatively
low-momentum tracks in the alignment, which helps to constrain the $z$
scale of the spectrometer. Another novel aspect is that tracks can be
combined in vertices, allowing for the use of primary vertex and mass
constraints~\cite{Amoraal:2012qn}.

All methods were used during the commissioning of the detectors
and in the initial $pp$ collisions and found to be in good
agreement~\cite{VeloTedArticle}.
The method using the Kalman track fit is used routinely for the
tracking alignment updates.

\subsubsection{Vertex locator alignment}
\label{veloalignment}

The most stringent alignment requirements apply to the vertex
detector. In order not to degrade impact parameter or decay time
resolutions, the \velo sensors need to be aligned with a precision of a few
microns in $x$ and $y$ and a few tens of microns in $z$.  Components of the
detector have been surveyed at various stages of the assembly at
ambient temperature.  The relative position of the \PhiSens{} sensor with
respect to the \RSens{} sensor in each module has been measured with an
accuracy of about $3$\mum for the $x$ and $y$ translation and with an
accuracy of about $20$\murad for rotations around the $x$ and $y$ axis.
The relative module position within each half of the detector has been
measured with a precision of about 10\mum for the translations along
$x$, $y$ and $z$.  The position of the two \velo{} halves has been
determined with an accuracy of 100\mum for the translations and
100\murad for the rotations.

The main degrees of freedom in the track alignment of the \velo{} sensors
and modules are the $x$ and $y$ translation and the rotation around
the $z$ axis.  The alignment for the $x$ and $y$ translation can be
evaluated at the sensor level, while the one for the rotation around
the $z$ axis can be determined only at the module level, as only the
\PhiSens{} sensors are sensitive to this degree of freedom.  The
misalignment due to the other three degrees of freedom (the $z$
translation and the rotations around the $x$ and $y$ axis) causes a
second-order effect.  To obtain the desired sensitivity, a track sample
with a wide distribution of the angle between the track and the strips
in the sensor plane is required. Consequently, the alignment for these
degrees of freedom can be evaluated only for the \RSens{} sensors.

The track-based alignment is insensitive to the overall $z$ scale, $xz$
and $yz$ shearing and to the global position and orientation of the
\velo{}~\cite{Viret:2008jq}. To constrain these degrees of freedom the position
of two modules in each half are fixed to their nominal survey position in the
\velo{} half frame. The average position and rotation of the two halves is also
fixed. After correcting for differences in temperature, the position of the
modules and sensors evaluated by the alignment with tracks is found to be in
good agreement with the metrology. 

The alignment of the two \velo{} halves relies on two constraints.
The first one is determined by tracks that cross both halves of the detector, in
particular those that traverse the region where the sensors in the two
halves overlap.  This gives sensitivity to misalignment due to $x$ and
$y$ translations and to rotation around the $z$ axis. In addition,
reconstructed primary vertices are used which adds sensitivity to
relative $x$, $y$ and $z$ translations and to rotations around the $x$
and $y$ axis.

\begin{figure}[!tb]
  \centerline{
    \includegraphics[width=0.91\textwidth]{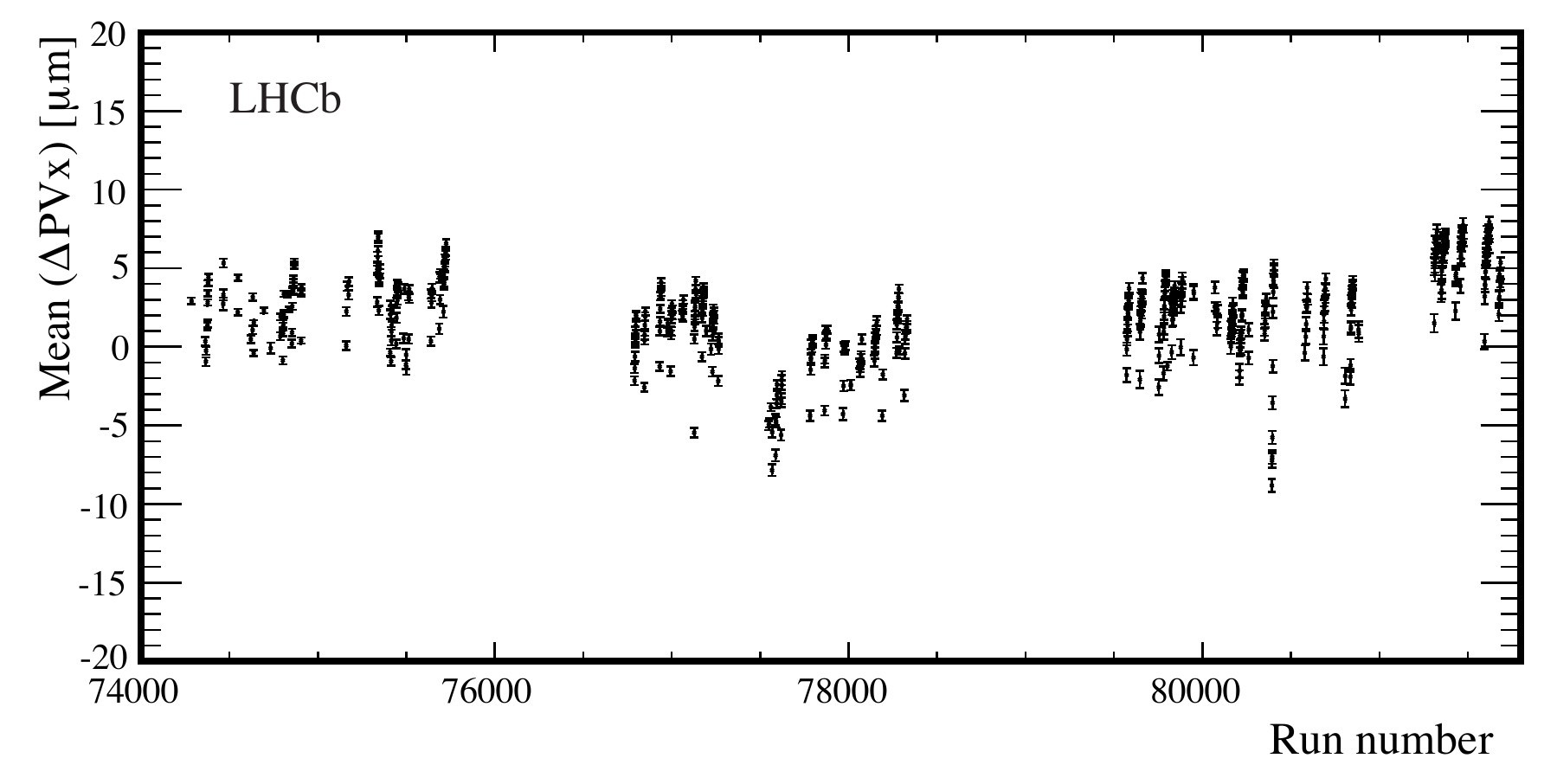}}
  \caption{Run dependence of the relative misalignment of the two
    \velo{} halves along the $x$ axis evaluated with primary
    vertices. }
  \label{fig:veloalignstability}
\end{figure}

The operating temperature was found to have an effect on the alignment
and hence is kept sufficiently stable such that variations can
be ignored. A more important issue is the fact that the \velo{} halves
are moved every fill in order to put them at a safe distance from the
beam during LHC injection. This movement corresponds to about $29\mm$ in
$x$. The \velo is closed only once stable beam conditions are
declared.  
The position of the \velo stepper motors is measured using resolvers mounted
on the motor axes and is reproducible with a precision better than 10\mum.
This measurement is then used as an alignment correction.
Figure~\ref{fig:veloalignstability} shows the distribution
of the difference between the $x$ position of primary vertices that
are separately reconstructed in the left and right detector halves as a
function of time. The variation illustrates that the resolver position
measurement is accurate to about $5\mum$.

The uncertainty in the $z$ scale of the \velo{} is important for
precision measurements of \bquark-hadron lifetimes and \BzBs mixing
frequencies.  At the
time of assembly the length of the \velo base plate was
measured with an accuracy of approximately $100\mum$ over the full
length of the \velo{}~\cite{LHCb-DP-2014-001}. This translates into a
length scale uncertainty of about $0.01$\%. To verify the
understanding of the survey, the measurements are compared to
the track-based alignment. In the latter, the length scale is fixed by
constraining two modules in each half to their nominal position. The
RMS of the differences in the $z$ positions of unconstrained modules
is $20\mum$, in agreement with the estimated survey uncertainty. To
interpret this as a length scale, the RMS of the distribution of the
$z$ positions of the first hits on typical \velo{} track segments  is
conservatively used. In combination with the number above this
leads to a total systematic uncertainty on the length scale of
$0.022\%$.

\begin{figure}[!tb]
  \centerline{
    \includegraphics[width=0.69\textwidth]{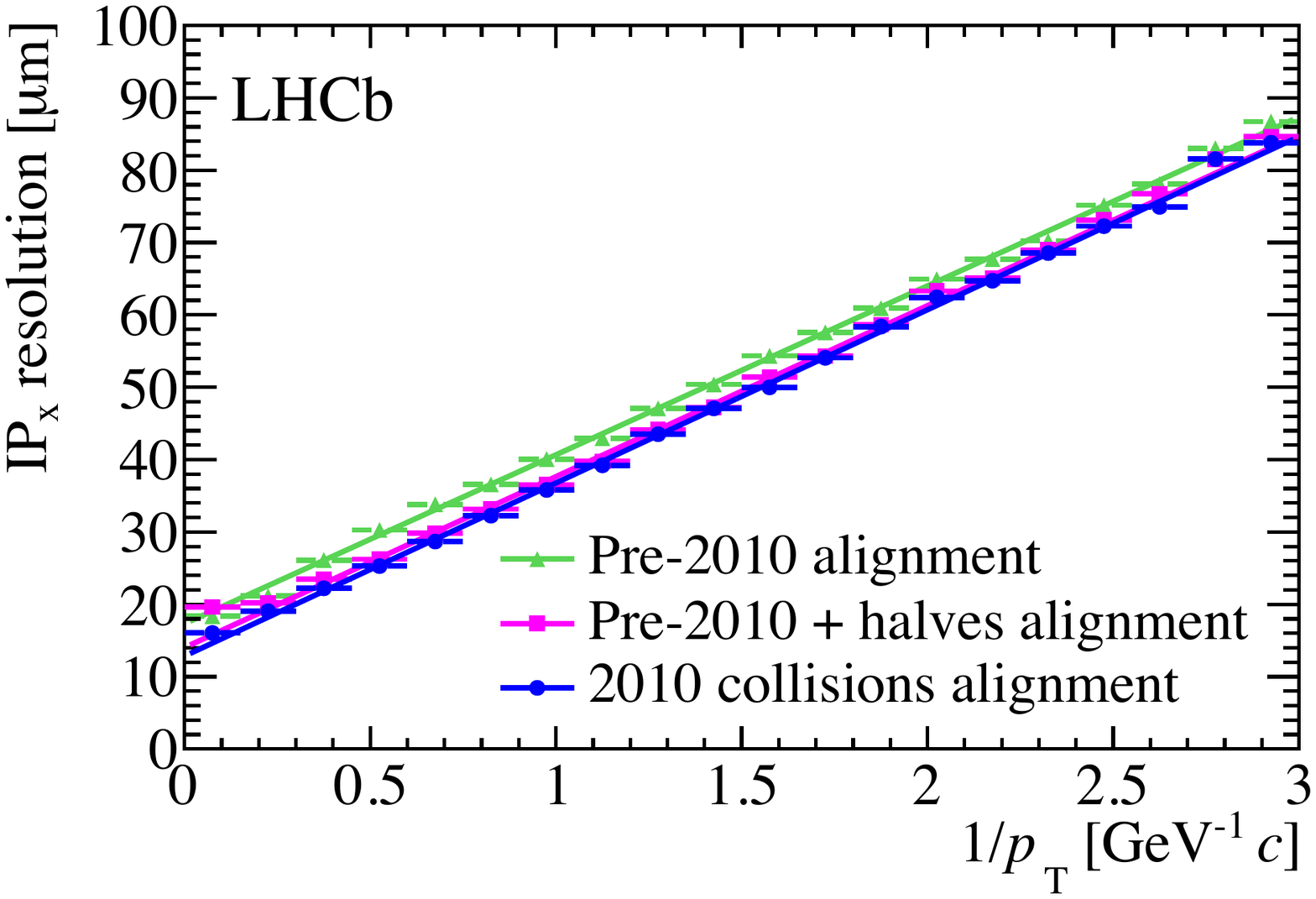}}
  \caption[]{IP$_x$ resolution as a function of \invpt, comparing different
    qualities of alignment, measured on 2010 data.} 
  \label{fig:IPXRes_CompareAlignments}
\end{figure}

To illustrate the effects of misalignment on the \velo{} performance, the
impact parameter (IP) resolution is examined (see Section~\ref{IPsection}).
Figure~\ref{fig:IPXRes_CompareAlignments} shows the IP$_x$ resolution
versus $1/\pt$ obtained at different stages of the alignment, namely
by using the alignment from the commissioning phase, after a
track-based alignment that only corrects for the relative alignment of
the two halves, and after the full alignment of the sensors. The
refinement of the alignment improves the IP resolution by about 25\%
at high transverse momentum.  As the remaining alignment uncertainties
are smaller than the corrections obtained in the last stage, the
residual misalignment has no significant effect on the IP resolution.

\subsubsection{Alignment of the silicon tracker and outer tracker}

The rest of the spectrometer is aligned relative to the \velo{} using
long tracks. The alignment is performed at different levels of
granularity, exploiting differences in the precision of survey
between `small' and `large' structures.  Typical alignment degrees of
freedom are displacements in $x$ and rotations around the $z$ axis for
the smallest structures (modules in the \ot and ladders in \intr and
\st) and displacements in $z$ for the layers.

Global deformations are a concern in a forward spectrometer with parallel 
detector planes, in particular $x$ scaling, $z$ scaling, $xz$ shearing
and curvature ($q/p$) bias. An $x$ scaling corresponds to a
displacement along $x$ of detector modules assembled in a single layer
proportional to the $x$ coordinate. In the \ttracker and \intr
detectors such a scaling is constrained by tracks that traverse
neighbouring ladders in the same layer. To profit from this constraint
the sample of tracks is enriched by preferentially selecting such
`overlap'~tracks.

In principle, the $z$ scaling of the spectrometer is fixed by the $z$
scale of the vertex detector, which comes from a survey. In practice,
this leads to a relatively poor constraint on the tracking layers
downstream of the magnet.  It has been verified that the last \ot detector
plane downstream of the magnet can be fixed to its survey position without
introducing a momentum bias.

A global $xz$ shearing can also be fixed using information from the \velo survey.
Whilst this leads to a relatively weak constraint on the tracking
layers downstream of the magnet, any remaining shearing between the
\velo and the rest of the tracking system is absorbed in the
curvature bias, which is a global deformation that is typical for
alignment with tracks in a non-zero
field \cite{Simioni:thesis,Amoraal:2012qn}. The curvature bias is
constrained by including mass constraints from cleanly selected $\DKpi$ or
$\jpsi\to\mup\mun$ candidates \cite{Amoraal:2012qn}.

Another concern are observed displacements in the $z$ coordinate, in
particular in \ttracker and \introne, the detectors closest to the magnet.
In the presence of a magnetic field, tracks are sensitive to the position of the
tracking detectors relative to the dipole field. An alignment performed with
early 2010 data indicated a displacement of approximately 1\cm
of the entire spectrometer along the $z$ axis.  In winter 2011 an
\textit{in-situ} measurement of the magnetic field map on a finite number of
points along the $x$, $y$ and $z$ axes in the centre of the magnet
confirmed this displacement.

The survey of the tracking detectors was performed with the dipole
magnet switched off. After anomalously large differences between
survey data and track-based alignment were observed in \introne, the
position of all \intr boxes was monitored before and after ramping the
field, revealing movements of up to 5\mm. This illustrates
that data collected in the absence of magnetic field are only of
limited value in the alignment.

As for the \velo a crucial aspect of the alignment is stability over
time. Detectors may be moved, for example, for maintenance during accelerator
technical stops. These occurred at least once every two months in the first
years of LHC running. The dipole field is reversed about twice per
month, which also affects alignment. Consequently, the detector is
realigned after every technical stop and every magnetic field reversal.
Remaining misalignments in the relative position of neighbouring
detector modules  are estimated from hit residual distributions
to be approximately 10\mum in \intr and 30\mum in \ttracker.

\begin{figure}[!tb]
  \begin{center}
    \includegraphics[width=0.9\textwidth]{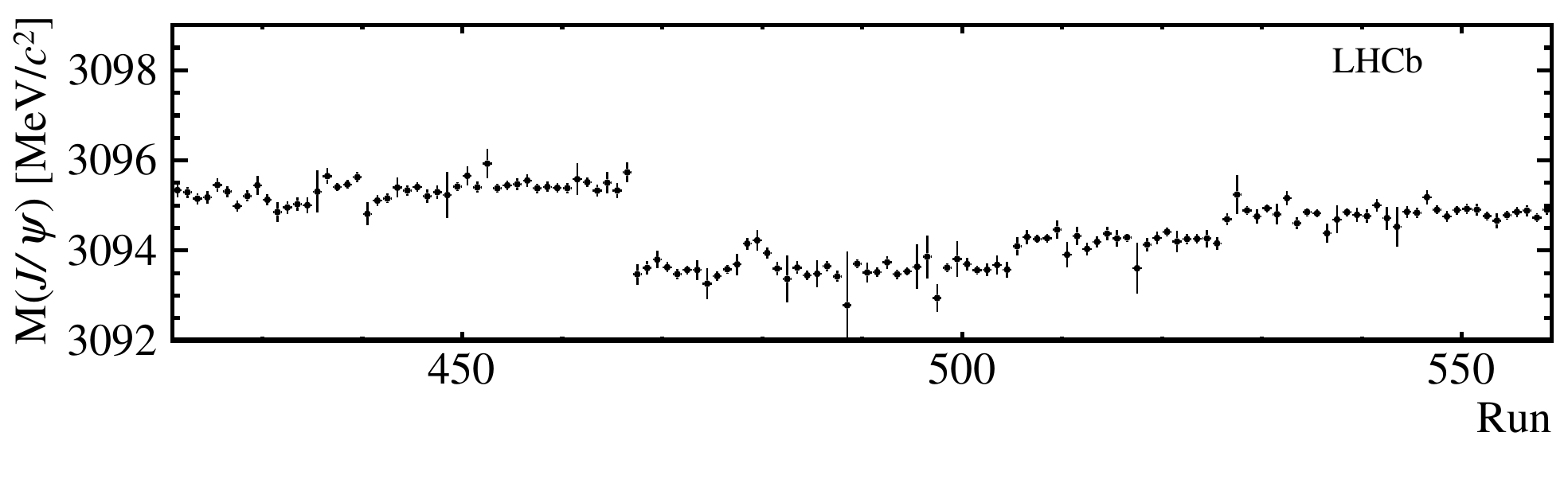}\\
    \includegraphics[width=0.9\textwidth]{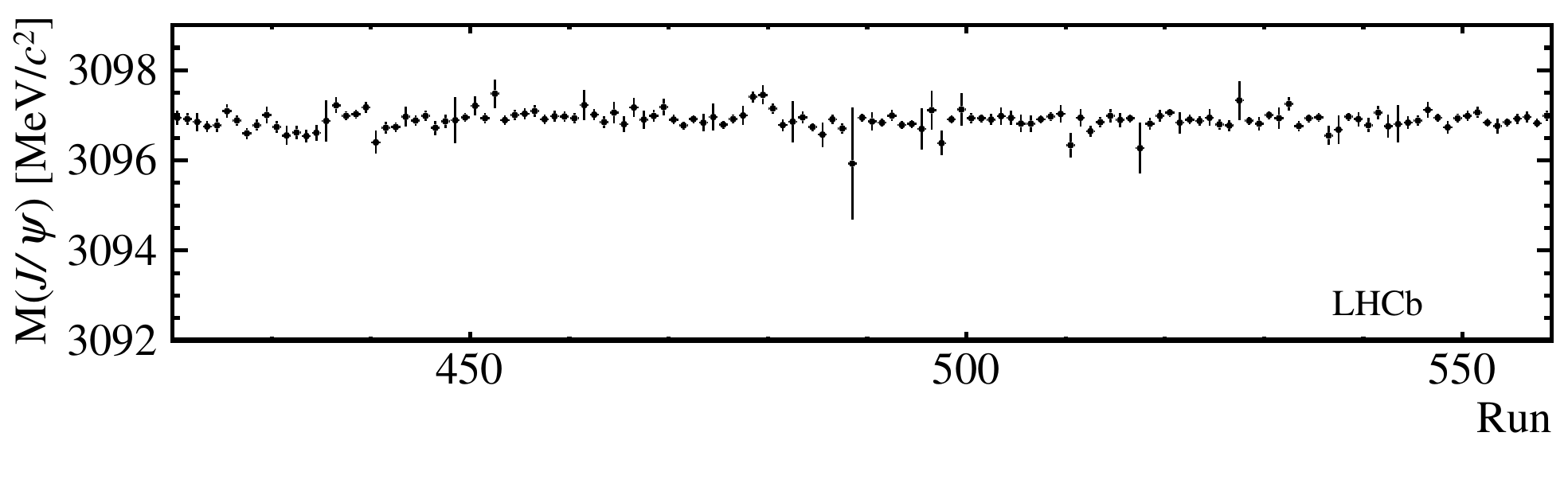}
    \caption{Fitted position of the peak of the $\jpsi\to\mumu$
      invariant mass distribution as a function of run number in a two-week
      period in which the operating temperature of \ttracker modules was varied.
      The mass is evaluated using the same alignment for the full period
      on the top and using a dedicated track alignment for each period
      with constant temperature on the bottom. }
    \label{fig:jpsivstime_TT_temperature.pdf}
  \end{center}
\end{figure}

Figure~\ref{fig:jpsivstime_TT_temperature.pdf} shows the position of the peak of
the $\jpsi\to\mumu$ invariant mass distribution as a function of time in a
period in which the operating temperature of the \ttracker modules was varied by
15\degrees C in order to study detector performance. The temperature change
causes the support structure on which the modules are mounted to contract by an
amount that is large enough to affect the curvature measurement, as shown by the
bias in the \jpsi mass. After a separate track-based alignment is performed for
each period with constant temperature, the bias in the mass
disappears~\cite{Marki:thesis}. This illustrates the importance of operating the
detector under stable conditions.

Although the average curvature bias is constrained with $\DKpi$
decays, other misalignment effects, uncertainties in modelling of the
magnetic field (evaluated to be about 0.1\%) and detector material still
affect the reconstructed invariant mass. In particular,  small variations in the
invariant mass as function of particle momentum are observed. To obtain precise
mass measurements corrections on the momentum scale are tabulated and calibrated
using samples of $\jpsi\to\mup\mun$ and $\Bp\to\jpsi\Kp$ decays (see
Section~\ref{sec:momentumresolution}). 

\subsubsection{Muon system alignment}

The read-out pads of the muon detector are less fine-grained than the read-out
channels  of the other tracking systems, leading to a coarser spatial
resolution.  Consequently the alignment of the muon chambers is in general less
critical. However, misalignments larger than a few \mm in the first two muon
stations can affect  the efficiency of the L0 muon trigger and introduce a
charge asymmetry. In the L0 trigger the muon momentum is estimated by the $x$
coordinate of hits in stations M1 and M2. Studies on simulated events have shown
that an alignment precision of $1\mm$ is enough to guarantee charge symmetry of
the trigger efficiency and momentum measurement to the 0.1\% level. The
alignment is even less critical for the other stations, which are not used for
momentum estimate and have lower spatial resolution.

\begin{figure}[!bt]
  \begin{center}
    \includegraphics[width = 0.95\textwidth]{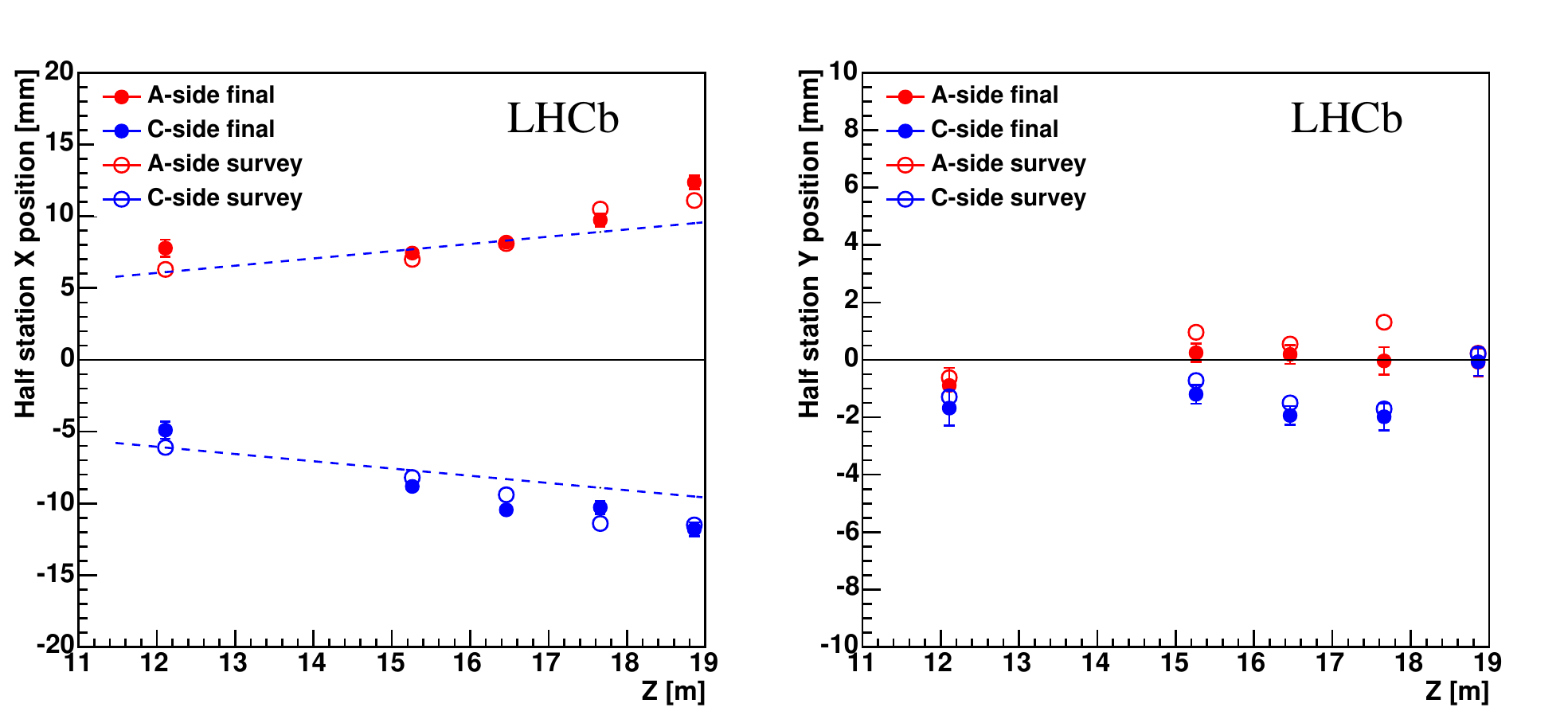}
    \caption{Alignment of the ten muon half stations for the  2012 run.
      The values of the inner edge in the $x$ position (left) and
      of the median $y$ position are shown as a function of the
      station position along $z$. The empty and full dots represent
      the results from the survey measurements and the software
      alignment respectively. The error bars, when visible, show the sum in
      quadrature of statistic and systematic uncertainties.The dashed lines
      represent the ideal projective alignment of the detector in the closed
      position.} 
    \label{fig:muon_alig}
  \end{center}
\end{figure}

The muon chambers are mounted in support structures called `half
stations'. The alignment accuracy of the chambers within a half station is
about 1\mm in the $x$ and $y$ direction.  Each half station can be
independently moved on rails in the $x$ direction. Due to mechanical
limitations observed during installation, the half stations could not
be put exactly in their nominal position. Moreover, in order to keep a safe
distance from the beam-pipe, a small separation between the two sides
was maintained, preserving as much as possible the symmetry and
projectivity with respect to the interaction point. The position of
the closed half stations was surveyed using four reference points on
each half station. The result of the survey is shown in
Figure~\ref{fig:muon_alig}. Displacements with respect to the ideal
projective position are found to be within 2\mm.

The alignment obtained from the survey is refined offline using
reconstructed tracks after the alignment of the tracking system.  Muon
hits are attached to reconstructed tracks matching standalone muon
segments with a good \chisq. The global \chisq of an ensemble of 
tracks is minimised using the same method as used for the rest of the tracking
system. As shown in Figure~\ref{fig:muon_alig} for 2012 data, small, but
significant, differences are found for the translational degrees of freedom,
while no significant rotations are found.  The result of this procedure is
confirmed, independently of the alignment of the other tracking detectors, using
standalone muon segment reconstruction from events selected without the muon
trigger.  The resulting differences with respect to the survey positions are
within $1.5$\mm in $x$ and $y$.  The accuracy of the alignment based on tracks
is $1$\mm or less, sufficient to avoid any detector efficiency effects that
could introduce charge asymmetries in the L0 trigger. The alignment results are
thus used by the L0 muon trigger for the computation of transverse momenta and
are accounted for in the subsequent offline reconstruction of muons.

\subsection{Vertexing and decay time resolution}
\label{sec:vertexing}

\begin{figure}[!tb]
 \vspace{3mm}
  \centering
  \includegraphics[width=.8\textwidth]{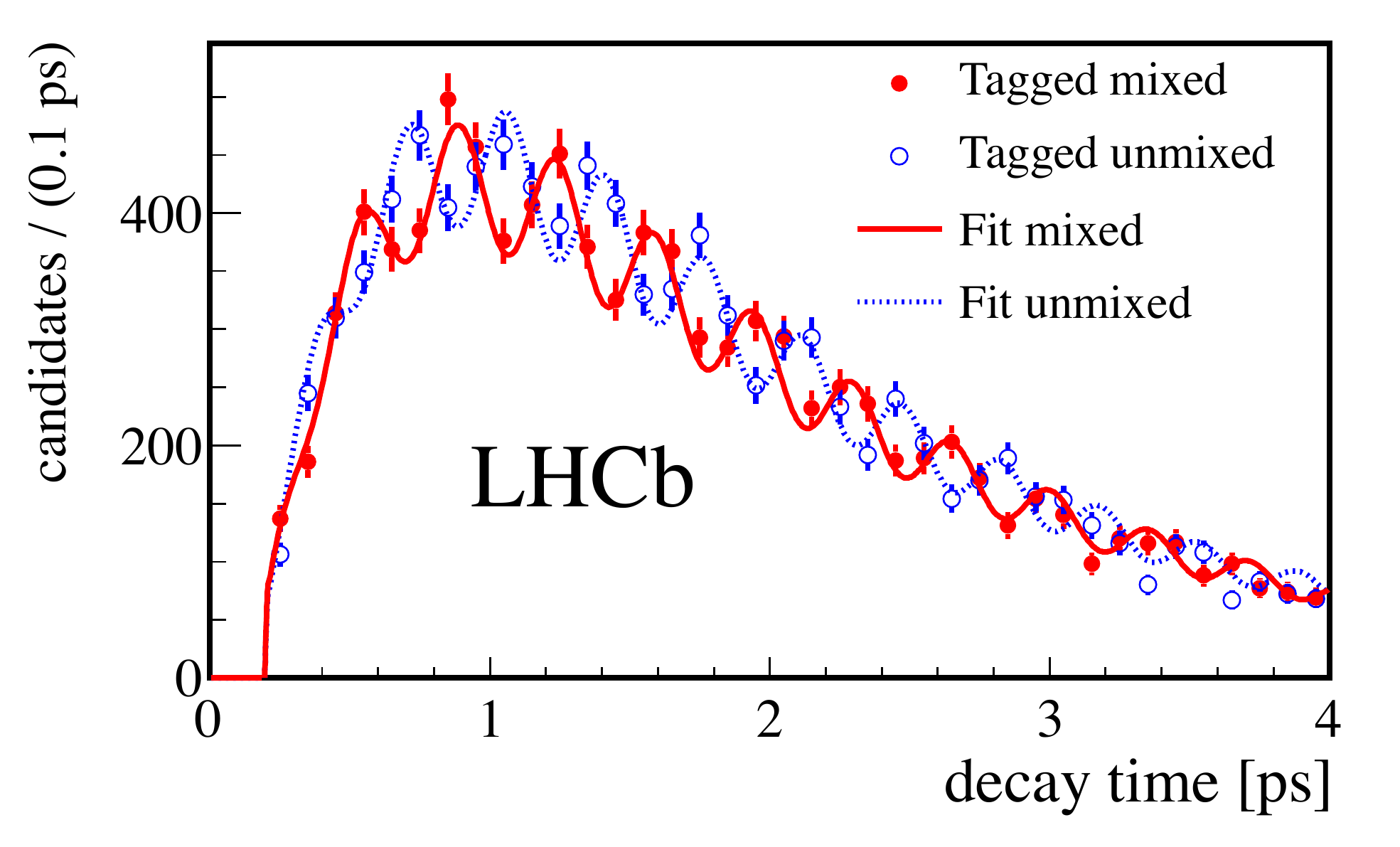}
 \caption{Decay time distribution for \decay{ \Bs} {D_s^- \pi^+} candidates
   tagged as mixed (different flavour at decay and production; red, continuous
   line) or unmixed (same flavour at decay and production; blue, dotted  line)
   \protect\cite{LHCb-PAPER-2013-006}.} 
\label{fig:BsOscillation}
\vspace{5mm}
\end{figure}

The study of \CP violation and rare decays in the heavy flavour sector requires
the accurate measurement of production and decay vertices and track impact
parameters, both for flavour tagging and for background rejection. The most
stringent demands on the vertex reconstruction arise from the decay time
resolution requirements to resolve the fast flavour  oscillations induced by
\Bs--\Bsb{} mixing. \lhcb has made the world's most precise measurement of the
\Bs oscillation frequency using the decay \decay{\Bs} {D_s^-
  \pi^+}~\cite{LHCb-PAPER-2013-006}. The decay time resolution in \lhcb is
sufficient to observe the oscillations in the flavour tagged decay time
distribution, as illustrated in Figure~\ref{fig:BsOscillation}.

\subsubsection{Primary vertex reconstruction}

\begin{figure}[!tb]
 \vspace{5mm}
  \centering
  \includegraphics[width=0.49\textwidth]{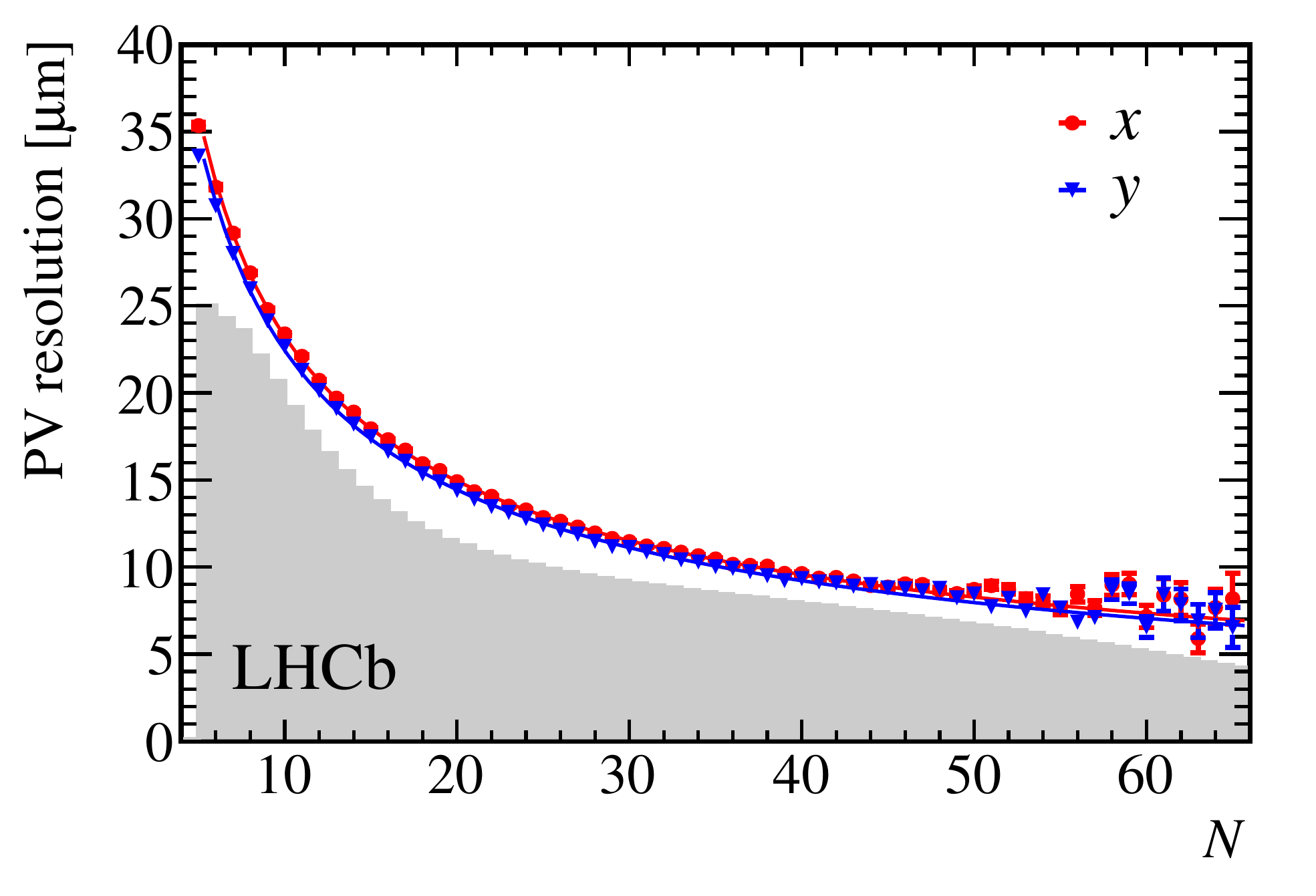}
  \includegraphics[width=0.49\textwidth]{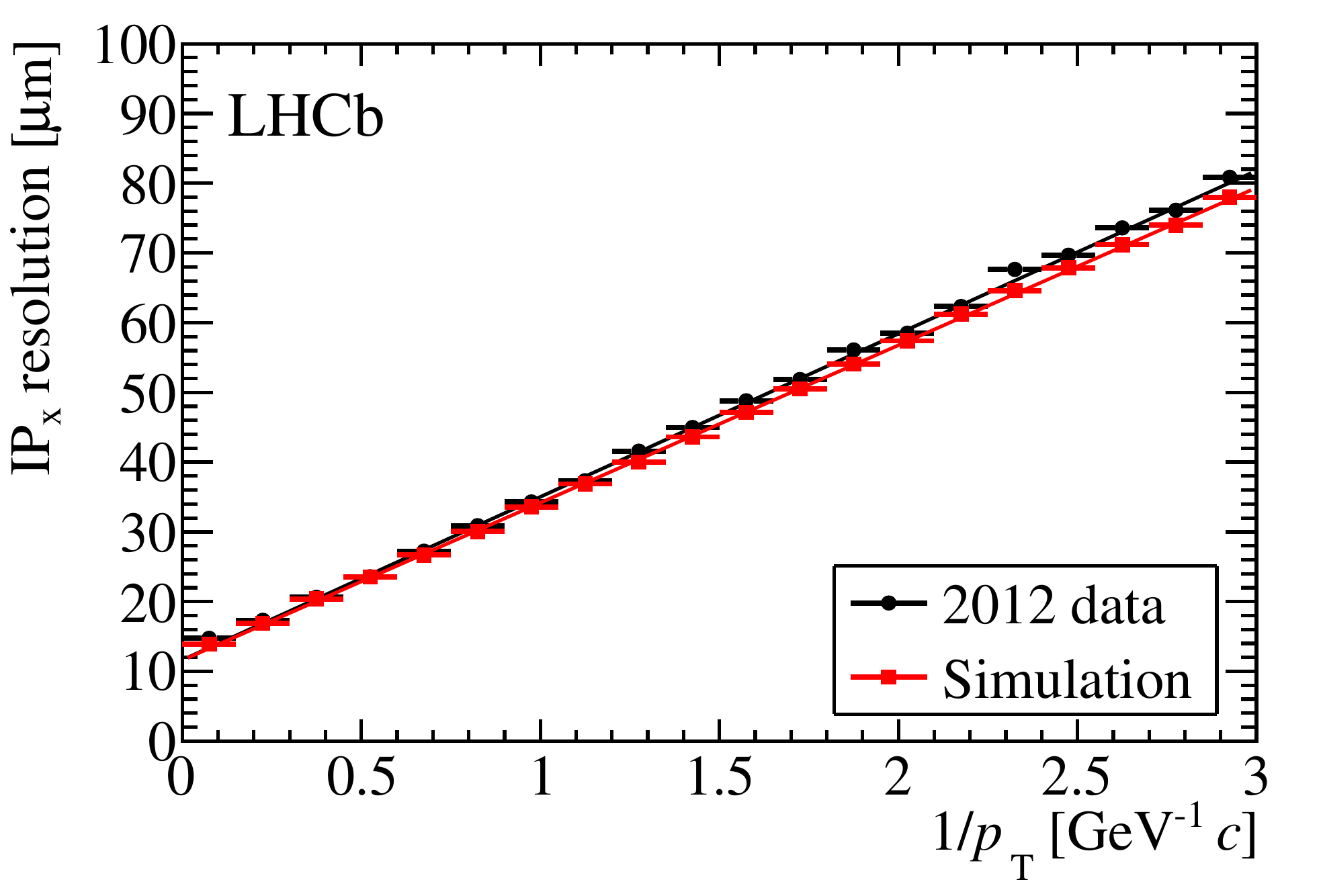}
  \caption[Primary vertex and impact parameter resolutions]{
    The primary vertex resolution (left), for events with one reconstructed
    primary vertex, as a function of track multiplicity. The $x$ (red) and $y$
    (blue) resolutions are separately shown and the superimposed histogram shows
    the distribution of number of tracks per reconstructed primary vertex for all
    events that pass the high level trigger. The impact parameter in $x$
    resolution as a function of \invpt (right). Both plots are made using data
    collected in 2012. }
  \label{fig:vertexing}
 \vspace{3mm}
\end{figure}

The primary vertex (PV) resolution is measured by comparing two
independent measurements of the vertex position in the same event.
This is achieved by randomly splitting the set of tracks in an event
into two and reconstructing the PVs in both sets. The width of the
distribution of the difference of the vertex positions is corrected
for a factor $\sqrt{2}$ to extract the vertex resolution.
The number of tracks making a vertex ranges from $5$ (the minimum required
by the PV reconstruction) to around $150$, and this technique allows the
resolution to be measured using up to around $65$ tracks. The PV resolution is
strongly correlated to the number of tracks in the vertex (the track
multiplicity). To determine the vertex resolution as a function of the track
multiplicity, only vertex pairs with exactly the same number of tracks
are compared. The result for the resolution in the $x$ and $y$
direction is shown in Figure~\ref{fig:vertexing}. A PV with $25$ tracks has a
resolution of $13$\mum in the $x$ and $y$ coordinates and $71$\mum in $z$.

\subsubsection{Impact parameter resolution}
\label{IPsection}

The impact parameter (IP) of a track is defined as its distance from the primary
vertex at its point of closest approach to the primary vertex. Particles
resulting from the decay of long lived \B or \D mesons
tend to have larger IP than those of particles produced at the primary vertex.
Selections on IP and IP~$\chi^2$ are extensively used in LHCb analyses to
reduce the contamination from prompt backgrounds. Consequently, an
optimal IP resolution and a good understanding of the effects
contributing to the IP resolution are of prime importance to LHCb
performance.

The IP resolution is governed by three main factors: multiple scattering of
particles by the detector material; the resolution on the position of hits in
the detector from which tracks are reconstructed; and the distance of
extrapolation of a track between its first hit in the detector and the
interaction point.  The minimisation of these factors is achieved in the design
of the \velo. The sensors are positioned close to the beams, separated from them
by only a thin aluminium foil. The first active strips are only 8\mm away from
the beams during physics collisions. The detector provides high-precision hit
position measurements as shown in Section \ref{sec:velohitresolution}.

As the IP is defined as the distance between a point and a line,
it is not a Gaussian distributed quantity. 
It is therefore customary to divide the IP in two quantities that
follow a normal distribution by projecting out two independent
components. In LHCb these are the components of the IP vector in the
transverse plane,
\begin{equation}
   \text{IP}_x \; = \;  x - x_\text{\textsc{PV}} -
   ( z - z_\text{\textsc{PV}} ) t_x
\end{equation}
and similarly for $y$, where $(x,y,z)_\textsc{PV}$ is the position of
the primary vertex, $(x,y,z)$ is the point on the track of closest approach to
the primary vertex and $(t_x,t_y,1)$ is the direction vector of the
track. Figure~\ref{fig:vertexing} shows the IP$_x$ resolution
as a function 1\pt{}. The IP$_y$ resolution is similar. The linear
dependence on 1/\pt{} is a consequence of multiple scattering and the
geometry of the vertex detector. At asymptotically high \pt{} the
IP$_x$ resolution is about 13\mum~\cite{LHCb-DP-2014-001}.

\subsubsection{Decay time resolution}

The distance between the production and secondary decay vertices of long
lived mesons is used to reconstruct the particle's decay time. This is
required for lifetime measurements and for resolving flavour
oscillations in time-dependent \CP violation measurements. Consequently, the
performance of the VELO is illustrated here with an analysis of the decay time
resolution of \BsToJPsiPhi decays \cite{LHCb-PAPER-2011-021}. 

Time dependent \CP violation effects are measured as the amplitude of
an oscillation in the $B$ decay time distribution. The size of the
observed amplitude is damped by a dilution factor from the finite
decay time resolution~\cite{Moser:1996xf}.  Hence, achieving optimal
decay time resolution is important and any bias in the estimated decay
time resolution leads to a bias in the measurement of the \CP violating effect.

The reconstructed decay time in the rest frame of the decaying
particle can be expressed in terms of the reconstructed decay length
$l$, momentum $p$ and mass $m$ of the particle in the LHCb frame as
$ t = ml/p$.
The decay time is computed with a vertex fit that constrains the
decaying particle to originate from the primary vertex.
The uncertainty on the decay length $l$ and on the momentum $p$ are
essentially uncorrelated in LHCb. Consequently, the decay time
uncertainty can be expressed in terms of the decay length uncertainty
$\sigma_l$ and the momentum uncertainty $\sigma_p$ as

\begin{equation}
\sigma_t^2 \; = \;
  \left(\frac{m}{p}\right)^2 \: \sigma_l^2 \; + \;
  \left(\frac{t}{p}\right)^2 \: \sigma_p^2 \; .
\end{equation}

This expression shows an explicit dependence on the decay time.
However, for decay times up to a few times the \B meson lifetime, the
uncertainty is dominated by the $\sigma_l$ term, motivating the use of
a `prompt' control channel to calibrate the decay time uncertainty.
The decay time resolution depends on the topology of the decay and is
calibrated for each final state on data. For \BsToJPsiPhi{} decays,
the calibration method uses prompt combinations that fake signal candidates.
Subtracting the small contribution from signal candidates and long-lived
background using the \sPlot technique~\cite{Pivk:2004ty}, the shape of the
decay time distribution is determined only by the resolution function.

\begin{figure}[!tb]
  \includegraphics[width=0.49\textwidth]{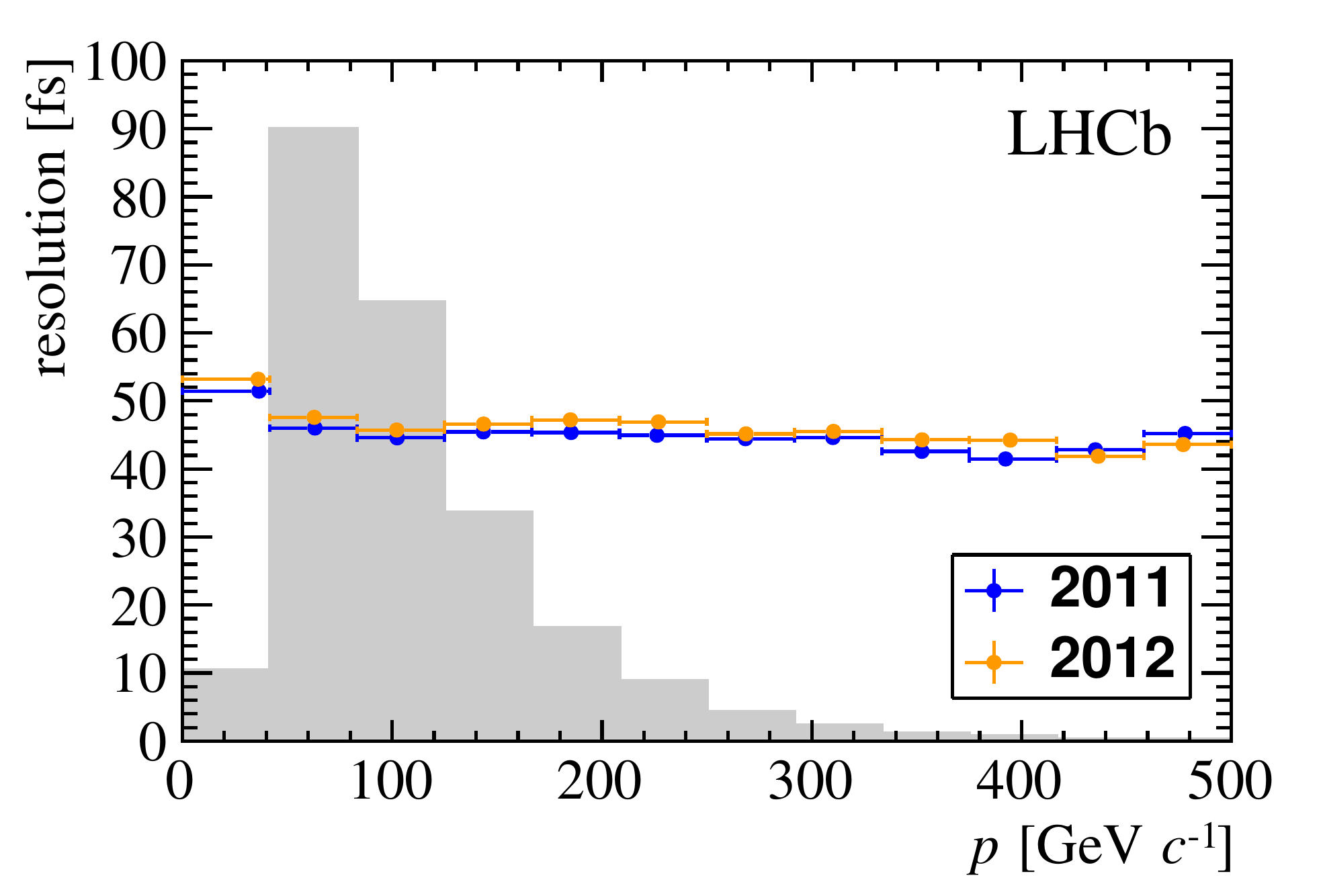}
  \includegraphics[width=0.49\textwidth]{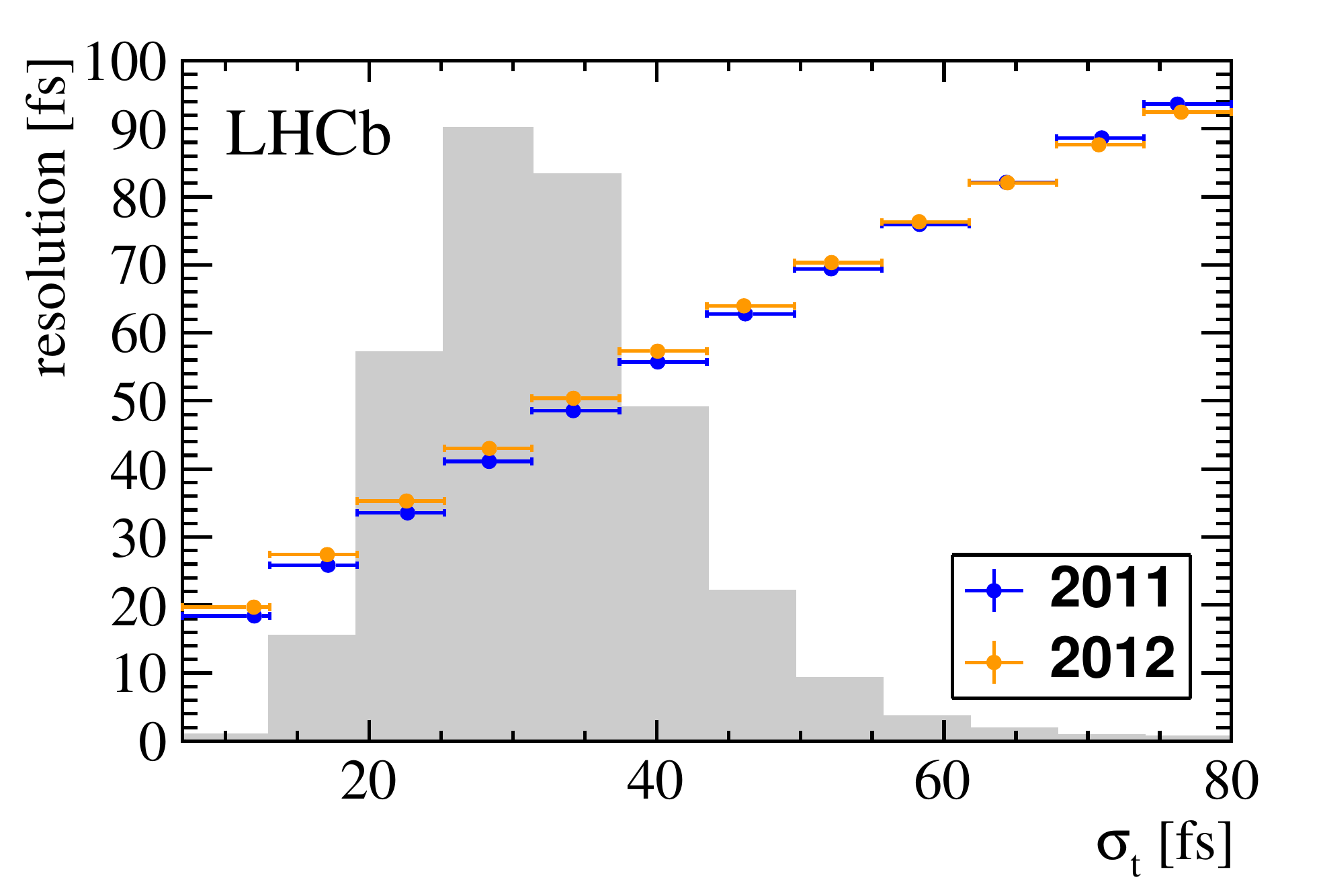}
  \caption{Decay time resolution as a function of momentum (left) and
    as a function of the estimated decay time uncertainty (right) of
    fake, prompt $\Bs\to\jpsi\phi\to\mumu\Kp\Km$ candidates in
    2011 and 2012 data. Only events with a single reconstructed primary vertex
    are used. The superimposed histogram shows the distribution of momentum
    (left) and estimated decay time uncertainty (right) on an arbitrary scale.}
  \label{fig:decaytimereso}
  \vspace{2mm}
\end{figure}

Figure~\ref{fig:decaytimereso} shows the resolution as a function of the (fake)
\B candidate momentum. It should be noted that the decay time resolution is
essentially independent of the \B momentum, illustrating that 
$\sigma_l \propto p$. This is a consequence of the fact that the larger the
momentum is, the smaller the opening angle, and hence the larger the uncertainty
on the position of the vertex in the direction of the boost. The resolution is
also shown as a function of the per-event estimated uncertainty in the decay
time, which is obtained from the vertex fit. As expected, the resolution is a
linear function of the estimated uncertainty. A decay time resolution of $\sim
50 \fs$ is obtained in \lhcb. For a mixing frequency of 17.7\invps, such as for
\Bs oscillations, this decay time resolution leads to a dilution of the
\CP asymmetry by a factor $\sim 0.7$.

\subsubsection[$V^0$ reconstruction]{$\mathbf{V^0}$ reconstruction}

\begin{figure}[!tb]
  \includegraphics[width=0.49\textwidth]{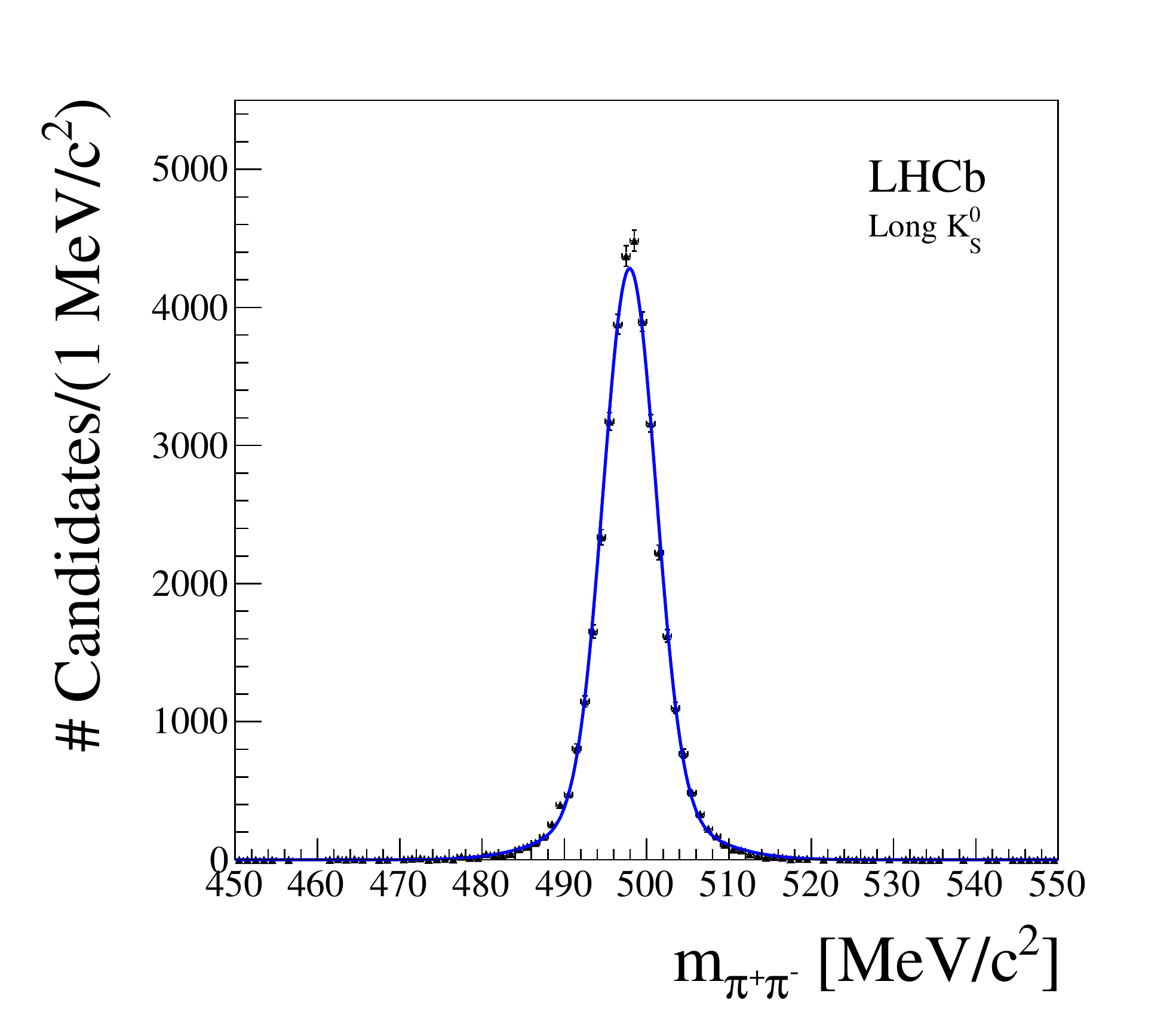}
  \includegraphics[width=0.49\textwidth]{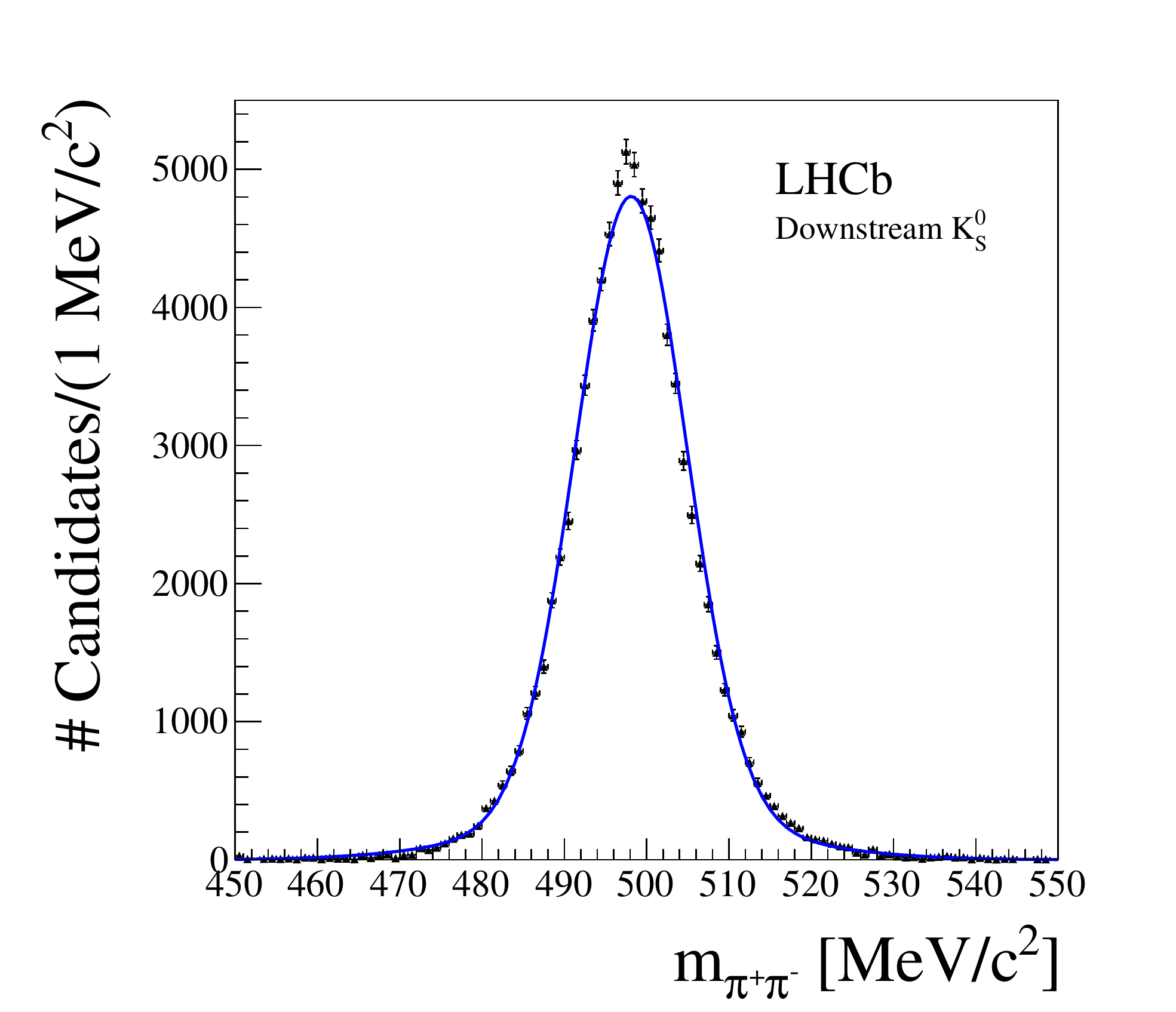}
  \caption{Distribution of the invariant mass of $\KS \to \pipi$ candidates with
  a decay vertex at a significant distance to the PV, for long tracks (left) and
  downstream tracks (right). A mass resolution of $3.5$\mevcc is achieved for
  the candidates reconstructed from long tracks and $7$\mevcc for those using
  downstream tracks.}
  \label{fig:ksmass}
\end{figure}

Reconstructed $V^0$ decays ($\KS\to\pip\pim$ and $\L\to \proton\pim$)
are an essential ingredient of many LHCb analyses. If the decay time
is sufficiently small, the daughter particles are reconstructed as
long tracks, and for these decays the invariant mass resolution is as good
as for short-lived resonances (see Section~\ref{sec:momentumresolution}
and Figure~\ref{fig:ksmass}). For $V^0$s that decay outside the
\velo{} acceptance, but before the magnet, the daughter particles are
reconstructed as downstream tracks from hits in the TT and T
stations. As the resolution on the track direction reconstructed in
the layers of the TT is not as good as in the \velo{}, the invariant mass
resolution for the downstream category is worse than for the
short-lived category, as shown in Figure~\ref{fig:ksmass}.
For \KS{} momenta typical of $B$ decay products,
about two thirds of the reconstructed \KS decays are found using
downstream tracks, illustrating the importance of the downstream
tracking for physics performance.

\section{Neutral particle reconstruction}
\label{sec:neutral-particle-id}

Neutral particle reconstruction is based on information provided by
the four systems (SPD, PS, ECAL and HCAL), which together form the
calorimeter. The SPD and the PS both consist of a plane of
scintillator tiles, separated from each other by a thin lead
layer, while the ECAL and HCAL have shashlik and sampling
constructions, respectively. In all four cases, the light produced in
the organic scintillators is transmitted to photomultiplier tubes
(PMT) by optical fibres~\cite{LHCb-TDR-002,Alves:2008zz}. In general, the
detected signal pulses are longer than the nominal read-out window of
25\ns, and this must be taken into account to minimise spill-over
effects. In the ECAL and HCAL detectors, this is performed by first
clipping the signal to fit within the read-out window. In the PS and
SPD detectors, the effects of spill-over are removed by subtracting a
fraction of the signal integrated in the previous clock cycle.

The SPD uses a single bit for each cell to indicate whether or not it
was traversed by a charged particle, with a discriminator comparing the
energy deposited in the given cell to half of that expected from a
minimum ionising particle (MIP). The signal from the PS detector is
digitised using a 10-bit ADC with a dynamic range of 0.1--100 times
the corresponding MIP energy deposit. The ECAL and the HCAL have the
same read-out electronics, which digitises signals with a 12-bit
precision and a dynamic range that results in a maximum detectable
transverse energy of 10\gev, optimised for the typical energy deposits
that occur in LHCb events.  The operational status and stability of
the ECAL and the HCAL detectors are examined using dedicated systems
based on light emitting diodes (LEDs).  The average PMT responses when
illuminated by pulses of known intensities are used to monitor the
behaviour of the corresponding read-out channels.

Reliable neutral particle reconstruction requires calibration
\cite{CaloTALP,CALOR-XCAL} of the
calorimeter system, which is therefore outlined briefly before
describing the methods used to reconstruct neutral particles

\subsection{Calibration of the calorimeter system}

The SPD calibration is performed by adjusting the discriminator
threshold for each channel to half of the expected MIP energy
deposit. These values are established at the beginning of each data
taking period, and the fraction of tracks pointing to SPD cells that
have an associated SPD hit is used to monitor performance during
data-taking \cite{CALOR-SPD}. Thresholds are adjusted to ensure high
and uniform efficiency.

The PS calibration consists of equalising the ADC response of all
channels, based on the most probable values of MIP energy
deposits. The MIP sample is composed of reconstructed tracks of
particles with momentum greater than 2\gevc to ensure that they reach
the PS~\cite{CALOR-PS}.  As for the SPD, the calibration is
established at the beginning of each data-taking period, checked
regularly and corrected as necessary during the run.

\begin{figure}[!tb]
\begin{center}
\includegraphics[width=0.65\textwidth]{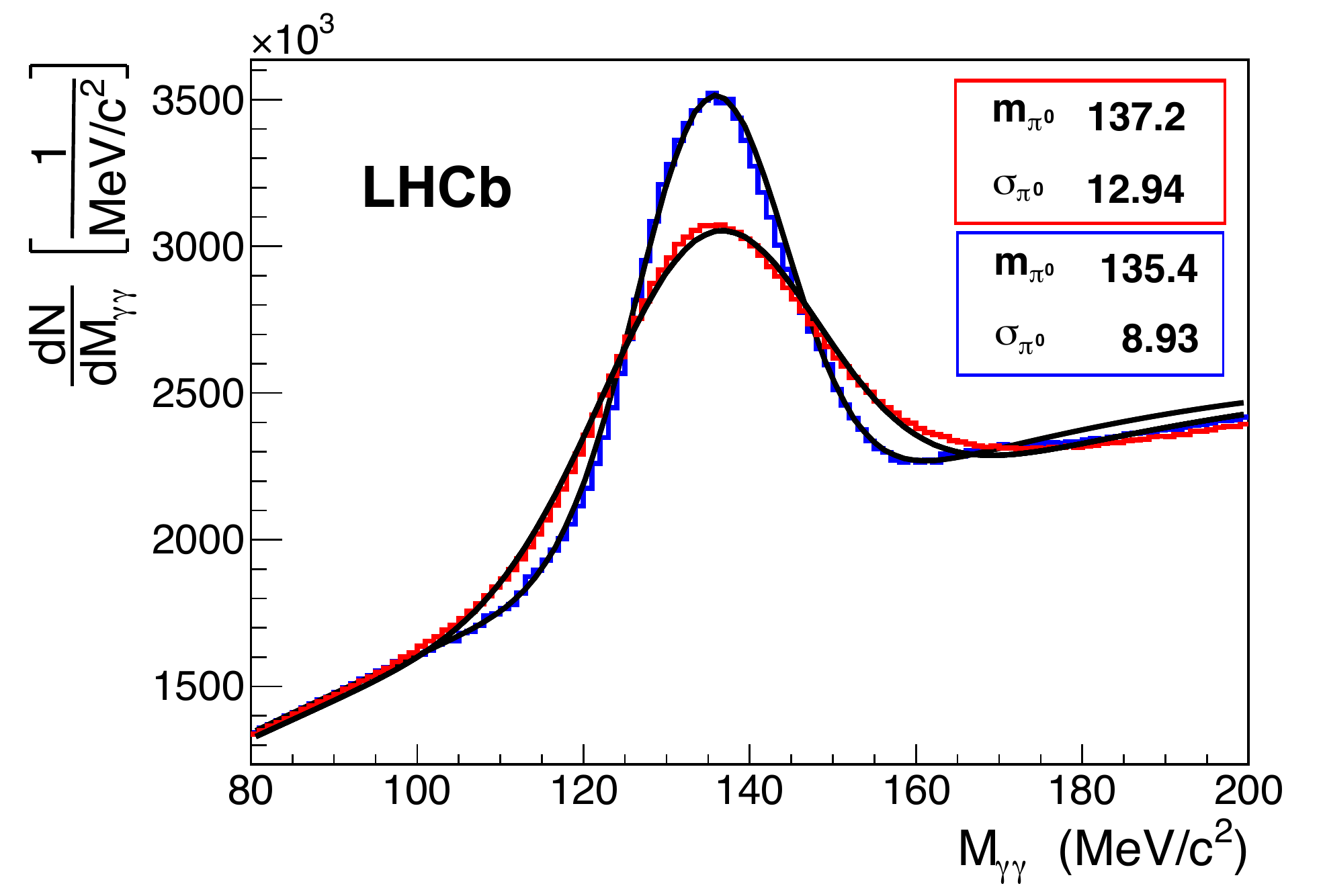}
\end{center}
\caption{\label{fig:pi0mass} Invariant mass distribution for
  $\piz\to\gamma\gamma$ candidates upon which the fine calibration
  algorithm is applied. The red curve corresponds to the distribution
  before applying the method, while the blue curve is the final
  one. Values in the red (blue) box are the mean and sigma of the
  signal peak distribution in \mevcc before (after) applying the fine
  calibration method.}
\end{figure}

The calibration of the ECAL requires two main steps: the first to set
an initial calibration, and the second to refine the values through an
iterative procedure. The first step was performed at the beginning of
LHCb commissioning by using test-beam measurements to reproduce the
design energy range. With these settings, a $\pi^0\to\gamma\gamma$
mass resolution at the level of 10\% was achieved for the first
collisions in 2009. In subsequent years, initial settings were obtained
using the LED system, achieving a similar accuracy of 8--10\%. In the
second step, initial calibration parameters are refined by studying
the energy deposited over many events and requiring continuity across
cell boundaries ~\cite{thesis_martens}.  
The calibration constants are improved by performing fits to the invariant
mass distribution of  
$\piz\to\gamma\gamma$ decays~\cite{CHEP-XCAL}, combining a photon
hitting the cell to be calibrated with another reconstructed
photon. The procedure is repeated until all coefficients are
stable. Figure~\ref{fig:pi0mass} shows the change in the fitted  \piz
invariant mass distribution, before and after the calibration
procedure. By applying the method to a sample of miscalibrated
simulated events, the final precision of the cell-to-cell
intercalibration is estimated to be approximately 2\%. In order to
apply the \piz mass fit to every cell in the ECAL, several hundred
thousand events are required.

The calibration of the HCAL uses two $^{137}{\rm Cs}$ sources of 
$\sim 10$\,{mCi}, one per detector side. 
This procedure takes about an hour during which
they are transported through all of the scintillator cells by a
hydraulic system. The response of the PMTs is measured by a dedicated
system of current integrators.  The relationship between the
integrated anode current and the particle energy was measured in test-beam 
and is used to set the values for the HCAL parameters, obtaining
a cell-to-cell intercalibration at the level of 5\%. 
The use of the \textit{in-situ} source limits the calibration procedure 
of the HCAL to technical stops, which occur bi-monthly.

The performance of the ECAL and the HCAL is monitored during the
data-taking periods using the built-in LED system: the response to the
LED flashes is found to agree well with the response to actual
particles. In addition, the distribution of \eOverP, for electrons in
the case of the ECAL and hadrons in the case of the HCAL, can be
compared to simulations and is used for monitoring purposes.

\begin{figure}[!tb]
\begin{center}
\includegraphics[width=1.0\textwidth]{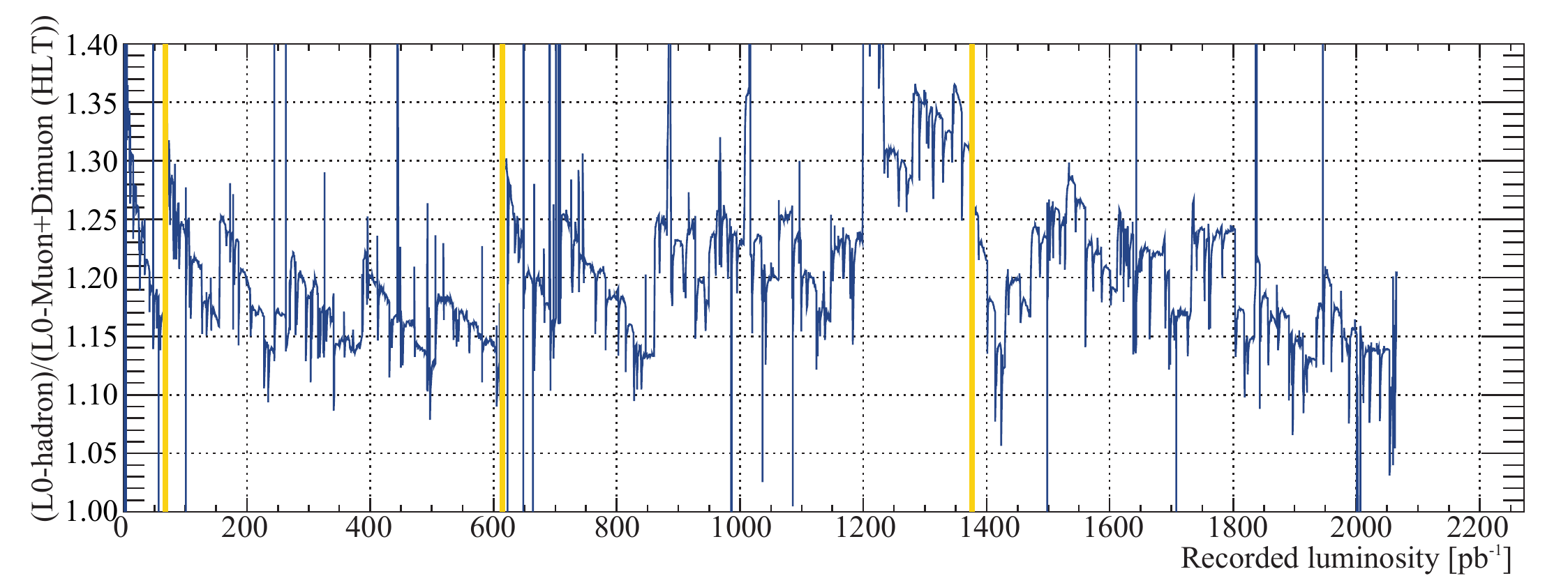}
\caption{\label{fig:l0trigger} Ratio between the rate of events
  triggered by the L0-hadron trigger, based on the HCAL, and
  Muon-based triggers.  Distinct increases in rate, \eg as at a
  recorded luminosity of around 50\invpb, correspond to the
application of a new set of PMT gains.}
\end{center}
\end{figure}

While no significant degradation of the SPD and the PS performance has
been seen, ageing has been observed in both the ECAL and the HCAL. 
The main cause of this is a decrease of
the PMT gains due to the degradation of the dynode system at high
integrated anode current. The gain losses are compensated by
increasing the voltage between the PMT dynodes.
The time period over which the gradual gain
reduction takes place is shorter than the interval between which
absolute calibrations can be performed for the ECAL (using $\piz$ mass
reconstruction) and the HCAL (using the $^{137}{\rm Cs}$ sources).
Therefore, the relative corrections required to maintain performance
are estimated using dedicated procedures.
For the ECAL, the \eOverP ratio is used whilst the HCAL is monitored using the
LED pulse system. The change in performance
with time is illustrated in Figure~\ref{fig:l0trigger}, which shows the
ratio between the L0-hadron trigger rate, based on the HCAL hardware, 
and a combination of muon-based triggers. As a result of the ageing, 
a gradual reduction of the rate is observed, with intermittent increases,
corresponding to the application of a new set of PMT gains.

\subsection{Selection of neutral energy deposits in ECAL}
\label{subsec:neutralreco}

ECAL cells with energy deposits are grouped together to form clusters
by applying a $3\times3$ cell pattern around local energy deposition
maxima.  Consequently, the centres of the reconstructed clusters are
always separated by at least one cell.  If a given cell is present in
more than one cluster, its energy is redistributed between the
clusters under consideration according to the total energies of the
clusters.  Although this process is iterative, it converges rapidly
because the effective Moli\`ere radius (3.5\cm) of the ECAL is smaller
than the size of the cells, which have lengths of side of 4.04\cm,
6.06\cm and 12.12\cm for the inner, middle and outer regions,
respectively.  Each cluster is characterised by its energy-weighted
moments up to second order, namely the total energy, the
energy-weighted position and the two-dimensional energy spread matrix.

Clusters corresponding to energy deposits of neutral particles are
identified as those without an associated charged track. This is done
using the procedure summarised below. First, all reconstructed tracks
in the event are extrapolated to the calorimeter. Next, all pairwise
combinations of extrapolated tracks and reconstructed clusters are
formed. The matching between tracks and clusters is evaluated using
the $\chi^2_\text{2D}$ metric,

\begin{equation}
\chi^2_\text{2D} \; = \left (\vec{r}_\text{tr} - \vec{r}_\text{cl}\right)^T \;
  \left( {\cal C}_\text{tr} + {\cal S}_\text{cl}\right)^{-1} \;
  \left(\vec{r}_\text{tr} - \vec{r}_\text{cl}\right),
\label{eq:chi2calo}
\end{equation}

where $\vec{r}_\text{tr}$ and $\vec{r}_\text{cl}$ represent the local
coordinates of tracks and clusters, respectively, at the $z$
barycentres of clusters, ${\cal C}_\text{tr}$ is the covariance matrix
of the $\vec{r}_\text{tr}$, and ${\cal S}_\text{cl}$ is the cluster
energy spread matrix. The $z$ barycentre of a cluster is the average
energy-weighted position of clusters in $z$, corrected assuming
logarithmic energy dependence.  A cluster generated by a neutral
particle is considered to be isolated, and hence a photon candidate,
if it has a minimum value of $\chi^2_\text{2D}$ with respect to any
extrapolated track of at least 4. This cut significantly suppresses the clusters
due to other charged particles while keeping high efficiency for photons
\cite{Olivier}.  

\subsection{Photon reconstruction}

The photon energy is determined from the total cluster energy in the
ECAL and the reconstructed energy deposit in
the PS~\cite{Olivier}. The photon direction is derived from an assumed
origin for the photon and the energy-weighted position of the photon
candidate: the transverse profile is corrected for the spread of the
cluster, and the $z$ barycentre calculated as for the $\chi^2_\text{2D}$
matching.

\begin{figure}[!tb]
  \vspace{4mm}
  \begin{center}
  \includegraphics[width = 0.65\textwidth]{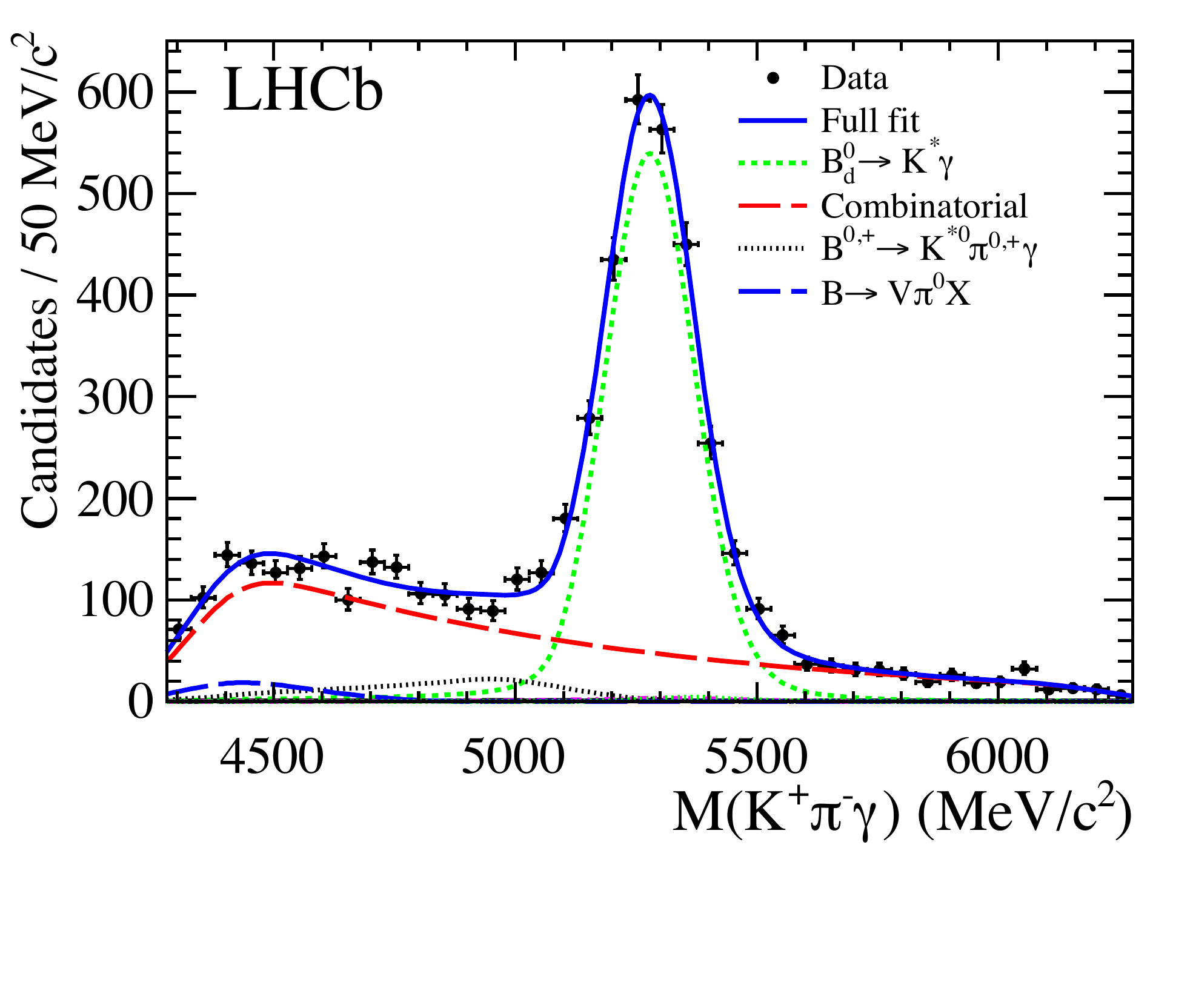}
  \vspace{-1cm} 
    \caption{Mass distribution of reconstructed 
      $B^0\to K^{*0}(K^+ \pi^-)\gamma$ candidates obtained in the 2011 data
      sample. The blue curve corresponds to the mass shape fit. The
      $K^{*0}\gamma$ signal (green dotted line) and the various background
      contaminations are
      shown~\protect\cite{LHCb-PAPER-2012-019}. \label{Kstg} }
  \end{center}
  \vspace{4mm}
\end{figure}

The performance of high-energy photon reconstruction is illustrated by
the reconstructed $B^0\to K^{*0}\gamma$ mass distribution shown in
Figure~\ref{Kstg}. The mass resolution obtained for this radiative decay
is dominated by the ECAL energy resolution and is found to be
93\mevcc~\cite{LHCb-PAPER-2012-019}.  A comparison of the data with
simulated samples shows that this corresponds to an accuracy of the
cell-to-cell intercalibration of around 2\%.

\subsection{Neutral pion reconstruction}

Neutral pions with low transverse momenta are mostly reconstructed as
pairs of well-separated photons (resolved \piz candidates). A mass
resolution of 8\mevcc is obtained for such neutral pions.  However,
due to the finite ECAL granularity, photon pairs from the decay of
sufficiently high momentum \piz{} cannot be resolved as individual
clusters. This essentially holds for all \piz{} meson decays with
transverse momentum above 2\gevc. To reconstruct such `merged' \piz{}
candidates, a procedure has been designed to identify overlapping
clusters.  The algorithm consists of splitting each single ECAL
cluster into two $3\times3$ subclusters built around the two highest
energy deposits of the original cluster.  The energy of the common
cells is then distributed between the two assumed subclusters by
fitting the energy distribution with that of two photons, using the
expected transverse profile obtained from simulations.  Since the
position of the two subcluster barycentres is a function of the energy
distribution, this procedure requires an iterative process.

\begin{figure}[!tb]
  \vspace{4mm}
  \begin{center}
    \resizebox{\textwidth}{!}{
      \includegraphics[width = 0.3\textwidth]{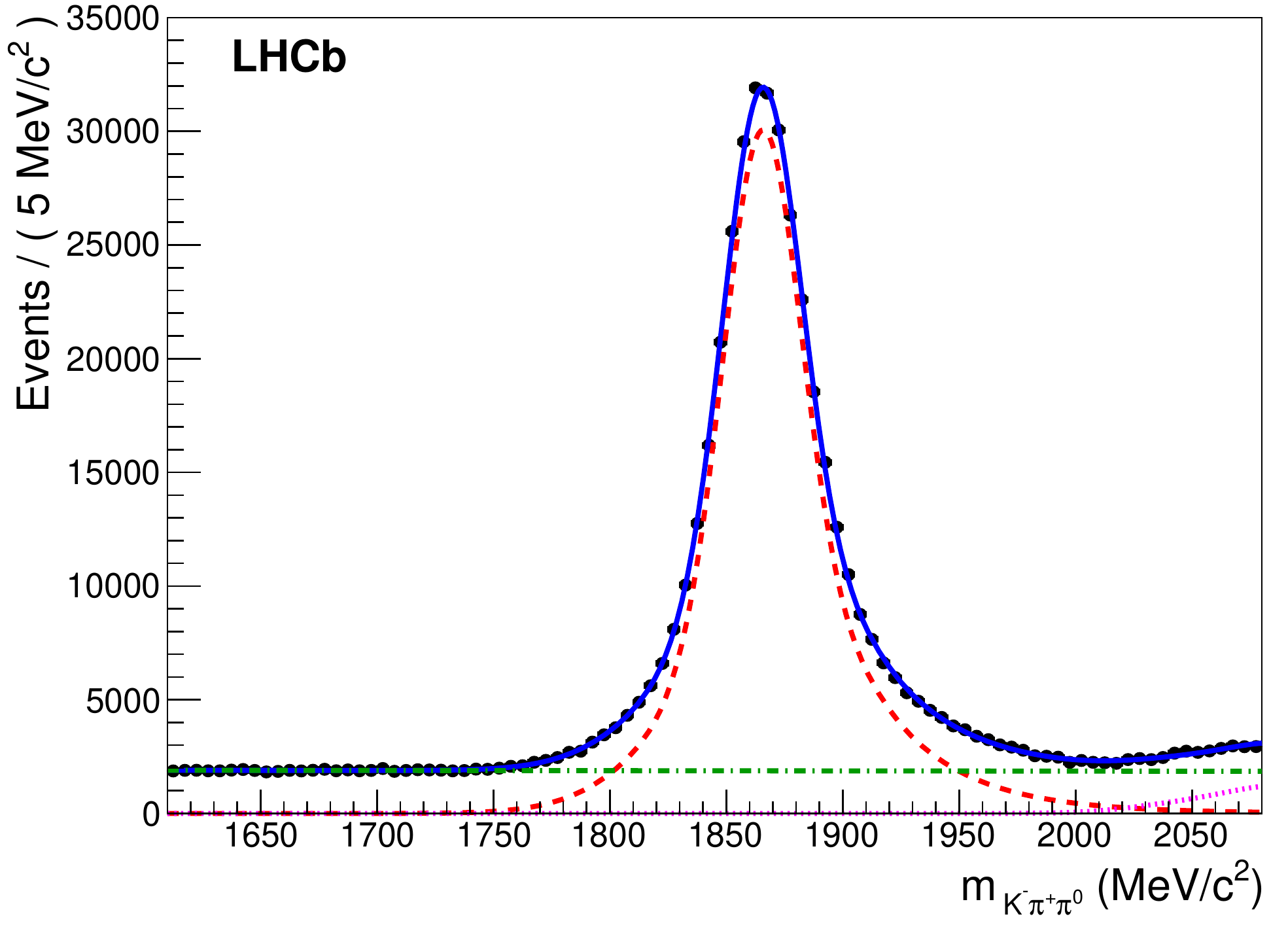}
      \includegraphics[width = 0.3\textwidth]{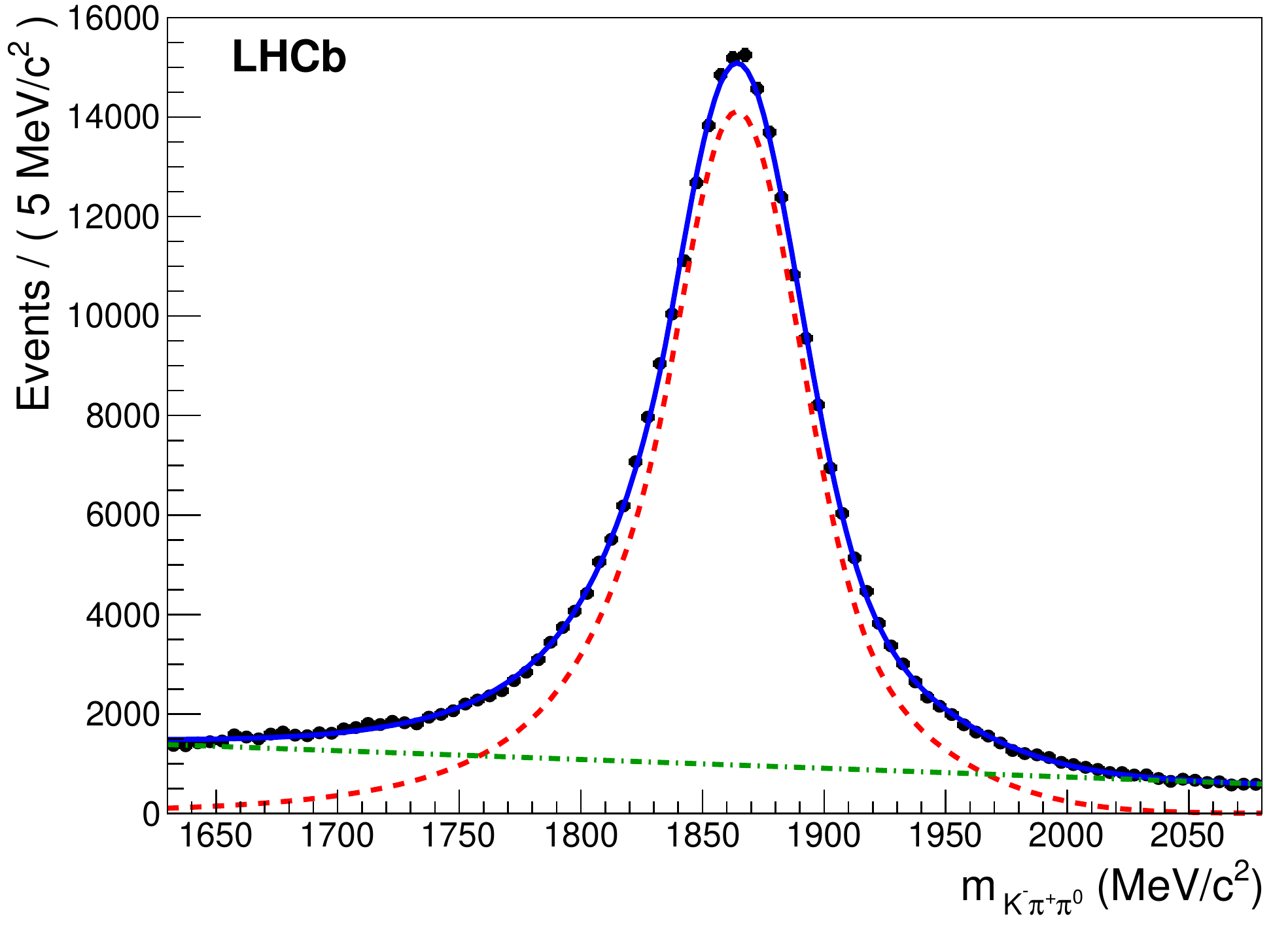}
    }
    \caption{Mass distributions of reconstructed {$D^0\to K^-\pi^+
        \pi^0$} candidates with resolved $\pi^0$ (left) and merged
      $\pi^0$ (right). Both are obtained from the 2011 data
      sample. The overall mass fit~\protect\cite{DiegoThesis} is
      represented by the blue curve, with the signal (red dashed line)
      and background (green dash-dotted line and purple dotted lines)
      contributions also shown.\label{Kpipi0} }
  \end{center}
  \vspace{4mm}
\end{figure}

The performance of neutral pion reconstruction is illustrated in
Figure~\ref{Kpipi0}, which shows the invariant mass distribution for
$D^0\to K^-\pi^+\piz$ candidates for resolved and merged \piz{}
candidates~\cite{DiegoThesis}. In this example, the estimated
invariant mass resolution is \unit[30]{\mevc} for the merged $\piz$
candidates and \unit[20]{\mevc} for the resolved ones.

\section{Particle identification}
\label{sec:charged-particle-id}

Particle identification in LHCb is provided by four different
detectors: the calorimeter system, the two RICH detectors and the muon
stations.  In the following sections the performance of the individual
sub-systems is presented first, followed by a description of the
methods used to combine the information for charged particles in a
single set of variables that provide optimal particle identification
performance.

\subsection{Calorimeter system based particle identification}

The main role of the calorimeters in terms of particle identification
is to provide for the recognition of photons, electrons and
\piz{} candidates. 
Distinguishing charged from neutral particles is performed by
studying the presence or absence of tracks in front of the energy
deposits using the techniques described in
Section~\ref{subsec:neutralreco}. For energy deposits related to
neutral particles, the shape of the cluster is used to distinguish
between photons and \piz{} candidates.
The photon hypothesis is established by
taking into account the possibility that photons convert when
interacting with the detector material upstream of the
calorimeter. When an energy deposit corresponds to a charged particle,
the electron hypothesis is constructed to distinguish electrons from
hadrons. Outline descriptions of how the photon and electron
hypotheses are built are given below.

\subsubsection[Photon and merged $\pi^0$ identification]
{Photon and merged $\protect\mathbf{\pi^0}$ identification}

\begin{figure}[!tb]
\begin{center}
\includegraphics[width=1.0\textwidth]{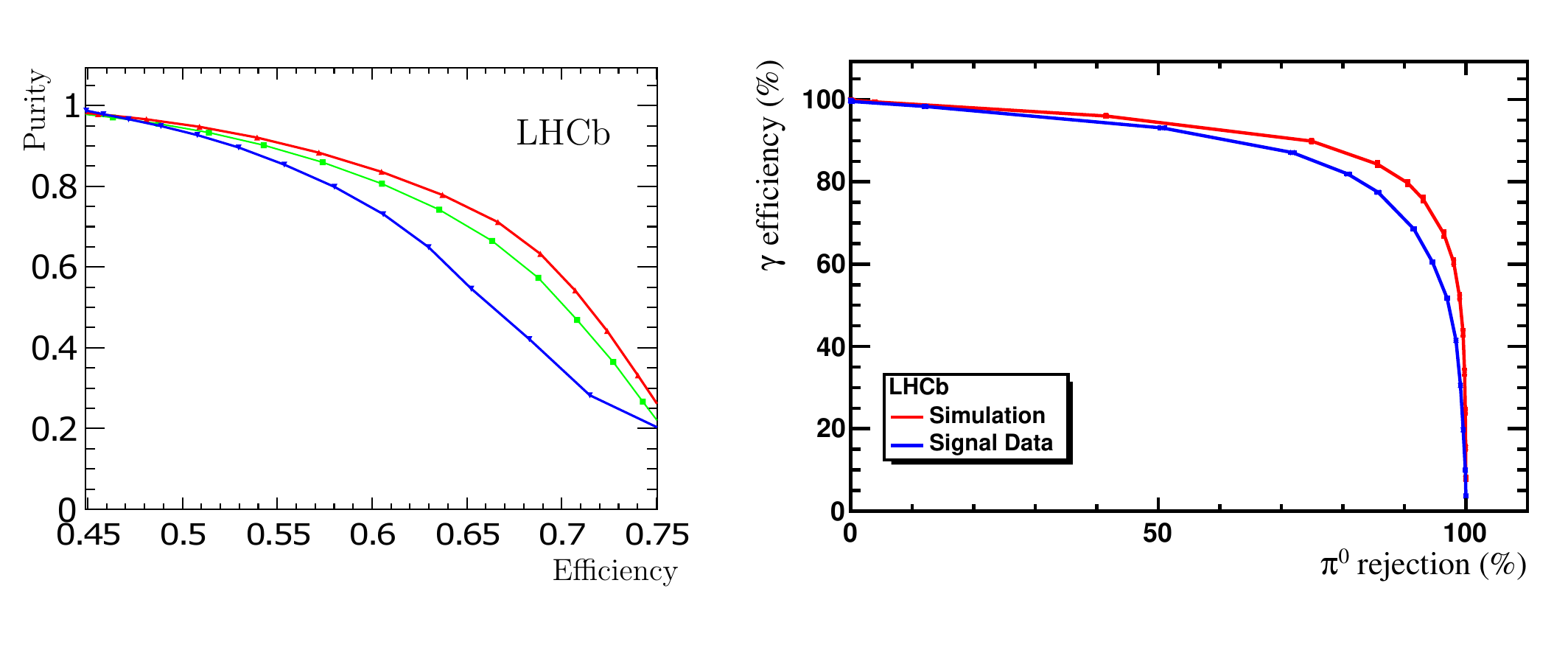}
\end{center}
\vspace{-7mm}
\caption{\label{fig:pidphoton} Performance of the photon
  identification. Purity as a function of efficiency for (green)
  the full photon candidate sample, (blue) converted candidates
  according to the SPD information and (red) non-converted
  candidates (left). Photon identification efficiency as a function of
  \piz rejection efficiency for the $\gamma-\piz$ separation tool for
  simulation, the red curve, and data, the blue curve (right).}
\end{figure}

Two independent estimators are built to establish the photon
hypothesis, one each for the converted and non-converted
candidates. Non-converted photons are identified by computing a photon
hypothesis likelihood from the signal and background probability
density functions of several variables, namely the PS energy deposited
in front of the ECAL cluster cells, the matching estimator
$\chi^2_\text{2D}$ between the cluster and any track defined for a
charged particle, and the ratio between the energy of the central cell
of the ECAL cluster and the total ECAL energy. Because the
non-converted photon estimator depends on the energy and the
calorimeter zone, several probability density functions are
constructed from simulations, for both signal and background,
corresponding to each of the zones. The difference in log-likelihood
between the photon and the background hypotheses ($\Delta \log {\cal
  L}$) is calculated and used to identify
photons. Figure~\ref{fig:pidphoton} shows the performance of the
photon identification in terms of the efficiency and purity obtained 
for candidates with $\pt > 200$\mevc. 

To avoid the misidentification of photons with high-$E_{\rm T}$ merged
\piz candidates, the difference between the distribution of the expected energy
deposit of a photon with respect to that of a \piz is used. This difference is
evaluated by a neural network classifier (specifically, a multi-layer
perceptron) trained with photons from a simulated $\Bz\to\Kstarz\gamma$ sample
as signal and \piz mesons from a mixture of $B$ decays as background. These \piz
mesons are reconstructed and selected as photons using the same
$\Bz\to\Kstarz\gamma$ preselection used for the signal sample. A photon
identification efficiency of 95\% can be obtained while rejecting 45\% of the
merged \piz meson background that are reconstructed as
photons. Figure~\ref{fig:pidphoton} shows the photon identification efficiency
with respect to the \piz rejection efficiency for simulation and data.

Photons converted before the magnet are reconstructed from
electron-positron tracks.  The electron (positron) is selected on the
basis of its electron PID variables and electron confidence level,
requiring a minimum \pt value and an \eOverP value within a
selected range. The algorithm only combines electron-positron pairs
for which the associated clusters have energy-weighted positions that
are closer than 3$\sigma$ of cluster extent (and 200\mm) in the
vertical plane, at the average $z$ barycentre of clusters.  Pairs are
selected on the basis of their transverse momenta, their di-electron
masses and their reconstructed vertex positions.  The electron energy
is corrected by including any bremsstrahlung photons measured by the
calorimeters that are compatible with the electron-positron pair.

\begin{figure}[bt]
\begin{center}
\includegraphics[width=0.6\textwidth]{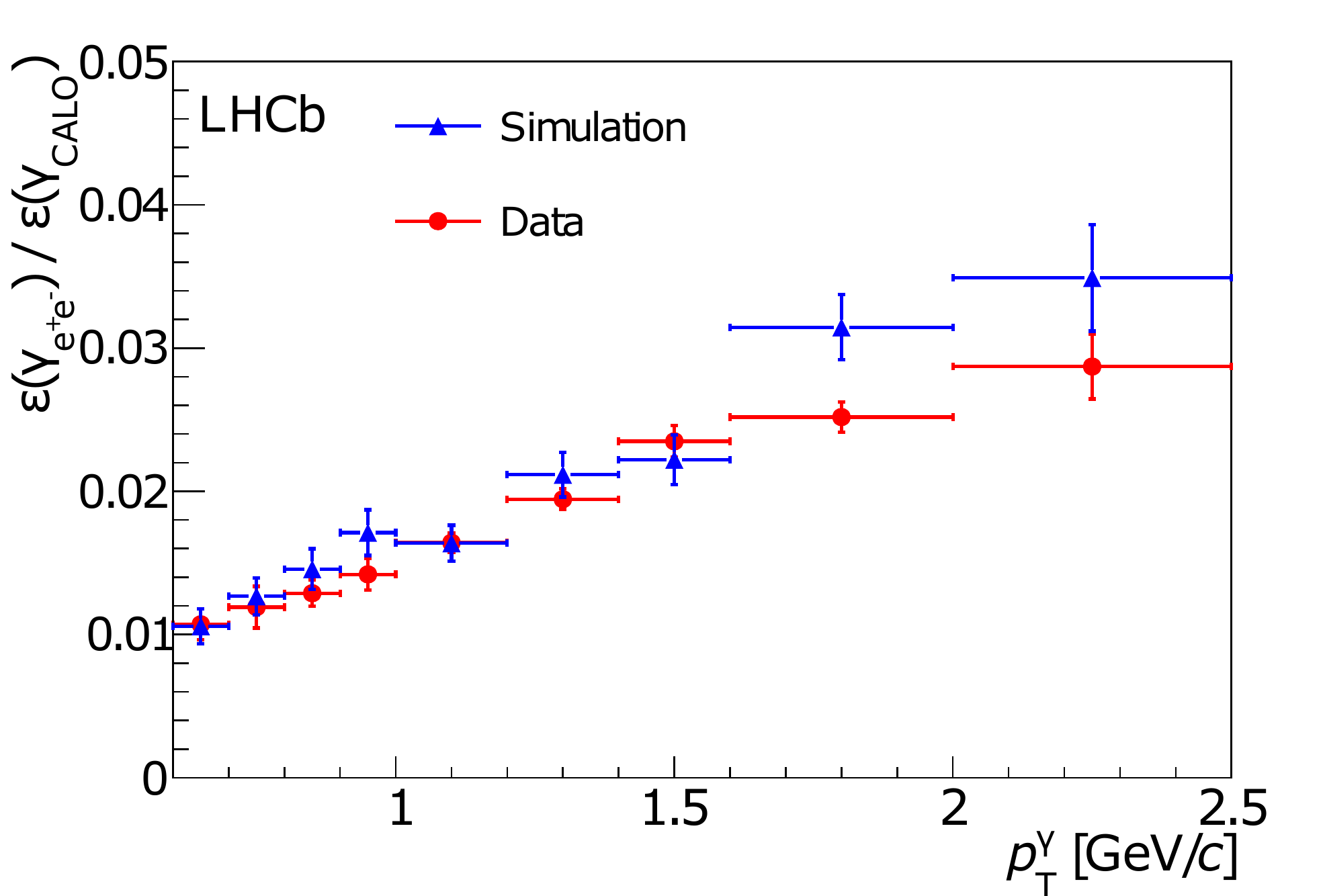}
\caption{Ratio of photon detection efficiencies $\epsilon (\gamma \to
  e^+e^-)/\epsilon(\gamma_\text{CALO})$ from the decay of \piz mesons in
  data (red) and simulations (blue).}
\label{pi0converted}
\end{center}
\end{figure}

Figure~\ref{pi0converted} shows the ratio of photon detection
efficiencies between converted and non-converted photons coming from
the decay of \piz mesons for both simulation and data.  The simulation
provides a good description of the photon reconstruction efficiency
implying that the detector material where the conversions occur is modelled
well, and that the reconstruction algorithms work equally well in data and
simulation. The level of performance is illustrated by analyses that benefit
from the good resolution obtained using converted photons, such as $\chi_c \to
\jpsi\gamma$~\cite{LHCb-PAPER-2013-028} or $\chi_b \to
\Upsilon\gamma$~\cite{LHCb-PAPER-2012-015}. In the case of the $\chi_c$, for
instance, the resolution on the mass difference $\Delta M =
M(\mu^+\mu^-\gamma) - M(\mu^+\mu^-)$ is about $5$\mevcc. With this
resolution, the $\chi_{c0}$, $\chi_{c1}$ and $\chi_{c2}$ states can be
disentangled from one another~\cite{LHCb-PAPER-2013-028}.

\subsubsection{Electron identification}
\label{sec:CALOPID}

\begin{figure}[!tb]
\vspace{3mm}
\begin{center}
\includegraphics[width=0.6\textwidth]{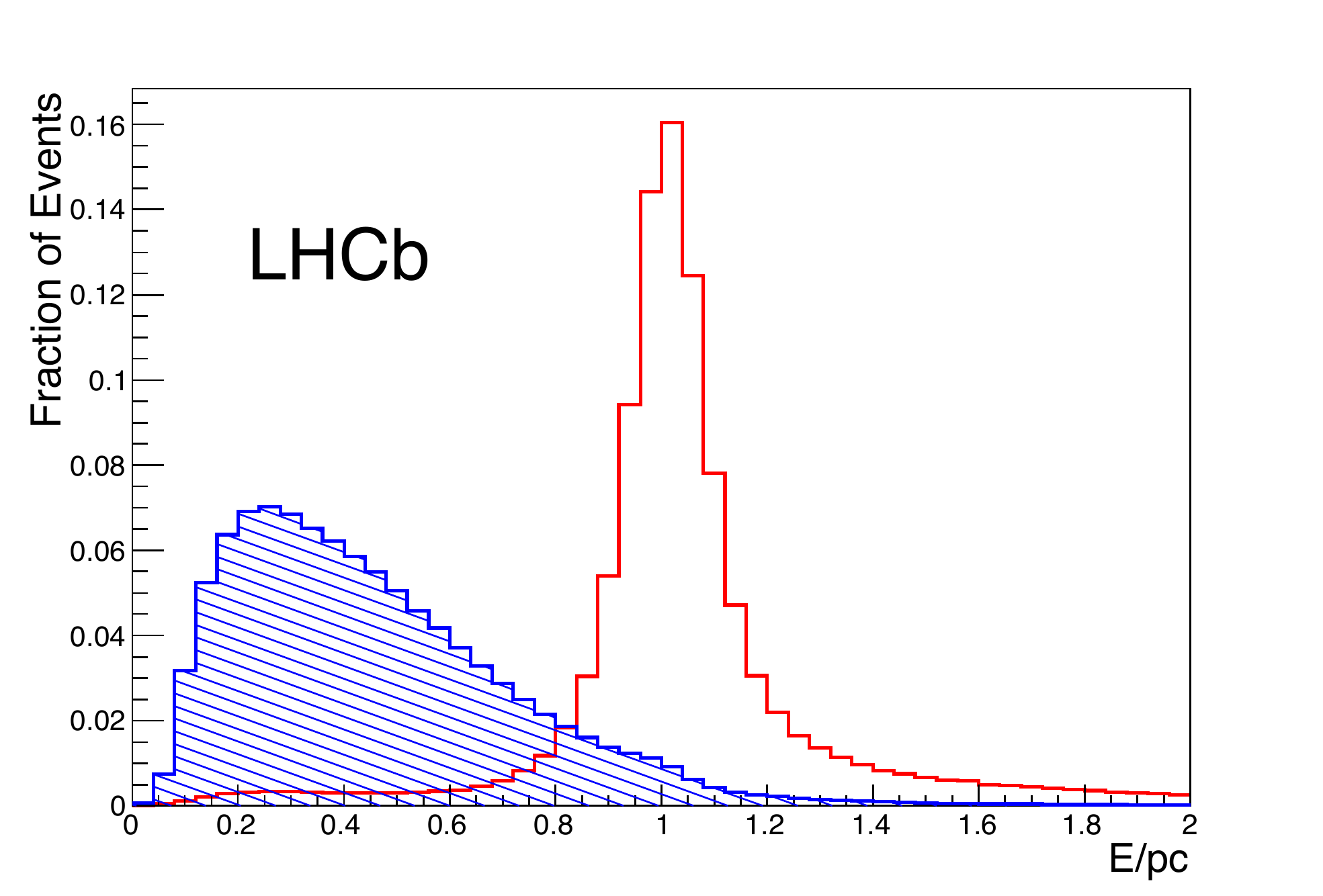}
\caption{Distribution for the ECAL of \eOverP for electrons (red) and hadrons
  (blue), as obtained from the first 340 \invpb recorded in 2011.}  
\label{fig:e2p}
\end{center}
\vspace{3mm}
\end{figure}

The identification of electrons in the calorimeter system uses
information derived from the ECAL, the PS and the HCAL. The procedure
to combine these different sources of information is based on signal
and background likelihood distributions constructed for each
sub-detector. In each case, reference histograms correlating the
energy measurement with the particle momentum are produced. For
example, Figure~\ref{fig:e2p} shows the \eOverP distribution in the ECAL
for electrons and hadrons, produced using the first 340\invpb recorded
in 2011. The electron distribution has been produced using
reconstructed electrons from photon conversions and the hadron
distribution using pions and kaons from $D^0$ meson decays.
From these distributions, the log-likelihood difference between electrons and
hadrons is derived.

For the ECAL, the log-likelihood difference for electron and hadron
hypotheses $\deltaLLecal$ is computed based on both \eOverP and the
$\chi^2_\text{2D}$ estimator defined in Section~\ref{subsec:neutralreco}. The
electron hypothesis likelihoods for the PS, $\deltaLLps$ and the HCAL
$\deltaLLhcal$ are built using the energy deposits in each sub-detector. A
combined estimator is then formed for the calorimeter system by taking the sum
of the individual estimators from the PS, the ECAL and the HCAL,

\begin{equation}
  \deltaLLcalo = \deltaLLecal + \deltaLLhcal + \deltaLLps \;\;\;.
\end{equation}

\begin{figure}[!tb]
 \vspace{3mm}
 \begin{center}
 \includegraphics[width=0.5\textwidth]{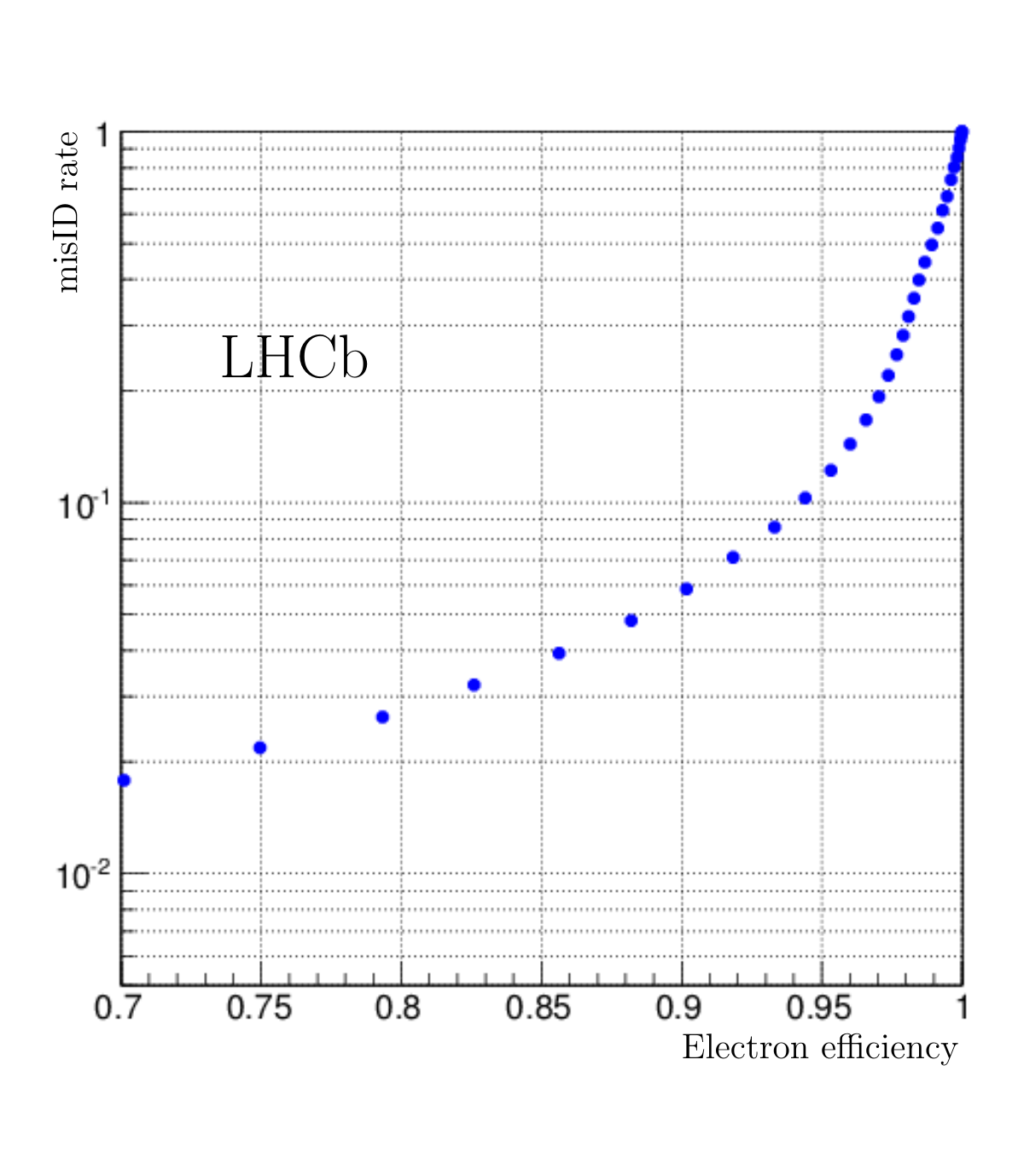}
 \caption{Electron identification efficiency versus misidentification rate.}
 \label{fig:CaloEffVMisID}
 \end{center}
 \vspace{3mm}
\end{figure}

Figure~\ref{fig:CaloEffVMisID} shows the combined electron identification
efficiency defined above versus the misidentification rate obtained by varying
the selection criteria applied to the likelihood difference.

The electron identification performance is evaluated using the data
recorded in 2011, which are sufficient for it to be measured using a
tag-and-probe method. This is applied to $B^{\pm} \rightarrow
J/\psi K^{\pm}$ candidates with $J/\psi \rightarrow e^+ e^-$, where
one of the electrons is required to be identified by its electron ID
($e_\text{tag}$) while the second electron is selected without using
any information from the calorimeter system ($e_\text{probe}$). This
second electron is then used to estimate the efficiency of the electron ID.

\begin{figure}[!tb]
\begin{center}
\includegraphics[width=\textwidth]{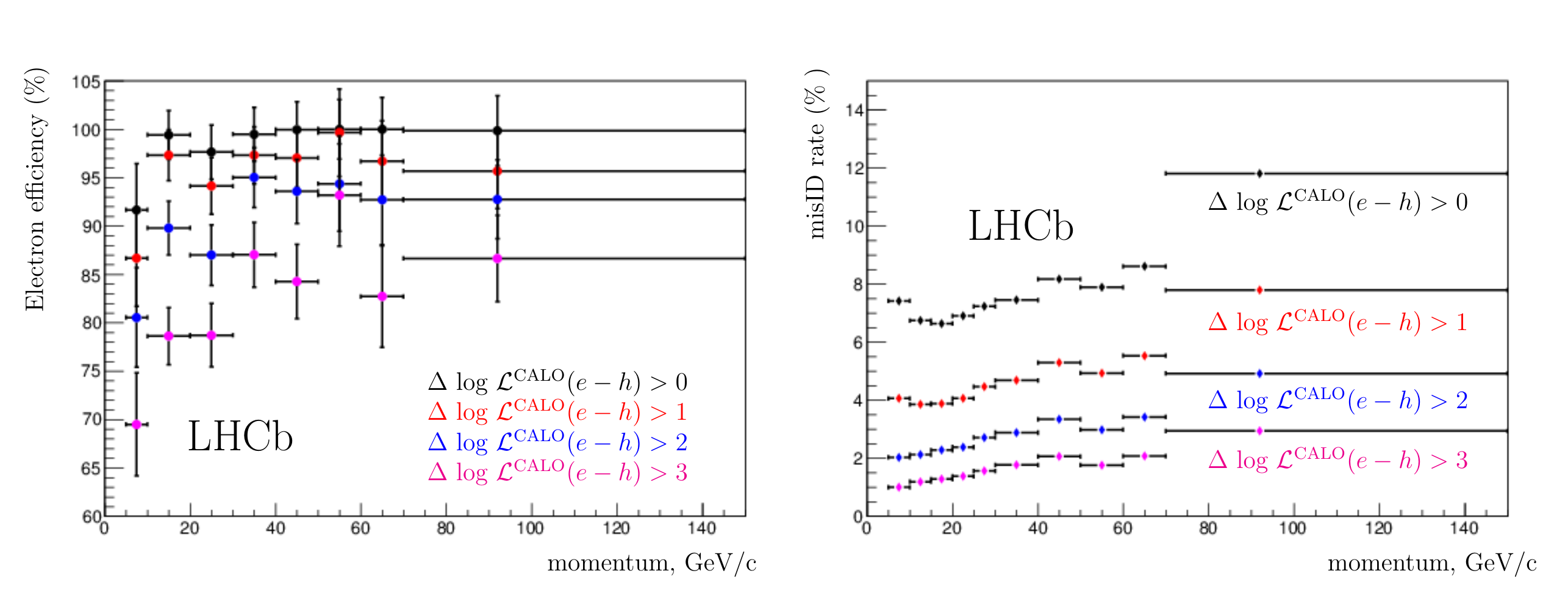}
 \caption{\small Electron identification performances for various
   $\Delta$ log ${\mathcal L}^\text{CALO}(e-h)$ cuts: electron
   efficiency (left) and misidentification rate (right) as functions
   of the track momentum.
 \label{perf_calo} }
 \end{center}
\end{figure}

The efficiency and the misidentification rate as a function of the
$e_\text{probe}$ momentum are presented in Figure~\ref{perf_calo} for
several cuts on $\deltaLLcalo$. The
electron identification efficiency is observed to be lower for $p <
10$ GeV/c. As expected, the higher momenta particles have higher
misidentification rates as illustrated in Figure~\ref{perf_calo}.
To quantify the typical identification performance of the
entire calorimeter system, the average identification efficiency of
electrons from the $J/\psi \rightarrow e^+ e^-$ decay in $B^{\pm}
\rightarrow J/\psi K^{\pm}$ events is $(91.9 \pm 1.3)\%$ for a
misidentification rate of $(4.54 \pm 0.02)\%$ after requiring 
$\deltaLLcalo > 2$.

\subsection{RICH system based particle identification}
\label{sec:rich-pid}

The primary role of the RICH system is the identification of charged hadrons
($\pi$, $K$, $p$). The information provided is used both at the final analysis
level, and as part of the software trigger (see Section~\ref{sec:trigger}).
In addition, the RICH system can contribute to the identification of charged
leptons ($e$, $\mu$), complementing information from the calorimeter and muon
systems, respectively. 

\subsubsection{Cherenkov angle resolution}
\label{sec:rich:AngularResolution}

\begin{figure}[!tb]
  \begin{center}
    \includegraphics[width=0.49  \textwidth]{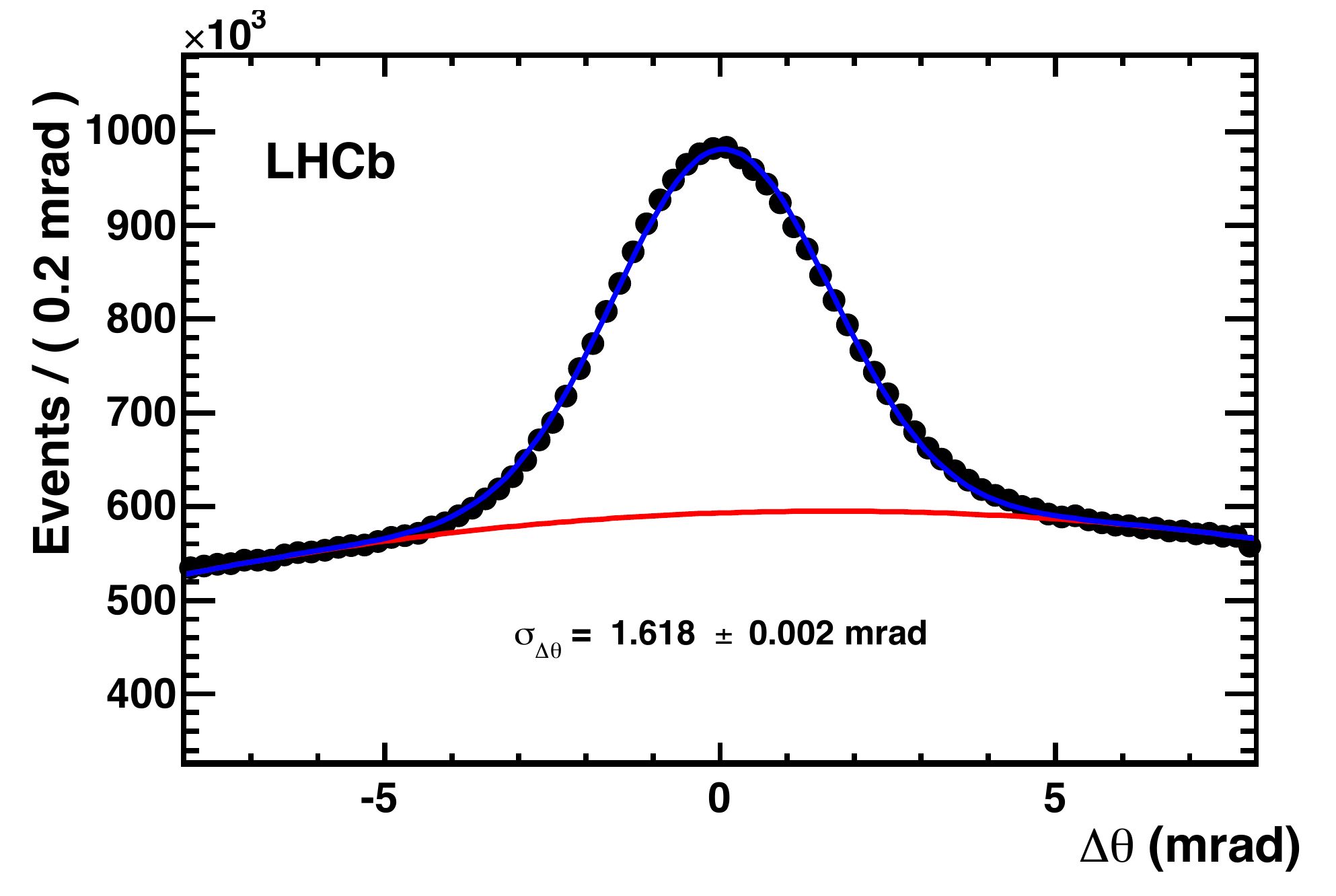}
    \includegraphics[width=0.495 \textwidth]{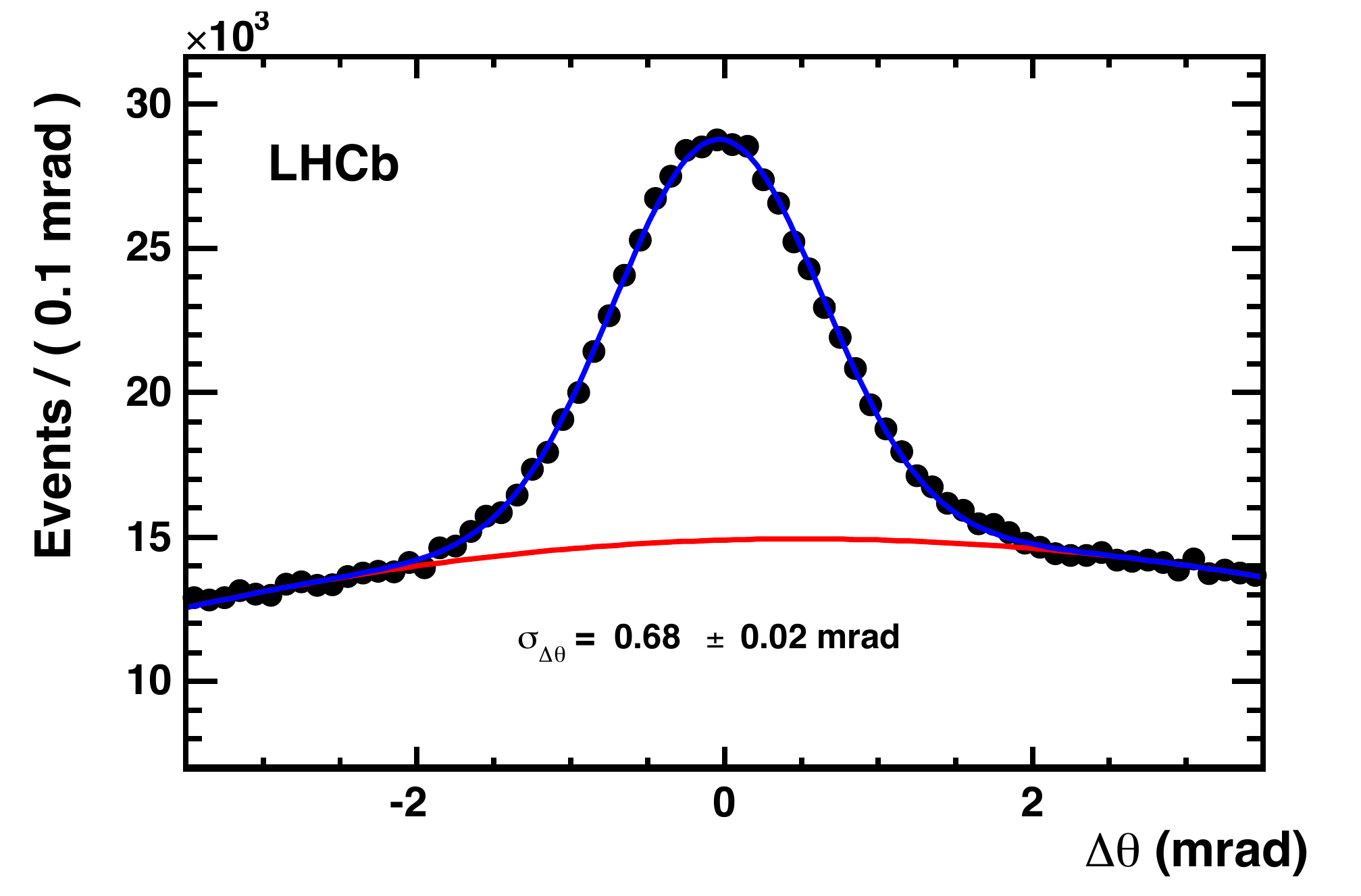}
    \includegraphics[width=0.45  \textwidth]{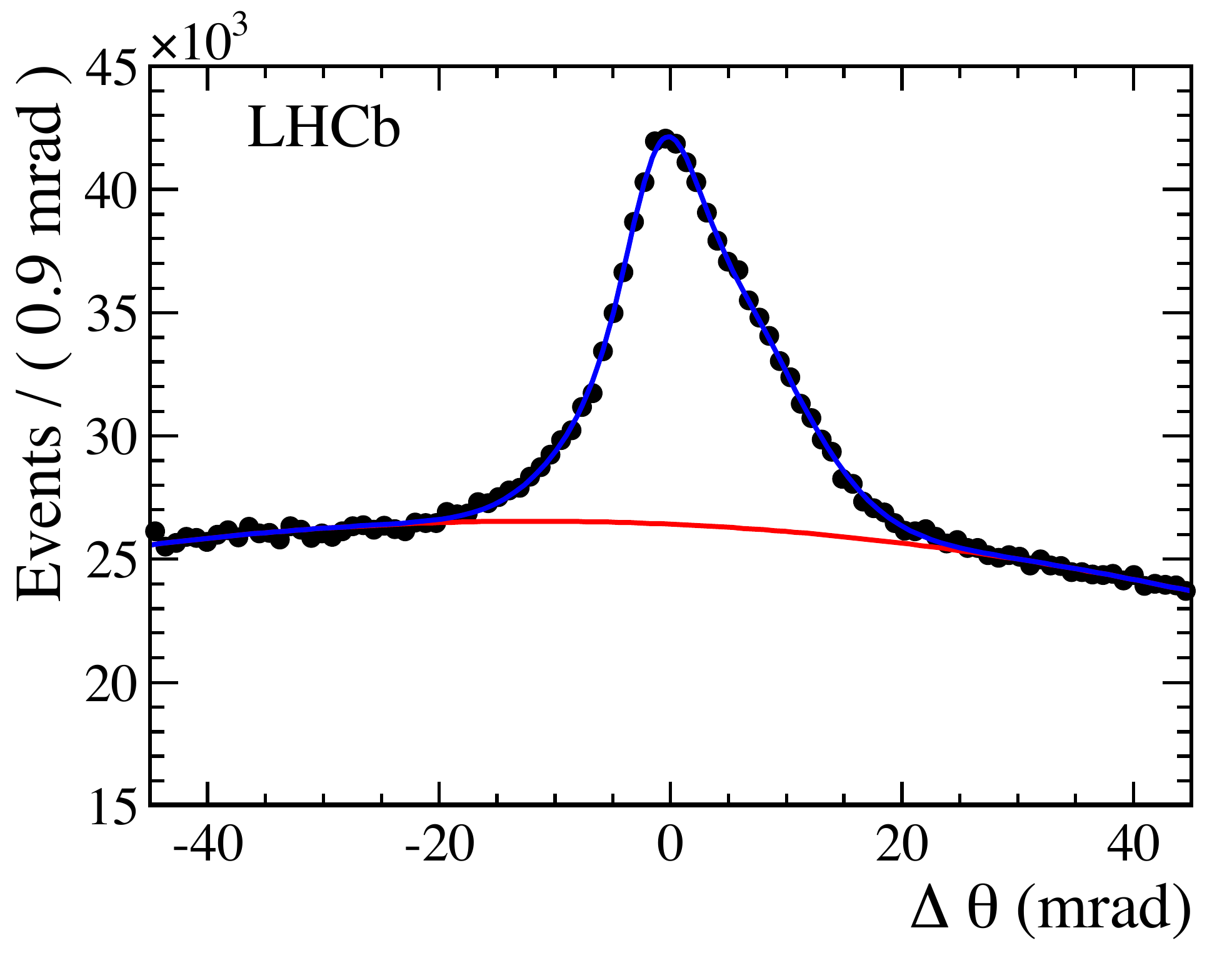}
    \caption{ \DeltaThetaC distributions for the RICH\,1 gas (top left), RICH\,2
      gas (top right) and Aerogel
      (bottom)\protect\cite{LHCb-DP-2012-003}. }
    \label{fig:rich:angle_res}
  \end{center}
\end{figure}

One of the primary measures of the RICH performance is \SigmaThetaC,
the resolution of the Cherenkov angle with which
the photons, radiated from the particles as they traverse the various
radiator volumes, can be reconstructed.
The distributions for \DeltaThetaC, the difference between the reconstructed
and expected photon Cherenkov angles, are shown in
Figure~\ref{fig:rich:angle_res} for 2011 data,
after all detector alignment and calibration procedures
have been performed \cite{LHCb-DP-2012-003}. 
The expected Cherenkov angles for each track are calculated using reconstructed
momenta and radiator refractive index information. Only high-momentum
tracks are selected, to ensure that the Cherenkov angle is close to saturation.

The values of \SigmaThetaC, extracted from a simple fit to the \DeltaThetaC
distributions, are determined to be   
$1.618 \pm 0.002$\mrad for RICH1 gas (\cfourften) and $0.68 \pm 0.02$\mrad for
RICH2 (\cffour),
comparable with the expectations from simulation of $1.52 \pm 0.02$\mrad and
$0.68 \pm 0.01$\mrad respectively. 
The disagreement seen between data and simulation for \cfourften
are largely attributed to imperfect corrections for distortions in the RICH
photon detector images caused by the residual magnetic field in the vicinity of
the RICH1 detector. Enhancements to the procedures used to compute these
corrections are foreseen for Run II, thus improving the resolutions achieved in
data. 

For the RICH1 aerogel radiator, where the distribution is not symmetric, the
standard deviation is estimated to be $5.6$\mrad. This value is about a factor
of 1.8 larger than the expectation from simulation. This discrepancy is, at
least partially, explained by the unmodelled absorption of \cfourften gas by the
very porous aerogel radiator, with which it is in contact.

\begin{figure}[!tb]
\begin{center}
\includegraphics[width=0.85\textwidth]{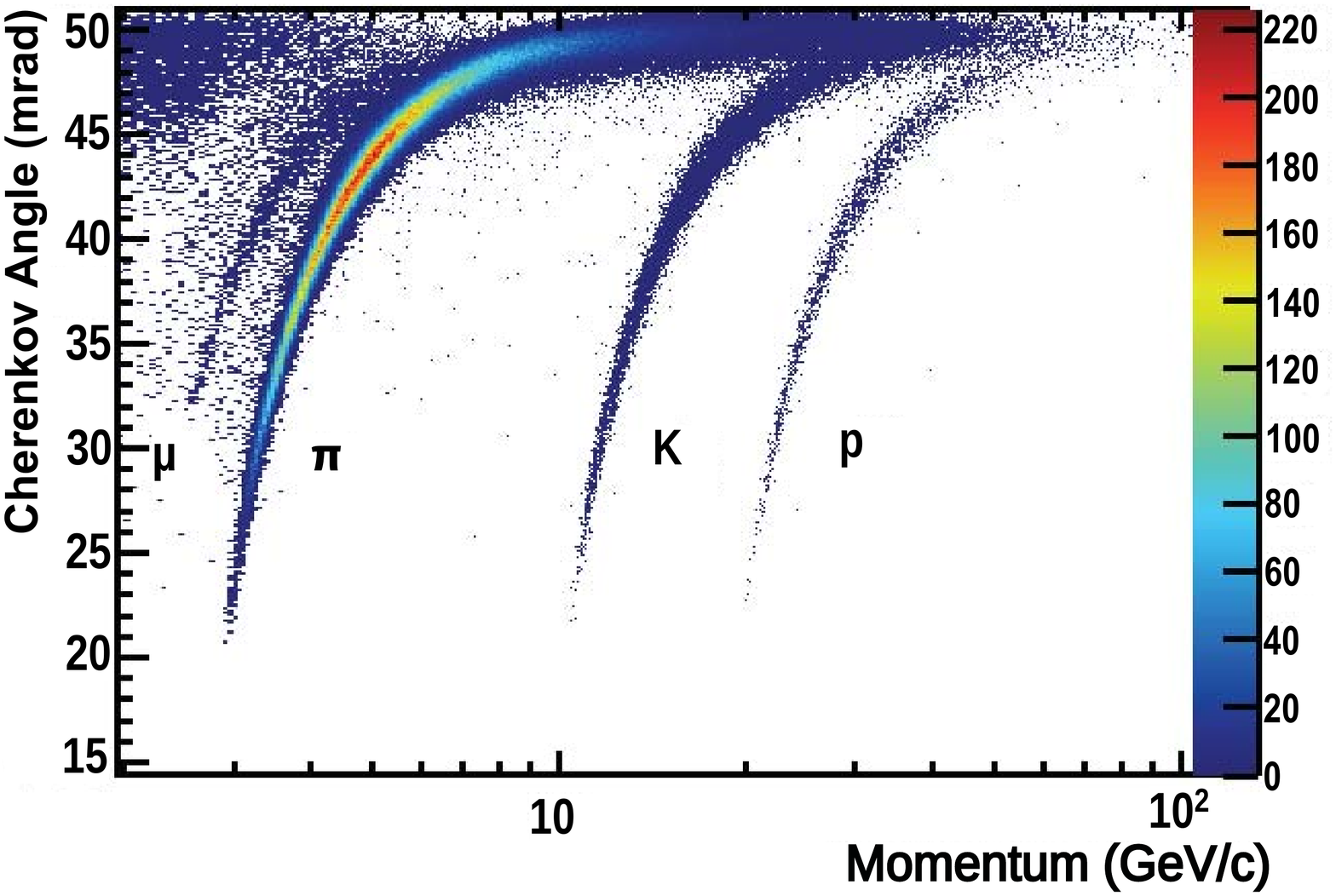}
\caption{Reconstructed Cherenkov angle for \emph{isolated} tracks,
 as a function of track momentum in the \cfourften radiator 
 \protect\cite{LHCb-DP-2012-003}. The
 Cherenkov bands for muons, pions, kaons and protons are clearly visible.}
\label{fig:CAngleVMom}
\end{center}
\end{figure}

Due to the high average track multiplicity in LHCb events, a reconstructed
Cherenkov ring will generally overlap with several neighbouring rings.
Solitary rings from isolated tracks, where no overlap is found,
provide a useful test of the RICH performance, since isolated rings can be
cleanly and unambiguously associated with a single track.
Figure~\ref{fig:CAngleVMom} shows the Cherenkov angle as a 
function of particle momentum using information from the \cfourften
radiator for isolated tracks selected in data ($\sim 2\%$ of all tracks). 
As expected, the events populate distinct bands according to their mass. 

\subsubsection{Photoelectron yield}
\label{sec:rich:Npe}

The average number of detected photons for each track traversing the Cherenkov
radiator media, called the photoelectron yield (\Npe), is another important
measure of the performance of a RICH detector. The yields for the three
radiators used in LHCb are measured in data
using two different samples of events \cite{LHCb-DP-2012-003}. 
The first sample is representative of normal LHCb data taking conditions, and
consists of the kaons and pions originating from the decay
\RichDzeroKPi, where the $\D^0$ is selected from \RichDstarDPi decays.
The second sample consists of low detector occupancy
\RichPPPPmumu events, which provide a clean track sample with
very low background levels. In both samples, only high-momentum tracks
are selected, to ensure that the Cherenkov angle is close to saturation.

\begin{table}[!tb]
\begin{center}
\tbl{Comparison of photoelectron yields (\Npe) determined from
\RichDstarDPi decays in simulation and data, and
\RichPPPPmumu events in data. }
{\begin{tabular}{c c c c c}
\hline
Radiator & \multicolumn{2}{ c }{\Npe from data} & \Npe from simulation \\ 
 & \RichDzeroKPi & \RichPPPPmumu & \RichDzeroKPi \\ [0.5ex]
\hline
Aerogel & $5.0 \pm 3.0$  & $4.3 \pm 0.9$& $8.0 \pm 0.6$ \\ 
C$_4$F$_{10}$  & $20.4 \pm 0.1$ & $24.5 \pm 0.3$& $28.3 \pm 0.6$ \\ 
CF$_4$ & $ 15.8 \pm 0.1$  & $ 17.6 \pm 0.2$& $22.7 \pm 0.6$ \\
\hline
\end{tabular}
\label{table:results}}
\end{center}
\end{table}

Table \ref{table:results} shows the results of the photoelectron yield
extraction, performed on both real and simulated data.
In data, the \RichDstarDPi events have values of \Npe that are less than those 
for \RichPPPPmumu events. This is mainly due to the higher charged track 
multiplicities of the \RichDstarDPi events, reducing the 
effective \Npe, and the track geometry cut that is applied to the \RichPPPPmumu
events increasing their \Npe yield.
The aerogel \Npe data values have a large uncertainty due to the significant 
background levels in the \DeltaThetaC distributions and the additional 
uncertainty in the shape of the signal peak~\cite{LHCb-DP-2012-003}.

The photoelectron yields for data are lower than those predicted by the
simulation. One reason for this is a small detector read-out inefficiency, which
was identified during high trigger rate data taking in Run I. 
The results presented include a retuning of the read-out settings, applied during
data taking to minimise the impact of the inefficiency.
A further optimisation will be performed for LHC Run II to reduce the effect to
the negligible level, and is expected to improve the yields further by a few percent.
The remaining discrepancy is accounted for by an over-estimate of the yield in
the simulation, which will be addressed by improved simulation tunings.
It must be stressed however, that the smaller yield measured in
data does not have a significant impact on the final particle identification 
performance, as described in Section \ref{sec:rich:PIDperf}.

\subsubsection{Particle identification performance}
\label{sec:rich:PIDperf}

To determine the RICH particle identification performance on data, large samples
of genuine $\pi$, $K$ and $p$ tracks are required. Such control samples must be
selected independent of RICH information which would otherwise bias the results.  
The strategy employed is to reconstruct exclusive decays purely from kinematic
selections. Only decay modes with large branching fractions, for which large
samples can be easily collected, are used to allow for precise calibration over
a range of track kinematics. 

\begin{figure}[!tb]
\begin{center}
\includegraphics[width=0.49\textwidth]{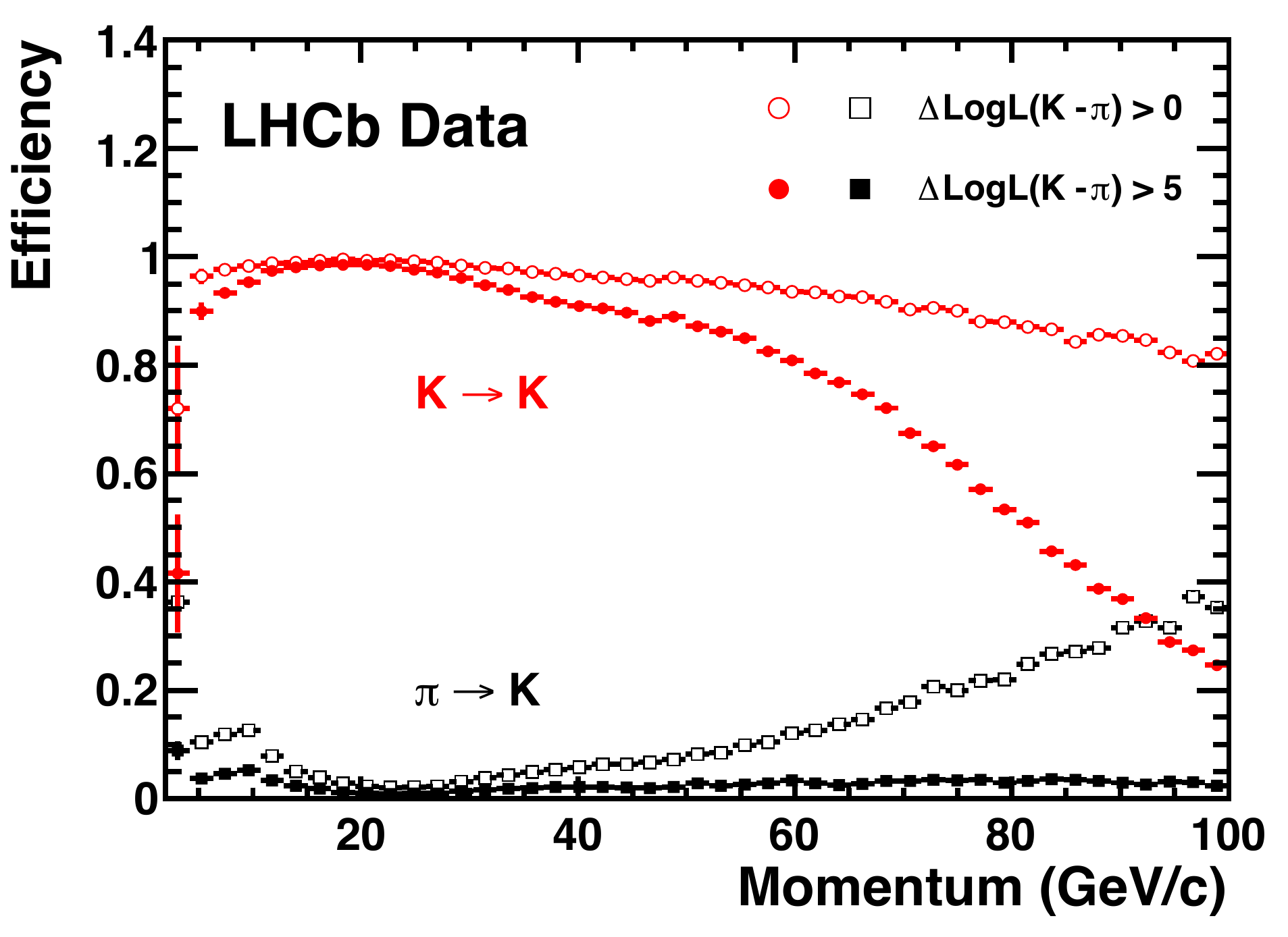}
\includegraphics[width=0.49\textwidth]{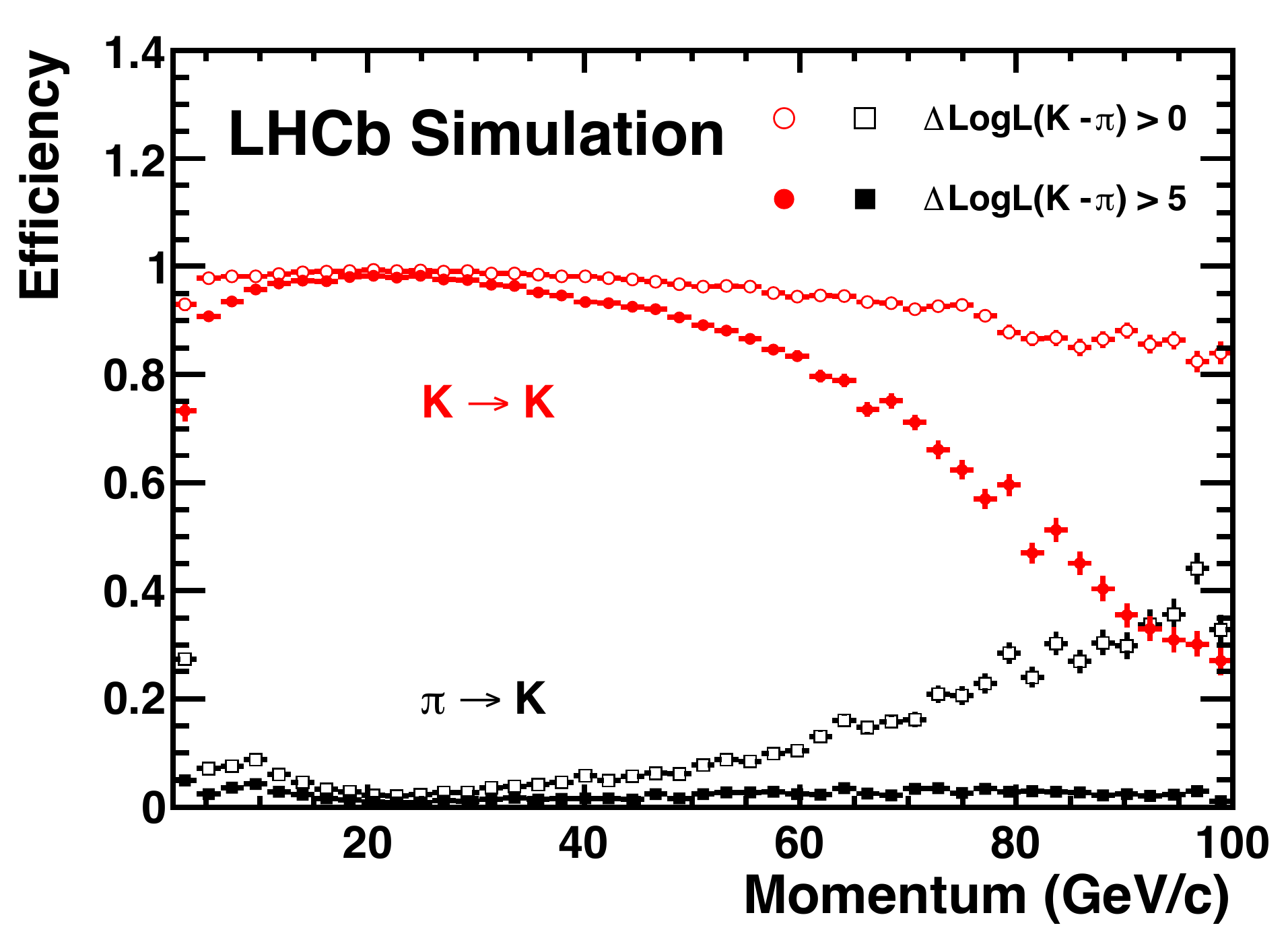}
\caption{Kaon identification efficiency and pion misidentification rate as
  measured using data (left) and from simulation (right)
as a function of track momentum\protect\cite{LHCb-DP-2012-003}. 
Two different \DeltaLLKPi requirements have been imposed on the samples,
resulting in the open and filled marker distributions, respectively. } 
\label{fig:rich:KPiSeparation}
\end{center}
\end{figure}

The following decays, and their charge conjugates, are identified: 
$\KS \to \pi^{+} \pi^{-}$, $\Lambda \to p \pi^{-}$ and 
$\Dstarp \to \Dz(K^{-}\pi^{+})\pi^{+}$.
This ensemble of final states provides a complete set of charged particle types
needed to assess comprehensively the hadron PID performance.
Utilising the track samples obtained from these exclusive control decay modes,
Figure~\ref{fig:rich:KPiSeparation} demonstrates the kaon efficiency (kaons
identified as kaons) and pion misidentification (pions misidentified as kaons)
fraction achieved in LHCb data, as a function of momentum.
For illustration the data is shown with two different
PID requirements, one optimising the efficiency, the other
minimising the misidentification rate.

For each track the likelihood that it is an electron, muon, pion, kaon or
proton is computed. 
In the first approach it is required that, for each track, the
likelihood for the kaon mass hypothesis is larger than that for the
pion hypothesis, i.e. \DeltaLLKPi$>0$.
When averaging over the momentum range 2 -- 100 GeV/$c$ one finds the
kaon efficiency to be $\sim95\%$ with a pion misidentification rate of $\sim10\%$.
A stricter PID requirement, \DeltaLLKPi$>5$,
reduces the pion misidentifiaction rate to $\sim 3\%$ at a modest loss in
kaon efficiency of $\sim 10\%$ on average.
Figure~\ref{fig:rich:KPiSeparation} also shows the performance in
simulation, for the same exclusive control channels and PID
requirements as above for data.
Good agreement with data is observed for both sets of PID requirements. 

\begin{figure}[!tb]
\begin{center}
\includegraphics[width=0.49\textwidth]{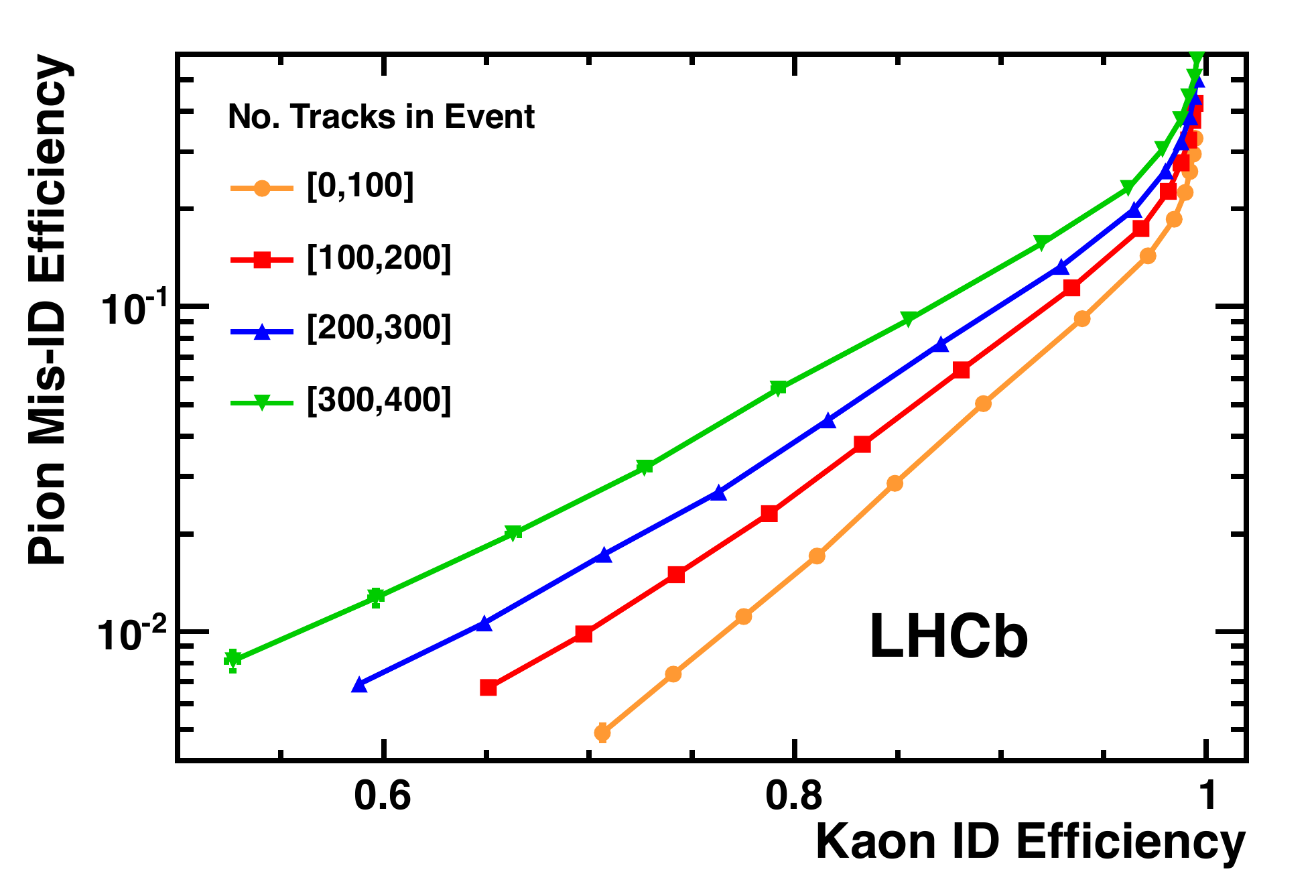}
\includegraphics[width=0.49\textwidth]{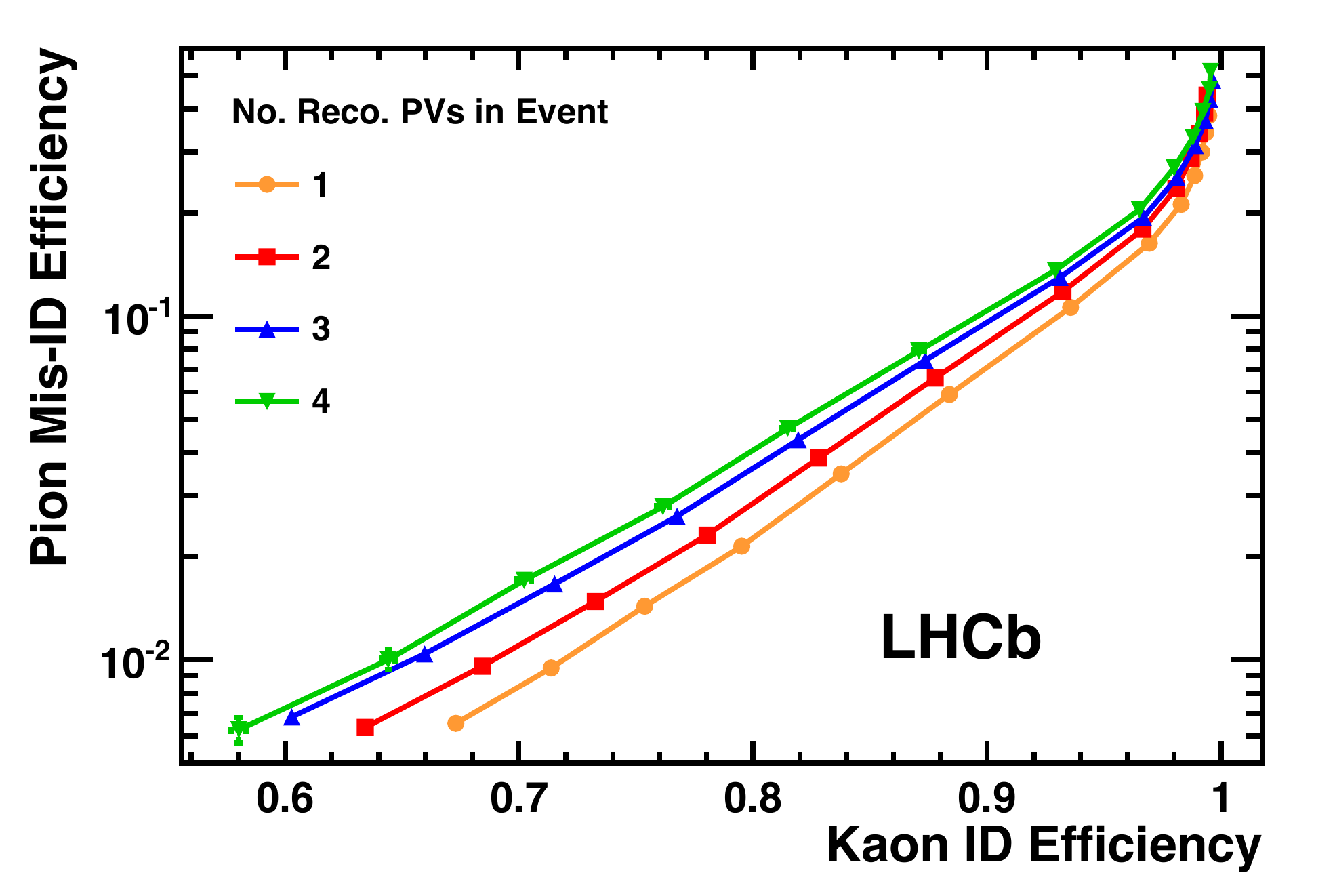}
\caption{Pion misidentification fraction versus kaon identification efficiency
as measured in 7\,TeV LHCb collisions: (left) as a function of track multiplicity,
and (right) as a function of the number of reconstructed primary vertices
\protect\cite{LHCb-DP-2012-003}. The
efficiencies are averaged over all particle momenta. }
\label{fig:rich:Multiplicity}
\end{center}
\end{figure}

The Run I conditions, with multiple interactions per bunch crossing
and the resulting high particle multiplicities, provide an insight into
the RICH performance at possible future higher luminosity running.
Figure~\ref{fig:rich:Multiplicity} shows the pion misidentification fraction
versus the kaon identification efficiency as a function of track
multiplicity and the number of reconstructed primary vertices, as the
requirement on the likelihood difference \DeltaLLKPi is
varied. The results demonstrate some degradation in PID
performance with increased interaction multiplicity. However, the performance is
still excellent and gives confidence that the RICH system will continue to
perform well during LHC Run II.

\subsection{Muon system based particle identification}
\label{sec:muon-pid}

The identification of a track reconstructed in the tracking system as a muon
is based on the association of hits around
its extrapolated trajectory in the muon system~\cite{LHCb-DP-2013-001}. 
A search is performed for hits within rectangular windows 
around the extrapolation points where the $x$ and $y$ dimensions of the windows 
are parameterised as a
function of momentum at each station and separately for each muon system
region. The parameters are optimised to maximise the efficiency and at the same
time provide low misidentification probabilities of pions as
muons. The same criterion is used to define the number of stations required to
have hits within a window 
as a function of momentum. A minimum momentum of 3\gevc is
necessary for a muon to traverse the calorimeters and reach the M2 and M3 stations, 
while above 6\gevc they traverse all five of the stations. 
For each muon candidate, likelihoods for the muon and non-muon hypotheses are
computed, based on the average squared distance of the hits that are closest to the
extrapolation points. 

\begin{figure}[!tb]
  \vspace{15mm}
  \begin{center}
    \includegraphics[width = 0.49\textwidth]{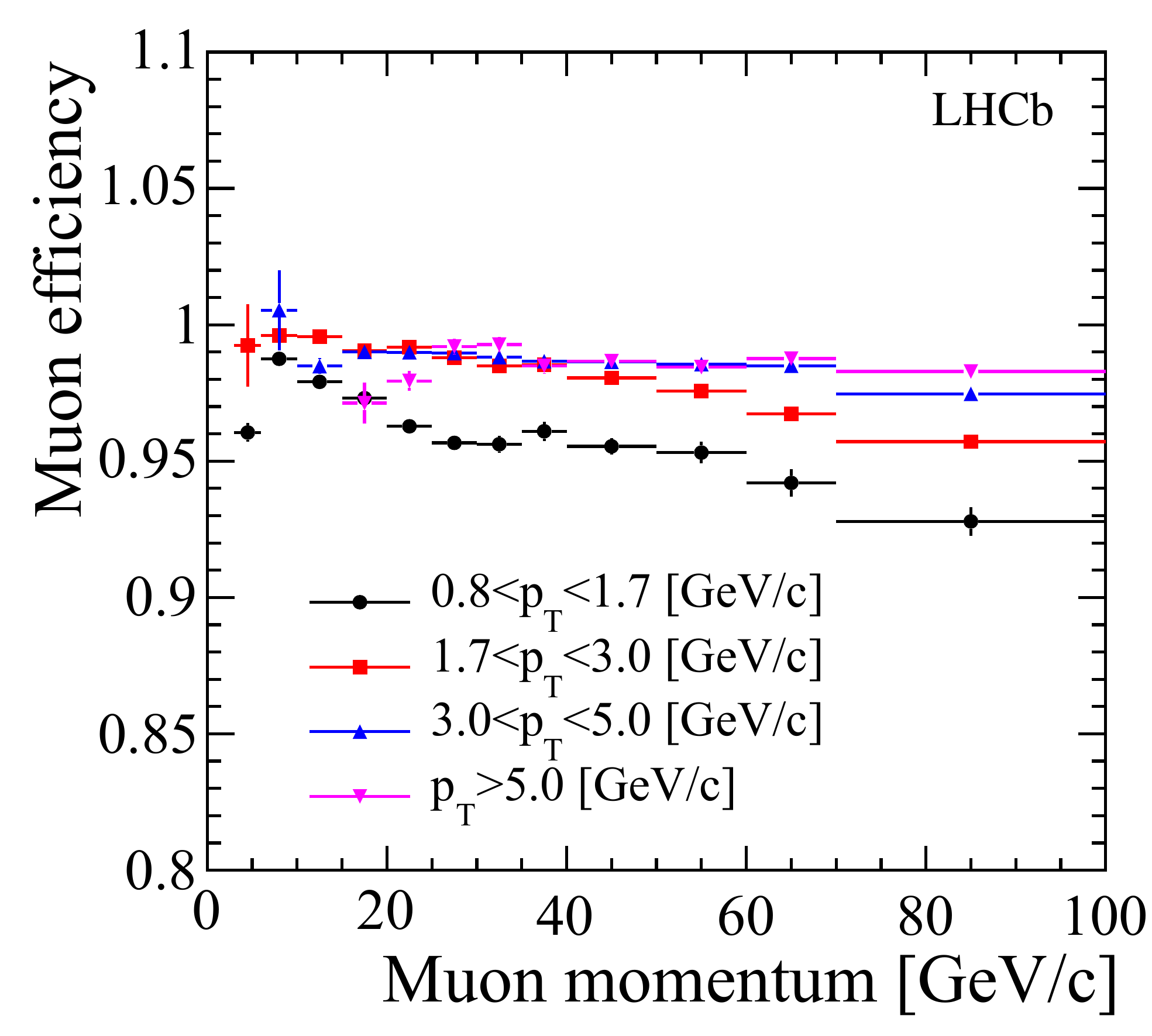}
    \includegraphics[width = 0.49\textwidth]{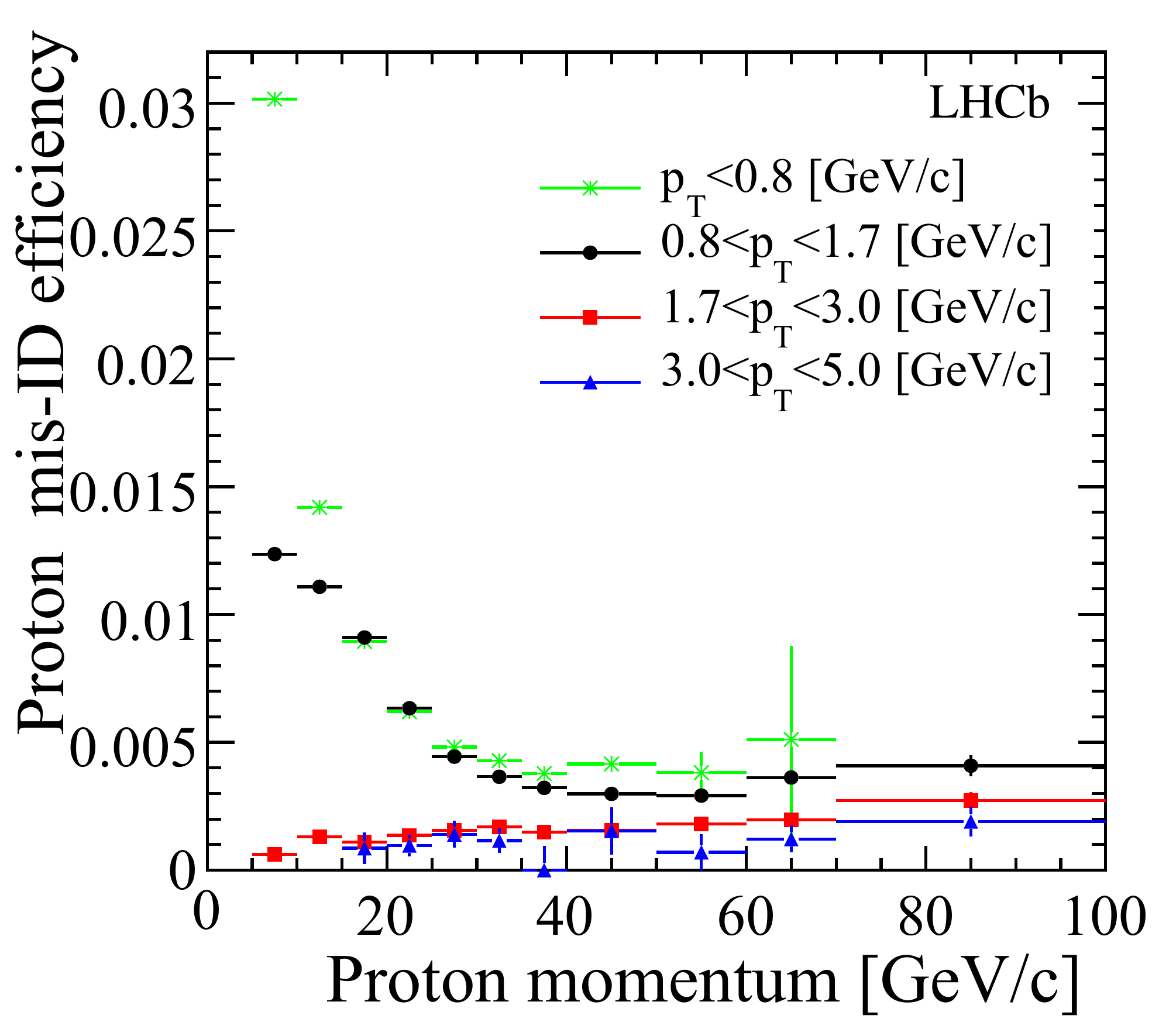}
  \end{center}
  \begin{center}
    \includegraphics[width = 0.49\textwidth]{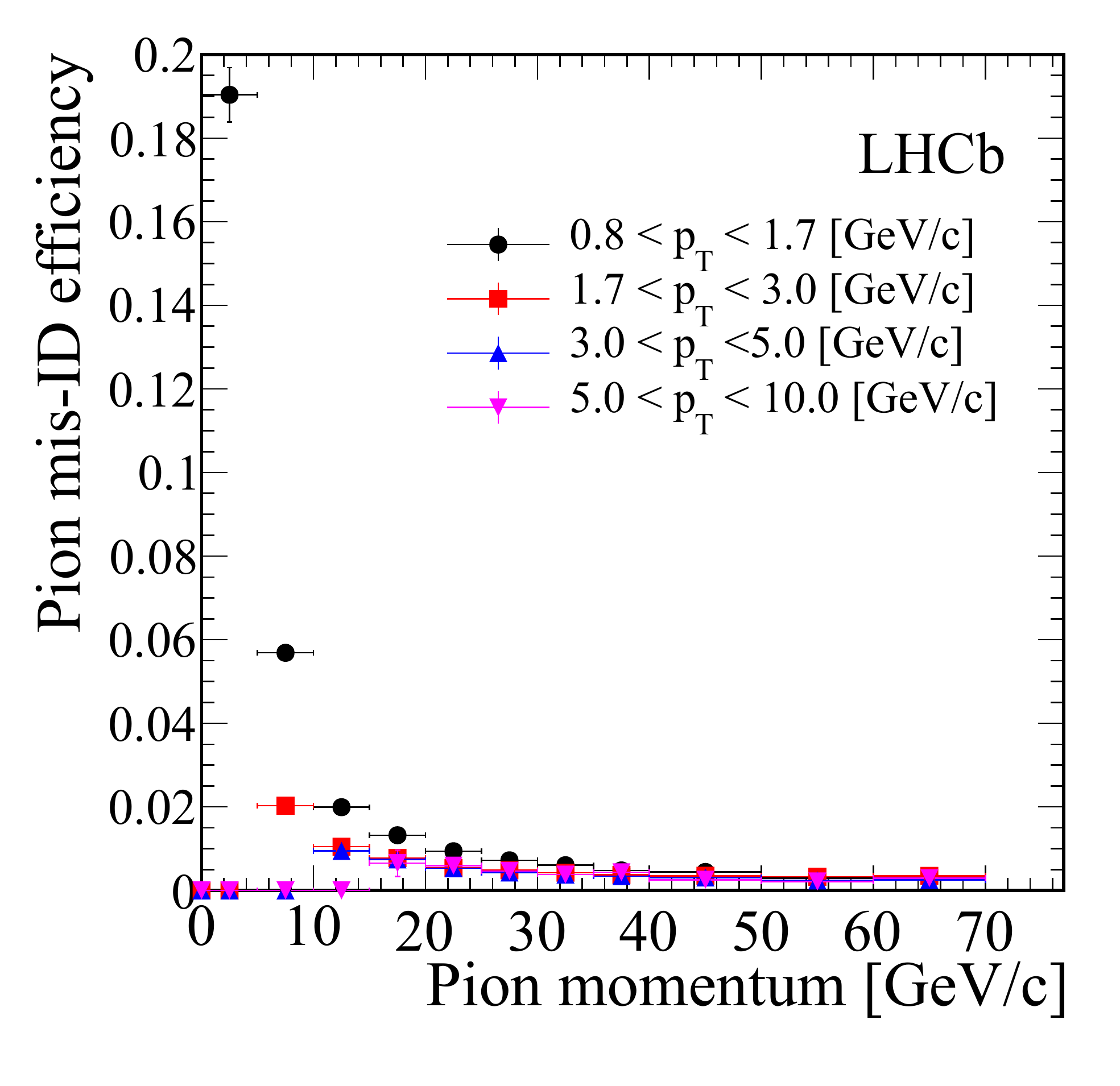} 
    \includegraphics[width = 0.49\textwidth]{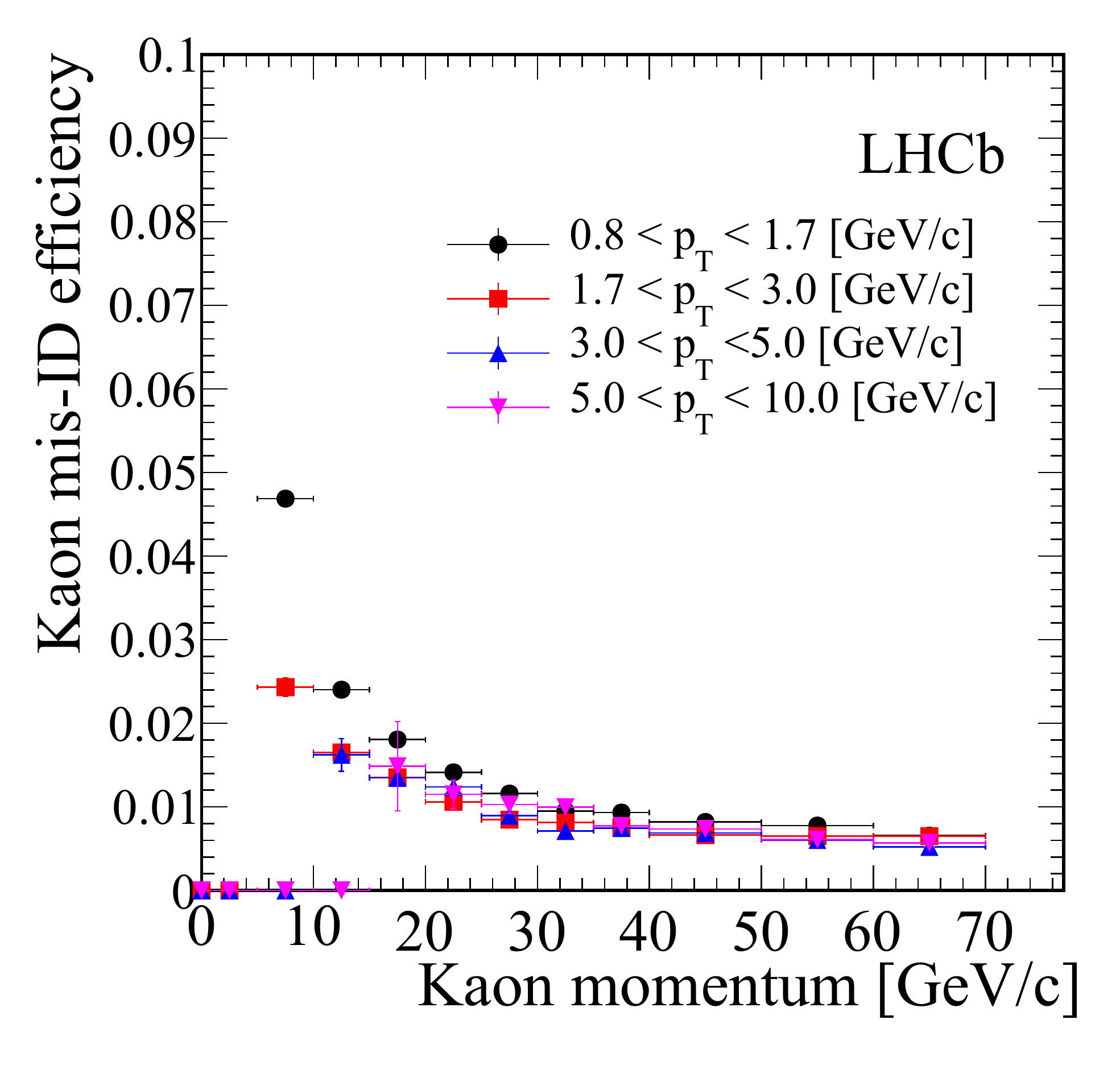} 
  \end{center}
  \caption{Top left: efficiency of the muon candidate selection based on the
    matching of hits in the muon system to track extrapolation, as a function
    of momentum for different \pt ranges. Other panels: misidentification
    probability of protons (top right), pions (bottom left), and kaons (bottom
    right) as muon candidates as a function of momentum, for different \pt ranges. 
    \label{fig:imeff}}
   \vspace{15mm}
\end{figure}

The performance of the muon identification is obtained from data using muons
from $J/\psi\to\mu^+\mu^-$ decays, protons from $\Lambda\to p\pi^-$ decays and
kaons and pions  from $\Dz \to K^-\pi^+$,  where the $D^0$ is selected from
$\Dstarp \to D^0\pi^+$ decays. These samples can be selected without using PID
information and are characterised by relatively high statistics and low
background. The latter is subtracted by fitting the appropriate invariant mass
distribution. Figure~\ref{fig:imeff} shows, as a function of the track momentum
and for different ranges of transverse momentum, the efficiency of the muon
candidate selection, and the probabilities of incorrect identification of
protons, pions and kaons as muons. 

The incorrect assignment of the muon identity to a proton occurs either 
due to a combination of spurious hits in the different muon stations that are 
aligned with the proton direction, or due to the existence of a true muon in the
event that points in the  same direction as the proton in the muon system. This
muon can be produced close to the interaction point or in the calorimeter shower.  
Since the window dimension decreases with momentum and an increasing number of
hits is required for tracks above 10\gevc,  
a strong reduction of the proton misidentification rate is seen in the interval
$3-30\gevc$ and for $\pt<1.7$\gevc. 
For higher \pt values, the protons have a high polar angle and therefore fall
outside of the high-occupancy part of the detector. 
Decays in flight are the main cause of misidentification of pions and kaons as
muons.  To a good approximation, the misidentification rate is the sum of the
contribution of decays in flight  and the proton misidentification probability. 

The background rejection power can be improved by the computation of a
likelihood for  the muon and non-muon hypotheses, based on the pattern
of hits around the extrapolation to the different muon stations of the charged
particles trajectories reconstructed with high precision in the tracking
system. The logarithm of the ratio between the muon and non-muon hypotheses,
$\rm\Delta log\mathcal{L}(\mu)$, is used as a discriminating variable. 
The likelihood for the non-muon hypothesis is calibrated using proton data,
since the other charged hadrons (pions or kaons) selected as muons will have a
component identical to the protons and a component very similar to the true
muons, due to decays in flight before the calorimeter.
The muon likelihood has been calibrated with muons from $J/\psi\to\mu^+\mu^-$
decays selected  from data, while the non-muon likelihood has been calibrated
with a simulated sample of $\Lambda\to p\pi^-$ decays.

\subsection{Combined particle identification performance}

The PID information obtained separately from the muon, RICH, and
calorimeter systems is combined to provide a single set of more
powerful variables.  Two different approaches are used. In the first
method the likelihood information produced by each sub-system is
simply added linearly, to form a set of combined likelihoods,
$\deltaLLCombXpi$, where $X$ represents either
the electron, muon, kaon or proton mass hypothesis. These variables
give a measure of how likely the mass hypothesis under consideration is,
for any given track, relative to the pion hypothesis.  A second
approach has been subsequently developed to improve upon the simple
log likelihood variables both by taking into account correlations
between the detector systems and also by including additional
information.  This is carried out using multivariate
techniques \cite{Hocker:2007ht}, combining PID information from each
sub-system into a single probability value for each particle
hypothesis.

\begin{figure}[!tb]
  \begin{center}
   \subfigure{\raisebox{1mm}{\includegraphics[width=0.47\textwidth]{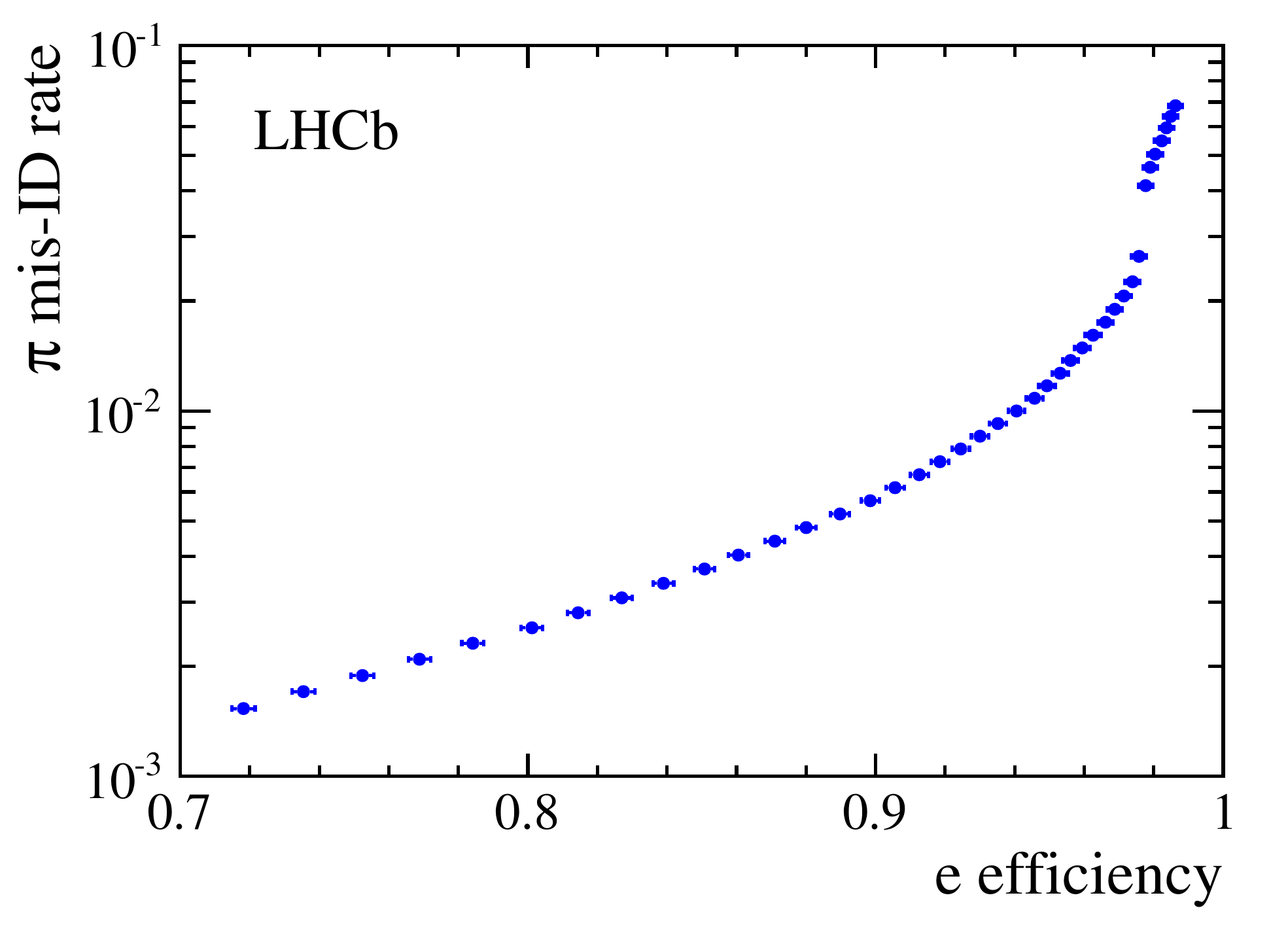}}}
   \subfigure{\raisebox{0mm}{\includegraphics[width=0.51\textwidth]{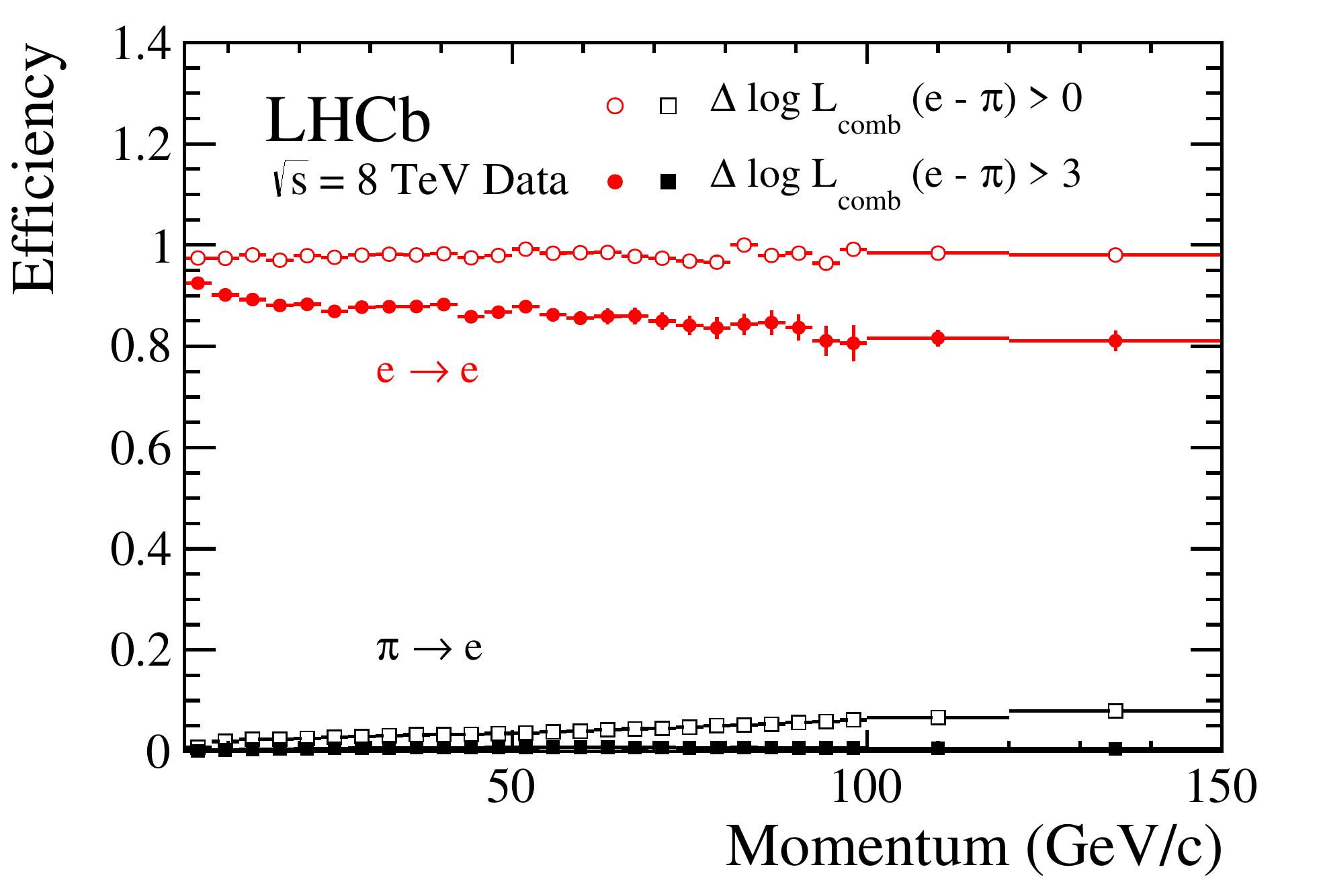}}}
  \end{center}
  \vspace{-3mm}
  \caption{Electron identification performance using the 
    $\deltaLLCombepi$ variable, as measured in 8\,TeV
    collision data, using a tag and probe technique with electrons
    from the decay $B^{\pm} \to (J/\psi \to e^+e^-) K^{\pm}$.  Left,
    pion misidentication rate versus electron identification
    probability when the cut value is varied.  Right, electron
    identification efficiency and pion misidentification rate as a
    function of track momentum, for two different cuts on $\deltaLLCombepi$.}
  \label{fig:CombinedEPID}
\end{figure}

To illustrate the improvement made by combining information from
different sub-detectors, the performance of the variable
$\deltaLLCombepi$ is first considered, using a similar tag and
probe technique to that of Section~\ref{sec:CALOPID} in which the
calorimeter-only performance is presented.  Figure~\ref{fig:CombinedEPID}
shows the pion misidentification versus electron identification probability, for
various cuts on $\deltaLLCombepi$. Compared to  Figure~\ref{fig:CaloEffVMisID}
the improvement in the misidentification rate can clearly be seen, \eg at
$\sim90\,\%$ electron efficiency the pion misidentification rate drops from
$\sim6\,\%$ to $\sim0.6\,\%$. 

An improvement in
performance can also be seen for the muon identification, as
illustrated by one of the most prominent LHCb results, the measurement
of the $B^0_s\to \mu^+\mu^-$ branching fraction and the search for
$B^0 \to \mu^+\mu^-$ decays~\cite{LHCb-PAPER-2013-046}. The $B^0_{(s)} \to
h^+h^-$ decay modes, where $h=(K,\pi)$,
can fake a signal if both hadrons are misidentified
as muons.  Therefore the minimisation of these backgrounds is of
paramount importance for this analysis.  This double misidentification
probability has been evaluated by folding the $K\to\mu$ and
$\pi\to\mu$ fake rates extracted from a $D^0\to K\pi$ sample from
$\Dstarp \to D^0\pi^+$ decays, in bins of $p$ and \pt, into the
spectrum for selected simulated $B^0_{(s)}\to h^+h^-$ events.  If
the tracks identified as muons are also required to satisfy a
selection using the combined PID information ($\deltaLLCombKpi<10$ and 
$\deltaLLCombmupi>-5$), the $B^0_{(s)}\to h^+h^-$
misidentification probability is reduced by a factor of $\sim6$,
whilst only $\sim3\%$ of the $B_s\to \mu^+\mu^-$ signal is lost.

\begin{figure}[!tb]
  \vspace{-2mm}
  \begin{center}
    \includegraphics[width = 0.49\textwidth]{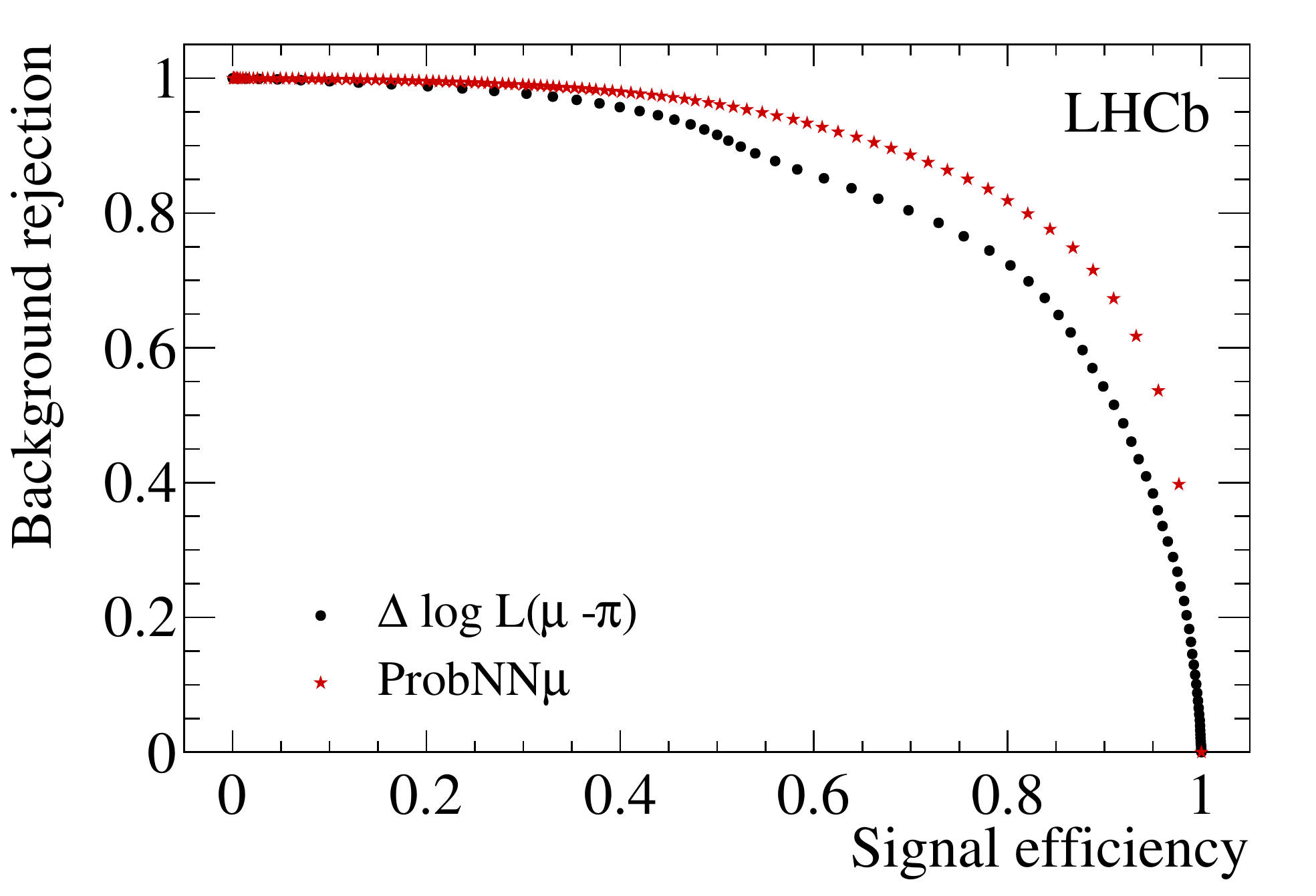}
    \includegraphics[width = 0.49\textwidth]{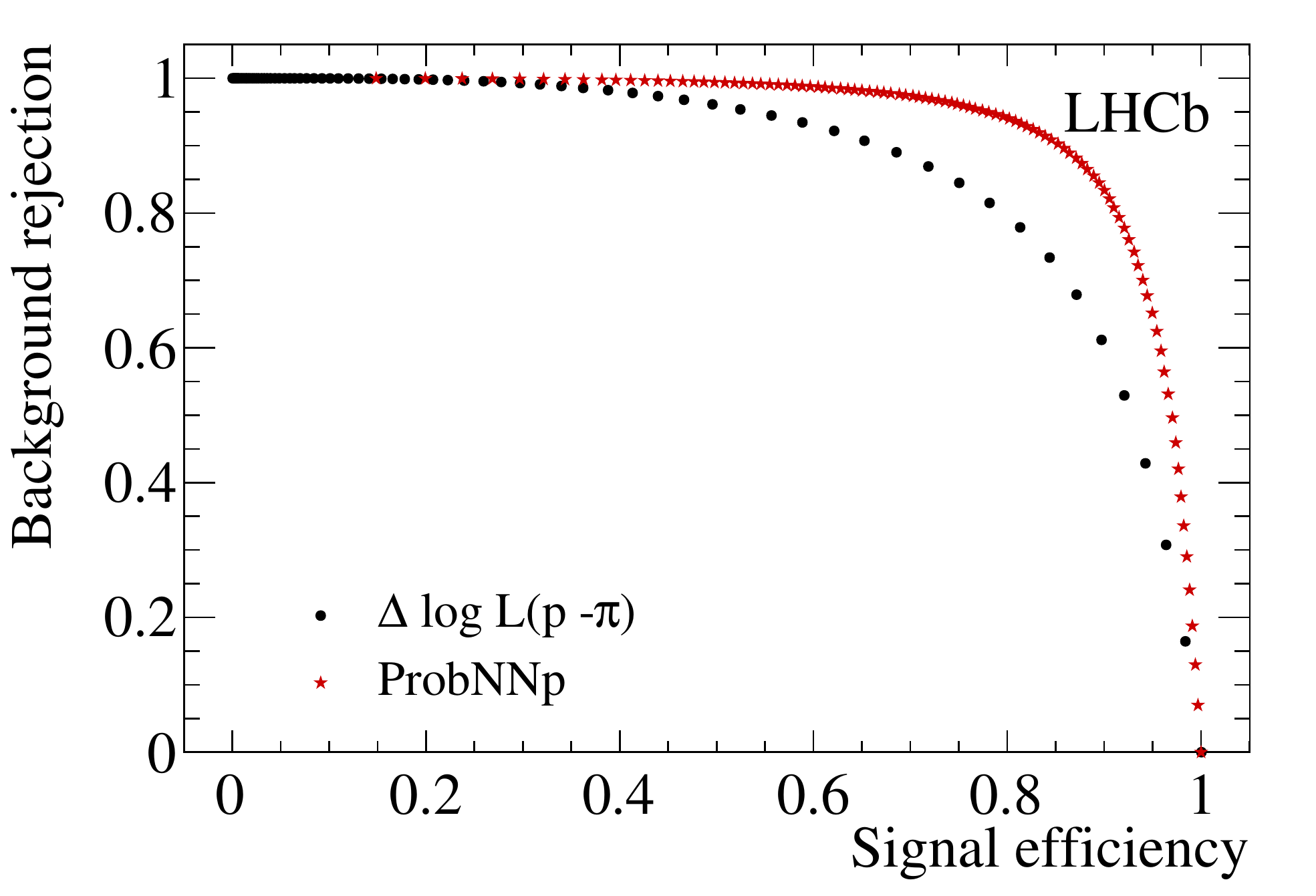}
  \end{center}
  \caption{Background misidentification rates versus muon (left) and
    proton (right) identification efficiency, as measured in the $\Sigma^+\to
    p\mu^+\mu^-$ decay study. 
    The variables $\deltaLLXpi$ (black) and ProbNN (red), 
    the probability value for each particle hypothesis, are compared for
    $5-10$\gevc muons and $5-50$\gevc protons, using data sidebands for
    backgrounds and Monte Carlo simulation for the signal. }
  \label{fig:Sigmapmm_rocs}
 \vspace{-2mm}
\end{figure}

The possible improvement of the multivariate approach with respect to
the simple log likelihood may also be illustrated by the ongoing
search for the flavour-changing neutral current decay 
$\Sigma^+\to p\mu^+\mu^-$. 
In Figure~\ref{fig:Sigmapmm_rocs} the misidentification rates versus
efficiency curves for 
the $\deltaLLXpi$ and the probability value for each particle hypothesis
variables are shown. 
The improvement is clearly visible
for both muons and protons. These multivariate variables will be
further developed and utilised more extensively during Run II.

\section{Trigger}
\label{sec:trigger}

The LHCb trigger consists of two levels; Level-0 (L0) and the High Level Trigger
(HLT). In Run I the trigger reduced the rate of events to be saved for physics
analysis to 2~--~5\khz. 

The L0 trigger is implemented in hardware and makes decisions based on
information from the calorimeter and muon systems in order to reduce
the event rate to below $1$\mhz, at which point the
whole detector can be read out. The HLT is a software application
running on the event filter farm (EFF).
A fraction of L0 accepted
events are deferred to disk for processing by the HLT during the
inter-fill time, optimising the use of available EFF resources.  

After the HLT, events are stored and later processed with a more
accurate alignment and calibration of the sub-detectors, and with
the reconstruction software described in Section~\ref{sec:tracking}.  
This part of the reconstruction and subsequent selection of
interesting events is referred to as the offline reconstruction and
selection for the remainder of this discussion.  

Section~\ref{sec:tistos} describes a data-driven method to determine
the efficiency and purity of the signals that are selected by the LHCb trigger.
The implementation of the L0 trigger~\cite{Alves:2008zz} is briefly summarised in
Section~\ref{sec:l0} and the performance of the HLT~\cite{TriggerPaper:2012xx}
is discussed in Section~\ref{sec:hlt}.
The selection criteria used in the two trigger levels during 2011 and 2012 are
described in detail in Reference~\cite{LHCb-PUB-2014-046}.
The deferral system is discussed in Section~\ref{sec:deferral} and a
short summary of the LHCb trigger is provided in Section~\ref{sec:summary}.   

The results presented here are based on the configuration and performance of the LHCb
trigger during 2012 when the deferral system was first used and the majority of the
Run~I dataset was collected.  

\subsection{Data driven trigger performance determination}
\label{sec:tistos}

The trigger efficiencies are evaluated using events reconstructed
with the full offline software, and are calculated with respect to candidates
selected by the full offline analysis of the respective channel, excluding the trigger. 
They therefore quantify the inefficiencies due to the simplified reconstruction,
possible misalignments and reduced resolution, as compared to the offline reconstruction.
They also account for any tighter selection requirements that are needed to satisfy 
the rate and processing time limitations.
The following decay channels are chosen to highlight the performance of the
trigger \footnote{Here, and in the following, charge conjugated
decays are implicitly included.} :

\begin{itemize}
\item \BuJpsiK decays, with $\Jpsi \to \mumu$. This decay evaluates the muon trigger
  efficiency and serves as a proxy for several key physics decay
  channels like $\BsToJPsiPhi$, $\BdToKstmm$ or $\Bsmm$.  
\item $\BdKpi$ as a typical two-body hadronic beauty decay.
\item $\BdDpi$ with $\DpKpipi$, as a typical four-body beauty decay.
\item $\DKpi$ as a two-body charm decay.
\item $\DpKpipi$ represents a three-body charm decay.
\item $\DstDpi$, followed by the four-body charm decay
  $D^{0} \ra K^{-} \pi^{+} \pi^{-} \pi^{+}$.
\end{itemize}

These decay channels and their selections are representative of the
trigger needs of the core physics analyses of the LHCb experiment. 
The selected charm modes cover the topologies that are most sensitive to \CP
violating effects. 
All samples used in this study carry a large signal to background ratio.
Nevertheless, the yields are determined by fitting the signal peaks
in the invariant mass distributions in order to subtract the residual background.

In the following, the term `signal' refers to a combination of tracks
forming the offline reconstructed and selected \bquark- or \cquark-hadron
candidates. To determine the trigger efficiency, trigger objects are associated with
signal tracks. The criteria used to associate a trigger object with a signal
track are as follows. An event is classified as Trigger on Signal (TOS) if the
trigger objects that are associated with the signal candidate are sufficient to
trigger the event. An event is classified as Trigger Independent of Signal (TIS)
if it has been triggered by trigger objects that are not associated with the
signal. Some events can be classified as TIS and TOS simultaneously 
(${\rm TIS ~\&~ TOS}$), 
which allows the determination of the trigger efficiency relative
to the offline reconstructed events from data alone. The efficiency to trigger
an event independently of the signal, 
\etis, is given by 
\begin{equation}
\etis = N^{\rm TIS ~\&~ TOS}/N^{\rm TOS}\, , \nonumber 
\end{equation}
where $N^{\rm TOS}$ is the number of events classified as TOS.
The efficiency to trigger an event on the signal alone, \etos, is
given by 
\begin{equation}
\etos = N^{\rm TIS ~\&~ TOS}/N^{\rm TIS}\, , \nonumber
\end{equation}
where $N^{\rm TIS}$ is the number of events classified as TIS. This method
of measuring trigger efficiencies is discussed in detail in Reference~\cite{LHCb-PUB-2014-039}.

\subsection{Level-0 hardware trigger}
\label{sec:l0}

The L0 trigger is divided into three independent units; 
the L0-Calorimeter trigger, the L0-Muon trigger and the L0-PileUp trigger, 
the latter being used only for the determination
of the luminosity \cite{LHCb-PAPER-2011-015}.
The L0 trigger system is fully synchronous with the 40\mhz bunch crossing
rate of the LHC. The latencies are fixed and are independent of the occupancy or
the bunch crossing history. 

\begin{table}[tb]
\begin{center}
\centering
\tbl{Typical L0 thresholds used in Run I \protect\cite{LHCb-PUB-2014-046}.} 
{\begin{tabular}{l@{\hspace{5mm}}*{7}{c}{c}}
\hline &\multicolumn{2}{c}{\pt or \et}& ~~~SPD~~~ \\
&\hspace{8mm} 2011 \hspace{14mm}&\hspace{8mm} 2012 \hspace{14mm}& 2011 and 2012\hspace{30mm}\\
\hline
single muon&1.48\gevc&1.76\gevc& 600\\
dimuon $\pt_1 \times \pt_2$ &$(1.30\gev/c)^2$&$(1.60\gev/c)^2$&900\\
hadron   & 3.50\gev&3.70\gev&600\\
electron & 2.50\gev&3.00\gev&600\\
photon   & 2.50\gev&3.00\gev&600\\
\hline
\end{tabular}
\label{l0cuts}}
\end{center}
\end{table}

The L0-Calorimeter system uses information from the SPD, PS, ECAL and
HCAL detectors, as described in Section~\ref{sec:neutral-particle-id}.
It computes the transverse energy, \et, deposited by incident particles in
clusters of $2\times 2$ cells.  
From these clusters, the following three types of candidates are built.
{\tt L0Hadron} is the highest \et HCAL cluster, which also contains the energy
of the matching ECAL cluster. 
{\tt L0Photon} is the highest  \et ECAL cluster with 1 or 2 PS hits in front of
the ECAL cluster and no hit in the SPD cells corresponding to the PS cells. 
{\tt L0Electron} has the same requirements as {\tt L0Photon}, with the
additional condition of at least one SPD cell hit in front of the PS cells. 
The \et of each candidate is compared to a fixed threshold 
and events containing at least one candidate above threshold fire the L0 trigger. 
The total number of hits
in the SPD is also determined, and is used to veto events that would take a
disproportionately large fraction of the available processing time in the HLT.
The SPD hit multiplicity requirements are listed in Table~\ref{l0cuts}.

The L0 muon processors look for the two highest \pt muon tracks in each quadrant.
The position of a track in the first two stations allows the determination of
its \pt with a measured momentum resolution of roughly $25\%$.
The trigger considers the eight candidates and sets a single threshold on either
the largest transverse momentum, $\pt^{\rm largest}$, or on the product,
$\pt^{\rm largest}\times \pt^{\rm 2nd~largest}$. These thresholds are listed in
Table~\ref{l0cuts}.  The tightening of L0 thresholds in the 2012 configuration
is a consequence of the increased luminosity and beam energy.

\begin{figure}[tb]
\vspace{1mm}
\includegraphics[width=0.49\textwidth]{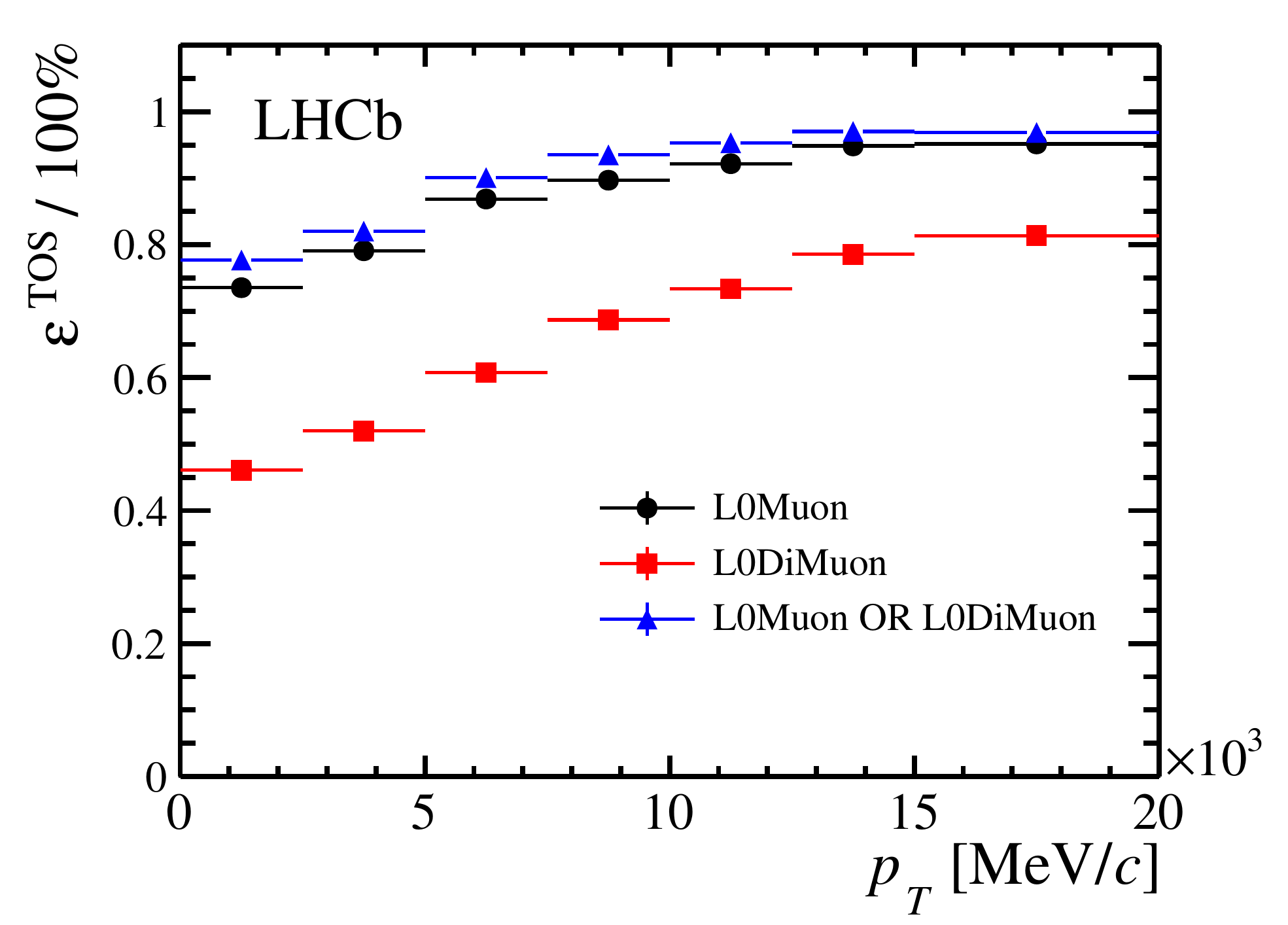}
\hfill
\includegraphics[width=0.49\textwidth]{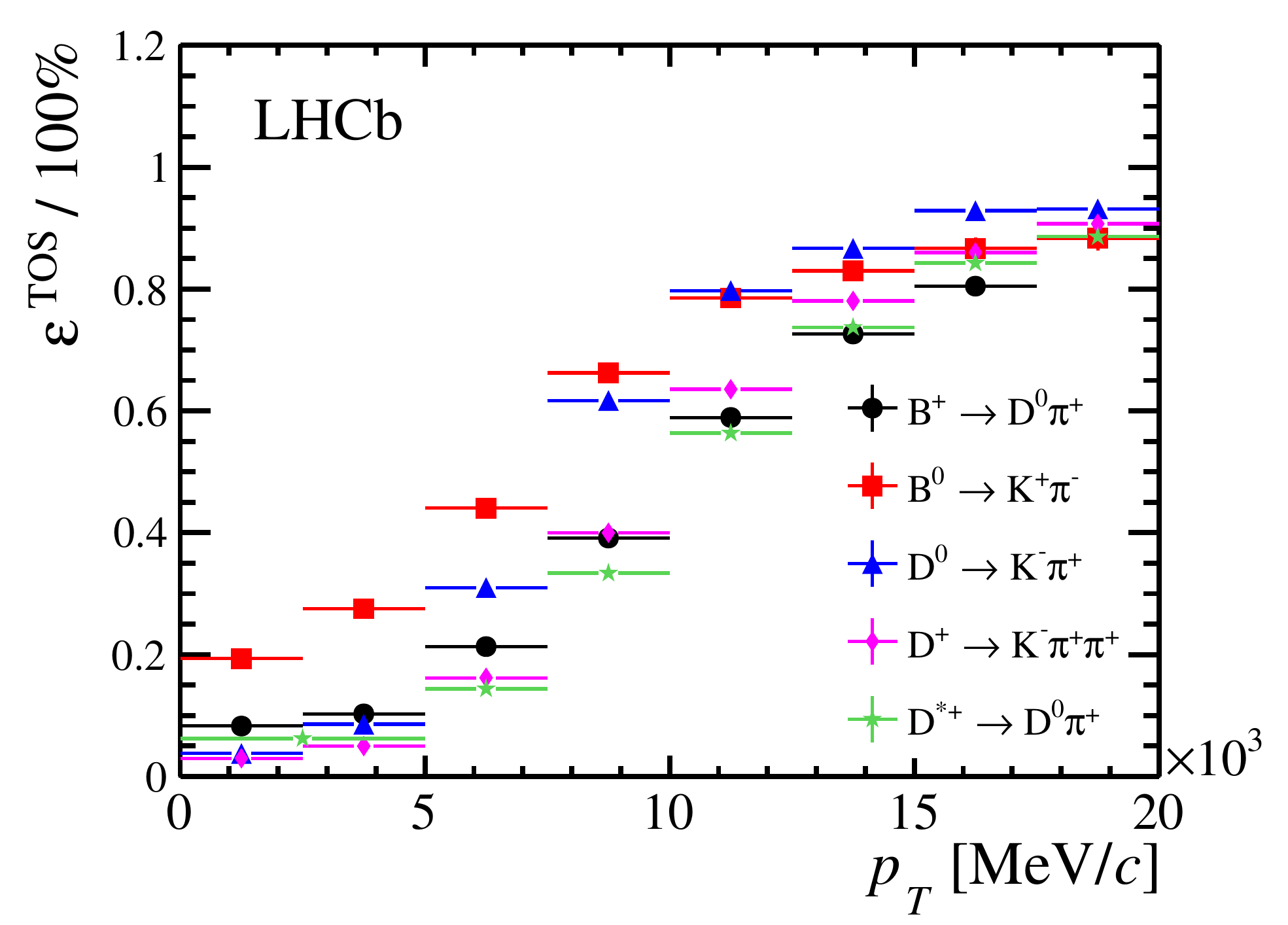}
\caption{\label{fig:l0hmu} (left) L0 muon trigger performance: TOS trigger
  efficiency for selected \BuJpsiK candidates.
(right) L0 hadron trigger performance: TOS trigger efficiency for different
  beauty and charm decay modes.}
\vspace{1mm}
\end{figure}

The trigger efficiencies are measured on offline selected events,
using the techniques described in Section~\ref{sec:tistos}. The efficiencies of
the L0 muon triggers evaluated on \BuJpsiK events are shown in
Figure~\ref{fig:l0hmu}. The majority of events are accepted by the single muon
trigger.  The largest inefficiency originates from the tight muon identification
requirements inside the L0 reconstruction algorithm. The L0 dimuon trigger
selects a small fraction of additional candidates at lower transverse
momenta. The combined efficiency for both L0 muon triggers is evaluated to be
$89\pm0.5\%$~\cite{LHCb-PUB-2014-039}. 

The L0 hadron efficiency is shown in Figure~\ref{fig:l0hmu} for the two-
and three-prong beauty decays \BdKpi and \BdDpi and the two-, three- and
four-prong charm decays \DKpi, \DpKpipi and \DstDpi, as a function of the $B$ or
$D$ meson \pt. The two-prong beauty decay is triggered with highest efficiency
by the L0 hadron \et criterion (\etos = 40\%) while the four-prong charm decay
\DstDpi is selected with the lowest efficiency (\etos = 22\%).  The other modes
lie in between, as shown in Figure~\ref{fig:l0hmu}. With the inclusion of TIS
triggers, the total efficiencies increase significantly, e.g. 
from 40\% to 53\% for \BdKpi.

The total output rate of the L0 trigger is limited to 1\mhz, which is the
maximum rate at which the LHCb detector can be read out. This output
rate consists of approximately 400\khz of muon triggers, 500\khz of
hadron triggers and 150\khz of electron and photon triggers, where the
individual triggers have an overlap of about 10\%.

\subsection{High Level Trigger}
\label{sec:hlt}

Events accepted by L0 are transported by the data acquisition network
to one of the processors of the EFF. The HLT is a software application, of which
29\,500 instances run on the EFF, and each instance consists of independently
operating trigger ``lines'', each of which consists of selection parameters for
a specific class of events.  

The HLT is based on the same software framework used throughout LHCb. 
Given the available resources in the EFF, the time per event is around fifty
times smaller than in the offline processing. The HLT is divided in two levels. 
In the first level (HLT1), a partial event reconstruction is performed. 
In the second level (HLT2), the complete event is reconstructed.
Where time allows, the HLT uses the same reconstruction algorithms as employed
offline, with some simplifications that are needed to satisfy the time
constraints. 

\subsubsection{First level}
\label{sec:HLT1}

The offline VELO reconstruction algorithm which performs a full 3D
pattern recognition is sufficiently fast to be run on all events entering the
HLT.  However, the offline algorithm makes a second pass on unused hits to
enhance the efficiency for tracks pointing far away from the beam-line, while in
the HLT this second pass is not used due to CPU constraints. Vertices are
reconstructed from a minimum of five intersecting VELO tracks. Vertices within a
radius of $300$\mum of the mean position of the $pp$-interaction envelope are
considered to be primary vertices.   

HLT1 limits the number of VELO tracks that are passed through the forward
tracking algorithm that searches for matching hits in the tracking stations. 
VELO tracks must have a significant IP with respect to all PVs,
or be matched to muon chamber hits by a fast muon identification algorithm.
This algorithm is only run in events that were triggered by a muon line in the
hardware step. To further limit the processing time, the forward track search
has a minimum momentum requirement that varied between 3 and 6\gevc during Run I.
VELO tracks without matching muon hits are also subject to a minimum \pt
threshold that varied between 0.5 and 1.25\gevc. 
The reconstructed forward tracks are fitted using a Kalman filter with 
a simplified detector geometry description and fewer iterations than in the
offline configuration. The invariant mass resolution of
$\jpsi\rightarrow\mu^+\mu^-$ candidates reconstructed with this procedure is
found to be $3\%$ worse than the 15.1\mevcc obtained offline. For tracks
identified as muon candidates, the basic offline muon identification
algorithm~\cite{LHCb-DP-2013-001} is applied to increase the purity of the muon
sample.  

\begin{figure}[tb]
\vspace{-2mm}
\includegraphics[width=0.49\textwidth]{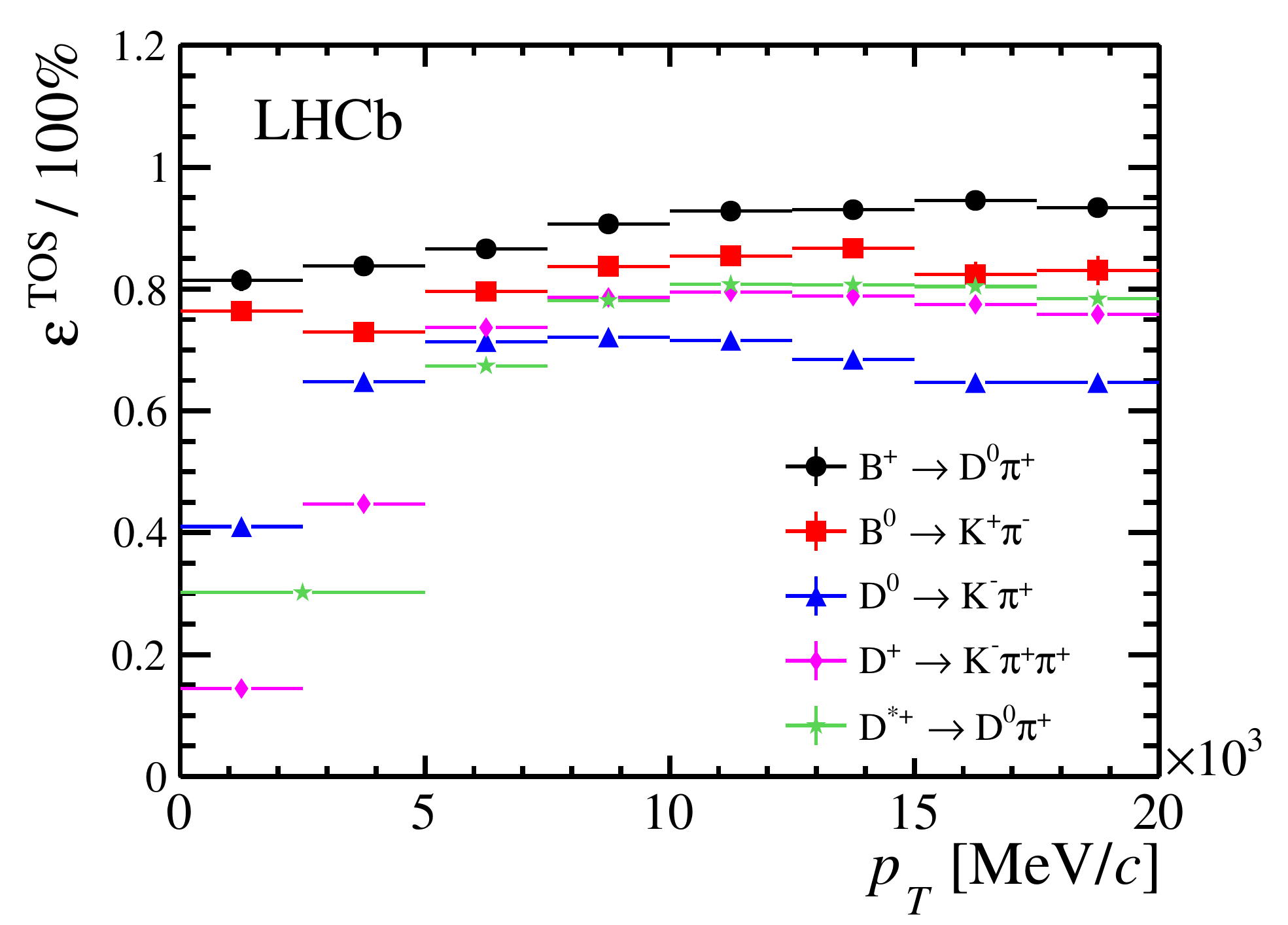}
\includegraphics[width=0.49\textwidth]{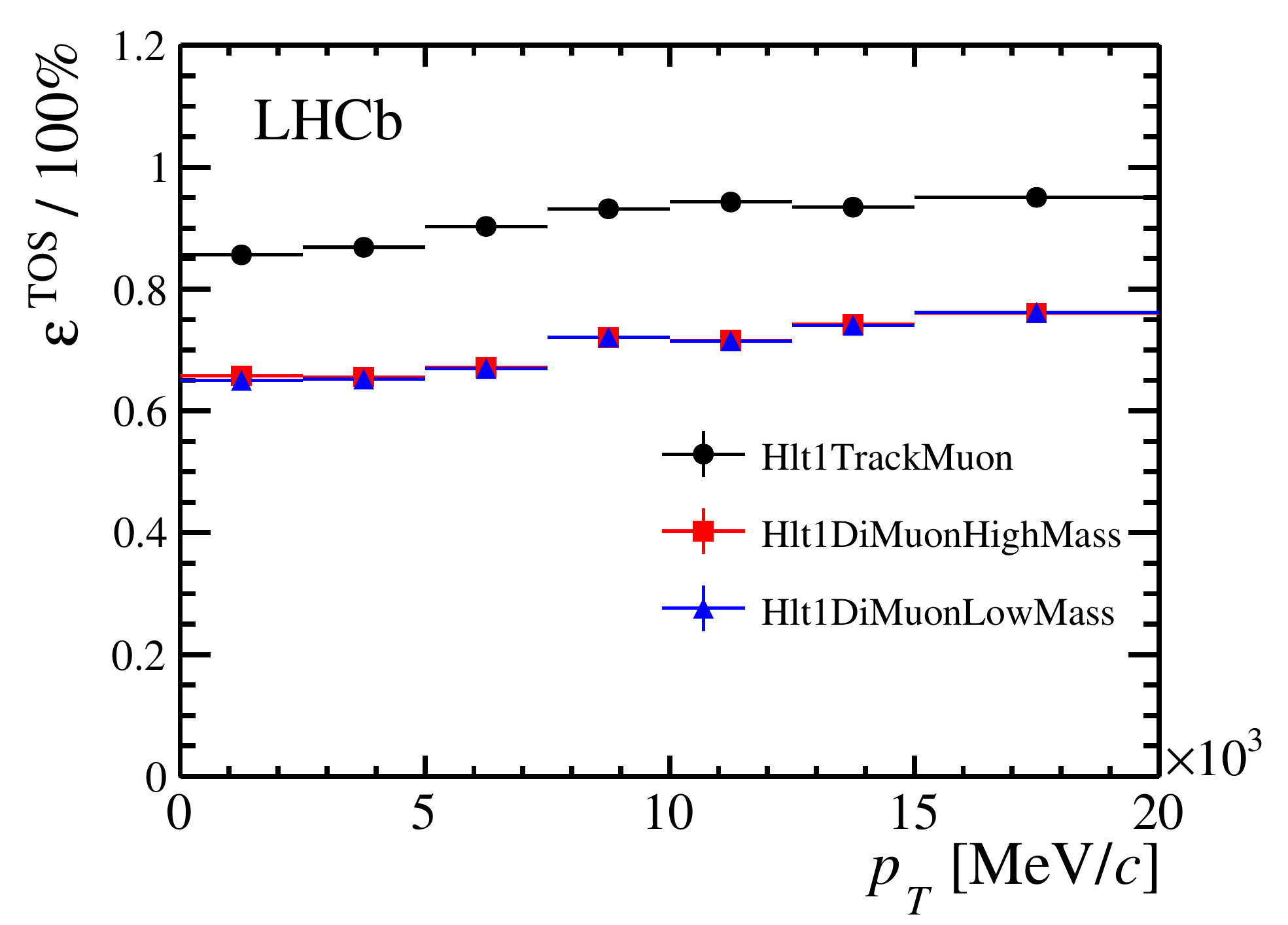}
\caption{\label{fig:hlt1hmu_pt} HLT1 inclusive track trigger performance: TOS
  efficiency for various channels as a function of \B or \D \pt (left) .
  HLT1 muon trigger performance : TOS
  efficiency for \BuJpsiK candidates as function of \Bu \pt (right).}
\vspace{-2mm}
\end{figure}

The inclusive beauty and charm trigger is based on the properties of
one good quality track candidate. The selection is based on track \pt
(typical value $\pt>1.6 - 1.7 \gevc$) and displacement from the primary vertex
(typical value IP $>0.1$~mm). This trigger line produces around 58\khz of
output, which is the largest contribution to the allocated HLT1 bandwidth.
It is the most efficient line for physics channels
that do not contain leptons in the final state. The performance of
HLT1 for hadronic signatures is shown in Figure~\ref{fig:hlt1hmu_pt} as a
function of the \pt of the resonance considered. The inclusive
one-track based trigger also exists in a version for electrons or photons
identified by L0, with reduced thresholds relative to the inclusive version. 
The output rate of these lines is around 7\khz.

A similar line exists for tracks that are matched to hits in the muon
chambers~\cite{Aaij:1384386}. This single muon
trigger line selects good quality muon candidates that are displaced from the
primary vertex and satisfy $\pt>1\gevc$. Single muon candidates that satisfy
$\pt>4.8\gevc$ are selected by a special trigger line without any
vertex separation requirements. 

Dimuon candidates are either selected based on their mass
($m_{\mu\mu}>2.5\gevcc$) without any displacement requirement, or based on their
displacement without the mass restriction.  The dominant inefficiency for these 
lines originates from the online muon identification algorithms. The
performance of HLT1 on muonic signatures as a function of \pt of the
\Bu parent is shown in Figure~\ref{fig:hlt1hmu_pt}. 
The single muon line has an efficiency of around 90\% to select \BuJpsiK decays,
while the dimuon lines have an efficiency of around 70\%.
The combination of muon trigger lines produces an output rate of around 14\khz.

In addition to the trigger lines discussed above, several dedicated
lines are implemented to enhance the trigger performance for events
containing candidates for high \pt electrons, di-protons, displaced
vertices or high \et jets. A set of technical lines including
selections for luminosity and beam-gas measurements completes the list
of HLT1 triggers.  

\subsubsection{Second level}

HLT1 reduces the event rate to about 80\khz, which is sufficiently low
to allow the forward tracking of all VELO tracks in HLT2. As described in
Section \ref{sec:tracking}, the offline reconstruction uses two  complementary
tracking algorithms. Due to the CPU constraints, HLT2 only searches for long
tracks based on VELO seeds. This simplification leads to a lower tracking
efficiency compared to the offline reconstruction of $1-2\%$ per track.

The processing time is further reduced by restricting the search to tracks with
$p>3$\gevc and $\pt>0.3$\gevc. Muon identification in HLT2 is performed using
the offline muon identification algorithm. Tracks are also associated to ECAL
clusters to identify electrons.  Photons and neutral pions are built starting
from the energy clusters reconstructed by the L0-Calorimeter system.

\subsubsection*{Generic beauty trigger}

A significant portion of the output rate of HLT2 is selected by
the `topological' lines, which are designed to trigger on partially
reconstructed $b$-hadron decays.  These topological lines cover all $b$-hadron
decays with at least two charged particles in the final state and a displaced
decay vertex. The inclusive nature of these lines makes them less susceptible to
the $1-2\%$ loss in efficiency per reconstructed track in HLT2. Tracks are
selected based on their track fit $\chi^2/\rm ndf$, IP and muon or electron
identification.  Two-, three- or four-body vertices are constructed from the
selected tracks with a requirement on their distance of closest approach
(DOCA).

\begin{figure}[!tb]
\begin{center}
\begin{minipage}{0.45\textwidth}
\includegraphics[width=\textwidth, height=0.79\textwidth]{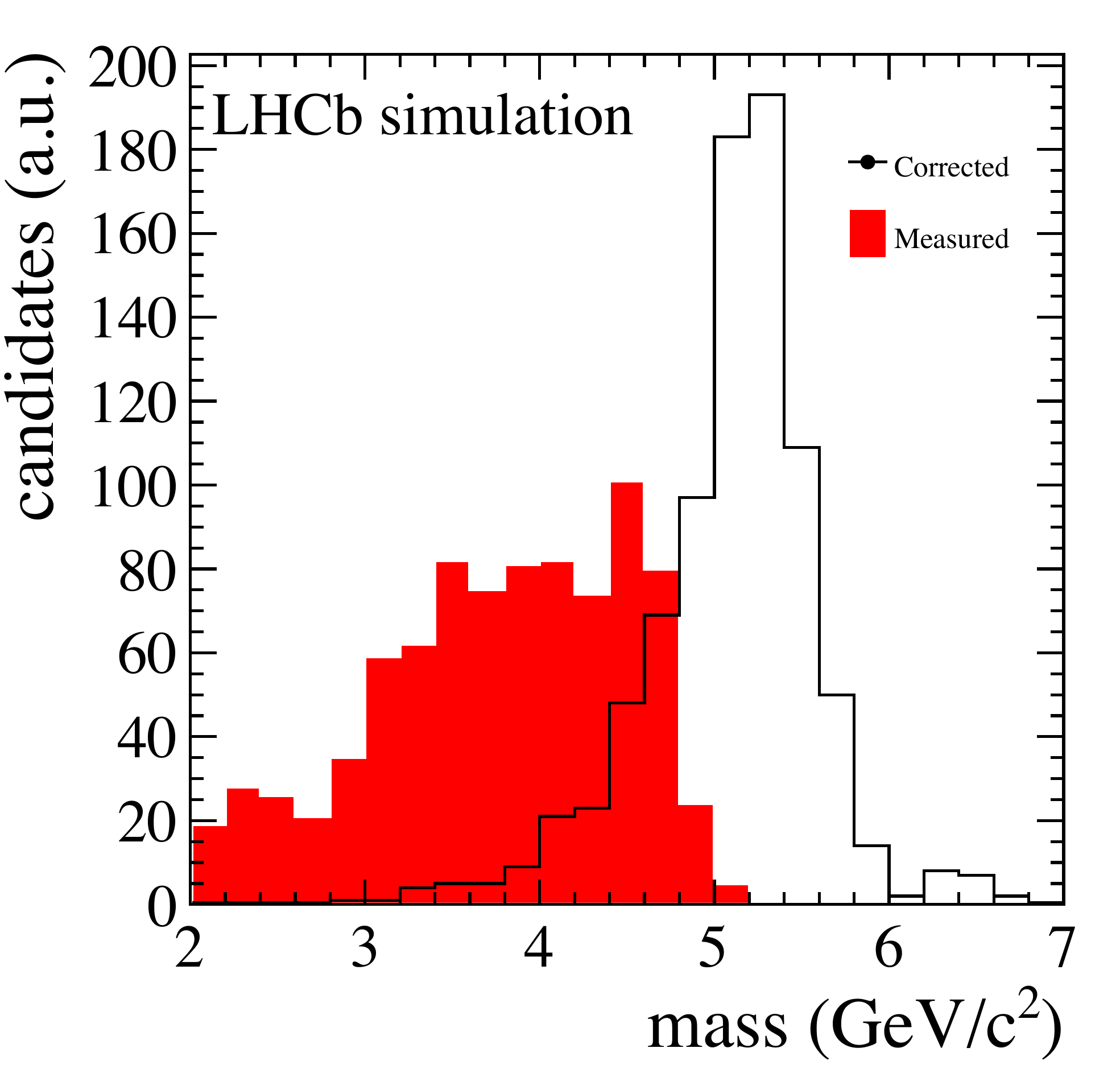}
\vspace{2mm}
\caption{\label{fig:mcorr} Simulated \BdToKstmm events. The reconstructed
  2-body mass is shown in red and the corrected mass ($m_{\rm
  corr}$, see text for
  definition) is shown in black.}
\end{minipage}
\hspace{1pc}
\begin{minipage}{0.45\textwidth}
\includegraphics[width=1.05\textwidth]{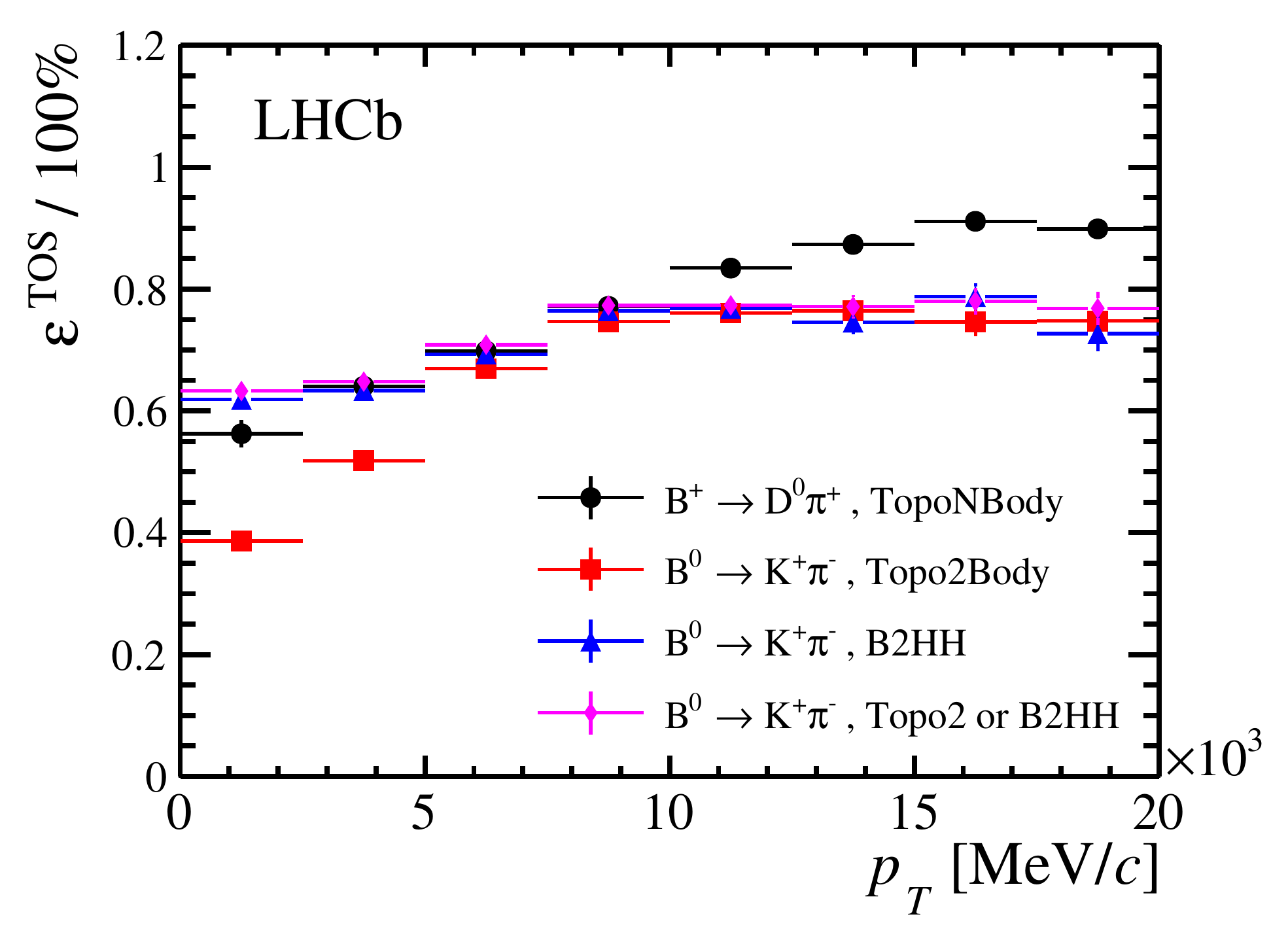}
\vspace{2mm}
\caption{\label{fig:topo} HLT2 inclusive beauty trigger performance as
a function of \B \pt. The efficiency for the exclusive \BdKpi trigger
line is also given.}
\end{minipage}
\end{center}
\end{figure}

Candidate $n$-body combinations are selected based on the following variables:
$\sum|\pt|,~ p^{\rm min}_{\rm T}$, $n$-body invariant mass ($m$),
DOCA, IP$\chi^2$ and flight distance (FD) $\chi^2$. 
In addition, the corrected mass is defined as $m_{\rm
  corr}=\sqrt{m^2+|p^\prime_{\rm T miss}|^2}+|p^\prime_{\rm T miss}|$, 
where $p^\prime_{\rm T miss}$ is the missing momentum transverse to the
line of flight between the $n$-body vertex and the PV to which it has the
smallest IP~\cite{Abe:1997sb}. 
Figure~\ref{fig:mcorr} shows the reconstructed 2-body and corrected mass
distributions for \BdToKstmm events. 
These variables are combined using a boosted decision tree trained on
simulated signal events and data taken in 2010~\cite{Gligorov:2012qt}.
An explicit veto on prompt charm is also applied to reduce its output rate.

Figure~\ref{fig:topo} shows the efficiency for the topological trigger
lines for \BdKpi and \BdDpi events as well as the additional
efficiency that can be gained by an exclusive selection for \BdKpi in
the low \pt regime. The output rate of the topological trigger is
2\khz, in which it efficiently selects beauty decays to charged
tracks. For example, the efficiencies for \BdKpi and
\BdDpi decays are approximately 78\% and 76\%, respectively.

If one of the tracks forming the generic beauty trigger decision is
identified as a muon, the selection on the boosted decision tree
classifier is loosened which enhances the efficiency for muonic beauty
decays like \BdToKstmm.

\subsubsection*{Muon triggers}

Several trigger lines select events that contain one or two muons. 
The muon identification procedure in HLT2 is identical to that which is used
offline \cite{LHCb-DP-2013-001}. Single muon candidate events are selected if
the muon passes a tight \pt requirement ($\pt>10\gevc$) or if candidates are
inconsistent with originating from the PV and they satisfy moderate \pt
($\pt>1.3\gevc$) and tight track quality requirements. Candidates selected with
the latter criteria are prescaled by factor of two. 

Dimuon candidate events are selected without a mass requirement if the dimuon
vertex is separated from the primary vertex. If the mass of the muon pair is
within $\pm 100$\mevcc ($\sim8\sigma$) of the \jpsi mass, three trigger
selections are considered.  Decay-time unbiased $\jpsi \rightarrow \mumu$
candidates are extensively used for calibration of the LHCb decay time acceptance. 
Therefore, all muon pairs with $\pt > 2\gevc$ are considered, as well as a
fraction of those without any \pt requirement. These two prompt selections are
complemented by an additional detached \jpsi trigger. The separation requirement
of this trigger is looser than that of the  generic detached dimuon trigger
described above. 

\begin{figure}[!tb]
\begin{center}
\includegraphics[width=0.49\textwidth]{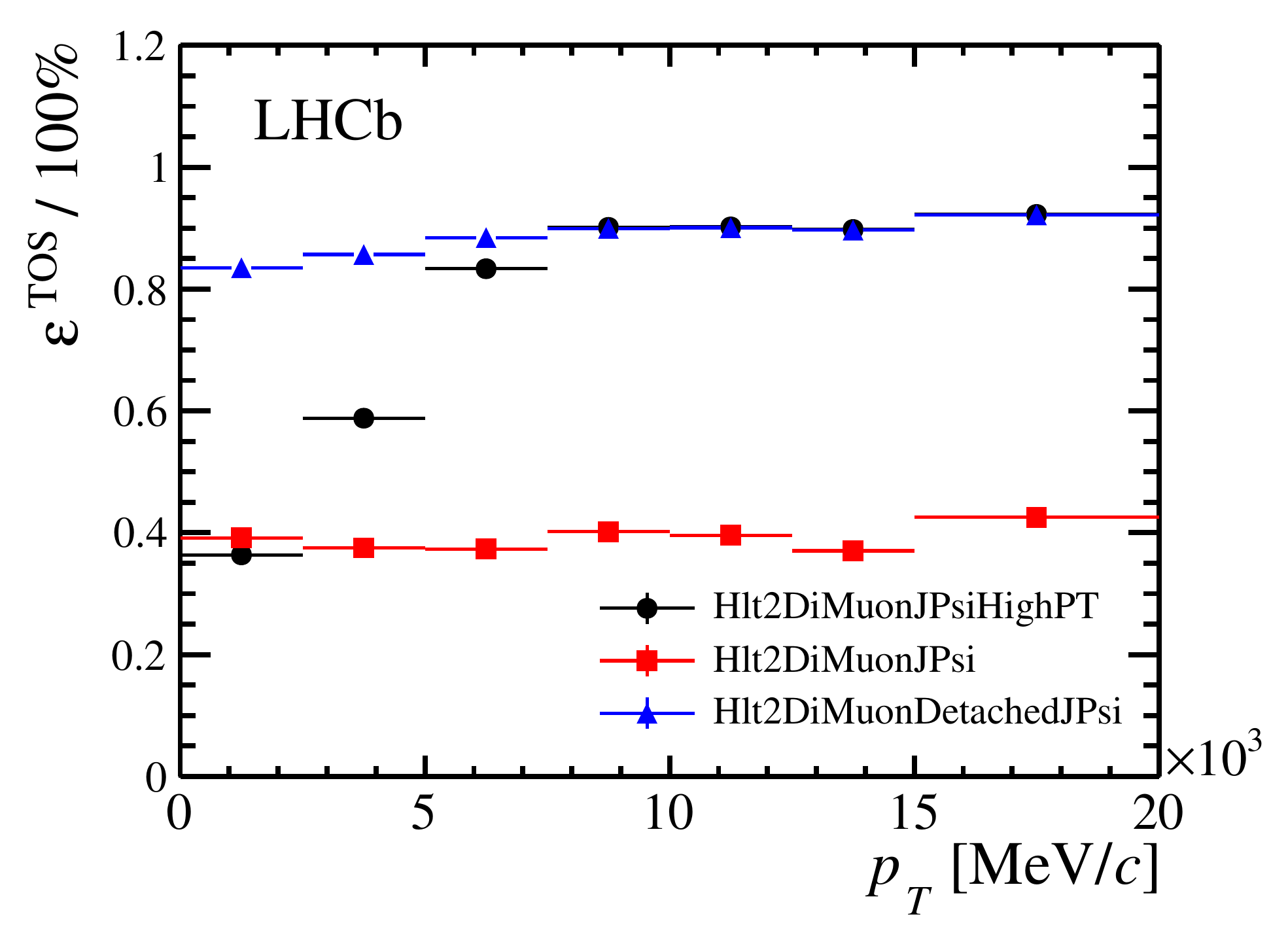}
\hfill
\includegraphics[width=0.49\textwidth]{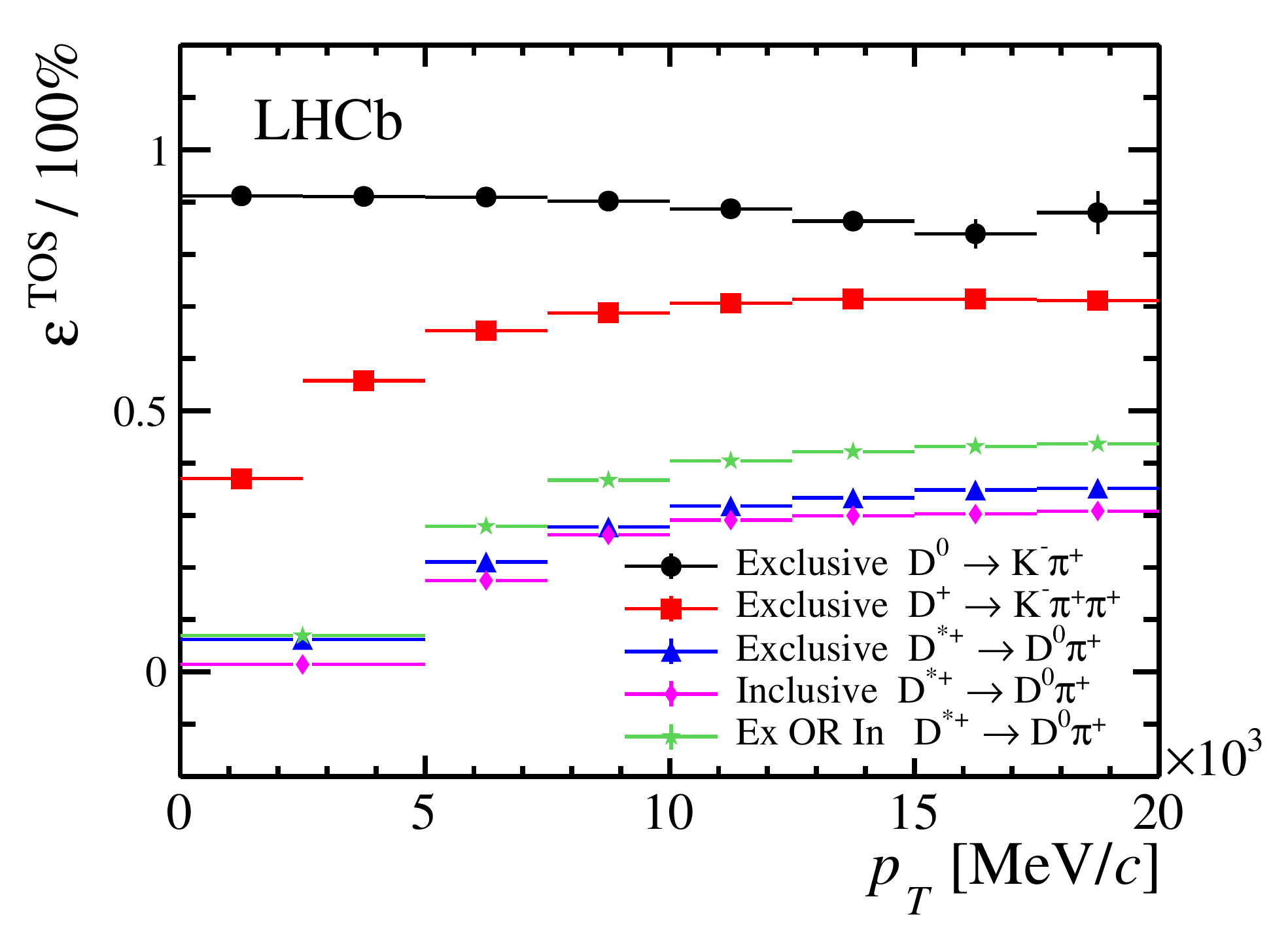}
\end{center}
\caption{\label{fig:hlt2cmu} HLT2 muon trigger performance for the
  \jpsi trigger lines  (left). The triggers \texttt{Hlt2DiMuonJPsi} and
  \texttt{Hlt2DiMuonJPsiHighPT} are the two prompt \jpsi triggers and
  \texttt{Hlt2DiMuonDetachedJPsi} is the trigger that selects \jpsi candidates
  that are inconsistent with coming from the primary vertex. 
  HLT2 charm trigger performance for inclusive and exclusive selections
  (right). The decay $D^{*+} \rightarrow D^0 \pi^+$ is followed by 
  $D^0 \rightarrow K^- \pi^+\pi^+\pi^-$.}
\end{figure}

This set of lines is optimised to fully exploit the large physics
potential of both prompt \jpsi and $\B \to \jpsi X$ decays. Relative to the
offline selection, the trigger efficiencies are typically above
90\%. Figure~\ref{fig:hlt2cmu} shows the performance of the \jpsi triggers,
where the effective prescale of about a factor of two on the prompt \jpsi line
is visible, as well as the \pt acceptance of the high \pt line. The total output
rate of all single and dimuon trigger lines is around $1\,\khz$. 

\subsubsection*{Charm triggers}

In the 2012 running conditions, roughly $600$\khz of $c\bar{c}$-events are
produced within the acceptance of LHCb. This large rate of charm production
means that tight exclusive selections are required in the trigger.
The exception is the decay chain \DstDpi, which can be selected inclusively,
{\it i.e.} only reconstructing two charged tracks from the \Dz decay matched to
a slow pion from the \Dst decay.  The mass difference between the \Dst and \Dz
candidates remains a good discriminating variable, enabling the rate to be
sufficiently reduced. The \Dz is partially reconstructed in all combinations
containing $\pi^\pm$, $K^\pm$, $p$, $\mu^\pm$, \KS or $\Lambda^0$  
enabling both rare decay and \CP violation measurements.

The dominant exclusive selections for prompt
charm are the hadronic two body and three body lines. The efficiency of
these triggers is summarised in Figure~\ref{fig:hlt2cmu}. 
Additional selections for hadronic, leptonic and semi-leptonic \D and \Lc
decays are implemented. 
The total output rate of all charm selections is $\sim2$\khz, but the
trigger efficiencies strongly depend on the offline selection: using a
pure sample of \DKpi events from \Dst decays, the HLT2 efficiency is 90\%, while
the trigger efficiency for the multibody decay \DstDpi is 26\%.

\subsubsection*{Exclusive and technical lines}

The HLT incorporates a large number of exclusive and technical trigger lines to
complement the signal selection by the inclusive lines discussed above. 
For example, 100\hz of random events are recorded throughout the full data
taking period. These events can be used to understand the trigger system,
for fast monitoring of the data and for luminosity determinations.
A number of trigger lines also exist that maximise the performance for
decays with electrons or photons in the final state \cite{LHCb-DP-2012-004}.
A single muon line with hard \pt requirements is designed
to select heavy particles decaying promptly to one or more muons,
like $W^\pm$ or $Z^0$ for electroweak measurements. 

The remaining lines are required for luminosity measurements, prescaled physics
lines with looser cuts, lines which trigger on low multiplicity events and lines
which look for large transverse momentum jets.
The trigger also contain lines that pass events for monitoring to allow
fast feedback on the quality of the data.

\subsection{Deferred trigger}
\label{sec:deferral}

The LHC delivers stable beams $\sim30\%$ of the available
time, and thus unless otherwise occupied the EFF would be idle the remaining
$\sim70\%$ of the time.
Therefore to maximise the use of available resources, LHCb developed 
a deferred triggering system for data taking in 2012\cite{deferal,MarkusCHEP}.
Around $20\%$ of the L0 accepted events are temporarily saved on the EFF
node local disks and these events are subsequently processed by the HLT during
the inter-fill periods.
The effective increase in CPU resources this provides was used to
reduce the \pt requirement in the
forward tracking algorithm and to include additional tracking
algorithms that allow the reconstruction of tracks from particles that
decay beyond the \velo{} acceptance. These modifications significantly
enhance the efficiency for charm decays.

An additional benefit of this system is that it provides redundancy 
against problems downstream in the dataflow. Prior to the adoption of
this scheme, such problems would quickly lead to back-pressure all the 
way up into the farm and lead to dead-time.
In the deferred trigger scheme, although the fraction of deferred
events would increase, no dead-time would be incurred. 
The deferred trigger requires a substantial amount of local
disk space, which was added to the farm before the 2012 run. The
requirements on the performance of the disks are
modest\footnote{The input data rate per farm-server is about 60 MB/s,
which is significantly lower than the typical disk performance of
about 120 MB/s.}, such that inexpensive desktop hard drives could be used.

\subsection{Trigger performance summary}
\label{sec:summary}

The LHCb trigger system has delivered a range of physics modes with
high efficiency in the first running phase of the LHC. It is primarily designed
to select charm and beauty hadrons over a large range of momentum and decay
time, and its efficiency can be determined directly  from the data.

The flexible design of the HLT, which is fully deployed in software,
allows for the rapid adaptation to changes in running conditions and physics
goals. Several innovative concepts have been developed that enable  inclusive
selections to be utilised in the full trigger chain and thus provide an
efficient trigger for nearly any beauty decay to charged particles. The deferred
triggering permits optimisation of computing resources for mean instead of peak
usage, leading to an effective increase in farm size in 2012 of 20 -- 30\%.
Multivariate selections allow the inclusive triggering of beauty
decays to charged tracks with high efficiency.  

\begin{table}[tb]
\begin{center}
\centering
\tbl{Efficiencies of selected channels for the whole
  trigger chain using the 2012 trigger configuration. The efficiencies
  are normalised to the number of events that are offline selected.}
{\begin{tabular}{lccc}
\hline
~~~~~channel~~~~~ & ~~~~~~L0~~~~~~ & ~~~~~~HLT1~~~~~~&~~~~~~HLT2~~~~~~ \\
\hline
\BuJpsiK, $\Jpsi \rightarrow \mumu$
& 89\% & 92\% & 87\%  \\
\BdKpi
& 53\% & 97\% &  80\% \\
\BdDpi, \DpKpipi
& 59\% & 98\% & 77\% \\

\DpKpipi
& 44\% &  89\% & 91\%  \\
\DstDpi, $D^{0} \ra K^{-} \pi^{+} \pi^{-} \pi^{+}$
& 49\% & 93\% & 30\% \\
\hline
\end{tabular}
\label{tab:HLTTotEff}}
\end{center}
\end{table}

Typical trigger efficiencies for selected signals are summarised in
Table~\ref{tab:HLTTotEff}. The trigger efficiency is high for
muonic \bquark-decays. For hadronic \bquark-decays, the hardware trigger stage
causes a significant loss of efficiency, but the software component remains 
efficient. The trigger efficiency for multibody charm decays is lower than for
\bquark-hadron decays, due to the softer momentum spectra of the final state
particles. 

The LHCb trigger system also efficiently covers physics beyond the core beauty
and charm programme. This includes $W^\pm$ or $Z^0$ production, inclusive particle
production, and exotic phenomena, for instance displaced vertices from
heavy long-lived particles.

\section{Conclusion and outlook}
\label{sec:Conclusion-Outlook}

During the first running period of the LHC between 2009 and 2013, the LHCb
experiment recorded a total of about 3.2\invfb of integrated luminosity with
$pp$ collisions at centre-of-mass energies of of 0.9, 7 and 8\tev, and 1.6\invnb
of integrated luminosity with proton-lead collisions. The majority of the data
were recorded under conditions corresponding to  a luminosity of $4 \times
10^{32}\mathrm{\,cm^{-2}s^{-1}}$, a bunch spacing of 50\ns and a pile-up of
1.8. Despite the fact that these are significantly more challenging than the
conditions originally foreseen for the experiment, it has been demonstrated that
the performance of each sub-system and the global performance of the detector
are in good agreement with the original expectations presented in the LHCb
detector paper~\cite{Alves:2008zz}. For some physics analyses the original
expectations are even exceeded, thanks to new ideas and well understood
backgrounds. 

During Run~II of the LHC, the LHCb experiment expects to collect an additional
5\invfb of data, which improves the prospects for observing physics beyond the
Standard Model using heavy quark flavours as a tool. However, the read-out and
trigger scheme of the current LHCb detector limit the luminosity that can be
recorded. To overcome these limitations, the LHCb experiment will be
upgraded~\cite{CERN-LHCC-2011-001,LHCb-TDR-012,LHCb-TDR-013,LHCb-TDR-014,LHCb-TDR-015,CERN-LHCC-2014-016}
to allow the detector to be read out at the maximum LHC bunch crossing rate of
40\mhz with a flexible software-based trigger. This will not only allow the data
rate to be increased substantially, but will also increase the trigger
efficiency, especially for channels currently triggered at the hardware level by
energy deposits in the calorimeters. In addition to the significant increase in
sensitivity for flavour physics, the experiment will continue to explore
other interesting signatures and thus act as a general purpose detector in the
forward region.

\clearpage
\section*{Acknowledgements}


\noindent We express our gratitude to our colleagues in the CERN
accelerator departments for the excellent performance of the LHC. We
thank the technical and administrative staff at the LHCb
institutes. We acknowledge support from CERN and from the national
agencies: CAPES, CNPq, FAPERJ and FINEP (Brazil); NSFC (China);
CNRS/IN2P3 (France); BMBF, DFG, HGF and MPG (Germany); SFI (Ireland); INFN (Italy); 
FOM and NWO (The Netherlands); MNiSW and NCN (Poland); MEN/IFA (Romania); 
MinES and FANO (Russia); MinECo (Spain); SNSF and SER (Switzerland); 
NASU (Ukraine); STFC (United Kingdom); NSF (USA).
The Tier1 computing centres are supported by IN2P3 (France), KIT and BMBF 
(Germany), INFN (Italy), NWO and SURF (The Netherlands), PIC (Spain), GridPP 
(United Kingdom).
We are indebted to the communities behind the multiple open 
source software packages on which we depend. We are also thankful for the 
computing resources and the access to software R\&D tools provided by Yandex LLC (Russia).
Individual groups or members have received support from 
EPLANET, Marie Sk\l{}odowska-Curie Actions and ERC (European Union), 
Conseil g\'{e}n\'{e}ral de Haute-Savoie, Labex ENIGMASS and OCEVU, 
R\'{e}gion Auvergne (France), RFBR (Russia), XuntaGal and GENCAT (Spain), Royal Society and Royal
Commission for the Exhibition of 1851 (United Kingdom).

\bibliographystyle{LHCb}
\bibliography{LHCb-PAPER,LHCb-DP,LHCb-CONF,LHCb-TDR,main,main-calo}

\clearpage
\centerline{\large\bf LHCb collaboration}
\begin{flushleft}
\small
R.~Aaij$^{41}$, 
B.~Adeva$^{37}$, 
M.~Adinolfi$^{46}$, 
A.~Affolder$^{52}$, 
Z.~Ajaltouni$^{5}$, 
S.~Akar$^{6}$, 
J.~Albrecht$^{9}$, 
F.~Alessio$^{38}$, 
M.~Alexander$^{51}$, 
S.~Ali$^{41}$
% ...

{\footnotesize \it
% ...
$ ^{5}$Clermont Universit\'{e}, Universit\'{e} Blaise Pascal, CNRS/IN2P3, LPC, Clermont-Ferrand, France\\
$ ^{6}$CPPM, Aix-Marseille Universit\'{e}, CNRS/IN2P3, Marseille, France\\
% ...
$ ^{9}$Fakult\"{a}t Physik, Technische Universit\"{a}t Dortmund, Dortmund, Germany\\
% ...
$ ^{37}$Universidad de Santiago de Compostela, Santiago de Compostela, Spain\\
$ ^{38}$European Organization for Nuclear Research (CERN), Geneva, Switzerland\\
% ...
$ ^{41}$Nikhef National Institute for Subatomic Physics, Amsterdam, The Netherlands\\
% ...
$ ^{46}$H.H. Wills Physics Laboratory, University of Bristol, Bristol, United Kingdom\\
% ...
$ ^{51}$School of Physics and Astronomy, University of Glasgow, Glasgow, United Kingdom\\
$ ^{52}$Oliver Lodge Laboratory, University of Liverpool, Liverpool, United Kingdom\\
% ...
}
\end{flushleft}

\end{document}